
\magnification=1200
\def\tt{{\bf t}}
\def\ll{{\rm left}}
\def\rr{{\rm right}}
\def\moduli{{\cal M}_{g,s}}
\def\dela{{\partial_1}}
\def\delb{{\partial_2}}

\def\Cc{{\bf C}}
\def\LE{{\cal L}_E}
\def\half{{1\over 2}}
\def\dalpha{\partial_\alpha}
\def\dbeta{\partial_\beta}
\def\dgamma{\partial_\gamma}
\def\res{\mathop{\rm res}}
\def\deli{\partial_i}
\def\delk{\partial_k}
\def\delj{\partial_j}
\def\deg{{\rm deg}\,}
\def\diag{{\rm diag}}
\font\Large=cmbx10 scaled \magstep2
\vskip 10 cm
\centerline{\Large GEOMETRY OF 2D TOPOLOGICAL FIELD THEORIES}
\bigskip
\centerline{\bf Boris DUBROVIN}
\medskip
\centerline{\sl SISSA, Trieste}
\bigskip
\centerline{Preprint SISSA-89/94/FM}
\vskip 5 cm
{\bf Abstract.} These lecture notes are devoted to
the theory of equations of associativity describing geometry of
moduli spaces of 2D topological field theories.
\vfill\eject

\vfill\eject
Contents.

Introduction.

Lecture 1. WDVV equations and Frobenius manifolds.

\noindent{Appendix A.}

Polynomial solutions of WDVV.

\noindent{Appendix B.}

Symmetriies of WDVV. Twisted Frobenius manifolds.

\noindent{Appendix C.}

WDVV and Chazy equation. Affine connections on curves
with projective structure.

Lecture 2. Topological conformal field theories and their moduli.

Lecture 3. Spaces of isomonodromy deformations as Frobenius manifolds.

\noindent{Appendix D.}

Geometry of flat pencils of metrics.

\noindent{Appendix E.}

WDVV and Painlev\'e-VI.

\noindent{Appendix F.}

Branching of solutions of the equations of isomonodromic
deformations and braid group.

\noindent{Appendix G.}

Monodromy group of a Frobenius manifold.

\noindent{Appendix H.}

Generalized hypergeometric equation associated to a Frobenius
manifold and its monodromy.

\noindent{Appendix I.}

Determination of a superpotential of a Frobenius manifold.

Lecture 4. Frobenius structure on the space of orbits of a Coxeter group.

\noindent{Appendix J.}

Extended complex crystallographic groups and twisted
Frobenius manifolds.

Lecture 5. Differential geometry of Hurwitz spaces.

Lecture 6. Frobenius manifolds and integrable hierarchies. Coupling to
topological gravity.

References.

\vfill\eject
\bigskip
{\bf Introduction.}
\medskip
In these lecture notes I consider one remarkable system of differential
equations that appeared in the papers of physisists on two-dimensional
topological field theory (TFT)
in the beginning of
'90 [148, 39]. Roughly speaking, the problem is to find
a quasihomogeneous
function $F=F(t)$ of the variables
$t=(t^1,\dots, t^n)$ such that the third derivatives
of it
$$c_{\alpha\beta\gamma}(t) := {\partial^3 F(t)\over \partial t^\alpha
\partial t^\beta \partial t^\gamma}
$$
for any $t$ are structure constants of an associative algebra
$A_t$ with a
$t$-independent
unity (the algebra will be automatically commutative) (see Lecture 1
for the precise formulation of the problem). For the function $F(t)$
one obtains a very complicated overdetermined system of PDEs. I call it
WDVV equations. In the physical setting the solutions of WDVV describe
moduli space of topological conformal field theories.
One of the projects of these lectures is to try
to reconstruct the building of 2D TFT one the base of WDVV equation.

{}From the point of view of a mathematician particular solutions of WDVV
with certain \lq\lq good" analytic properties are generating functions
for the Gromov - Witten invariants of K\"ahler (and, more generally,
of symplectic) manifolds [149]. They play the crucial role in the
formulation (and, may be, in the future explanation) of the phenomenon
of mirror symmetry of Calabi - Yau 3-folds [151]. Probably, they also play
a central role in understanding of relations between matrix integrals,
integrable hierarchies, and topology of moduli spaces of
algebraic curves [149, 71, 89-91, 57].

We discuss briefly the \lq\lq physical" and \lq\lq topological"
motivations of WDVV equations in Lecture 2. The other lectures
are based mainly on the papers [44-46, 48-50] of the author. The material
of Appendices was not published before besides Appendix E and
Appendix D (this was a part of the preprint [50]).
\medskip
In the abbreviated form our contribution to the theory of WDVV
can be encoded by the following key words:
\smallskip
WDVVF as Painlev\'e equations
\smallskip
Discrete groups and their invariants and WDVV
\smallskip
Symmetries of WDVV
\smallskip
To glue all these together we employ an amazingly rich (and nonstandard)
differential geometry of WDVV. The geometric reformulation
of the equations is given in Lecture 1. We observe then (on some simple
but important examples) that certain analyticity conditions work
as a very strong  rule for selection of solutions. Probably, solutions
with good analytic properties (in a sense to be formulated in a more
precise way) are isolated points in the see of all solutions of WDVV.

The main geometrical playing characters - the deformed affine connection
and the deformed Euclidean metric - are introduced in Lecture 3. In this
Lecture we find the general solution of WDVV for which the algebra $A_t$
is semisimple for generic $t$. This is expressed via certain transcendental
functions of the Painlev\'e-VI type and their higher order generalizations.
The theory of linear differential operators with rational coefficients
and of their monodromy preserving deformations plays an important role
in these considerations.

In Appendix G we introduce another very important object: the monodromy
group that can be constructed for any solution of WDVV. This the
monodromy group of some holonomic system of differential equations
describing the deformation of Euclidean structure on the space
of parameters $t$. We are tempted to conjecture that in the
problem of mirror symmetry this construction fills the gap from
the quantum cohomologies of a Calabi - Yau 3-fold $X$ to the
Picard - Fuchs equation of the mirror dual of $X$. However, this
should be first checked for the known examples of mirror pairs.

More generally, our conjecture is that, for a solution of
WDVV with good analytic properties our monodromy group is a discrete
group. Some general properties of the monodromy group are obtained
in Appendices G and H. We give many examples where the group is a finite
Coxeter group, an extension of an affine Weyl group (Lecture 4) or
of a complex crystallographic group (Appendix J). The solutions
of WDVV for these monodromy groups are given by simple formulae
in terms of the invariants of the groups.
To apply this technique to the topological
problems, like mirror symmetry, we need to find some natural groups related
to, say, Calabi - Yau 3-folds (there are interesting results in this
direction
in the recent preprints [135]).

B\"acklund-type symmetries of the equations of associativity play an important
technical role in our constructions. It turns out that the group of
symmetries of WDVV is rich enough: for example, it contains elements
that transform, in the physical notations, a solution with the given
$d = \hat c = {c\over 3}$ to a solution with $d' =2-d$ (I recall that for
topological sigma-models $d$ is the complex dimension of the target space).
Better understanding of the structure of the group of symmetries
of WDVV could be useful in the mirror problem (may be, the mirror
map is also a symmetry of WDVV).

In the last Lecture we briefly discuss relation of the equations of
associativity to integrable hierarchies and their semiclassical limits.
Some of these relations were discussed also in [45, 81, 87-88, 97-98,
133] for the
dispersionless limits of various integrable hierarchies of KdV type
(the dispersionless limit corresponds to the tree-level approximation in TFT).
Some of these observations were known also in the theory of Gauss - Manin
equations [111]. Our approach is principally different: we construct
an integrable hierarchy (in a semi-classical approximation) for
{\it any} solution of WDVV. We obtain also for our hierarchies the
semi-classical analogue of Lax representation. The problem of
reconstruction of all the hierarchy (in all orders in the small dispersion
expansion) is still open (see the recent papers [90, 57] where this problem
was under investigation).
\bigskip
{\bf Acknowledgement.} I wish to thank V.I.Arnol'd, V.V.Batyrev,
A.B.Givental, D.R. Morrison,
A.N.Todorov, C.Vafa, A.N.Varchenko, M.Verbitski
for helpful discussions.

\vfill\eject
\centerline{\bf Lecture 1.}
\smallskip
\centerline{\bf WDVV equations and Frobenius manifolds.}
\medskip
I start with formulation of the main subject of these lectures:
a system of differential equations arising originally in the
physical papers on two-dimensional field theory (see below Lecture 2).
We look for
a function $F=F(t)$, $t=(t^1,\dots, t^n)$ such that the third derivatives
of it
$$c_{\alpha\beta\gamma}(t) := {\partial^3 F(t)\over \partial t^\alpha
\partial t^\beta \partial t^\gamma}
$$
obey the following equations

1) Normalization:
$$\eta_{\alpha\beta} := c_{1\alpha\beta}(t)
\eqno{(1.1)}$$
is a constant nondegenerate matrix. Let
$$(\eta^{\alpha\beta}) := (\eta_{\alpha\beta})^{-1}.
$$
We will use the matrices $(\eta_{\alpha\beta})$ and $(\eta^{\alpha\beta})$
for raising and lowering indices.

2) Associativity: the functions
$$c_{\alpha\beta}^\gamma (t):= \eta^{\gamma\epsilon}c_{\epsilon\alpha\beta}
(t)
\eqno{(1.2)}$$
(summation over repeated indices will be assumed in these lecture notes)
for any $t$ must define in the $n$-dimensional space with a basis
$e_1$, ..., $e_n$ a structure of an assosciative algebra $A_t$
$$e_\alpha\cdot e_\beta = c_{\alpha\beta}^\gamma (t) e_\gamma.
\eqno{(1.3)}$$
Note that the vector $e_1$ will be the unity for all the algebras
$A_t$:
$$c_{1\alpha}^\beta (t) = \delta_\alpha^\beta.
\eqno{(1.4)}$$

3) $F(t)$ must be quasihomogeneous function of its variables:
$$F(c^{d_1}t^1,\dots, c^{d_n}t^n) = c^{d_F}F(t^1,\dots, t^n)
\eqno{(1.5)}$$
for any nonzero $c$ and for some numbers $d_1$, ..., $d_n$, $d_F$.

It will be convenient to rewrite the quasihomogeneity condition (1.5) in
the infinitesimal form introducing the {\it Euler vector field}
$$E = E^\alpha(t)\dalpha
$$
as
$${\cal L}_EF(t) := E^\alpha(t)\dalpha F(t) = d_F\cdot F(t).
\eqno{(1.6)}$$
For the quasihomogeneity (1.5) $E(t)$ is a linear vector field
$$E = \sum_\alpha d_\alpha t^\alpha \dalpha
\eqno{(1.7)}$$
generating the scaling transformations (1.5). Note that for the Lie
derivative ${\cal L}_E$ of the unity vector field $e = \partial_1$
we must have
$${\cal L}_E e = - d_1 \, e.
\eqno{(1.8)}$$

Two  generalizations of the quasihomogeneity condition will be
important in our considerations:

1.  We will consider the functions $F(t^1,\dots, t^n)$
up to adding of a (nonhomogeneous) quadratic function in $t^1,\dots, t^n$.
Such an addition does not change the third derivatives. So the algebras
$A_t$ will remain unchanged. Thus the quasihomogeneity condition (1.6) could
be modified  as follows
$$\LE F(t) = d_F F(t)
+ A_{\alpha\beta}t^\alpha t^\beta + B_\alpha t^\alpha +C.
\eqno{(1.9)}$$
This still provides quasihomogeneity of the functions $c_{\alpha\beta
\gamma}(t)$.Moreover, if
$$d_F\neq 0,~ d_F-d_\alpha\neq 0,~ d_F-d_\alpha -d_\beta\neq 0~
{\rm for~ any}~\alpha,~\beta
\eqno{(1.10)}$$
then the extra terms in (1.9) can be killed by adding of a quadratic form
to $F(t)$.

2. We will consider more general linear nonhomogeneous Euler vector
fields
$$E(t) = \left( q_\beta^\alpha t^\beta + r^\alpha \right) \dalpha.
\eqno{(1.11)}$$
If the roots of $E(t)$ (i.e., the eigenvalues of the matrix
$Q=(q_\beta^\alpha)$
) are simple and nonzero then $E(t)$ can be reduced to the form (1.7)
by a linear
change of the variables $t$. If some of the roots of $E(t)$ vanish then,
in general the linear nonhomogeneous terms in (1.11) cannot be killed by linear
transformations of $t$. In this case for a diagonalizable matrix $Q$ the Euler
vector field can be reduced to the form
$$E(t) = \sum_\alpha d_\alpha t^\alpha \dalpha + \sum_{\alpha | d_\alpha = 0}
r^\alpha \dalpha
\eqno{(1.12)}$$
(here $d_\alpha$ are the eigenvalues of the matrix $Q$). The numbers $r^\alpha$
can be changed by linear transformations in the kernel
${\rm Ker} Q$. However, in important examples the function $F(t)$ will be
periodic (modulo quadratic terms) w.r.t. some lattice of periods
in ${\rm Ker} Q$ (note that periodicity of $F(t)$ can happen only along
the directions with zero scaling dimensions). In this case the vector
$(r^\alpha)$ is defined modulo the group of automorphisms of the
lattice of periods. Particularly, in topological sigma models with
non-vanishing first Chern class of the target space the vector $(r^\alpha)$
is always nonzero (see below Lecture 2).
\medskip

The degrees $d_1$, ..., $d_n$, $d_F$ are well-defined up to a nonzero
factor. We will consider only the case
$$d_1\neq 0
$$
(the variable $t^1$ is marked due to (1.1)). It is convenient in this case
to normalize the degrees $d_1$, ..., $d_n$, $d_F$ in such a way that
$$d_1 = 1.
\eqno{(1.13a)}$$
In the physical literature the normalized degrees usually are parametrized
by some numbers $q_1=0$, $q_2$,..., $q_n$ and $d$ such that
$$d_\alpha = 1-q_\alpha, ~~d_F = 3-d.
\eqno{(1.13b)}$$
If the coordinates are normalized as in (1.18) then
$$q_n = d, ~~q_\alpha + q_{n-\alpha +1} = d.
\eqno{(1.13c)}$$

Associativity imposes the following system of nonlinear PDE for
the function $F(t)$
$${\partial^3F(t)\over \partial t^\alpha \partial t^\beta \partial t^\lambda}
\eta^{\lambda\mu}
{\partial^3F(t)\over \partial t^\gamma \partial t^\delta \partial t^\mu}
=
{\partial^3F(t)\over \partial t^\gamma \partial t^\beta \partial t^\lambda}
\eta^{\lambda\mu}
{\partial^3F(t)\over \partial t^\alpha \partial t^\delta \partial t^\mu}
\eqno{(1.14)}$$
for any $\alpha$, $\beta$, $\gamma$, $\delta$ from 1 to n. The
quasihomogeneity (1.9) determines the  scaling reduction of the system.
The normalization (1.1) completely specifies the dependence of the function
$F(t)$ on the marked variable $t^1$. The resulting system of equations will be
called {\it Witten - Dijkgraaf - E.Verlinde - H.Verlinde (WDVV)} system: it
was first found in the papers [148, 39] in topological field theory (see
Lecture
2 below). A solution of the WDVV equations will be called
{\it (primary) free energy}.

{\bf Remark 1.1.} More general reduction of (1.14) is given by {\it conformal}
transformations of the metric $\eta_{\alpha\beta}$. The generator $E$
of the correspondent one-parameter group of diffeomorphisms of the $t$-space
for $n\geq 3$ due to Liouville theorem [55] must have the form
$$ E = a\left\{ t_1 t^\gamma\partial_\gamma - \half t_\sigma t^\sigma
\partial_1 \right\} + \sum_{\epsilon =1}^n d_\epsilon t^\epsilon
\partial_\epsilon
\eqno{(1.15a)}$$
for some constants $a, d_1, \dots, d_n$ with
$$d_\alpha = 1-q_\alpha, ~~q_1 = 0, ~~q_\alpha + q_{n-\alpha +1} = d
$$
(in the normalization (1.17)). The function $F(t)$ must obey the equation
$$\sum_{\epsilon = 1}^n (at_1 + d_\epsilon )t^\epsilon \partial_\epsilon
F = (3-d+2at_1)F +{a\over 8}\left( t_\sigma t^\sigma\right)^2
\eqno{(1.15b)}$$
modulo quadratic terms. The equations of associativity (1.14) for $F$
satisfying (1.15) also can be reduced to a system of ODE. I will not
consider this system here. However, a {\it discrete} group of conformal
symmetries of WDVV plays an important role in our considerations
(see below Appendix B).

Observe that the
system is invariant w.r.t. linear changes of the coordinates $t^1,\dots,
t^n$.
To write down
WDVV system more explicitly we use the following
\smallskip
{\bf Lemma 1.1} {\it The scaling transformations generated by
the Euler vector field $E$ (1.9)
act as linear conformal transformations of the metric $\eta_{\alpha\beta}$
$$\LE\eta_{\alpha\beta}=({d_F - d_1}) \eta_{\alpha\beta}
\eqno{(1.16)}$$
where the numbers $d_F$ and $d_1$ are defined in (1.6) and (1.8).
}

Proof. Differentiating the equation (1.9) w.r.t. $t^1$, $t^\alpha$ and
$t^\beta$
and using $\partial_1 E^\rho = \delta_1^\rho$ (this follows from (1.8))
we obtain
$$q_\alpha^\rho \eta_{\rho\beta} +q_\beta^\rho \eta_{\rho\alpha}
= (2-d) \eta_{\alpha\beta}.
$$
The l.h.s. of this equality coincides with the Lie derivative
$\LE \, \eta_{\alpha\beta}$ of the metric $<~,~>$. Lemma is proved.
\medskip
{\bf Corollary 1.1.} {\it If $\eta_{11}=0$ and all the roots of $E(t)$
are simple then
by a linear change of coordinates $t^\alpha$ the matrix
$\eta_{\alpha\beta}$ can be reduced to the antidiagonal form
$$\eta_{\alpha\beta} = \delta_{\alpha + \beta, n+1}.
\eqno{(1.17)}$$
In these coordinates
$$F(t) = {1\over 2} (t^1)^2 t^n +
{1\over 2} t^1\sum_{\alpha =2}^{n-1} t^\alpha t^{n-\alpha + 1}
+f(t^2,\dots, t^n)
\eqno{(1.18)}$$
for some function $f(t^2,\dots, t^n)$, the sum
$$d_\alpha + d_{n-\alpha +1}
\eqno{(1.29)}$$
does not depend on $\alpha$
and
$$d_F = 2d_1 + d_n.
\eqno{(1.20)}$$
If the degrees are normalized in such a way that $d_1 =1$ then they can be
represented in the form
$$d_\alpha = 1-q_\alpha, ~~ d_F = 3-d
\eqno{(1.21a)}$$
where the numbers $q_1, \dots, d_n, d$ satisfy
$$q_1 = 0, ~~ q_n = d, ~~ q_\alpha + q_{n-\alpha +1} = d.
\eqno{(1.21b)}$$
}

{\bf Exercise 1.1.} Show that if $\eta_{11}\neq 0$
(this can happen only  for $d_F = 3 d_1$) the function $F$ can be reduced by a
linear
change of $t^\alpha$ to the form
$$F = {c\over 6} (t^1)^3 + \half t^1\sum_{\alpha = 1}^{n-1}
t^\alpha t^{n-\alpha + 1} + f(t^2, \dots, t^n)
\eqno{(1.22)}$$
for a nonzero constant $c$
where the degrees satisfy
$$d_\alpha + d_{n-\alpha + 1} = 2 d_1.
\eqno{(1.23)}$$

Proof of Corollary. If $<e_1, e_1>= 0$ then one can chose the basic vector
$e_n$ such that
$<e_1, e_n>=1$ and $e_n$ is still an eigenvector
of $Q$.
On the orthogonal
complement of the span of $e_1$ and $e_n$ we can
reduce the bilinear form $<~,~>$
to the antidiagonal form using only eigenvectors of
the scaling transformations.
In these coordinates (1.18) follows from (1.1). Independence
of the sum $d_\alpha + d_{n-\alpha +1}$ of $\alpha$ follows from (1.16).
The formula for $d_F$ follows from (1.16). Corollary is proved.
\medskip
I will mainly consider  solutions of WDVV of the type
(1.18). I do not know physical examples of the solutions of the second
type (1.22). However, we are to take into account also the solutions
with $\eta_{11}\neq 0$ for completeness of the mathematical theory
of WDVV (see, e.g., Appendix B and Lecture 5).
\medskip
{\bf Example 1.1.} $n=2.$ Equations of associativity are empty. The other
conditions specify the following general solution of WDVV
$$\eqalignno{F(t_1, t_2) &= {1\over 2} t_1^2 t_2 + t_2^k, ~~k={3-d\over 1-d},
{}~d\neq -1, 1, 3 &{(1.24a)}\cr
F(t_1, t_2) &= {1\over 2} t_1^2 t_2 + t_2^2\log t_2, ~~d=-1 &{(1.24b)}\cr
F(t_1, t_2) &= {1\over 2} t_1^2 t_2 + \log t_2, ~~d=3 &{(1.24c)}\cr
F(t_1, t_2) &= {1\over 2} t_1^2 t_2 + e^{t_2}, ~~d=1 &{(1.24d)}\cr}
$$
(in concrete formulae I will label the coordinates
$t^\alpha$ by subscripts for the sake of graphical simplicity).
In the last case $d=1$ the Euler vector field is
$E= t_1 \partial_1 + 2 \partial_2$.
\medskip
{\bf Example 1.2.} $n=3.$ In the three-dimensional algebra $A_t$ with
the basis $e_1=1$, $e_2$, $e_3$ the law of multiplication is
determined by the following table
$$\eqalign{e_2^2 &= f_{xxy} e_1 + f_{xxx} e_2 + e_3
\cr
e_2 e_3 &= f_{xyy} e_1 + f_{xxy} e_2
\cr
e_3^2 &= f_{yyy} e_1 + f_{xyy} e_2\cr}
\eqno(1.25)$$
where the function $F$ has the form
$$F(t) = {1\over 2} t_1^2 t_3 +{1\over 2} t_1 t_2^2 + f(t_2, t_3)
\eqno(1.26)$$
for a function $f=f(x,y)$ (the subscripts denote the correspondent partial
derivatives).
The associativity condition
$$(e_2^2) e_3 = e_2 (e_2 e_3)
\eqno(1.27)$$
implies the following PDE for the function $f=f(x,y)$
$$f_{xxy}^2 = f_{yyy} + f_{xxx} f_{xyy}.
\eqno(1.28)$$
It is easy to see that this is the only one equation of associativity
for $n=3$.
The function $f$ must satisfy also the following scaling condition
$$\eqalignno{\left( 1-{d\over 2}\right) x f_x + (1-d) y f_y &= (3-d) f,
{}~~d \neq 1,
\, 2,\, 3 &(1.29a)
\cr
\half x f_x + r f_y &= 2f , ~~ d =1 &(1.29b)\cr
rf_x - yf_y &= f, ~~ d =2 &(1.29c)\cr
\half x f_x + 2 y f_y &= c, ~~ d = 3 &(1.29d)\cr}
$$
for some constants $c$, $r$.
The correspondent scaling reductions of the equation (1.28)
$$\eqalignno{f(x,y)& = x^{4+q} \phi(yx^q),~ q= -{1-d\over 1-\half d},
{}~~ d \neq 1,\, 2,\, 3 &(1.30a)\cr
f(x,y) &= x^4 \phi( y-2 r \log x), ~~ d =1 &(1.30b)\cr
f(x,y) &=y^{-1}\phi(x+r\log y),  ~~d =2 &(1.30c)\cr
f(x,y) &= 2 c \log x + \phi (yx^{-4}) , ~~ d = 3 &(1.30d)\cr}
$$
are the following
third order ODEs for the function $\phi = \phi(z)$,
$$[(12 + 14q + 4q^2)\phi' + (7q+5q^2) z\phi''
+q^2z^2 \phi''' ]^2 = \phi''' +
$$
$$[(2+q)(3+q)(4+q)\phi + q(26 + 27q + 7q^2) z\phi'
+ q^2(9+6q)z^2\phi'' + q^3z^3 \phi''']
[(4+3q)\phi'' + qz \phi''']
\eqno(1.31a)$$
for $d\neq 1, \, 2$,
$$\phi''' [r^3 + 2\phi' - r \phi''] -(\phi'')^2 - 6r^2 \phi''
+ 11 r \phi' - 6\phi = 0
\eqno(1.31b)$$
for $d=1$,
$$ -144\,{{\phi '}^2} +
      96\,\phi \,\phi '' +
      128\,r\,\phi '\,\phi '' -
      52\,{r^2}\,{{\phi ''}^2} +
      \phi''' -
      48\,r\,\phi \,\phi''' +
      8\,{r^2}\,\phi '\,\phi''' +
      8\,{r^3}\,\phi ''\,\phi''' =0
      \eqno(1.31c)$$
for $d=2$,
$$\phi''' =
400\,{{\phi '}^2} +
      32\,c\,\phi '' +
      1120\,z\,\phi '\,
       \phi '' +
      784\,{z^2}\,{{\phi ''}^2} +
      16\,c\,z\,\phi''' +
      160\,{z^2}\,\phi '\,
       \phi ^{(3)} +
      192\,{z^3}\,\phi ''\,
       \phi ^{(3)}
\eqno(1.31d)$$
for $d=3$.
Any solution of our main problem for $n=3$ can be obtained
from a solution of these ODEs. Later these will be shown to be reduceable
to a particular
case of the Painlev\'e-VI equation.

{\bf Remark 1.2.} Let us compare (1.28) with the WDVV equations for the
prepotential $F$ of
the second type (1.22). Here we look for a solutions of (1.14) in the form
$$F = {1\over 6} t_1^3 + t_1 t_2 t_3 + f(t_2, t_3).
\eqno(1.32)$$
The three-dimensional algebra with a basis $e_1 = e$, $e_2$, $e_3$ has
the form
$$\eqalign{ e_2^2 &= f_{xxy} e_2 + f_{xxx} e_3 \cr
e_2 e_3 &= e_1 + f_{xyy} e_2 + f_{xxy} e_3 \cr
e_3^2 &= f_{yyy}e_2 + f_{xyy} e_3. \cr}
\eqno(1.33)$$
The equation of associativity has the form
$$f_{xxx} f_{yyy} - f_{xxy}f_{xyy} = 1.
\eqno(1.34)$$
It is interesting that this is the condition of unimodularity
of the Jacobi matrix
$$\det \left( \matrix{{\partial P\over \partial x} &
{\partial P \over \partial y} \cr
{\partial Q \over \partial x} & {\partial Q \over \partial y} \cr}
\right) = 1
$$
for
$$P = f_{xx}(x,y), ~~Q = f_{yy}(x,y).
$$
The function $f$ must satisfy the quasihomogeneity condition
$$f(c^{1-a}x, c^{1+a}y) = c^3 f(x,y).
\eqno(1.35)$$
\medskip
{\bf Example 1.3.} $n=4$. Here we have a system of 5 equations for
the function $f=f(x,y,z)$
where
$$F(t_1, t_2, t_3, t_4) = {1\over 2} t_1^2 t_4 + t_1 t_2 t_3
+ f(t_2, t_3, t_4)
\eqno(1.36)$$
$$ \eqalign{-2\,f_{xyz} -
      f_{xyy}\,
       f_{xxy} +
      f_{yyy}\,
       f_{xxx} &=0 \cr
     -f_{xzz} -
      f_{xyy}\,
       f_{xxz} +
      f_{yyz}\,
       f_{xxx} &=0 \cr
     -2\,f_{xyz}\,
       f_{xxz} +
      f_{xzz}\,
       f_{xxy} +
      f_{yzz}\,
       f_{xxx} &=0 \cr
f_{zzz} -
      {{f_{xyz}}^2} +
      f_{xzz}\,
       f_{xyy} -
      f_{yyz}\,
       f_{xxz} +
      f_{yzz}\,
       f_{xxy} &=0 \cr
   f_{yyy}\,
       f_{xzz} -
      2\,f_{yyz}\,
       f_{xyz} +
      f_{yzz}\,
       f_{xyy} &=0 .\cr}
\eqno(1.37)$$
It is a nontrivial exercise even to verify compatibility of
this overdetermined system of equations.
\medskip
I am going now to give a coordinate-free formulation of the main
problem. Let me give first more details about the algebras $A_t$.
\smallskip
{\bf Definition 1.1.} An algebra  $A$ over {\Cc}  is called
(commutative) {\it Frobenius
algebra} if:

1) It is a commutative associative {\Cc}-algebra with a unity $e$.

2) It is supplied with a {\Cc}-bilinear symmetric nondegenerate
inner product
$$A\times A \to \Cc , ~~a,\,b\mapsto <a,b>
\eqno(1.38)$$
being invariant in the following sense:
$$<ab,c> = <a,bc>.
\eqno(1.39)$$
\medskip
{\bf Remark 1.3.} Let $\omega\in A^*$ be the linear functional
$$\omega (a) := <e,a>.
\eqno(1.39)$$
Then
$$<a,b> = \omega (ab).
\eqno(1.40)$$
This formula determines a bilinear symmetric invariant inner product
for arbitrary linear functional $\omega$. It will be nondegenerate
(for finite-dimensional Frobenius algebras) for generic $\omega\in A^*$.
Note that we consider a Frobenius algebra with a {\it marked} invariant
inner product.
\medskip
{\bf Example 1.4.} $A$ is the direct sum of $n$ copies of one-dimensional
algebras. This means that a basis $e_1,\dots, e_n$ can be chosen in the
algebra with the multiplication law
$$e_i \, e_j = \delta_{ij} \, e_i,~~i,\, j =1,\dots, n.
\eqno(1.41)$$
Then
$$<e_i,e_j> = 0 ~{\rm for}~ i\neq j,
\eqno(1.42)$$
the nonzero numbers $<e_i,e_i>$, $i=1,\dots, n$ are the parameters
of the Frobenius algebras of this type. This algebra is semisimple
(it has no nilpotents).
\smallskip
{\bf Exercise 1.2.} Prove that any Frobenius algebra without nilpotents
over {\Cc} is of the above form.
\medskip
An operation of {\it rescaling} is defined for an algebra with a unity
$e$: we modify the multiplication law and the unity as folllows
$$a\cdot b \mapsto k\, a\cdot b, ~~e \mapsto k\, e
\eqno(1.43)$$
for a given nonzero constant $k$. The rescalings preserve
Frobenius property of the algebra.
\medskip
Back to the main problem: we have a family of Frobenius algebras
depending on the parameters $t=(t^1,\dots, t^n)$. Let us denote by $M$
the space of the parameters. We have thus a fiber bundle
$$\matrix{ & \big\downarrow & A_t \cr
	  t\in & M & \cr}
\eqno(1.44)$$
The basic idea is to identify this fiber bundle with the tangent
bundle $TM$ of the manifold $M$.

We come thus to our main definition. Let $M$ be a $n$-dimensional
manifold.
\smallskip
{\bf Definition 1.2.} $M$ is {\it Frobenius manifold} if a structure
of Frobenius algebra is specified on any tangent plane $T_tM$ at
any point $t\in M$ smoothly depending on the point such that
\item{1.} The invariant inner product $<~,~>$ is a flat metric
on $M$.
\item{2.} The unity vector field $e$ is covariantly constant
w.r.t. the Levi-Civit\'a connection $\nabla$ for the metric $<~,~>$
$$\nabla e = 0.
\eqno(1.45)$$
\item{3.} Let
$$c(u,v,w) := <u\cdot v,w>
\eqno(1.46)$$
(a symmetric 3-tensor). We require the 4-tensor
$$(\nabla_zc)(u,v,w)
\eqno(1.47)$$
to be symmetric in the four vector fields $u$, $v$, $w$, $z$.
\item{4.} A vector field $E$ must be determined on $M$ such that
$$\nabla (\nabla E) =0
\eqno(1.48)$$
and that
the correspondent one-parameter group of diffeomorphisms acts
by conformal transformations of the metric $<~,~>$ and by
rescalings on the Frobenius algebras $T_tM$.
\medskip
In these lectures the word `metric' stands for a \Cc -bilinear quadratic
form on $M$.

Note that the requirement 4 makes sense since we can locally identify
the spaces of the algebras $T_tM$ using the Euclidean parallel transport
on $M$. We will call $E$ {\it Euler vector field} (see formula (1.9) above)
of the Frobenius manifold. The covariantly constant operator
$$Q= \nabla E(t)
\eqno(1.49)$$
on the tangent spaces $T_tM$
will be called {\it the grading operator} of the Frobenius manifold.
The eigenvalues
of the operator $Q$ are constant functions on $M$. The eigenvalues $q_\alpha$
of
${\rm id} - Q$
will be called
{\it scaling dimensions} of $M$. Particularly, as it follows
from (1.45), the unity vector field $e$ is an eigenvector of $Q$ with the
eigenvalue 1.

The infinitesimal form of the requirement 4 reads
$$\eqalignno{\nabla_\gamma\left(\nabla_\beta E^\alpha\right) &=0
&(1.50a)\cr
{\cal L}_E c_{\alpha\beta}^\gamma & = c_{\alpha\beta}^\gamma
&(1.50b)\cr
{\cal L}_E e & = -e &(1.50c)\cr
{\cal L}_E \eta_{\alpha\beta} & =  D \eta_{\alpha\beta}
&(1.50d)\cr}
$$
for some constant $D= 2-d$. Here ${\cal L}_E$ is the Lie derivative along the
Euler vector field. In a coordinate-free way  (1.50b) and (1.50d) read
$$\eqalignno{\LE (u\cdot v) - \LE u\cdot v - u\cdot \LE v &= u\cdot v
&(1.50b')\cr
\LE <u,v> - <\LE u,v> - <u,\LE v> &= D\, <u,v> &(1.50d')\cr}
$$
for arbitrary vector fields $u$ and $v$.
\smallskip
{\bf Exercise 1.3.} Show that the operator
$$\hat V := \nabla E - \half (2-d) {\rm id}
\eqno(1.51)$$
is skew-symmetric w.r.t. $<~,~>$
$$<\hat Vx,y> = - <x,\hat Vy>.
\eqno(1.52)$$
\medskip
{\bf Lemma 1.2.} {\it Any solution of WDVV equations with $d_1\neq 0$
defined in a domain
$t\in M$ determines in this domain the structure of a Frobenius manifold
by the formulae
$$\dalpha\cdot\dbeta := c_{\alpha\beta}^\gamma (t)\dgamma
\eqno(1.53a)$$
$$<\dalpha ,\dbeta > := \eta_{\alpha\beta}
\eqno(1.53b)$$
where
$$\dalpha := {\partial \over \partial t^\alpha}
\eqno(1.53c)$$
etc.,
$$e := \partial_1
\eqno(1.53d)$$
and the Euler vector field (1.9).

Conversely, locally any
Frobenius manifold has the
structure (1.53), (1.11) for some solution of WDVV equations.
}

Proof. The metric (1.53b) is manifestly flat being constant in
the coordinates $t^\alpha$. In these coordinates the covariant
derivative of a tensor coincides with the correspondent partial
derivatives. So the vector field (1.53d) is covariantly constant.
For the covariant derivatives of the tensor (1.46)
$$c_{\alpha\beta\gamma}(t) = {\partial^3F(t)\over
\partial t^\alpha \partial t^\beta \partial t^\gamma}
\eqno(1.54)$$
we have a completely symmetric expression
$$\partial_\delta c_{\alpha\beta\gamma}(t) = {\partial^4F(t)\over
\partial t^\alpha \partial t^\beta \partial t^\gamma \partial t^\delta}.
\eqno(1.55)$$
This proves the property 3 of our definition. The property 4
is obvious since the one-parameter group of diffeomorphisms for
the vector field (1.9) acts by rescalings (1.43).

Conversely, on a Frobenius manifold locally one can chose flat
coordinates $t^1$, ..., $t^n$ such that the invariant metric
$<~,~>$ is constant in these coordinates. The symmetry condition
(1.47) for the vector fields $u=\dalpha$, $v=\dbeta$, $w=\dgamma$
and $z=\partial_\delta$ reads
$$\partial_\delta c_{\alpha\beta\gamma} (t) ~{\rm is~symmetric~
in}~\alpha,~\beta,~\gamma,~\delta
$$
for
$$c_{\alpha\beta\gamma}(t) = <\dalpha\cdot\dbeta ,\dgamma >.
$$
Together with the symmetry of the tensor $c_{\alpha\beta\gamma}(t) $
this implies local existence of a function $F(t)$ such that
$$c_{\alpha\beta\gamma}(t) = {\partial^3 F(t)\over
\partial t^\alpha \partial t^\beta \partial t^\gamma.}
$$
Due to covariant constancy of the unity vector field $e$
we can do a linear change of coordinates in such a way that
$e = \partial_1$. This gives (1.53d).

We are to prove now that the function $F(t)$ satisfies (1.9). Due to (1.48)
in the flat coordinates  $E(t)$ is a linear vector field
of a form (1.11).
{}From the definition of rescalings we have
in the coordinates $t^\alpha$
$$[\partial_1, E] = \partial_1.
\eqno(1.56)$$
Hence $\partial_1$ is an eigenvector of the
operator $Q = \nabla E$
with the eigenvalue $1$. I.e. $d_1 = 1$. The constant matrix
$\left(Q_\beta^\alpha\right)$
must obey
the equation
$$Q_{\alpha\beta} = D\, \eta_{\alpha\beta}
\eqno(1.57)$$
(this follows from (1.50d))
for some constant $D$. The last step is to use the condition
(1.50b) (the definition of rescalings). From (1.50b) and (1.50d) we obtain
$${\cal L}_E c_{\alpha\beta\gamma} = (1+D) c_{\alpha\beta\gamma}.
$$
Due to (1.54) this can be rewritten as
$$\dalpha\dbeta\dgamma \left[E^\epsilon \partial_\epsilon
F - (1+D)F \right] = 0.
$$
This gives (1.9). Lemma is proved.
\medskip
{\bf Exercise 1.4.} In the case $d_1 = 0$ show that a two-dimensional
commutative group of diffeomorphisms acts locally  on the space of
parameters $t$ preserving the multiplication (1.53a) and the metric (1.53b).

This symmetry provides integrability in quadratures of
the equation (1.31a) for $d_1 = 0$ (observation of [30]). Indeed,
(1.31a) for $d_1 = 0$ ($q=-2$) reads
$$12z^2 (\phi'')^2 + 8z^3 \phi''\phi''' - \phi''' = 0.
$$
The integral of this equation is obtained in elliptic  quadratures from
$$\phi'' = {1\over 8z^3} \pm{\sqrt{ cz^3 + {1\over 64}}\over z^3}.
$$
\medskip
{\bf Definition 1.3.} Two Frobenius manifolds $M$ and $\tilde M$
are {\it equivalent} if there exists
a diffeomorphism
$$\phi : M\to\tilde M
\eqno(1.58a)$$
being a linear conformal transformation of the correspondent
invariant metrics $ds^2$ and $d\tilde s^2$
$$\phi^* d\tilde s^2 = c^2\, ds^2
\eqno(1.58b)$$
($c$ is a nonzero constant)
with the differential $\phi_*$ acting as an
isomorphism on the tangent algebras
$$\phi_*: T_tM \to T_{\phi(t)}\tilde M.
\eqno(1.58c)$$
If $\phi$ is a local diffeomorphism with the above properties
then it will be called {\it local equivalence}.
\medskip
Note that for an equivalence $\phi$ not necessary
$F = \phi^* \tilde F$. For example,
if the coordinates $(t^1,\dots, t^n)$
and $(\tilde t^1,\dots, \tilde t^n)$ on $M$ and $\tilde M$ resp. are normalized
as in (1.18) then the map
$$\tilde t^n = c^{2} t^n, ~~\tilde t^\alpha
= t^\alpha ~{\rm for}~\alpha \neq n
\eqno(1.59a)$$
$$\tilde F(\tilde t^1,
\dots, \tilde t^n) = c^2 F(t^1,\dots, t^n).
\eqno(1.59b)$$
for a constant $c\neq 0$ is an equivalence. Any equivalence is a superposition
of (1.59) and of a linear $\eta$-orthogonal transformation of the coordinates
$t^1, \dots, t^n$.
\medskip
{\bf Examples of Frobenius manifolds.}
\smallskip
{\bf Example 1.5.} Trivial Frobenius manifold. Let $A$ be a graded
Frobenius algebra. That means that some weights $q_1$, ..., $q_n$
are assigned to the
basic vectors $e_1$, ..., $e_n$ such that
$$c_{\alpha\beta}^\gamma = 0 ~{\rm for }~ q_\alpha + q_\beta \neq
q_\gamma
\eqno(1.60a)$$
and also
$$\eta_{\alpha\beta} = 0 ~{\rm for }~ q_\alpha + q_\beta \neq d
\eqno(1.60b)$$
for some $d$. Here
$$e_\alpha e_\beta = c_{\alpha\beta}^\gamma e_\gamma
$$
$$\eta_{\alpha\beta} = <e_\alpha,e_\beta >
$$
in the algebra $A$.
These formulae define a structure of Frobenius manifold on
$M=A$. The correspondent free energy $F$ is a cubic function
$$F(t) = {1\over 6} c_{\alpha\beta\gamma} t^\alpha
t^\beta t^\gamma = {1\over 6} <\tt^3, e>
\eqno(1.61)$$
for
$$c_{\alpha\beta\gamma} = \eta_{\alpha\epsilon}c_{\alpha\beta}^\epsilon,
$$
$$\tt = t^\alpha e_\alpha, ~~e = e_1 ~~{\rm is ~the ~unity}.
$$
Here the degrees of the coordinates $t^\alpha$ and of the function $F$
are
$$d_\alpha = 1-q_\alpha,~~ d_F = 3 - d.
$$

For example, the cohomology ring $A= H^*(X)$ of a $2d$-dimensional oriented
closed manifold $X$ satisfying
$$H^{2i+1}(X) = 0 ~~{\rm for ~any~} i
\eqno(1.62)$$
is a graded Frobenius algebra w.r.t. the cup product and the Poincar\' e
duality pairing. The degree of an element $x\in H^{2q}(X)$ equals $q$.
\smallskip
{\bf Remark 1.4.} To get rid of the restriction (1.62) one is to generaalize
the notion of Frobenius manifold to supermanifolds, i.e. to admit
anticommuting coordinates $t^\alpha$. Such a generalization was done
by Kontsevich and Manin [85].
\medskip
{\bf Example 1.6.} The direct product $M'\times M''$ of two Frobenius manifolds
of the dimensions $n$ and $m$ resp. carries a natural structure of a
Frobenius manifold if the scaling dimensions satisfy the constraint
$${d_1'\over d_1''} = {d_F'\over d_F''}.
\eqno(1.63)$$
If the flat coordinates ${t^1}', \dots ,{t^n}'$, ${t^1}'', \dots ,
{t^m}''$ are normalized as in (1.18), and $\deg\, {t^1}' = \deg\, {t^1}''
= 1$, then (1.63) reads $\deg\, {t^n}' = \deg\, {t^m}''$. Thus only the
case $m>1, ~n>1$ is of interest. The prepotential $F$ for the direct
product has the form
$$F\left( t^1, \hat t^1, {t^2}', \dots, {t^{n-1}}',
{t^2}'', \dots, {t^{m-1}}'', \hat t^N, t^N\right) =
$$
$$= \half {t^1}^2 t^N + t^1 \hat t^1 \hat t^N + \half
t^1 \sum_{\alpha =2}^{n-1}{t^\alpha}' {t^{n-\alpha +1}}'
+ \half t^1\sum_{\beta =2}^{m-1} {t^\beta}'' {t^{m-\beta +1}}'' +
$$
$$+ f'\left( {t^2}', \dots, {t^{n-1}}', \half (t^N + \hat t^N)\right)
+ f''\left( {t^2}'', \dots, {t^{m-1}}'', \half (t^N - \hat t^N)\right)
\eqno(1.64a)$$
where the functions $f'({t^2}', \dots, {t^{n}}')$ and
$f''({t^2}'', \dots, {t^{m}}'')$ determine the prepotentials
of $M'$ and $M''$ in the form (1.18). Here $N= n+m$,
$$\eqalign{ t^1 &= {{t^1}' + {t^1}''\over 2}, \cr
t^N &= {t^n}' + {t^m}'', \cr}
\eqalign{
{}~~\hat t^1 &=
{{t^1}' - {t^2}''\over 2} )\cr
{}~~\hat t^N &= {t^n}' - {t^m}''. )\cr
}
\eqno(1.64)$$
Observe that only trivial Frobenius manifolds can be multiplied by
a one-dimensional Frobenius manifold.
\medskip
{\bf Example 1.7.} [39] $M$ is the space of all polynomials of
the form
$$M = \{ \lambda (p) = p^{n+1} + a_n p^{n-1} + \dots
+ a_1 | a_1,\dots , a_n \in \Cc\}
\eqno(1.65)$$
with a nonstandard affine structure. We identify the tangent
plane to $M$ with the space of all polynomials of the degree
less than $n$. The algebra $A_\lambda$ on $T_\lambda M$ by definition
coincides with the algebra of truncated polynomials
$$A_\lambda = \Cc [p]/(\lambda'(p))
\eqno(1.66a)$$
(the prime denotes $d/dp$).
The invariant inner product is defined by the formula
$$<f,g>_\lambda = \res_{p=\infty}{f(p)g(p)\over \lambda'(p)}.
\eqno(1.66b)$$
The unity vector field $e$ and the Euler vector field $E$
read
$$e := {\partial\over \partial a_1},~~E := {1\over n+1}\sum_i (n-i+1)a_i
{\partial\over\partial a_i}.
\eqno(1.66c)$$
We will see that this is an example of a Frobenius manifold
in Lecture 4.
\smallskip

{\bf Remark 1.5.} The notion of Frobenius manifolfd admits algebraic
formalization in terms of the ring of functions on a manifold.
More precisely, let $R$ be a commutative associative algebra
with a unity over a field $k$ of characteristics $\neq 2$. We
are interested in structures of Frobenius algebra over $R$
in the $R$-module of $k$-derivations $Der(R)$ (i.e. $u(\kappa
)
=0$ for $\kappa \in k, u\in Der(R)$) satisfying
$$\tilde\nabla_u(\lambda )\tilde\nabla_v(\lambda )-
\tilde\nabla_v(\lambda )\tilde\nabla_u(\lambda )=
\tilde\nabla_{[u,v]}(\lambda )~ ~{\rm
identicaly~in}~\lambda\eqno(1.67a)$$
$${\rm for}~\tilde\nabla_u(\lambda )v=\nabla_uv+\lambda u\cdot
v,\eqno(1.67b)$$
(see below Lemma 3.1)
$$\nabla_ue=0~{\rm for ~all}~u\in Der(R)\eqno(1.67c)$$
where $e$ is the unity of the Frobenius algebra $Der(R)$.
Non-degenerateness of the symmetric inner product
$$<~,~>:Der(R)\times Der(R)\to R$$
means that it provides an isomorphism ${\rm
Hom}_R(Der(R),R)\to Der(R)$. I recall that the covariant
derivative is a derivation $\nabla_uv\in Der(R)$ defined for
any $u,~v\in Der(R)$ being determined from the equation
$$<\nabla_uv,w>=$$
$${1\over
2}[u<v,w>+v<w,u>-w<u,v>+<[u,v],w>+<[w,u],v>+<[w,v],u>]
\eqno(1.68)$$
for any $w\in Der(R)$ (here $[~,~]$ denotes the commutator of
derivations).

To reformulate algebraically the scaling invariance (1.50) we need
to introduce gradings to the algebras $R$ and $Der(R)$. In the
case of algebras of functions the gradings are determined by the
assumptions
$$\deg t^\alpha = 1-q_\alpha, ~~ \deg \dalpha = q_\alpha
\eqno(1.69)$$
where the numbers $q_\alpha$ are defined in (1.21).
\vfill\eject
\centerline{\bf Appendix A.}
\smallskip
\centerline{\bf Polynomial solutions of WDVV.}
\medskip
Let all the structure constants of a Frobenius manifold be analytic
in the point $t=0$. Then the germ of the Frobenius manifold near
the point $t=0$ can be considered as a deformation of the Frobenius
algebra $A_0 :=T_{t=0}M$. This is a graded Frobenius algebra with
a basis $e_1$, \dots, $e_n$ and
with
the structure constants $c_{\alpha\beta}^\gamma(0)$. The degrees
of the basic vectors are
$$\deg e_\alpha = q_\alpha
$$
where the numbers $q_\alpha$ are defined in (1.21). The algebras $T_tM$
for $t\neq 0$ can be considered thus as {\it deformations} of the
graded Frobenius algebra $T_0M$. In the physical setting (see Lecture
2 below) $T_0M$ is the primary chiral algebra of the correspondent
topological conformal field theory. The algebras $T_tM$ are operator
algebras of the perturbed topological field theory. So the problem of
classification of analytic deformations of graded Frobenius
algebras looks to be also physically motivated. (Probably, analytic
deformability in the sense that the graded Frobenius algebra
can be
the tangent algebra at the origin
of an analytic Frobenius manifold imposes a strong constraint
on the algebra.)

We consider here the case where
all the degrees $\deg t^\alpha$ are {\it real} positive
numbers and not all of them are equal.
In the normalization (1.18) that means that $0<d<1$.
\smallskip
{\bf Problem.} To find all the solutions of WDVV being analytic in the
origin $t=0$ with real positive degrees of the flat coordinates.
\medskip
Notice that for the positive degrees analyticity in the origin and
the quasihomogeneity (1.5) implies that the function $F(t)$ is a polynomial
in $t^1, \dots, t^n$. So the problem coincides with the problem of
finding of the {\it polynomial solutions} of WDVV.

For $n=2$ all the noncubic polynomial solutions have the form (1.24a)
where $k$ is
an integer and $k\geq 4$. Let us consider here the next case $n=3$.
Here we have a function $F$ of the form (1.26) and
$$\deg t^1 = 1, ~\deg t^2 = 1-{d\over 2}, ~\deg t^3 = 1-d, ~
\deg f = 3-d.
\eqno(A.1)$$
The function $f$ must satisfy the equation (1.28). If
$$f(x,y) = \sum a_{pq} x^p y^q
$$
then the condition of quasihomogeneity reads
$$a_{pq} \neq 0 ~{\rm only ~for}~ p+q-3 = (\half p + q - 1) d.
\eqno(A.2)$$
Hence $d$ must be a rational number. Solving the quasihomogeneity
equation (A.2) we obtain the following two possibilities for the
function $f$:
{1.}
$$f = \sum_k a_k x^{4-2km}y^{kn-1}
\eqno(A.3a)$$
$$d = {n-2m\over n-m}
\eqno(A.3b)$$
for some natural numbers $n, ~m$, $n$ is odd, and
{2.}
$$f = \sum_k a_k x^{4-km} y^{kn-1}
\eqno(A.4a)$$
$$d = {2(n-m)\over 2n -m}
\eqno(A.4b)$$
for some natural numbers $n, ~m$, $m$ is odd. Since the powers
in the expansions of $f$ must be nonnegative, we obtain the following
three possibilities for $f$:
$$f = a x^2 y^{n-1} + b y^{2n-1}, ~~n\geq 3
\eqno(A.5a)$$
$$f = a y^{n-1}, ~~n\geq 5
\eqno(A.5b)$$
$$f = a x^3 y^{n-1} + b x^2 y^{2n-1} + c x y^{3n-1}
+ d y^{4n-1}, ~~n\geq 2.
\eqno(A.5c)$$

\noindent The inequalities for $n$ in these formulae can be assumed since the
case when $f$ is at most cubic polynomial is not of interest. (It is easy to
see
that for a cubic solution with $n=3$ necessary $f=0$.)

Substituting (A.5) to (1.28) we obtain resp.
$$\eqalign{a(n-1)(n-2)(n-3) &= 0 \cr
-4a^2 (n-1)^2 + b (2n-1)(2n-2)(2n-3) &=0 \cr}
\eqno(A.6a)$$
$$a(n-1)(n-2)(n-3) = 0
\eqno(A.6b)$$
$$\eqalign{a(n-1)(n-2)(n-3) &= 0 \cr
18 a^2\left( (n-1)(n-2) - 2(n-1)^2\right) + b(2n-1)(2n-2)(2n-3) &=0 \cr
c(3n-1)(3n-2)(3n-3) &=0 \cr
- 4b^2 (2n-1)^2 + 6ac(3n-1)(3n-2) + d(4n-1)(4n-2)(4n-3) &= 0. \cr}
\eqno(A.6c)$$
It is clear that $a$ must not vanish for a nonzero $f$. So $n$ must equal
$3$ in the first case (A.6a), $2$ or $3$ in the third one, and in the
second case (A.6b) there is no non-zero solutions. Solving the system
(A.6a) for $n=3$ and the system (A.6c) for $n=2$ and $n=3$ we obtain
the following 3 polynomial solutions of WDVV:
$$\eqalignno{F& =
{t_1^2t_3+t_1t_2^2\over 2} + {t_2^2t_3^2\over 4} + {t_3^5\over 60}
&(A.7)\cr
F &= {t_1^2t_3+t_1t_2^2\over 2} + {t_2^3t_3\over 6} + {t_2^2t_3^3\over 6}
+{t_3^7\over 210}
&(A.8)\cr
F &= {t_1^2t_3+t_1t_2^2\over 2} + {t_2^3t_3^2\over 6}
+{t_2^2t_3^5\over 20} +{t_3^{11}\over 3960}.
&(A.9)\cr}
$$
These are unique up to the equivalence noncubic polynomial solutions of WDVV
with $n=3$ with positive degrees of $t^\alpha$.
The polynomial (A.7) coincides with the prepotential of Example 1.7 with $n=3$.

The crucial observation to understand the nature of the other polynomials
(A.8) and (A.9) was done by V.I.Arnold. He observed that the degrees $5 = 4+1$
$7=6+1$ and $11 = 10+1$ of the polynomials have a simple relation to the
Coxeter numbers of the groups of symmetries of the Platonic solids
(the tetrahedron, cube and icosahedron resp.).
In Lecture 4 I will show how to explain this observation using a hidden
symmetry of WDVV (see also Appendix G).
\vfill\eject
\bigskip
\centerline{\bf Appendix B.}
\smallskip
\centerline{\bf Symmetries of WDVV. Twisted Frobenius manifolds.}
\medskip
By definition {\it symmetries} of WDVV are the transformations
$$\eqalign{t^\alpha &\mapsto \hat t^\alpha,\cr
\eta_{\alpha\beta} &\mapsto \hat\eta_{\alpha\beta},\cr
F &\mapsto \hat F\cr}
\eqno(B.1)$$
preserving the equations.
First examples of the symmetries have been
introduced above: they are equivalencies
of Frobenius manifolds and shifts along vectors belonging
to the kernel of the grading operator $Q$.

Here we describe two types of less trivial symmetries for which the map
$t^\alpha\mapsto\hat t^\alpha$ preserves the multiplication of the
vector fields.
\smallskip
Type 1. Legendre-type
transformation $S_\kappa$ for a given $\kappa = 1, \dots, n$
$$\eqalignno{\hat t_\alpha &= \dalpha\partial_\kappa F(t)
&(B.2a)\cr
{\partial^2\hat F\over \partial \hat t^\alpha \partial \hat t^\beta}
&= {\partial^2 F\over \partial t^\alpha \partial t^\beta}
&(B.2b)\cr
\hat\eta_{\alpha\beta} &= \eta_{\alpha\beta}.
&(B.2c)\cr}
$$
We have
$$\dalpha =\partial_\kappa\cdot\hat\dalpha.
\eqno(B.3)$$
So the transformation $S_\kappa$ is invertible where $\partial_\kappa$
is an invertible element of the Frobrnius algebra of vector fields.
Note that the unity vector field
$$e = {\partial\over \partial\hat t^\kappa}.
\eqno(B.2d)$$

The transformation $S_1$ is the identity; the transformations $S_\kappa$
commute for different $\kappa$.

To describe what happens with the scaling degrees (assuming diagonalizability
of the degree
operator $Q$)
we shift
the degrees putting
$$\mu_\alpha := q_\alpha -{d\over 2},
{}~~\alpha = 1, \dots, n.
\eqno(B.4)$$
Observe that the spectrum consists of the eigenvalues of the
operator $-\hat V$ (see (1.51)).
The shifted degrees are symmetric w.r.t. zero
$$\mu_\alpha + \mu_{n-\alpha +1} = 0.
\eqno(B.5)$$
We will call the numbers $\mu_1, \dots, \mu_n$ {\it spectrum} of
the Frobenius manifold. Knowing the spectrum we can uniquely
reconstruct the degrees putting
$$q_\alpha = \mu_\alpha - \mu_1, ~~d= - 2\mu_1.
\eqno(B.6)$$

It is easy to see that the transformations $S_\kappa$ preserve the spectrum
up to permutation of the numbers $\mu_1, \dots, \mu_n$:
for $\kappa \neq {n\over 2}$
it interchanges the pair $(\mu_1, \mu_n)$ with the pair
$(\mu_\kappa , \mu_{n-\kappa +1})$. For $\kappa = {n\over 2}$
the transformed Frobenius manifold is of the second type (1.22).

To prove that (B.2) determines a symmetry of WDVV we introduce
on the Frobenius manifold $M$ a new metric $<~,~>_\kappa$ putting
$$<a,b>_\kappa := <\partial_\kappa^2, a\cdot b>.
\eqno(B.7)$$
\smallskip
{\bf Exercise B.1.} Prove that the variables $\hat t^\alpha$ (B.2a)) are the
flat
coordinates of the metric (B.7). Prove that
$$<\hat\dalpha\cdot\hat\dbeta, \hat\dgamma >_\kappa
= \hat\dalpha\hat\dbeta\hat\dgamma \hat F(\hat t).
\eqno(B.8)$$
\medskip
{\bf Example B.1.} For
$$F = \half {t^1}^2 t^2 + e^{t^2}
$$
($d=1$) the transformation $S_2$ gives
$$\eqalign{\hat t^1 &= e^{t^2}\cr
\hat t^2 &= t^1.\cr}
$$
Renumbering $\hat t^1\leftrightarrow \hat t^2$ (due to (B.2d)) we
obtain
$$\hat F = \half (\hat t^1)^2 \hat t^2 + \half (\hat t^2)^2
\left( \log t^2 - {3\over 2}\right).
\eqno(B.9)$$
This coincides with (1.24b) (now $d= -1$). See also Example 5.5 below.
\medskip
If there are coincidences between the degrees
$$q_{\kappa_1} = \dots = q_{\kappa_s}
\eqno(B.10)$$
then we can construct more general transformation $S_c$ putting
$$\hat t_\alpha = \sum_{i=1}^s c^i \dalpha\partial_{\kappa_i} F(t)
\eqno(B.11)$$
for arbitrary constants $(c^1, \dots, c^s) =: c$. This is invertible
when the vector field
$$\sum c^i\partial_{\kappa_i}
$$
is invertible. The transformed metric on $M$ depends quadratically
on $c^i$
$$<a,b>_c := <\left(\sum c^i\partial_{\kappa_i}\right)^2, a\cdot b>.
\eqno(B.12)$$
\medskip
Type 2. The inversion $I$:
$$\eqalign{\hat t^1 &= \half {t_\sigma t^\sigma\over t^n}
\cr
\hat t^\alpha &= {t^\alpha\over t^n} ~{\rm for ~}\alpha \neq 1,\, n\cr
\hat t^n &= -{1\over t^n}\cr}
\eqno(B.13a)$$
(the coordinates are normalized as in (1.18)),
$$\eqalign{\hat F(\hat t) &=
(t^n)^{-2}\left[ F(t) - \half t^1 t_\sigma t^\sigma\right]
= (\hat t^n)^2 F +\half \hat t^1 \hat t_\sigma \hat t^\sigma ,\cr
\hat\eta_{\alpha\beta} &= \eta_{\alpha\beta}.\cr}
\eqno(B.13b)$$

Note that the inversion acts as a conformal transformation of the metric
$<~,~>$
$$\eta_{\alpha\beta} d\hat t^\alpha d\hat t^\beta = (t^n)^{-2}
\eta_{\alpha\beta} dt^\alpha dt^\beta.
\eqno(B.14)$$

The inversion {\it changes} the spectrum $\mu_1, \dots, \mu_n$.
\smallskip
{\bf Lemma B.1.} {\it If
$$E(t) = \sum (1-q_\alpha)t^\alpha \dalpha
\eqno(B.15)$$
then after the transform one obtains
$$\hat E(\hat t) = \sum (1-\hat q_\alpha ) \hat t^\alpha \hat \dalpha
\eqno(B.16a)$$
where
$$\hat \mu_1 = -1+\mu_n, ~\hat\mu_n = 1+\mu_1, ~
\hat\mu_\alpha = \mu_\alpha ~{\rm for~} \alpha \neq 1,\, n.
\eqno(B.16b)$$
Particularly,
$$\hat d = 2-d.
\eqno(B.16d)$$
If
$$E(t) = \sum(1-q_\alpha) t^\alpha \dalpha + \sum_{q_\sigma =1}
r^\sigma \partial_\sigma
$$
and $d\neq 1$, or $d=1$ but $r^n = 0$, then
$$\hat E(\hat t) = \hat E^\alpha(\hat t)\hat \dalpha
\eqno(B.17a)$$
where
$$\eqalign{\hat E^1 &= \hat t^1 + \sum_{q_\sigma=1} r^\sigma\hat t^{n-\sigma
+1}
\cr
\hat E^\alpha &= (d-q_\alpha)\hat t^\alpha ~~{\rm for ~any }~\alpha ~
{\rm s.t.} ~ q_\alpha \neq 1\cr
\hat E^\sigma &= (d-1) \hat t^\sigma - r^\sigma \hat t^n ~~
{\rm for ~any}~ \sigma ~{\rm s.t.} ~ q_\sigma =1\cr
\hat E^n &= (d-1) \hat t^n.\cr}
\eqno(B.17b)$$}

Proof is straightforward.
\medskip
The transformation of the type 2 looks more misterious (we will see
in Lecture 3 that this is a Schlesinger transformation of WDVV in
the sense of [127]). We leave to the reader to verify that the inversion
preserves the multiplication of vector fields. Hint: use the formulae
$$\eqalign{\hat c_{\alpha\beta\gamma} &= t_1 c_{\alpha\beta\gamma}
- t_\alpha \eta_{\beta\gamma} - t_\beta\eta_{\alpha\gamma}
-t_\gamma\eta_{\alpha\beta}\cr
\hat c_{\alpha\beta n} &= t_1 c_{\alpha\beta\sigma}t^\sigma -
\half \eta_{\alpha\beta} t_\sigma t^\sigma - 2 t_\alpha t_\beta\cr
\hat c_{\alpha nn} &= t_1 c_{\alpha\lambda\mu}t^\lambda t^\mu
-2 t_\alpha t_\sigma t^\sigma\cr
\hat c_{nnn} &= t_1 c_{\lambda\mu\nu} t^\lambda t^\mu t^\nu
-{3\over 2} (t_\sigma t^\sigma)^2\cr}
\eqno(B.18a)$$
(here $\alpha, \beta, \gamma \neq 1, \, n$)
together with (B.14) to prove that
$$\hat c_{\alpha\beta\gamma} = (t^n)^{-2} {\partial t^\lambda
\over \partial \hat t^\alpha} {\partial t^\mu\over\partial
\hat t^\beta} {\partial t^\nu\over \partial \hat t^\gamma}
c_{\lambda\mu\nu}.
\eqno(B.18b)$$
\smallskip
{\bf Exercise B.2.} Show that the solution (1.24c) is the $I$-transform
of the solution (1.24b).
\smallskip
{\bf Exercise B.3.} Prove that the group $SL(2, {\bf C})$ acts on the space
of solutions of WDVV with $d=1$ by
$$\eqalign{t^1 &\mapsto t^1 + \half {c\over ct^n+d} \sum_{\sigma\neq 1}
t_\sigma t^\sigma\cr
t^\alpha &\mapsto {t^\alpha \over ct^n +d}\cr
t^n &\mapsto {at^n + b\over ct^n+d},\cr}
\eqno(B.19)$$
$$ad -bc =1.
$$
[Hint: consider superpositions of $I$ with the shifts along $t^n$.]
\medskip
The inversion is an involutive transformation up to an equivalence
$$\eqalign{I^2 : (t^1, t^2, \dots, t^{n-1}, t^n) &\mapsto
(t^1, -t^2, \dots, -t^{n-1}, t^n)\cr
F &\mapsto F.\cr}
\eqno(B.20)$$
\medskip
\smallskip
{\bf Proposition B.1.} {\it Assuming invertibility of the transformations
$S_\kappa$, $I$ we can reduce by these transformations
any solution of WDVV to a solution with
$$0\leq {\rm Re}\, q_\alpha \leq {\rm Re}\, d \leq 1.
\eqno(B.21)$$
}
\medskip
{\bf Definition B.1.} A Frobenius manifold will be called {\it reduced}
if it satisfies the inequalities (B.21).
\medskip
Particularly, the transformations $S_\kappa$ are invertible near those points
$t$
of a Frobenius manifold $M$ where the algebra $T_tM$ has no nilpotents.
In the next Lecture we will obtain complete local classification of such
Frobenius manifolds.
\medskip
Using the above transformations $I$, $S$ we can glue together a few
Frobenius manifolds to obtain a more complicated geometrical object
that will be called {\it twisted Frobenius manifold}. The multiplication
of tangent vector fields is globally well-defined on a twisted Frobenius
manifold. But the invariant inner product (and therefore, the function $F$)
is defined only locally. On the intersections of the coordinate charts
these are to transform according to the formula (B.12) or (B.14). We will
construct examples of twisted Frobenius manifolds in Appendices C, J.
Twisted Frobenius manifolds could also appear as the moduli spaces
of the topological sigma models of the B-type [151] where the flat
metric is well-defined only locally.
\vfill\eject
\bigskip
\centerline{\bf Appendix C.}
\smallskip
\centerline{\bf WDVV and Chazy equation.}
\smallskip
\centerline{\bf Affine connections on curves with
projective structure.}
\medskip
Here we consider three-dimensional Frobenius manifolds with $d=1$.
The degrees of the flat variables are
$$\deg t^1 = 1,~ \deg t^2 = 1/2, ~ \deg t^3 = 0.
\eqno(C.1)$$
Let us look for a solution of WDVV being periodic in $t^3$ and analytic
in the point $t^1 = t^2 = 0, ~ t^3 = i\infty$.
The function $F$ must have the form
$$F = \half (t^1)^2 t^3 + \half t^1 (t^2)^2 - {(t^2)^4\over 16} \gamma (t^3)
\eqno(C.2)$$
for some unknown $2\pi$-periodic
function $\gamma = \gamma(\tau)$ analytic at $\tau = i\infty$
$$\gamma(\tau) = \sum_{n\geq 0} a_n q^n, ~~q = e^{2\pi i \tau}.
\eqno(C.3)$$
The coefficients $a_n$ are determined up to a shift along $\tau$,
$$\tau \mapsto \tau + \tau_0, ~~ a_n \mapsto a_n e^{2\pi i n\tau_0}.
\eqno(C.4)$$

For the function we
obtain from (1.28)
$$\gamma''' = 6 \gamma\gamma'' - 9 {\gamma'}^2.
\eqno(C.5)$$
\smallskip
{\bf Exercise C.1.} Prove that the equation (C.5) has a unique (modulo
the ambiguity (C.4)) nonconstant solution of the form (C.3),
$$\gamma(\tau) = {\pi i\over 3}\left[ 1 - 24 q - 72 q^2 - 96 q^3
- 168 q^4 -\dots
\right] ,~~ q = e^{2\pi i \tau}.
\eqno(C.6)$$
\medskip
The equation (C.5) was considered by J.Chazy [29] as an example of ODE
with the general solution having moving natural boundary. It arose as
a reduction of the self-dual Yang - Mills equation in [1]. Following [1,
134]
I will call (C.5) {\it Chazy equation}.
\smallskip
{\bf Exercise C.2.} Show that the roots $\omega_1(\tau)$, $\omega_2(\tau)$,
$\omega_3(\tau)$ of the cubic equation
$$\omega^3 + {3\over 2} \gamma(\tau)\omega^2 +{3\over 2} \gamma'(\tau)
\omega +{1\over 4}\gamma''(\tau) = 0
\eqno(C.7)$$
satisfy the system
$$\dot\omega_1 =- \omega_1(\omega_2+\omega_3) + \omega_2\omega_3
$$
$$\dot\omega_2 =- \omega_2(\omega_1+\omega_3) + \omega_1\omega_3
$$
$$\dot\omega_3 =- \omega_3(\omega_1+\omega_2) + \omega_1\omega_2.
\eqno(C.8)$$

The system (C.8) was integrated by Halphen [70]. It was rediscovered in the
context of the self-dual Einstein equations by Atiyah and Hitchin [11].
\medskip
The main property [29] of Chazy equation is the invariance
w.r.t. the group $SL(2,\Cc )$
$$\tau \mapsto \tilde \tau = {a\tau + b\over c\tau + d}, ~~
ad - bc = 1
\eqno(C.9a)$$
$$\gamma (\tau) \mapsto \tilde \gamma (\tilde \tau ) =
(c\tau + d)^2 \gamma (\tau ) + 2 c (c\tau + d).
\eqno(C.9b)$$
The invariance (C.9) follows immediately from the invariance
of WDVV w.r.t. the transformation (3.19).
Observe that (C.9b) coincides with the transformation law of one-dimensional
affine connection w.r.t. the M\"obius transformations (C.9a) (cf. [134]).

We make here a digression about one-dimensional affine connections.
One-dimensional real or complex manifolds will be considered; in the complex
case only holomorphic connections will be of interest. The connection is
determined by a function
$$\gamma(\tau) := \Gamma_{11}^1(\tau)
$$
(holomorphic in the complex case, and $\Gamma_{\bar 1 \bar 1}^{\bar 1}
= \overline{\gamma(\tau)}$) for any given local coordinate $\tau$. The
covariant
derivative of a $k$-tensor $f(\tau)\, d\tau^k$ by definition is a
$(k+1)$-tensor of the form
$$\nabla f(\tau)\, d\tau^{k+1} := \left( {df\over d\tau}
- k \gamma(\tau) f(\tau)\right) \, d\tau^{k+1}.
\eqno(C.10)$$
This implies that under a change of coordinate
$$\tau \mapsto \tilde \tau = \tilde \tau (\tau )
\eqno(C.11a)$$
(holomorphic in the complex case) the connection must transform as follows
$$\tilde \gamma (\tilde \tau) = {1\over d\tilde \tau /d\tau}
\gamma (\tau) - {d^2 \tilde \tau /d\tau^2 \over d \tilde \tau /d \tau}.
\eqno(C.11b)$$

One-dimensional affine connection has no local invariants: it can be
reduced to zero by an appropriate change of coordinate. To find the
{\it flat} local parameter $x$ one is to look for a 1-form $\omega
= \phi(\tau)\, d\tau$ such that $\nabla\omega = 0$, i.e.
$${d\phi\over d\tau} - \gamma \phi = 0
\eqno(C.12a)$$
and then put
$$\omega = dx.
\eqno(C.12b)$$
The covariant derivative of a $k$-tensor $f\, dx^k$ coincides with the
usual derivative w.r.t. the flat coordinate $x$
$$\nabla f\, dx^{k+1} \equiv {df\over dx}\, dx^{k+1}.
\eqno(C.13)$$
In arbitrary coordinate $\tau$ the covariant derivative can be written in
the form
$$\nabla f\, d\tau^{k+1} = \phi^k {d\over d\tau} \left( f \phi^{-k}\right)
\, d\tau^{k+1}.
\eqno(C.14)$$

Let us assume now that there is fixed a projective structure on the
one-dimensional
manifold. That means that the transition functions (C.11a) now are not
arbitrary
but they are the M\"obius transformations (C.9a). Then the transformation law
(C.11b) coincides with (C.9b).

When is it possible to reduce the connection to zero by a M\"obius
transformation? What is the complete list of differential-geometric
invariants of an affine connection on one-dimensional manifold with
a projective structure?

The following simple construction gives the answer to the questions.
\smallskip
{\bf Proposition C.1.} {\it \item{1.} For a one-dimensional connection
$\gamma$ the quadratic differential
$$\Omega\, d\tau^2, ~~\Omega := {d\gamma\over d\tau} - \half \gamma^2
\eqno(C.15)$$
is invariant w.r.t. the M\"obius transformations.
\item{2.} The connection $\gamma$ can be reduced to zero by a M\"obius
transformation {\rm iff} $\Omega = 0$.
}

Proof. The verification of the invariance of $\Omega\, d\tau^2$ is
straightforward.
{}From this it follows that $\Omega = 0$ when $\gamma$ is reducible to $0$
by a M\"obius transformation. Conversely, solving the equation $\Omega = 0$
we obtain
$$\gamma = - {2\over \tau - \tau_0}.
$$
After the inversion
$$\tilde \tau = {1\over \tau - \tau_0}
$$
we obtain $\tilde \gamma(\tilde \tau ) = 0$. Proposition is proved.
\medskip
{\bf Remark C.1.} For arbitrary change of coordinate $\tilde \tau =
\tilde \tau (\tau)$ the \lq\lq curvature" $\Omega$ transforms like
projective connection
$$\tilde\Omega = \left( {d\tau \over d\tilde \tau}\right)^2
\Omega + S_{\tilde\tau}(\tau)
\eqno(C.16)$$
where $S_{\tilde\tau}(\tau)$ stands for the Schwartzian derivative
$$S_z(w) := {d^3w/dz^3\over dw/dz} - {3\over 2}
\left( {d^2w/dz^2\over dw/dz}\right)^2.
\eqno(C.17)$$
We obtain a map
$${\rm affine~ connections~} \to {\rm ~projective~ connections.}
$$
This is the appropriate differential-geometric interpretation of the
well-known Miura transformation.
\medskip
{\bf Exercise C.3.} Let $P = P(\gamma, d\gamma /d\tau, d^2\gamma /d\tau^2,
\dots )$ be a polynomial such that for any affine connection $\gamma$
the tensor $P\, d\tau^k$ for some $k$ is invariant w.r.t. M\"obius
transformations. Prove that $P$ can be represented as
$$P = Q(\Omega , \nabla\Omega, \nabla^2\Omega, \dots )
\eqno(C.18)$$
where $Q$ is a graded homogeneous polynomial of the degree $k$ assuming
that $\deg \nabla^l\Omega = l+2$.
\medskip
In other words, the \lq\lq curvature" $\Omega$ and the covariant derivatives
of it provide the complete set of differential-geometric invariants
of an affine connection on a one-dimensional manifold with a projective
structure.
\smallskip
{\bf Example C.1.} Consider a Sturm - Liouville operator
$$L = - {d^2\over dx^2} + u(x), ~~x \in D, ~~ D\subset S^1 ~
{\rm or}~ D\subset {\bf CP}^1.
\eqno(C.19)$$
It determines a projective structure on $D$ in the following standard
way. Let $y_1(x)$, $y_2(x)$ be two linearly independent solutions of the
differential equation
$$L\, y = 0.
\eqno(C.20)$$
We introduce a new local coordinate $\tau$ in $D$ putting
$$\tau = {y_2(x)\over y_1(x)}.
\eqno(C.21)$$
This specifies a projective structure in $D$. If $D$ is not simply
connected then a continuation of $y_1(x), ~y_2(x)$ along a closed curve
gives a linear substitution
$$y_1(x) \mapsto cy_2(x) + dy_1(x)
$$
$$y_2(x) \mapsto ay_2(x) + by_1(x)
\eqno(C.22)$$
for some constants $a, ~b, ~c, ~d$, $ad-bc=1$ (conservation of the Wronskian
$y_1y_2' - y_2y_1'$). This is a M\"obius transformation of the local
parameter $\tau$. Another choice of the basis $y_1(x), ~y_2(x)$ produces
an equivalent projective structure in $D$.

We have also a natural affine connection in $D$. It is uniquely
specified by saying that $x$ is the flat coordinate for the connection.
What is the \lq\lq curvature" $\Omega$ of this affine connection w.r.t.
the projective structure (C.21)? The answer is
$$\Omega d\tau^2 = 2 u(x) dx^2
\eqno(C.23)$$
(verify it!).

Let us come back to the Chazy equation. It is natural to consider
the general class of equations of the form
$$P(\gamma, d\gamma /d\tau, \dots, d^{k+1}\gamma /d\tau^{k+1}) = 0
\eqno(C.24)$$
for a polynomial $P$ invariant w.r.t. the transformations (C.9).
Due to (C.18) these can be rewritten in the form
$$Q(\Omega, \nabla\Omega, \dots, \nabla^k\Omega) = 0
\eqno(C.25a)$$
for
$$\Omega = {d\gamma\over d\tau} - \half \gamma^2,
\eqno(C.25b)$$
$Q$ is a graded homogeneous polynomial with $\deg \nabla^l\Omega = l+2$.
Putting
$$u := \half {\Omega d\tau^2\over \omega^2}
\eqno(C.26)$$
(cf. (C.23)) where $\nabla\omega = 0$, $\omega =: dx$, we can represent (C.25)
as
$$Q(2u, 2u', \dots, 2u^{(k)}) = 0
\eqno(C.27)$$
for $u' = du/dx$ etc. Solving (C.27) we can reconstruct $\gamma(\tau)$
via two independent solutions $y_1(x), ~y_2(x)$ of the Sturm -
Liouville equation (C.20) normalized by $y_2'y_1 - y_1'y_2 = 1$
$$\tau = {y_2(x)\over y_1(x)}, ~~\gamma = {d(y_1^2)\over dx}.
\eqno(C.28)$$

Let us consider examples of the equations of the form (C.25). I will consider
only the equations linear in the highest derivative $\nabla^k\Omega$.

For $k=0$ we have only the conditions of flateness. For $k=1$ there
exists only one invariant differential equation $\nabla\Omega =0$
or
$$\gamma'' - 3\gamma\gamma' + \gamma^3 = 0.
\eqno(C.29)$$
We have $u(x) = c^2$ (a constant); a particular solution of (C.29) is
$$\gamma = -{2\over \tau}.
\eqno(C.30)$$
The general solution can be obtained from (C.30) using the invariance (C.9).

For $k=2$ the equations of our class must have the form
$$\nabla^2\Omega + c \Omega^2 = 0
\eqno(C.31a)$$
for a constant $c$ or more explicitly
$$\gamma''' - 6 \gamma \gamma'' + 9{\gamma'}^2 + (c-12)
\left( \gamma' - \half \gamma^2\right)^2 = 0.
\eqno(C.31b)$$
For $c=12$ this coincides with the Chazy equation (C.5). The corresponding
equation (C.27)
$$u'' + 2cu^2 = 0
\eqno(C.32)$$
for $c\neq 0$ can be integrated in elliptic functions
$$u(x) = -{3\over c} \wp_0(x)
\eqno(C.33)$$
where $\wp_0(x)$ is the equianharmonic Weierstrass elliptic function,
i.e. the inverse to the elliptic integral
$$x = \int_\infty^{\wp_0} {dz\over 2\sqrt{z^3 -1}}.
\eqno(C.34)$$
[All the solutions of (C.32) can be obtained from (C.33) by shifts and
dilations
along $x$. There is also a particular solution $u =-3/cx^2$ and the
orbit of this w.r.t. (C.9).] So the solutions of (C.31) can be expressed as in
(C.28) via the solutions of the Lam\'e equation with the equianharmonic
potential
$$y'' +{3\over c}\wp_0(x) y = 0
\eqno(C.35)$$
(for $c=0$ via Airy functions). It can be reduced to the hypergeometric
equation
$$t(t-1){d^2y\over dt^2} +\left( {7\over6}t - {1\over 2}\right)
{dy\over dt} + {1\over 12c} y = 0
\eqno(C.36)$$
by the substitution
$$t = 1 - \wp^3_0(x).
\eqno(C.37)$$
{}From (C.28) we express the solution of (C.31) in the form
$$\tau = {y_2(t)\over y_1(t)}, ~~\gamma = {d\log y_1^2\over d\tau}
\eqno(C.38)$$
for two linearly independent solutions $y_1(t)$, $y_2(t)$ of the hypergeometric
equation.

Particularly, for the Chazy equation one obtains [29]
 the hypergeometric
equation
$$t(t-1){d^2y\over dt^2} +\left( {7\over6}t - \half \right)
{dy\over dt} + {1\over 144} y = 0.
\eqno(C.39)$$
Note that the function $t = t(\tau)$ is the
Schwartz triangle function $S(0, \pi /2, \pi /3; \tau)$.
\smallskip
{\bf Remark C.2.} In the theory of the Lam\'e equation (C.35) the values
$${3\over c} = - m(m+1)
\eqno(C.40)$$
for an integer $m$ are of particular interest [52]. These look not to be
discussed from the point of view of the theory of projective structures.

Chazy considered also the equation
$$\gamma''' - 6\gamma \gamma'' + 9 {\gamma'}^2 +
{432\over n^2 - 36} \left( \gamma' - \half \gamma^2\right)^2 = 0
\eqno(C.41)$$
for an integer $n>6$. The correspondent Lam\'e equation
$$y'' - m(m+1)\wp_0(x) y =0
\eqno(C.42)$$
has
$$m = {3\over n} -\half .
\eqno(C.43)$$
Particularly, for the equation (C.5) $m=-\half$. The solutions of
(C.41), according to Chazy, can be expressed via the Schwartz triangle
function $S(\pi /n, \pi /2, \pi /3; \tau)$. This can be seen from (C.36).
\smallskip
{\bf Exercise C.4.} \item{1.} Show that the equation of the class (C.25)
of the order $k=3$ can be integrated via solutions of the Lam\'e
equation $y'' + A\wp(x) y = 0$ with arbitrary Weierstrass elliptic
potential. \item{2.} Show that for $k=4$ $\gamma$ can be expressed
via the solutions  of
the equation (C.20) with the potential $u(x)$ satisfying
$$u^{IV} + a u^3 + b uu'' + c {u'}^4 = 0
\eqno(C.44)$$
for arbitrary constants $a, ~b, ~c$. Observe that for $a= -b = 10$,
$c= -5$ the equation (C.44) is a particular case of the equation determining
the genus two algebraic-geometrical (i.e. \lq\lq two gap") potentials
of the Sturm - Liouville operator [52].
\medskip
Let me explain now the geometrical meaning of the solution (C.6) of the Chazy
equation (C.5).  The underlined complex one-dimensional manifold $M$ here
will be the modular curve
$$M := \left\{ Im \tau >0\right\} /SL(2, {\bf Z}).
\eqno(C.45)$$
(This is not a manifold but an orbifold. So I will drop away
the \lq\lq bad" points $\tau = i\infty , ~\tau = e^{2\pi i /3},
{}~\tau = i$ and the $SL(2, {\bf Z})$-images of them.) A construction
of a natural affine connection on $M$ essentially can be found in the
paper [63] of Frobenius and Stickelberger. They described an elegant
approach to the problem of differentiating of elliptic functions
w.r.t. their periods. I recall here the basic idea of this not very
wellknown paper because of its very close relations to the subject
of the present lectures.

Let us consider a lattice on the complex plane
$$L = \left\{ 2m\omega + 2n\omega' |m, n \in {\bf Z}\right\}
\eqno(C.46)$$
with the basis $2\omega$, $2\omega'$ such that
$$Im\left(\tau = {\omega'\over \omega}\right) >0.
\eqno(C.47)$$
Another basis
$$\omega',~\omega \mapsto \tilde\omega' = a\omega' + b\omega,~
\tilde\omega = c\omega' + d\omega,
\eqno(C.48)$$
$$\left( \matrix{a & b \cr c & d \cr}\right) \in SL(2, {\bf Z})
\eqno(C.49)$$
determines the same lattice.

Let ${\cal L}$ be the set of all lattices. I will drop away (as above)
the orbifold points of ${\cal L}$ corresponding to the lattices with
additional symmetry. So ${\cal L}$ is a two-dimensional manifold.

By $E_L$ we denote the complex torus (elliptic curve)
$$E_L := \Cc /L.
\eqno(C.50)$$
We obtain a natural fiber bundle
$$\matrix{ & \downarrow & E_L \cr
       {\cal M}= & {\cal L} & \cr}.
\eqno(C.51)$$
The space of this fiber bundle will be called {\it universal torus}.
(Avoid confusion with universal elliptic curve: the latter is two-dimensional
while our universal torus is three-dimensional. The points of the universal
torus corresponding to proportional lattices give isomorphic elliptic curves.)
Meromorphic functions on ${\cal M}$ will be called {\it invariant
elliptic functions}.
They can be represented as
$$f = f(z; \omega , \omega')
\eqno(C.52a)$$
with $f$ satisfying the properties
$$f(z+2m\omega + 2n\omega' ; \omega, \omega') = f(z; \omega, \omega'),
\eqno(C.52b)$$
$$f(z; c\omega' + d\omega,\, a\omega' + b\omega ) = f(z; \omega, \omega')
\eqno(C.52c)$$
for
$$\left( \matrix{a & b \cr c & d \cr}\right) \in SL(2, {\bf Z}).
\eqno(C.52d)$$
An example is the Weierstrass elliptic function
$$\wp \equiv \wp (z; \omega, \omega') = {1\over z^2}
+ \sum_{m^2 + n^2 \neq 0}
\left( {1\over (z- 2m\omega - 2n\omega')^2} - {1\over
(2m\omega + 2n\omega')^2}  \right).
\eqno(C.53)$$
It satisfies the differential equation
$$\left( \wp'\right)^2 = 4\wp^3 - g_2 \wp -g_3
\eqno(C.54)$$
with
$$g_2 \equiv g_2(\omega, \omega') = 60 \sum_{m^2 + n^2 \neq 0}
{1\over
(2m\omega + 2n\omega')^4},
\eqno(C.55)$$
$$g_3 \equiv g_3(\omega, \omega') = 140 \sum_{m^2 + n^2 \neq 0}
{1\over
(2m\omega + 2n\omega')^6}.
\eqno(C.56)$$

Frobenius and Stickelberger found two vector fields on the universal torus
${\cal M}$. The first
one is the obvious Euler vector field
$$\omega {\partial\over\partial\omega} + \omega'
{\partial\over\partial\omega'} + z {\partial\over\partial z}.
\eqno(C.57)$$
In other words, if $f$ is an invariant elliptic function then so is
$$\omega {\partial f\over\partial\omega} + \omega'
{\partial f\over\partial\omega'} + z {\partial f\over\partial z}.
$$
(There is even more simple example of a vector field on the universal
torus: $\partial /\partial z$.)
To construct the second vector field we need the Weierstrass $\zeta$-function
$$\zeta \equiv \zeta(z; \omega, \omega') = {1\over z}
+ \sum_{m^2 + n^2 \neq 0} \left( {1\over z- 2m\omega - 2n\omega'} +
{1\over 2m\omega + 2n\omega'} + {z\over (2m\omega +2n\omega')^2}
\right),
\eqno(C.58)$$
$${d\zeta\over dz} = - \wp .
\eqno(C.59)$$
The $\zeta$-function depends on the lattice $L$ (but not on the particular
choice of the basis $\omega, ~\omega'$) but it is not an invariant
elliptic function
in the above sense since
$$\zeta(z+ 2m\omega + 2n\omega'; \omega, \omega') =
\zeta (z;\omega, \omega') + 2m\eta + 2n\eta'
\eqno(C.60a)$$
where
$$\eta \equiv \eta(\omega, \omega') := \zeta (\omega; \omega, \omega'),
\eqno(C.60b)$$
$$\eta' \equiv \eta'(\omega, \omega') := \zeta (\omega'; \omega, \omega').
\eqno(C.60c)$$
The change (C.48) of the basis in the lattice acts on $\eta, ~\eta'$ as
$$\tilde\eta' = a\eta' + b\eta, ~\tilde\eta = c\eta' + d\eta.
\eqno(C.61)$$
\smallskip
{\bf Lemma C.1.} [63] {\it If $f$ is an invariant elliptic function then so is
$$\eta {\partial f\over\partial\omega} + \eta'
{\partial f\over\partial\omega'} + \zeta {\partial f\over\partial z}.
\eqno(C.62)$$
}

Proof is in a simple calculation using (C.60) and (C.61).
\medskip
{\bf Exercise.} \item{1).}
For $f= \wp (z; \omega, \omega')$ obtain [63, formula 11.]
$$\eta {\partial \wp\over\partial\omega} + \eta'
{\partial \wp\over\partial\omega'} + \zeta {\partial \wp\over\partial z}
= - 2 \wp^2 + {1\over 3} g_2.
\eqno(C.63)$$
\item{2).} For $f= \zeta(z;\omega, \omega')$ (warning: this is not
an elliptic function!) obtain [63, formula 29.]
$$\eta {\partial \zeta\over\partial\omega} + \eta'
{\partial \zeta\over\partial\omega'} + \zeta {\partial
\zeta\over\partial z} = \half \wp' - {1\over 12 } g_2 z.
\eqno(C.64)$$
\medskip
Consider now the particular class of invariant
elliptic functions not depending on
$z$.
\smallskip
{\bf Corollary C.1.} {\it If
$f= f(\omega, \omega')$ is a homogeneous function on the
lattice of the weight $(-2k)$,
$$f(c\omega, c\omega') = c^{-2k}f(\omega, \omega')
\eqno(C.65)$$
then
$$\eta {\partial f\over\partial\omega} + \eta'
{\partial f\over\partial\omega'}
\eqno(C.66)$$
is a homogeneous function of the lattice of the degree $(-2k-2)$.
}
\medskip
{\bf Exercise C.5.} Using (C.63) and (C.64) prove that
$$\eta {\partial g_2\over\partial\omega} + \eta'
{\partial g_2\over\partial\omega'} = - 6 g_3
\eqno(C.67)$$
$$\eta {\partial g_3\over\partial\omega} + \eta'
{\partial g_3\over\partial\omega'} = -{g_2^2\over 3}
\eqno(C.68)$$
[63, formula 12.] and
$$\eta {\partial \eta\over\partial\omega} + \eta'
{\partial \eta\over\partial\omega'} = -{1\over 12} g_2 \omega
\eqno(C.69)$$
[63, formula 31.].
\medskip
Any homogeneous function $f(\omega, \omega')$ on ${\cal L}$ of the
weight $(-2k)$ determines a $k$-tensor
$$\hat f(\tau) d\tau^k
\eqno(C.70a)$$
on the modular curve $M$ where
$$f(\omega, \omega') = \omega^{-2k}\hat f(\tau), ~~\tau =
{\omega'\over\omega}.
\eqno(C.70b)$$
In the terminology of the theory of automorphic functions $\hat f$ is
an automorphic form of the modular group of the weight $2k$. [Also
some assumptions about behaviour of $\hat f$ in the orbifold points
are needed in the definition of an automorphic form; we refer the
reader to a textbook in automorphic functions (e.g., [80]) for the details.]
Due to
Corollary we obtain a map
$$k-{\rm tensors ~on}~M \to (k+1)-{\rm tensors ~on}~M,
\eqno(C.71a)$$
$$\hat f (\tau) \mapsto \nabla\hat f(\tau) :=-{2\over \pi i}
\omega^{2k+2} \left(\eta{\partial f\over \partial\omega} +
\eta'{\partial f\over\partial\omega'}\right).
\eqno(C.71b)$$
(Equivalently: an automorphic form of the weight $2k$ maps to an
automorphic form of the weight $2k+2$.)
This is the affine connection on $M$ we need. We call it {\it
FS-connection}.

Explicitly:
$$\nabla\hat f = - {2\over \pi i} \left( \eta{\partial\over\partial\omega}
+\eta'{\partial\over\partial\omega'}\right) \left[ \omega^{-2k}
\hat f\left({\omega'\over\omega}\right)\right]
$$
$$= -{2\over \pi i} \left[ (-\eta\omega' +\eta'\omega) {d\hat f\over d\tau}
- 2k \omega\eta \hat f\right] = {d\hat f\over d\tau} - k\gamma\hat f
\eqno(C.72a)$$
(I have used the Legendre identity $\eta\omega' - \eta'\omega = \pi i/2$)
where
$$\gamma \equiv \gamma(\tau) := -{4\over \pi i} \omega \eta (
\omega, \omega').
\eqno(C.72b)$$

The FS-connection was rediscovered in the theory of automorphic forms
by Rankin [119] (see also [80, page 123]). From [63, (13.10)] we obtain
$$\gamma(\tau) = {1\over 3\pi i} {\theta_1'''(0;\tau)\over
\theta_1'(0;\tau)}.
\eqno(C.72c)$$
{}From (C.60) it follows the representation of $\gamma(\tau)$ via
the normalized Eisenstein series $E_2(\tau)$ (this is not an
automorphic form!)
$$\gamma(\tau)= {i\pi\over 3} E_2(\tau)
\eqno(C.72d)$$
$$E_2(\tau) = 1+ {3\over \pi^2} \sum_{m\neq 0} \sum_{n=-\infty}^\infty
{1\over (m\tau + n)^2} = 1 - 24 \sum_{n=1}^\infty \sigma(n) q^n.
\eqno(C.73)$$
Here $\sigma(n)$ stands for the sum of all the divisors of $n$.
\smallskip
{\bf Proposition C.2.} {\it The FS-connection on the modular curve
satisfies the Chazy equation (C.5).
}

Proof (cf. [134]). Put
$$\hat g_2 \equiv \hat g_2(\tau) = \omega^4 g_2(\omega, \omega'),
{}~~\hat g_3 = \hat g_3(\tau) = \omega^6 g_3 (\omega, \omega').
\eqno(C.74)$$
{}From (C.69) we obtain that
$$\Omega \equiv \gamma' - \half \gamma^2 = {2\over 3(\pi i)^2} \hat g_2.
\eqno(C.75)$$
Substituting to (C.67), (C.68) we obtain
$$\nabla^2\Omega + 12 \Omega^2 = 0.
$$
Proposition is proved.
\medskip
{}From (C.72d), (C.73)
 we conclude that the solution (C.6) specified by the
analyticity at $\tau = i\infty$ coincides with the FS-connection.
\smallskip
{\bf Exercise C.6.} Derive from (C.5) the following recursion relation for
the sums of divisors of natural numbers
$$\sigma(n) = {12\over n^2(n-1)} \sum_{k=1}^{n-1} k(3n-5k)\sigma(k)
\sigma(n-k).
\eqno(C.76)$$
\medskip
To construct the flat coordinate for the FS-connection we observe
that [63] for the discriminant
$$\Delta \equiv \Delta(\omega, \omega') = g_2^3 - 27 g_3^2
\eqno(C.77)$$
we have from (C.67), (C.68)
$$\eta{\partial\Delta\over\partial\omega} +
\eta'{\partial\Delta\over\partial\omega'} = 0.
\eqno(C.78)$$
So
$$\nabla\hat\Delta(\tau) = 0
\eqno(C.79)$$
where $\hat\Delta(\tau)$ is a 6-tensor
$$\hat\Delta(\tau) = (2\pi)^{12} q\prod_{n=1}^\infty (1-q^n)^{24}.
\eqno(C.80)$$
The sixth root of $(2\pi)^{-12}\hat\Delta(\tau)d\tau^6$ gives the
covariantly constant 1-form $dx$
$$dx := \eta^4(\tau) d\tau
\eqno(C.81)$$
where $\eta(\tau)$ is the Dedekind eta-function
$$\eta(\tau) = q^{1\over 24} \prod_{n\geq 1} (1-q^n)
\eqno(C.82)$$
(avoid confusions with the function $\eta = \zeta(\omega; \omega, \omega')$!).
We obtain particularly that the FS covariant derivative of a $k$-tensor
$\hat f(\tau)$ can be written as
$$\nabla\hat f = \eta^{4k}(\tau) {d\over d\tau} \left[
{\hat f\over \eta^{4k}(\tau)} \right] .
\eqno(C.82)$$
Another consequence is the following formula for the FS-connection
$$\gamma(\tau) = {1\over 6} {d\over d\tau}
\log \hat\Delta(\tau) = 4 {d\over d\tau} \log \eta(\tau) =
{8\pi i}\left( {1\over 24} - \sum_{n=1}^\infty
{nq^n\over 1-q^n}\right) .
\eqno(C.83)$$
\smallskip
{\bf Remark C.3.}
Substituting (C.84) in the Chazy equation we obtain a 4-th order
differential equation for the modular discriminant. It is a consequence
of the third order equation of Jacobi [77, S. 103]
$$\left[ 12\psi^3 {d^2\psi\over dz^2}\right]^3 -
27 \left[ {1\over 8} \psi^4 {d^3(\psi^2)\over dz^2} \right]^2
= 1,
\eqno(C.85a)$$
$$\psi = \eta^{-2}(\tau),~~ z = 2\pi i \tau.
\eqno(C.85b)$$
Notice also the paper [74] of Hurwitz where it is shown that any
automorphic form of the modular group
satisfies certain algebraic equation of the third
order.
\medskip
Consider now the Frobenius structure on the space
$$\hat{\cal M} := \{ t^1, t^2, t^3 | Im t^3 > 0\}
\eqno(C.86)$$
specified by the FS solution (C.72) of the Chazy equation. So
$$\eqalign{F &= \half (t^1)^2 t^3 + \half t^1 (t^2)^2 -
{\pi i\over 2} (t^2)^4 \left( {1\over 24} - \sum_{n=1}^\infty
{nq^n\over 1-q^n}\right)\cr
&= \half (t^1)^2 t^3 + \half t^1 (t^2)^2
- {\pi i\over 2}(t^2)^4
\left( {1\over 24} - \sum_{n=1}^\infty \sigma(n) q^n\right)\cr
}
\eqno(C.87)$$
where $q = \exp{2\pi i t^3}$.
Here we have
$\tilde \gamma = \gamma$, i.e. the solution $\gamma(\tau)$ obeys
the transformation rule
$$\gamma\left( {a\tau + b \over c\tau + d}\right)
= (c\tau + d)^2 \gamma(\tau) + 2c (c\tau + d), ~~
\left( \matrix{a & b \cr c & d \cr}\right) \in SL(2, {\bf Z}).
\eqno(C.88)$$
The formulae (B.19) for integer $a$, $b$, $c$, $d$
determine a realisation of the group $SL(2, {\bf Z})$
as a group of symmetries of the Frobenius
manifold (C.87). Factorizing $\hat{\cal M}$ over the transformations (B.19)
we obtain a first example of twisted Frobenius manifold in the sense of
Appendix B.
\medskip
The invariant metric is a section of a line bundle over the manifold.
This is the
pull-back of the tangent bundle of the modular curve under the natural
projection
$$(t^1, t^2, t^3) \mapsto t^3.
$$
Indeed, the object
$$\left( (dt^2)^2 + 2 dt^1 dt^3 \right) \otimes {\partial\over\partial\tau}
\eqno(C.89)$$
is invariant w.r.t. the transformations (B.19) (this follows from (B.14)).
\smallskip
{\bf Exercise C.7.} Show that the formulae
$$\eqalign{ t^1 &=- {1\over 2\pi i} \left[ \wp (z; \omega, \omega')
+ \omega^{-1} \eta(\omega; \omega, \omega')\right] \cr
t^2 &= {\sqrt{2}\over \omega} \cr
t^3 &= \tau = \omega' /\omega \cr}
\eqno(C.90)$$
establish an isomorphism of the twisted Frobenius manifold (C.87)
with the universal torus
${\cal M}$.
\medskip
In Appendix J I will explain the relation of this example
to geometry of complex crystallographic group.
\smallskip
{\bf Remark C.4.} The triple correlators $c_{\alpha\beta\gamma}(t)$ can be
represented like a \lq\lq sum over instanton corrections" [149] in topological
sigma models (see the next lecture). For example,
$$c_{333} = 4\pi^4 (t^2)^4 \sum_{n\geq 1} n^3 A(n) {q^n\over 1-q^n}
\eqno(C.91)$$
where
$$A(n) = n^{-3} \prod [ p_i^{k_i} (p_i^3 + p_i^2 + p_i + 1)
- (p_i^2 + p_i + 1)]
\eqno(C.92a)$$
for
$$n = \prod_i p_i^{k_i}
\eqno(C.92b)$$
being the factorization of $n$ in the product of powers of different
primes $p_1, p_2, \dots$.
\vfill\eject
\centerline{\bf Lecture 2.}
\smallskip
\centerline{\bf Topological conformal field theories}
\smallskip
\centerline{\bf and their moduli.}
\medskip
A quantum field theory (QFT) on a $D$-dimensional manifold $\Sigma$
consists of:

1). a family of local fields $\phi_\alpha (x),~x\in \Sigma$ (functions or
sections of a fiber bundle over $\Sigma$). A metric $g_{ij}(x)$ on $\Sigma$
usualy is one of the fields (the gravity field).

2). A Lagrangian $L=L(\phi , \phi_x, ...)$. Classical field theory is
determined by the Euler -- Lagrange equations
$${\delta S\over \delta\phi_\alpha (x)} = 0,~~S[\phi ]=\int_\Sigma L(\phi ,
\phi_x, ...).
\eqno(2.1)$$
As a rule, the metric $g_{ij}(x)$ on $\Sigma$ is involved explicitly
in the Lagrangian even if it is not a dynamical variable.

3). Procedure of quantization usualy is based on construction of an
appropriate  path integration
measure $[d\phi ]$. The partition function is a result of the path integration
over the space of all fields $\phi (x)$
$$Z_\Sigma=\int [d\phi ]e^{-S[\phi ]}.
\eqno(2.2)$$
Correlation functions (non normalized) are defined by a similar path
integral
$$<\phi_\alpha (x)\phi_\beta (y)\dots >_\Sigma = \int [d\phi ]\phi_\alpha
(x)\phi_\beta (y)\dots e^{-S[\phi ]}.\eqno(2.3)$$
Since the path integration measure is almost never well-defined (and
also taking into account that different Lagrangians could give
equivalent QFTs) an old idea of QFT is to construct a self-consistent
QFT by solving a system of differential equations for correlation
functions. These equations were scrutinized in 2D
conformal field theories where D=2 and Lagrangians are invariant with
respect to conformal transformations
$$\delta g_{ij}(x)=\epsilon g_{ij}(x), ~~\delta S = 0.$$
This theory is still far from being completed.

Here I will consider another class of solvable 2-dimensional
QFT: {\it topological field
theories}. These theories admit {\it topological invariance}: they are
invariant with respect to arbitrary change of the metric $g_{ij}(x)$
on the 2-dimensional surface $\Sigma$
$$\delta g_{ij}(x)={\rm arbitrary},~~\delta S=0.
\eqno(2.4)$$
On the quantum level that means that the partition function $Z_\Sigma$
depends only on topology of $\Sigma$. All the correlation functions also
are topological creatures: they depend only on the labels of operators
and on topology of $\Sigma$ but not on the positions of the operators
$$<\phi_\alpha (x)\phi_\beta (y)\dots >_\Sigma \equiv
<\phi_\alpha\phi_\beta\cdots >_g
\eqno(2.5)$$
where $g$ is the genus of $\Sigma$.
The simplest example is 2D gravity with the Hilbert -- Einstein action
$$S = \int R\sqrt{g} d^2x={\rm Euler~characteristic~of~}\Sigma.
\eqno(2.6)$$
There are two ways of quantization of this functional. The first one
is based on an appropriate discrete version of the model ($\Sigma \to $
polihedron). This way leads to considering matrix integrals of the
form [22]
$$Z_N(t) = \int_{X^*=X}\exp\{-{\rm tr}(X^2+t_1X^4+t_2X^6+\dots \} dX
\eqno(2.7)$$
where the integral should be taken over the space of all $N\times N$
Hermitean matrices $X$. Here $t_1$, $t_2$ ... are called coupling
constants. A solution of 2D gravity is based on the observation that
after an appropriate limiting procedure $N\to\infty$
(and a renormalization of $t$) the limiting
partition function coincides with $\tau$-function of a particular solution of
the
KdV-hierarchy (see details in [149]).

Another approach called {\it topological 2D gravity} is based on an appropriate
supersymmetric extension of the Hilbert -- Einstein Lagrangian [147 - 149].
This
reduces the path integral over the space of all metrics $g_{ij}(x)$ on
a surface $\Sigma$ of the given genus $g$ to an integral over the
finite-dimensional space of conformal classes of these metrics, i.e.
over the moduli space ${\cal M}_g$ of Riemann surfaces of genus $g$.
Correlation functions of the model are expressed via intersection
numbers of certain cycles on the moduli space [149, 38]
$$\sigma_p\leftrightarrow c_p\in H_*({\cal M}_g),~~p = 0, 1, \dots
\eqno(2.8a)$$
$$<\sigma_{p_1}\sigma_{p_2}\dots >_g= \# (c_{p_1}\cap c_{p_2}\cap\dots
)
\eqno(2.8b)$$
(here the subscript $g$ means correlators on a surface of genus $g$).
This approach is often called {\it cohomological field theory}.

More explicitly, let $g$, $s$ be integers satisfying the conditions
$$g\geq 0, ~~s>0, ~~2-2g-s<0.
\eqno(2.9)$$
Let
$$\moduli = \left\{ \left( \Sigma , x_1, \dots, x_s
\right) \right\}
\eqno(2.10)$$
be the moduli space of smooth
algebraic curves $\Sigma$ of genus $g$ with $s$
ordered distinct
marked points $x_1, \dots, x_s$
(the inequalities (2.9) provide that the curve with the marked
points is {\it stable}, i.e. it admits no infinitesimal automorphisms).
By $\overline\moduli$
we will denote the Deligne -- Mumford compactification of $\moduli$.
Singular curves with double
points obtained by a degeneration of $\Sigma$ keeping the marked
points off the singularities
are to be added to compactify
$\moduli$. Any of the components of $\Sigma\setminus ({\rm singularities})$
with the marked and the singular
points on it is required to be stable.
Natural line bundles $L_1, \dots, L_s$
over $\overline\moduli$ are defined. By definition,
$${\rm fiber ~of ~} L_i |_{(\Sigma, x_1, \dots, x_s)}
= T^*_{x_i}\Sigma.
\eqno(2.11)$$
The Chern classes $c_1(L_i)\in H^*(\overline\moduli )$
of the line bundles and their products are {\it Mumford - Morita - Miller
classes} of the moduli space [105].
The genus $g$ correlators of the topological gravity are defined
via the intersection numbers of these cycles
$$<\sigma_{p_1} \dots \sigma_{p_s}>_g := \prod_{i=1}^s(2p_i+1)!!
\int_{\overline\moduli}
c^{p_1}_1(L_1)\wedge \dots \wedge c^{p_s}_1(L_s).
\eqno(2.12)$$
These numbers could be nonzero only if
$$\sum (p_i - 1) = 3g-3.
\eqno(2.13)$$
These are nonnegative
rational numbers but not integers since $\overline\moduli$
is not a manifold but an orbifold.

It was conjectured by Witten that the both approaches to 2D quantum
gravity should give the same results. This conjecture was proved by
Kontsevich [82 - 83] (another proof was obtained by Witten [152]).
He showed that the generating function
$$\eqalign{F(t) &= \sum_{g,n}\sum_{p_1 < \dots <p_n}
\sum_{k_1, \dots, k_n=0}^\infty {T_{p_1}^{k_1}\dots
T_{p_n}^{k_n} \over {k_1}!\dots {k_n}!}
<\sigma_{p_1}^{k_1}\dots\sigma_{p_n}^{k_n} >_g\cr
&= \sum_{g=0}^\infty \big\langle \exp\sum_{p=0}^\infty T_p \sigma_p
\big\rangle_g\cr}
\eqno(2.14)$$
(the free energy of 2D gravity) is logarythm of $\tau$-function of a
solution of the KdV hierarchy where $T_0=x$ is the spatial variable
of the hierarchy, $T_1$, $T_2$, \dots are the times
(this was the original form of the
Witten's conjecture). The $\tau$-function is specified by the string
equation (see eq. (6.54b)
below). Warning: the matrix gravity and the topological
one correspond to two different $\tau$-functions of KdV (in the terminology
of Witten these are {\it different phases} of 2D gravity).

Other examples of 2D TFT's
(see below)
proved out to have important
mathematical applications, probably being the best tool for treating
sophisticated topological objects.
For some of these 2D TFT's a description in terms of integrable
hierarchies was conjectured.

This gives rise to the following
\medskip
{\bf Problem.} To find a rigorous mathematical foundation of 2D topological
field theory. More concretely, to elaborate a system of axioms providing
the description (if any) of 2D TFT's in terms of integrable hierarchies
of KdV type.
\medskip
A first step on the way to the solution of the problem was done
by Atiyah [10] (for any dimension
D) in the spirit of G.Segal's axiomatization of
conformal field theory. He proposed simple axioms specifying
properties of correlators of the fields in the
{\it matter sector} of a 2D topological
field theory. In the matter sector
the set of local fields $\phi_1(x),\dots
, \phi_n(x)$
(the so-called {\it primary fields} of the model)
does not contain the metric on $\Sigma$. (Afterwards one should
integrate over the space of metrics. This should give rise to a
procedure of {\it coupling to topological gravity} that will be
described below in Lecture 6. In the above example of topological
gravity the matter sector consists only of the identity operator.)
Then the correlators of the fields $\phi_1(x)$, ...,
$\phi_n(x)$ obey very simple algebraic axioms.
According to these axioms the matter sector of a 2D TFT is specified by:
\item{1.} The space of the local physical states $A$. I will consider
only finite-dimensional spaces of the states
$${\rm dim}\, A = n <\infty.
$$
\item{2.} An assignment
$$(\Sigma, \partial\Sigma ) \mapsto v_{(\Sigma, \partial\Sigma ) }
\in A_{(\Sigma, \partial\Sigma ) }
\eqno(2.15)$$
for any oriented 2-surface $\Sigma$ with an oriented boundary $\partial\Sigma$
that depends only on the topology of the pair $(\Sigma, \partial\Sigma ) $
\footnote{$^{*)}$}{We can modify this axiom assuming that the assignment
(2.15)
is covariant w.r.t. some representation in $A_{(\Sigma, \partial\Sigma)}$
of the mapping class group $(\Sigma, \partial\Sigma)\to (\Sigma,
\partial\Sigma)$. The simplest generalisation of such a type is that,
where the space of physical states is ${\bf Z_2}$-graded
$$A = A_{even}\oplus A_{odd}.
$$
A homeomorphism $(\Sigma, \partial\Sigma) \to (\Sigma, \partial\Sigma)$
permuting the co-oriented components $C_1$, \dots, $C_k$ of $\partial\Sigma$
$$C_1, \dots, C_k \to C_{i_1}, \dots, C_{i_k}
$$
acts trivially on $A_{even}$ but it multiplies the vectors of $A_{odd}$
by the sign of the permutation $(i_1, \dots, i_k)$. In this lectures
we will not consider such a generalisation.}.
Here the linear space $A_{(\Sigma, \partial\Sigma ) }$ is defined as follows:
$$\eqalign{A_{(\Sigma, \partial\Sigma ) } &=
\Cc {\rm ~if~} \partial\Sigma =\emptyset\cr
&= A_1\otimes \dots \otimes A_k\cr}
\eqno(2.16)$$
if the boundary $\partial\Sigma$ consists of $k$ components $C_1, \dots,
C_k$ (oriented cycles) and
$$A_i := \cases{ A & if the orientation of $C_i$ is coherent to the orientation
of $\Sigma$\cr
A^* & (the dual space) otherwise. \cr}
\eqno(2.17)$$
\medskip
Drawing the pictures I will assume that the surfaces are oriented via the
external normal vector; so only the orientation of the boundary will be shown
explicitly.
\medskip
The assignment (2.15) is assumed to satisfy the following three axioms.
\smallskip
\noindent 1. {\it Normalization}:
\vskip 2 cm
\centerline{Fig.1}
\vskip 2 cm
\noindent 2. {\it Multiplicativity}: if
$$(\Sigma, \partial\Sigma )  = (\Sigma_1, \partial\Sigma_1 ) \cup
(\Sigma_2, \partial\Sigma_2 )
\eqno(2.18a)$$
(disjoint union) then
$$v_{(\Sigma, \partial\Sigma ) } = v_{(\Sigma_1, \partial\Sigma_1 ) }
\otimes v_{(\Sigma_2, \partial\Sigma_2 ) }\in A_{(\Sigma, \partial\Sigma ) }
= A_{(\Sigma_1, \partial\Sigma_1 ) }\otimes A_{(\Sigma_2, \partial\Sigma_2 ) }.
\eqno(2.18b)$$
3. {\it Factorization.} To formulate this axiom I recall the operation of
contraction defined in tensor products like (2.16),
(2.17). By definition, $ij$-contraction
$$A_1\otimes\dots\otimes A_k \to
A_1\otimes\dots\otimes \hat A_i\otimes\dots\otimes\hat A_j\otimes\dots
\otimes A_k
\eqno(2.19)$$
(the $i$-th and the $j$-th factors are omitted in the r.h.s.) is defined
when $A_i$ and $A_j$ are dual one to another using the standard pairing
$$A^*\otimes A \to \Cc
$$
of the $i$-th and $j$-th factors and the identity on the other factors.

Let $(\Sigma, \partial\Sigma ) $ and $(\Sigma', \partial\Sigma' ) $
coincide outside of a ball; inside the ball the two have the form
\vskip 2 cm
\centerline{Fig.2}
\vskip 2 cm
\noindent (I draw an oriented cycle on the neck of $\Sigma$ to emphasize
that it is obtained from $\Sigma'$ by gluing together the cycles
$C_{i_0}$ and $C_{j_0}$.) Then we require that
$$v_{(\Sigma, \partial\Sigma ) } = i_0j_0{\rm -contraction ~of}~
v_{(\Sigma', \partial\Sigma' ) }.
\eqno(2.20)$$

Particularly let us redenote by $v_{g,s}$ the vector
\vskip 2 cm
\centerline{Fig.3}
\vskip 2 cm
\noindent This is a symmetric polylinear function on the space of
the states. Choosing a basis
$$\phi_1, \dots, \phi_n\in A
\eqno(2.21)$$
we obtain the components of the polylinear function
$$v_{g,s}\left( \phi_{\alpha_1}\otimes \dots \phi_{\alpha_s}\right)
=: <\phi_{\alpha_1} \dots \phi_{\alpha_s}>_g
\eqno(2.22)$$
that by definition are called the genus $g$ correlators of the
fields $\phi_{\alpha_1}, \dots, \phi_{\alpha_s}$.

We will prove, following [37], that the space of the states $A$ carries
a natural structure of a Frobenius algebra. All the correlators can be
expressed in a pure algebraic way in terms of this algebra.

Let
\vskip 2 cm
\centerline{Fig.4}
\vskip2 cm
\vskip 2 cm
\centerline{Fig.5}
\vskip 2 cm
\vskip 2 cm
\centerline{Fig.6}
\vskip 2 cm
{\bf Theorem 2.1.} {\it \item{1.} The tensors $c$, $\eta$ specify on $A$
a structure of a Frobenius algebra with the unity $e$.
\item{2.} Let
\vskip 2 cm
\centerline{Fig.7}
\vskip 2cm
Then
$$<\phi_{\alpha_1}\dots \phi_{\alpha_k}>_g =
<\phi_{\alpha_1}\cdot\dots \cdot\phi_{\alpha_k}, H^g>
\eqno(2.23)$$
(in the r.h.s. $\cdot$ means the product in the algebra $A$).
}

Proof. Commutativity of the multiplication is obvious since we can
interchange the legs of the pants
on Fig. 4 by a homeomorphism. Similarly, we obtain
the symmetry of the inner product $<~,~>$. Associativity follows from
Fig. 8
\vskip 2cm
\centerline{Fig.8}
\vskip 2cm
\noindent
Particularly, the $k$-product is determined by the $k$-leg pants
\vskip 2cm
\centerline{Fig.9}
\vskip 2cm
Unity:
\vskip 2cm
\centerline{Fig.10}
Nondegenerateness of $\eta$. We put
\vskip 2cm
\centerline{Fig.11}
\vskip 2cm
and prove that $\tilde\eta = \eta^{-1}$. This follows from Fig. 12
\vskip 2cm
\centerline{Fig.12}
\vskip 2cm
Compatibility of the multiplication with the inner product is proved
on the next picture:
\vskip 2cm
\centerline{Fig.13}
\vskip 2cm
The first part of the theorem is proved.

The proof of the second part is given on the following picture:
\vskip 2cm
\centerline{Fig.14}
\vskip 2cm
Theorem is proved.
\medskip
{\bf Remark 2.1.} If the space of physical states is ${\bf Z_2}$-graded
(see the footnote on page ?) then we obtain a ${\bf Z_2}$-graded
Frobenius algebra. Such a generalization was considered by
Kontsevich and Manin in [85].
\medskip
The Frobenius algebra on the space $A$ of local physical observables
will be called {\it primary chiral algebra} of the TFT. We always
will choose a basis $\phi_1, \dots, \phi_n$ of $A$ in such a way
that
$$\phi_1 = 1.
\eqno(2.24)$$
Note that the tensors $\eta_{\alpha\beta}$ and $c_{\alpha\beta\gamma}$
defining the structure of the Frobenius algebra are the following genus
zero correlators of the fields $\phi_\alpha$
$$\eta_{\alpha\beta} = <\phi_\alpha\phi_\beta >_0, ~~~c_{\alpha\beta\gamma}
= <\phi_\alpha \phi_\beta\phi_\gamma >_0.
\eqno(2.25)$$
The handle operator $H$ is the vector of the form
$$H = \eta^{\alpha\beta}\phi_\alpha\cdot\phi_\beta \in A.
\eqno(2.26)$$
Summarizing, we can reformulate the Atiyah's axioms saying that the matter
sector of a 2D TFT is encoded by a Frobenius algebra. No additional
restrictions
for the Frobenius algebra can be read out of the axioms.
\medskip
On this way
\medskip

Topologicaly invariant Lagrangian $\to$ correlators of local physical
fields
\medskip
\noindent we lose too much relevant information.
 To
capture more information on a topological Lagrangian we will consider
a topological field theory together with its deformations preserving
topological invariance
$$L\to L + \sum t^\alpha L_\alpha^{(pert)}
\eqno(2.27)$$
($t^\alpha$ are coupling constants). To construct these moduli of a TFT
we are to say a few more words about the construction of a TFT.

A realization of the topological invariance is provided by QFT with
a nilpotent
symmetry. We have a Hilbert space ${\cal H}$ where the operators
of the QFT act and an endomorphism (symmetry)
$$Q: {\cal H}\to{\cal H}, ~~Q^2 = 0.
\eqno(2.28)$$
In the classical theory
$$\{ {\rm physical ~observables}\} = \{ {\rm invariants ~ of ~ symmetry}\} .
$$
In the quantum theory
$$\{ {\rm physical ~observables}\} = \{ {\rm operators ~commuting ~with~} Q
\}.
$$
I will denote by $\{ Q, \phi\}$ the commutator/anticommutator of $Q$
with the operator $\phi$ (depending on the statistics of $\phi$).
\smallskip
{\bf Lemma 2.1.} $\{ Q, \{ Q, \cdot \}\} = 0.$

Proof follows from $Q^2 = 0$ and from the Jacobi identity.
\medskip
Hence the operators of the form
$$\phi = \{ Q, \psi\}
\eqno(2.29)$$
are always physical. However, they do not contribute to the correlators
$$<\{ Q,\psi\} \phi_1 \phi_2 \dots > = 0
\eqno(2.30)$$
if $\phi_1$, $\phi_2$ \dots are physical fields. So the space
of physical states can be identified with the {\it cohomology} of
the operator $Q$
$$A= {\rm Ker} Q/{\rm Im} Q.
\eqno(2.31)$$
The operators in $A$ are called {\it primary states}.

The topological symmetry will follow if we succeed to construct
operators $\phi_\alpha^{(1)}$, $\phi_\alpha^{(2)}$ for
any primary field $\phi_\alpha = \phi_\alpha(x)$ such that
$$d\phi_\alpha(x) = \{ Q, \phi_\alpha^{(1)} \}, ~~
d\phi_\alpha^{(1)}(x) = \{ Q, \phi_\alpha^{(2)}\} .
\eqno(2.32)$$
(We assume here that the fields $\phi_\alpha(x)$ are scalar functions
of $x\in\Sigma$. So $\phi_\alpha^{(1)}(x)$ and $\phi_\alpha^{(2)}(x)$
will be 1-forms and 2-forms on $\Sigma$ resp.) Indeed,
$$d_x<\phi_\alpha(x)\phi_\beta(y)\dots > =
<\{ Q, \phi_\alpha^{(1)}(x)\} \phi_\beta(y)\dots > = 0.
\eqno(2.33)$$

The operators $\phi_\alpha^{(1)}$ and $\phi_\alpha^{(2)}$ can be constructed
for a wide class of QFT obtained by a procedure of {\it twisting} [93] from
a $N=2$ supersymmetric quantum field theory (see [38]). Particularly, in this
case the primary chiral algebra is a {\it graded} Frobenius
algebra in the sense of Lecture 1. The degrees $q_\alpha$ of the
fields $\phi_\alpha$ are the correspondent eigenvalues of the $U(1)$-charge
of the $N=2$ algebra; $d$ is just the label of the $N=2$ algebra (it
is called {\it dimension} since in the case of topological sigma-models
it coincides with the complex dimension of the target space). The class
of TFT's obtained by the twisting procedure from $N=2$ superconformal
QFT is called {\it topological conformal field theories} (TCFT).

{}From
$$\{ Q , \oint_C \phi_\alpha^{(1)}\} = \oint_C d\phi_\alpha = 0
\eqno(2.34)$$
we see that $\oint_C \phi_\alpha^{(1)}$ is a physical observable
for any closed cycle $C$ on $\Sigma$. Due to (2.30)
this operator depends only on the homology class of the cycle.

Similarly, we obtain that
$${\int\!\int}_\Sigma \phi_\alpha^{(2)}
\eqno(2.35)$$
is also a physical observable. (Both the new types of observables
are non-local!)

Using the operators (2.35) we can construct a very important class of
perturbations of the TCFT modifying the action as follows
$$S \mapsto \tilde S(t) := S - \sum_{\alpha =1}^n t^\alpha
{\int\!\int}_\Sigma \phi_\alpha^{(2)}
\eqno(2.36)$$
where the parameters $t=(t^1, \dots t^n)$ are called {\it coupling constants}.
The perturbed correlators will be functions of $t$
$$<\phi_\alpha(x)\phi_\beta(y)\dots >(t) := \int [d\phi ] \phi_\alpha (x)
\phi_\beta(y) \dots e^{-\tilde S(t)}.
\eqno(2.37)$$
\smallskip
{\bf Theorem 2.2.} [39] {\it \item{1.} The perturbation (2.36) preserves
the topological
invariance.
\item{2.} The perturbed primary chiral algebra $A_t$ satisfies the WDVV
equations.}
\medskip
Due to this theorem the construction (2.36) determines a {\it canonical
moduli space} of dimension $n$ of a TCFT with $n$ primaries. And this
moduli space carries a structure of Frobenius manifold.
\medskip
I will not reproduce here the proof of this (physical) theorem (it looks
like the statement holds true under more general assumptions
than those were used in the proof [39] - see below). It would be
interesting to derive the theorem directly from Segal's-type axioms
(see in [142]) of TCFT.

The basic idea of my further considerations is to add the statement
of this theorem as {\it a new axiom} of TFT. In other words, we will
axiomatize not an isolated TCFT but the TCFT together with its canonical
moduli space (2.36). Let me repeat that the axioms of TCFT now read:
\smallskip
{\it The canonical moduli space of a TCFT is a Frobenius manifold.}
\medskip
The results of Appendices A, C above shows that the axiom together
with certain analiticity assumptions could give rise to a reasonable
classification of TCFT. Particularly, the formula (A.7) gives the
free energy of the $A_3$ topological minimal model (see below).
More general relation of Frobenius manifolds to discrete
groups will be established in Appendix G.
In lecture 6 we will show that
the axioms of coupling (at tree-level)
to topological gravity of Dijkgraaf and Witten
can be derived from geometry of Frobenius manifolds. The description of
Zamolodchikov metric (the $t\,t^*$ equations of Cecotti and Vafa [26])
is an additional differential-geometric structure on the Frobenius
manifold [47].

For the above example of topological gravity the matter sector is rather
trivial: it consists only of the unity operator. The correspondent
Frobenius manifold (the moduli space) is one-dimensional,
$$F = {1\over 6} (t^1)^3.
$$
All the nontrivial fields $\sigma_p$ for $p>0$ in the topological
gravity come from the integration over the space of metrics
(coupling to topological gravity that we will discuss in Lecture 6).

We construct now other examples of TCFT describing their
matter sectors. I will skip to describe the Lagrangians of these
TCFT giving only the \lq\lq answer": the description of the primary
correlators in topological terms.
\smallskip
{\bf Example 2.1.} Witten's algebraic-geometrical description [150] of the
$A_n$
topological minimal models [39]
(due to K.Li [93] this is just the topological counterpart
of the $n$-matrix model). We will construct some coverings
over the moduli spaces $\moduli$. Let us fix numbers ${\bf \alpha} = (\alpha_1,
\dots, \alpha_s)$ from $1$ to $n$ such that
$$n(2g-2) + \sum_{i=1}^s(\alpha_i -1) = (n+1) l
\eqno(2.38)$$
for some integer $l$. Consider a line bundle
$${\cal L}\to \Sigma
\eqno(2.39)$$
of the degree $l$ such that
$${\cal L}^{\otimes (n+1)} = K_\Sigma^n \bigotimes_{i} O(x_i)^{\alpha_i -1}.
\eqno(2.40)$$
Here $K_\Sigma$ is the canonical line bundle of the curve $\Sigma$ (of the
genus $g$). The sections of the line bundle in the r.h.s. are
$n$-tensors on $\Sigma$ having poles only at the marked points $x_i$
of the orders less than $\alpha_i$.

We have $(n+1)^{2g}$ choices of the line bundle ${\cal L}$. Put
$$\moduli'({\bf\alpha}) :=
\left\{ \left( \Sigma, x_1, \dots, x_s, {\cal L}\right)
\right\} .
\eqno(2.41)$$
This is a $(n+1)^{2g}$-sheeted covering over the moduli space of stable
algebraic curves. An important point [150] is that the covering can be
extended onto the compactification $\overline{\moduli}$.
Riemann -- Roch implies that,
generically
$${\rm dim}\, H^0\left( \Sigma, {\cal L}\right) =
l+1 - g = d(g-1) + \sum_{i= 1}^s q_{\alpha_i} =: N({\bf\alpha})
\eqno(2.42)$$
where we have introduced the notations
$$d := {n-1\over n+1}, ~~~q_\alpha := {\alpha -1\over n+1}.
\eqno(2.43)$$
Let us consider now the vector bundle
$$\matrix{\downarrow & \!\! V({\bf\alpha})\cr
\moduli'({\bf\alpha}) & \cr}
\eqno(2.44)$$
where
$$V({\bf\alpha}) := H^0\left( \Sigma, {\cal L}\right).
\eqno(2.45)$$
Strictly speaking, this is not a vector bundle since the dimension
(2.42) may jump on the curves where $H^1\left( \Sigma, {\cal L}\right)\neq 0$.
However, the top Chern class $c_N(V({\bf\alpha}))$, $N = N({\bf\alpha})$
of the bundle is well-defined (see [150]
for more detail explanation). We define the primary correlators by the
formula
$$<\phi_{\alpha_1} \dots \phi_{\alpha_s}>_g :=
(n+1)^{-g} \int_{\overline{\moduli'}({\bf\alpha})} c_N\left( V({\bf\alpha})
\right)
\eqno(2.46)$$
(the $\overline{\moduli'}$ is an appropriate compactification of
$\moduli'$).
These are nonzero only if
$$\sum_{i=1}^s (q_{\alpha_i} - 1) = (3-d)(g-1).
\eqno(2.47)$$
The generating function of the genus zero primary correlators
$$F(t) := <\exp{\sum_{\alpha =1}^n t^\alpha \phi_\alpha}>_0
\eqno(2.48)$$
due to (2.47)
is a quasihomogeneous polynomial of the degree $3-d$ where
the degrees of the coupling constants $t^\alpha$ equal
$1-q_\alpha$. One can verify directly that $F(t)$ satisfies
WDVV [150]. It turns out that the Frobenius manifold (2.48)
coincides with the Frobenius manifold of polynomials
of Example 1.7!

One can describe in algebraic-geometrical terms also the result
of \lq\lq coupling to topological gravity" (whatever it means)
of the above matter sector. One should take into consideration
an analogue of Mumford -- Morita -- Miller classes (see above
in the construction of topological gravity) $c_1(L_i)$
(I recall that the fiber of the line bundle $L_i$ is the
cotangent line $T^*_{x_i}\Sigma$). After coupling to
topological gravity of the matter sector (2.46) we will
obtain an infinite number of fields $\sigma_p(\phi_\alpha)$,
$p = 0, 1, \dots$. The fields $\sigma_0(\phi_\alpha)$ can be
identified with the primaries $\phi_\alpha$. For $p>0$
the fields $\sigma_p(\phi_\alpha)$ are called {\it
gravitational descendants} of $\phi_\alpha$. Their correlators
are defined by the following intersection number
$$<\sigma_{p_1}(\phi_{\alpha_1})\dots \sigma_{p_s}(\phi_{\alpha_s})
>_g :=
$$
$$={\left({\alpha_1\over n+1}\right)_{p_1}
\dots \left({\alpha_s\over n+1}\right)_{p_s}\over (n+1)^{g-p_1-\dots -p_s}}
\int_{\overline{\moduli'}({\bf\alpha})}
c_1^{p_1}(L_1)\wedge\dots\wedge c_1^{p_s}(L_s)
\wedge c_N(V({\bf\alpha})).
\eqno(2.49)$$
Here we introduce the notation
$$(r)_p := r(r+1)\dots (r+p-1), ~~(r)_0 := 1.
\eqno(2.50)$$
The correlator is nonzero only if
$$\sum_{i=1}^s (p_i + q_{\alpha_i} - 1) = (3-d)(g-1).
\eqno(2.51)$$
The generating function of the correlators
$${\cal F}(T) := \sum_g\big\langle \exp{\sum_{p=0}^\infty \sum_{\alpha =1}^n
T_{\alpha, p}\sigma_p(\phi_\alpha)}\big\rangle_g
\eqno(2.52)$$
is conjectured by Witten [150] to satisfy the $n$-th generalized KdV
(or Gelfand - Dickey) hierarchy. Here $T$ is the infinite
vector of the indeterminates $T_{\alpha,p}$, $\alpha = 1,\dots, n$,
$p = 0, 1, \dots$. We will come back to the discussion of the
conjecture in Lecture 6.

\smallskip
{\bf Example 2.2.} Topological sigma-models [147] (A-models in the terminology
of Witten [151]). Let $X$ be a compact K\"ahler manifold of (complex) dimension
$d$ with non-positive canonical class.
The fields in the matter sector of the TFT will be in 1-1-correspondence
with the cohomologies $H^*(X,\Cc )$. I will describe only the genus zero
correlators of the fields. For simplicity I will consider
only the case when the odd-dimensional cohomologies of $X$ vanish
(otherwise one should consider ${\bf Z_2}$-graded Frobenius manifolds
[85]).

The two-point correlator of two cocycles $\phi_\alpha$, $\phi_\beta\in
H^*(X,\Cc )$ coincides with the intersection number
$$<\phi_\alpha, \phi_\beta > = \int_X \phi_\alpha\wedge\phi_\beta.
\eqno(2.53)$$
The definition of multipoint correlators is given [151] in terms
of the intersection theory on the moduli spaces of \lq\lq instantons",
i.e. holomorphic maps of the Riemann sphere to $X$.

Let
$$\psi : CP^1 \to X
\eqno(2.54)$$
be a holomorphic map of a given homotopical type $[\psi ]$. Also
some points $x$, $y$, \dots are to be fixed on $CP^1$. Let
${\cal M}[\psi]$ be the moduli space of all such maps for the
given homotopical type $[\psi]$. We consider the \lq\lq universal
instanton": the natural map
$$\Psi : CP^1\times {\cal M}[\psi] \to X.
\eqno(2.55)$$
We define the primary correlators putting
$$<\phi_\alpha \phi_\beta\dots >_0 :=
\int_{{\cal M}[\psi]} \Psi^*(\phi_\alpha)|_{x\times{\cal M}[\psi]}\wedge
\Psi^*(\psi_\beta)|_{y\times{\cal M}[\psi]}
\wedge\dots .
\eqno(2.56)$$
This TFT can be obtained by twisting of a N=2 superconformal theory if
$X$ is a Calabi - Yau manifold, i..e. if the canonical class of $X$ vanishes.
So one could expect to describe this class
of TFT's by Frobenius manifolds only for Calabi - Yau $X$. However, we still
obtain a Frobenius manifold for more general K\"ahler manifolds
$X$ (though it has been proved rigorously only for some particular classes
of K\"ahler manifolds [104, 120]). But
the scaling invariance (1.7) must be modified to (1.12).

To define a generating function we are to be more careful in choosing
of a basis in $H^{1,1}(X,\Cc )$. We choose this basis $\phi_{\alpha_1},
\dots, \phi_{\alpha_k}$
of integer K\"ahler
forms $\in\, H^{1,1}(X,\Cc)\cap H^2(X,{\bf Z}) $. By definition,
integrals of the form
$$d_i \equiv d_i[\psi] := \int_{CP^1}\psi^*(\phi_{\alpha_i}), ~~i = 1, \dots, k
\eqno(2.57)$$
are all nonnegative. They are homotopy invariants of the map (2.54).
The generating function
$$F(t) := \big\langle \sum_{[\psi]}\sum_{\phi_\alpha\in H^*(X)} t^\alpha
\phi_\alpha \big\rangle_0
\eqno(2.58)$$
will be formal series in $e^{- t^{\alpha_i}}$, $i=1, \dots, k$
and a formal power series in other coupling constants. So the Frobenius
manifold
coincides with the cohomology space
$$M = \oplus H^{2i}(X,\Cc ).
\eqno(2.59)$$
The free energy $F(t)$ is $2\pi i$-periodic in $t^{\alpha_i}$ for
$\phi_{\alpha_i}\in H^{1,1}(X,\Cc)\cap H^2(X,{\bf Z})$. In the limit
$$t^{\alpha_i}\to +\infty ~{\rm for} ~\phi_{\alpha_i}\in
H^{1,1}(X,\Cc)\cap H^2(X,{\bf Z}),
{}~t^\alpha \to 0 ~{\rm for ~ other ~} \phi_\alpha
\eqno(2.60)$$
the multiplication on the Frobenius manifold coincides with the multiplication
in the cohomologies. In other words, the cubic part of the corresponding
free energy is determined by the graded Frobenius algebra $H^*(X)$
in the form (1.61). Explicitly, the free energy has the structure of (formal)
Forier series
$$F(t,\tilde t) = {\rm cubic ~part} +
\sum_{k_i, \, [\psi ]}
N_{[\psi ]}(k_1, \dots, k_m)
{(\tilde t^{\beta_1})^{k_1} \dots (\tilde t^{\beta_m})^{k_m}\over
k_1! \dots k_m!}
e^{- t^{\alpha_1} d_1 - \dots -t^{\alpha_k} d_k}.
\eqno(2.61)$$
Here the coupling constants $t^{\alpha_1}, \dots t^{\alpha_k}$ correspond
to a basis of K\"ahler forms as above, the coupling constants
$\tilde t^{\beta_j}$ correspond to a basis in the rest part of cohomologies
$$\tilde t^{\beta_j} \leftrightarrow \tilde\phi_{\beta_j}\in
\oplus_{k>1}H^{2k}(X,{\bf Z}) ({\rm modulo ~torsion}).
\eqno(2.62)$$
The coefficients $N_{[\psi ]}(k_1, \dots, k_m)$ are defined as the
Gromov - Witten invariants of $X$\footnote{$^*$}{One needs to make some
technical genericity assumptions about  the maps $\psi$ in order to prove
WDVV for the free energy (2.61). One possibility is to perturb the complex
structure on $X$ and to consider pseudoholomorphic curves $\psi$. This was
done (under some restrictions on $X$) in [104, 120]. Another scheme was
proposed
recently by Kontsevich [84]. It based on considerations of {\it stable maps},
i.e. of such maps that the restrictions (2.63)
allow no infinitesimal deformations
of them.}. By definition, they count the numbers of rational maps
(2.54)
of the homotopy type $[\psi ] = (d_1, \dots, d_k)$ intersecting with
the Poincar\'e-dual cycles
$$\#\left\{\psi(CP^1) \cap {\cal D}(\tilde \phi_{\beta_j})\right\} = k_j, ~
j= 1, \dots, m.
\eqno(2.63)$$
(We put $N_{[\psi ]}(k_1, \dots, k_m) = 0$ if the set of maps satisfying
(2.63) is not discrete.)

The Euler vector field $E$ has the form
$$E= \sum (1-q_\alpha ) t^\alpha \partial_\alpha -
\sum_{i=1}^k r_{\alpha_i}\partial_{\alpha_i}
\eqno(2.64)$$
where $q_{\alpha}$ is the {\it complex} dimension of the cocycle $\phi_\alpha$
and the integers $r_{\alpha_i}$ are defined by the formula
$$c_1(X) = \sum_{i=1}^k r_{\alpha_i} \phi_{\alpha_i}.
\eqno(2.65)$$

Particularly,
suppressing all the couplings but those corresponding to the K\"ahler forms
$\in\, H^{1,1}(X,\Cc)\cap H^2(X,{\bf Z})$ we reduce the perturbed primary
chiral ring (2.61) to the {\it quantum cohomology ring} of C.Vafa [138].
For a Calabi - Yau manifold $X$ (where $c_1(X) = 0$) this is the only
noncubic part of the Frobenius structure. The quantum multiplication on
Calabi - Yau manifolds is still defined on $H^*(X)$. It has a structure
of a graded algebra with the same gradings as in usual cohomology algebra,
but this structure depends on the parameters $t^{\alpha_1}, \dots,
t^{\alpha_k}$.

An equivalent reformulation of the multiplication (depending on the coupling
constants $t^{\alpha_1}, \dots , t^{\alpha_k}$)
in the quantum
cohomology ring reads as follows [138]
$$<\phi_\alpha\cdot\phi_\beta, \phi_\gamma>
:= \sum_{[\psi]}e^{-t^{\alpha_1}d_1 - \dots t^{\alpha_k}d_k}
\eqno(2.66)$$
where the summation is taken over all the homotopy classes of the maps
$$\psi :CP^1 \to X ~~{\rm such ~that ~} \psi(0)\in {\cal D}(\phi_\alpha),
{}~\psi(1)\in {\cal D}(\phi_\beta),~\psi(\infty)\in {\cal D}(\phi_\gamma)
\eqno(2.67)$$
(here ${\cal D}$ is the Poincar\'e duality operator ${\cal D}: H^i(X)
\to H_{d-i}(X)$). By definition we put zero in the r.h.s. of (2.66) for
those classes $[\psi]$ when the set of maps satisfying (2.67) is not
discrete.
In the limit
$$t^{\alpha_i}\to +\infty
\eqno(2.68)$$
the quantum cohomology ring coincides with $H^*(X)$.

The most elementary example is the quantum cohomology ring of $CP^d$
$$\Cc [x]/(x^{d+1} = e^{-t}).
\eqno(2.69)$$
Here $t= t_2$ corresponds to the K\"ahler class of the standard metric
on $CP^d$.

The trivial case is that $d=1$ (complex projective line). We obtain two
coupling parameters $t_1 \leftrightarrow e_1\in H^0(CP^1,{\bf Z})$,
$t_2 \leftrightarrow -e_2\in H^2(CP^1,{\bf Z})$ (I change the sign
for convenience). The Frobenius structure on $H^*(CP^1)$ is completely
determined
by the quantum multiplication (2.66). So
$$F(t_1, t_2) = \half t_1^2t_2 +e^{t_2}.
\eqno(2.70)$$
The Euler vector field has the form, according to (2.64)
$$E= t_1 \partial_1 +2\partial_2.
\eqno(2.71)$$

In the case of the projective plane ($d=2$) we choose again the basic elements
$e_1$,
$e_2$, $e_3$ in $H^0$, $H^2$, $H^4$ resp. The classical cohomology ring has the
form
$$e_2^2=e_3
\eqno(2.72a)$$
$$e_2e_3 = 0.
\eqno(2.72b)$$
In the quantum cohomology ring instead of (2.72b) we have
$$e_2e_3 = e^{t_2}e_1.
\eqno(2.73)$$
So we must have
$$\partial_2\partial_3^2 F|_{t_1=t_3=0} = e^{t_2}.
\eqno(2.74)$$
Hence the Frobenius structure on $H^*(CP^2)$ must have
the free energy of the form
$$F(t_1,t_2,t_3) = \half t_1^2t_3 + \half t_1t_2^2 +\half
t_3^2 e^{t_2} + o(t_3^2\, e^{t_2}).
\eqno(2.75)$$
The Euler vector field for the Frobenius manifold reads
$$E = t_1 \partial_1 + 3 \partial_2 - t_3 \partial_3.
\eqno(2.76)$$
So we obtain finally the following structure of the free energy
$$F = \half t_1^2t_3+\half t_1t_2^2 + t_3^{-1}\phi
(t_2+3\log t_3)
\eqno(2.77)$$
where the function $\phi = \phi (x)$ is a solution of the differential
equation
$$9\phi'''-18 \phi'' + 11 \phi' - 2\phi =
\phi''\phi'''-{2\over 3}\phi'\phi''' +{1\over 3}{\phi''}^2
\eqno(2.78)$$
of the form
$$\phi = \sum_{k\geq 1} A_ke^{kx}.
\eqno(2.79)$$
Due to (2.74) one must have
$$A_1 = \half .
\eqno(2.80)$$
For the coefficients of the Fourier series (2.79) we have the recursion
relations for $n>1$
$$A_n = {1\over (n-1)(9n^2-9n+2)}\sum_{k+l=n,~0<k,l}
kl\left[\left(n+{2\over 3}\right) kl -{2\over 3}(k^2+l^2)\right]
A_k A_l.
\eqno(2.81)$$
So the normalization (2.80) uniquely specifies the solution (2.79) (observation
of [85]). The coefficients $A_k$ are identified by Kontsevich and
Manin [85] as
$$A_k = {N_k\over (3k-1)!}
$$
where $N_k$ is the number of rational curves of the degree $k$
on $CP^2$ passing through generic $3k-1$ points.

Let us prove convergence of the series (2.79)\footnote{$^{*)}$}{As
I learned very
recently from D.Morrison, a general scheme of proving convergence of the series
for the free energy was recently proposed by J.Koll\'ar.}.
\smallskip
{\bf Lemma 2.2.} {\it $A_k$ are positive numbers satisfying
$$A_k \leq {3\over 5k^4} \left( {5\over 6}\right)^k
\eqno(2.83)$$
for any $k$.}

Proof. Positivity of $A_n$ follows from positivity of the coefficients in
(2.81).
The inequality (2.83) holds true
for $k=1,~ 2$ ($A_2 = 1/120$). We continue the proof by induction
in $k$. For $k\geq 3$ we have
$$(n-1)(9n^2 - 9n + 2) \geq 3n^3.
\eqno(2.84)$$
Assuming the inequality (2.83) to be valid for any $k<n$, we obtain
from (2.83), (2.84)
$$A_n\leq {3\over 50 n^3}\cdot
\left( {5\over 6}\right)^n
\sum_{k+l=n}\left[ {\left( n+{2\over 3}\right)\over
k^2\, l^2} + {2\over 3} \left( {1\over k\, l^3} + {1\over k^3\, l}
\right)\right]
$$
$$={3\over 50 n^3} \left({5\over 6}\right)^n
\left[ {4(n+1)\over n^3} \sum_1^{n-1}{1\over k}
+{2n+{8\over 3}\over n^2} \sum_1^{n-1}{1\over k^2}
+{4\over 3n}\sum_1^{n-1} {1\over k^3}\right] <{3\over 5n^4}\left(
{5\over 6}\right)^n
$$
where we use the following elementary inequalities
$$\eqalign{{4(n+1)\over n^3 }\sum_1^{n-1} {1\over k} &< {3\over n}\cr
{2n+{8\over 3}\over n^2}\sum_1^{n-1}{1\over k^2}
&<{5\over n}\cr
{4\over 3n}\sum_1^{n-1}{1\over k^3} &< {2\over n}.\cr}
$$
Lemma is proved.
\medskip
{\bf Corollary 2.1.} {\it The function $\phi(x)$ is analytic in
the domain
$${\rm Re}\, x< \log {6\over 5}.
\eqno(2.85)$$
}
\medskip
It would be interesting to give more neat description of the analytic
properties of the function $\phi(x)$. An analytic expression for this function
in a closed form is still not available, although there are
some very interesting formulae in the recent preprint of Kontsevich [84].
Such an expression could give a solution of the following open
\smallskip
{\bf Problem.} To compute the asymptotic for big $k$ of the numbers
$N_k$ of rational curves of degree $k$ on the projective plane
passing through generic $3k-1$ points.
\medskip
More complicated example is the quantum cohomology ring of the quintic
in $CP^4$ (the simplest example of a Calabi - Yau three-fold). Here
${\rm dim}\, H^{1,1}(X) = 1$. We denote by $\phi$ the basic element in
$H^{1,1}(X)$ (the K\"ahler class) and by $t$ the correspondent
coupling constant.
The only
nontrivial term in the quantum cohomology ring is
$$<\phi\cdot\phi, \phi > = 5 + \sum _{n=1}^\infty A(n)n^3 {q^n\over
1-q^n}, ~~q = e^{-t}
\eqno(2.86)$$
where $A(n)$ is the number of rational {\it curves} in $X$
of the degree $n$. The function (2.86) has been found in [24] in the setting
of mirror conjecture. Quantum cohomologies for other
manifolds were calculated in [9, 14, 28, 67, 79]. In [67] quantum cohomologies
of flag
varieties were found. A remarkable description of them in terms of
the ring of functions on a particular Lagrangian manifold of the
Toda system was discovered.
Also a relation of quantum cohomologies and
Floer cohomologies were elucidated in these papers. A general approach
to calculating higher genera corrections to (2.86) was found in [16].
\medskip
{\bf Example 2.3.} Topological sigma-models of B-type [151]. Let $X$
be a compact Calabi - Yau manifold (we consider only 3-folds,
i.e. $d = {\rm dim}_{\Cc} X = 3$). The correlation functions
in the model are expressed in terms of periods of some differential
forms on $X$. The structure of the Frobenius manifold $M = M(X)$
is decribed only on the sublocus $M_0(X)$ of complex structures
on $X$. I recall that ${\rm dim} M_0(X) = {\rm dim} H^{2,1}(X)$ while
the dimension of $M$ is equal to the dimension of the full cohomology
space of $X$.
The bilinear form $\eta_{\alpha\beta}$ coincides with the intersection
number
$$<\phi',\phi''> = \int_X\phi'\wedge\phi''.
\eqno(2.87)$$
To define the trilinear form on the tangent space to $M_0(X)$ we fix a
holomorphic
3-form
$$\Omega \in H^{3,0}(X)
\eqno(2.88)$$
and normalize it by the condition
$$\oint_{\gamma_0} \Omega = 1
\eqno(2.89)$$
(I recall that dim$\, H^{3,0}(X)= 1$ for a Calabi - Yau 3-fold) for an
appropriate cycle $\gamma_0 \in H_3(X,{\bf Z})$. Then the trilinear form
reads
$$<\partial, \partial', \partial''> := \int_X
\partial \partial' \partial'' \Omega \wedge \Omega
\eqno(2.90)$$
for three tangent vector fields $\partial$, $\partial'$, $\partial''$
on $M_0(X)$ (I refer the reader to [107] regarding the details of the
construction
of holomorphic vector fields on $M_0(X)$ using technique of variations of Hodge
structures). From the definition it follows that the Frobenius  structure
(2.90) is well-defined only locally. Globally we would expect to obtain a
twisted
Frobenius manifold in the sense of Appendix B because of the ambiguity
in the choice of the normalizing cycle $\gamma_0$.

The mirror conjecture claims that for any Calabi - Yau manifold $X$
there exists another Calabi - Yau manifold $\hat X$ such that the Frobenius
structure of A-type determined by the quantum multiplication on the
cohomologies
of $X$ is locally isomorphic to the Frobenius structure of the B-type
defined by the periods of $\hat X$. See [138, 151] concerning motivations of
the
conjecture and [8, 24, 28, 73, 108] for consideration of the particular
examples of
mirror dual manifolds $X$ and $\hat X$.
\medskip
{\bf Example 2.4.} Topological Landau - Ginsburg (LG) models [137]. The bosonic
part of the LG-action $S$ has the form
$$S = \int d^2z \left( |{\partial p\over \partial z}|^2 + |\lambda'(p)|^2
\right)
\eqno(2.91)$$
where the holomorphic function $\lambda(p)$ is called {\it superpotential}
and $S$ is considered as a functional of the holomorphic {\it superfield}
$p = p(z)$. The classical states are thus in the one-to-one correspondence
with the critical points of $\lambda(p)$
$$p(z) \equiv p_i, ~~\lambda'(p_i)=0, ~~i=1, \dots, n.
\eqno(2.92)$$
Quantum correlations can be computed [137, 27] in terms of solitons propagating
between the classical vacua (2.92). If the critical points of the
superpotential
are non-degenerate and the critical values are pairwise distinct then masses
of the solitons are proportional to the differences of the critical values.
In this case we obtain a {\it massive} TFT.

The moduli space of a LG theory can be realized as a family of LG models
with an appropriately deformed superpotential
$$\lambda = \lambda (p; t^1, \dots, t^n).
\eqno(2.93)$$
The Frobenius structure on the space of parameters is given by the
following formulae [137, 27]
$$\eqalignno{<\partial, \partial'>_\lambda &=
\sum \res_{\lambda' = 0} {\partial(\lambda dp) \partial'(\lambda dp)\over
d\lambda(p)}
&(2.94a)\cr
<\partial, \partial', \partial''>_\lambda &= \sum\res_{\lambda' =0}
{\partial(\lambda dp) \partial'(\lambda dp) \partial''(\lambda dp)\over
d\lambda(p) dp.}
&(2.94b)\cr}
$$
By definition, in these formulae the vector fields $\partial$,
$\partial'$, $\partial''$ on the space of parameters act trivially on $p$.

Particularly, for the superpotential
$$\lambda(p) = p^{n+1}
$$
the deformed
superpotential coincides with the generic polynomial of the degree $n+1$
of the form (1.65). The Frobenius structure (2.94) coincides with the one
of Example 1.7 (see Lecture 4 below).

We will not discuss here interesting relations between topological
sigma-models and topological LG models (see [153]).
\vfill\eject
\centerline{\bf Lecture 3.}
\smallskip
\centerline{\bf Spaces of isomonodromy deformations as Frobenius manifolds.
}
\bigskip
Let us consider a linear differential operator
of the form
$$\Lambda = {d\over dz} - U - {1\over z} V
\eqno(3.1)$$
where $U$ and $V$ are two complex $n\times n$ $z$-independent matrices,
$U$ is a diagonal matrix with pairwise distinct diagonal entries,
and the matrix $V$ is skew-symmetric. The solutions of the differential
equation with
rational coefficients
$$\Lambda \psi = 0
\eqno(3.2)$$
are analytic multivalued vectorfunctions in $z\in \Cc\setminus 0$. The
monodromy
of these solutions will be called monodromy of the differential operator
$\Lambda$ (we will give the precise definition of the monodromy
below). The main result of this Lecture is the following
\smallskip
{\bf Main Theorem.} {\it There is a natural Frobenius structure on the space
of all the operators of the form (3.1) with a given monodromy. This structure
is determined uniquely up to a symmetry described above in Appendix B.
Conversely, any Frobenius manifold satisfying to a semisimplisity assumption
can be obtained by such a construction.}
\medskip

The following two creatures are the principal playing characters
on a Frobenius manifold.

1. Deformed Euclidean connection

$$\tilde\nabla_u(z)v := \nabla_uv + z u\cdot v.
\eqno(3.3)$$
It is a symmetric connection depending on the parameter $z$.
\smallskip
{\bf Lemma 3.1.} {\it The connection $\tilde\nabla (z)$ is flat
identically in $z$ } iff {\it the algebra $c_{\alpha\beta}^\gamma(t)$
is associative and a function $F(t)$ locally exists such that
$$c_{\alpha\beta\gamma}(t) = \dalpha\dbeta\dgamma F(t).
\eqno(3.4)$$}

Proof. In the flat coordinates vanishing of the curvature of the deformed
connection reads
$$[\tilde\nabla_\alpha(z),\tilde\nabla_\beta(z)]^\epsilon =
[z(\dbeta c_{\alpha\gamma}^\epsilon - \dalpha c_{\beta\gamma}
^\epsilon) + z^2 (c_{\alpha\gamma}^\sigma c_{\beta\sigma}^\epsilon
- c_{\beta\gamma}^\sigma c_{\alpha\sigma}^\epsilon )]
= 0.
$$
Vanishing of the coefficients of the quadratic polynomial in $z$
together with the symmetry of the tensor $c_{\alpha\beta\gamma}
:= \eta_{\gamma\epsilon} c_{\alpha\beta}^\epsilon$ is equivalent
to the statements of the lemma.
\medskip
{\bf Remark 3.1.} Vanishing of the curvature is equivalent to the compatibility
of the linear system
$$\dalpha \xi_\beta = z c_{\alpha\beta}^\gamma (t) \xi_\gamma.
\eqno(3.5)$$
This gives a \lq\lq Lax pair" for the associativity equations ($z$ plays the
role of the spectral parameter). For any given $z$ the system
has $n$-dimensional space of solutions. The solutions are closely
related to the flat coordinates of the deformed connection
$\tilde\nabla(z)$, i.e to the independent functions
$\tilde t^1(t,z)$, ...,
$\tilde t^n(t,z)$ such that in these new coordinates the deformes
covariant derivatives coincide with partial derivatives

$$\tilde\nabla_\alpha (z) = {\partial\over\partial\tilde t^\alpha
(t,z)}.
\eqno(3.6)$$
\smallskip
{\bf Exercise 3.1.} Prove that 1) any solution of the system (3.5) is
the gradient of some function
$$\xi_\alpha = \dalpha \tilde t;
$$
2) if $\xi_\alpha^1$, ..., $\xi_\alpha^n$ is a fundamental
system of solutions of the system (3.5) for a given $z$
then the correspondent
functions $\tilde t^1$, ..., $\tilde t^n$ are flat coordinates
for the deformed connection $\tilde\nabla(z)$.
\smallskip
{\bf Exercise 3.2.} Derive from (3.5) the following Lax pair for the Chazy
equation
(C.5)
$$\left[ \partial_z + U, \partial_x +V \right] = 0
\eqno(3.8)$$
where the matrices $U$ and $V$ have the form
$$U = \left( \matrix{ 0 & -1 & 0 \cr
{3\over 4} z^2 \gamma' & {3\over 2} z \gamma & -1 \cr
{1\over 4} z^3 \gamma'' & {3\over 4} z^2 \gamma' & 0 \cr}
\right) , ~~
V = \left( \matrix{ 0 & 0 & -1 \cr
{1\over 4} z^3 \gamma'' & {3\over 4} z^2 \gamma' & 0 \cr
{1\over 16} z^4 \gamma''' & {1\over 4} z^3 \gamma'' & 0 \cr}
\right).
\eqno(3.9)$$

Observe that the more strong commutativity condition
$$\left[ U, V \right] = 0
\eqno(3.10)$$
holds true for the matrices $U$, $V$. It is easy to see that one can put
arbitrary three unknown functions in (3.9) instead of $\gamma'$, $\gamma''$,
$\gamma'''$. Then the commutativity (3.8) together with (3.10) is still
equivalent to the Chazy equation.

The commutation representation (3.8), (3.10) looks to be intermediate one
between Lax pairs with a derivative w.r.t. the spectral parameter $z$
(those being typical in the theory of isomonodromic deformations)
and \lq\lq integrable algebraic systems" of [51], i.e. the equations
of commutativity of matrices depending on the spectral parameter.

A Lax pair for the Chazy equation was obtained also in [1]. But
instead of finite dimensional matrices some differential
operators with partial derivatives are involved. This gives no possibility
to apply the machinery of the theory of integrable systems.

I learned recently from S.Chakravarty that he has found another
finite-dimensional Lax pair for Chazy equation. This looks similar
to the Lax pair of the Painlev\'e-VI equation with a nontrivial
dependence of the poles on the both dependent and independent variables.
\medskip
The further step is to consider WDVV as the scaling reduction of the equations
of associativity (1.14) as of an integrable system. The standard machinery
of integration of scaling reductions of integrable systems [60, 76, 127]
suggests
to add a differential equation in the spectral parameter $z$ for the
auxiliary function $\xi = \xi(t,z)$.
\smallskip
{\bf Proposition 3.1.} {\it WDVV is equivalent to compatibility of the system
of equations (3.5) together with the equation
$$z\partial_z \xi_\alpha = z E^\gamma(t) c_{\gamma\alpha}^\beta(t)
\xi_\beta + Q_\alpha^\gamma \xi_\gamma
\eqno(3.11)$$
where $Q_\alpha^\gamma = \nabla_\alpha E^\gamma$.}

Proof. Due to (1.50b) the system (3.5) is invariant w.r.t. the group of
rescalings (1.43) together with the transformations $z \mapsto kz$. So
the system (3.5) for the covector $\xi$ is compatible with the equation
$$x\partial_z \xi = \LE \xi.
\eqno(3.12)$$
On the solutions of (3.5) the equation (3.12) can be rewritten in the form
(3.11). Proposition is proved.
\medskip
The compatibility of (3.5) and (3.11) can be reformulated [85] as vanishing
of the curvature of the connection on $M\times CP^1$ (the coordinates
are $(t,z)$) given by the operators (3.5) and (3.11).

The further step of the theory (also being standard [60, 76,
127]) is to parametrize
the solutions of WDVV by the monodromy data of the operator (3.11) with
rational coefficients. We are not able to do this in general. The problem
is to {\it define} the monodromy of the operator. The main problem
is to choose a trvialization in the space of solutions of the
equation (3.11) for big $z$. This problem looks not to be purely
technical: one can see that WDVV does not satisfy the Painlev\'e
property of absence of movable critical points in the $t$-coordinates
(see, e.g., the discussion of the analytic properties of solutions
of Chazy equation in Appendix C above). Our main strategy will be to
find an appropriate coordinate system on $M$ and to do a gauge transform
of the operator (3.11)
providing applicability of the isomonodromy deformations technique to
WDVV. This can be done under certain semi-simplicity assumptions
imposed onto $M$.

\medskip
Another important playing character is a new metric on a
Frobenius manifold. It is convenient to define it as a
metric on the cotangent bundle $T^*M$
i.e. as an inner product of 1-forms. For two 1-forms
$\omega_1$ and $\omega_2$ we put
$$(\omega_1,\omega_2)^* := i_E(\omega_1\cdot\omega_2)
\eqno(3.13)$$
(I label the metric by $^*$ to stress that this is an inner
product on $T^*M$).
Here $i_E$ is the operator of contraction of a 1-form with
the vector field $E$; we multiply two 1-forms using
the operation of multiplication of tangent vectors on the
Frobenius manifold and the duality between tangent and
cotangent spaces established by the invariant inner product.
\smallskip
{\bf Exercise 3.3.} Prove that the inner product $(u,v)$ of
two vector fields w.r.t. the new metric isrelated to the old inner
product $<u,v>$ by the equation
$$(E\cdot u,v) = <u,v>.
\eqno(3.14)$$
Thus the new metric on the tangent bundle is welldefined
in the points $t$ of $M$ where $E(t)$ is an invertible element
of the algebra $T_tM$.
\medskip
In the flat coordinates $t^\alpha$ the metric $(~,~)^*$
has the components
$$g^{\alpha\beta}(t) := (dt^\alpha,dt^\beta)^* =
E^\epsilon (t) c_\epsilon
^{\alpha\beta}(t)
\eqno(3.15)$$
where
$$c_\epsilon^{\alpha\beta} (t)
:= \eta^{\alpha\sigma}c_{\sigma\epsilon}^\beta (t).
\eqno(3.16)$$
If the degree operator is diagonalizable
then
$$g^{\alpha\beta}(t) = (d+1 -q_\alpha -q_\beta )
F^{\alpha\beta}(t) + A^{\alpha\beta}
\eqno(3.17)$$
where
$$F^{\alpha\beta}(t) := \eta^{\alpha\lambda} \eta^{\beta\mu}
{\partial^2 F(t)\over \partial t^\lambda \partial t^\mu}
\eqno(3.18)$$
(I recall that we normalise the degrees $d_\alpha$ in such a way
that $d_1 = 1$) and the matrix $A_{\alpha\beta}= \eta_{\alpha\alpha'}\eta
_{\beta\beta'}A^{\alpha'\beta'}$ is defined in (1.9).
\smallskip
{\bf Lemma 3.2.} {\it The metric (3.13) does not
degenerate identically near the $t^1$-axis for
sufficiently small $t^1\neq 0$.}

Proof. We have
$$c_1^{\alpha\beta} (t) \equiv \eta^{\alpha\beta}.
$$
So for small $t^2$, ..., $t^n$
$$g^{\alpha\beta}(t) \simeq t^1 c_1^{\alpha\beta}+A^{\alpha\beta} = t^1
\eta^{\alpha\beta}+A^{\alpha\beta}.
$$
This cannot be degenerate identically in $t^1$.
Lemma is proved.
\medskip
It turns out that the new metric also is flat. In fact I will prove
a more strong statement, that any linear combination
of the metrics $(~,~)^*$ and $<~,~>^*$ is a flat metric (everywhere
when being nondegenerate). To formulate the precise statement
I recall some formulae of Riemannian geometry.
\smallskip

Let $(~,~)^*$ be a symmetric nondegenerate
bilinear form on the cotangent bundle $T^{*} M$
to a manifold $M$. In a local coordinate system $x^1$, ...,
$x^n$ the metric is given by its components
$$g^{ij} (x) := (dx^i,dx^j)^{*}
\eqno(3.19)$$
where $(g^{ij})$ is an invertible symmetric matrix. The inverse
matrix $(g_{ij}) := (g^{ij})^{-1}$ specifies a
metric on the manifold i.e. a
nondegenerate inner product on the tangent bundle
$TM$
$$(\partial_i,\partial_j) := g_{ij}(x)
\eqno(3.20)$$
$$\partial_i := {\partial \over \partial x^i}.
\eqno(3.21)$$
The {\it Levi-Civit\`a connection} $\nabla_k$ for the metric is uniquely
specified by the conditions
$$\nabla_kg_{ij}: = \partial_k g_{ij} -\Gamma_{ki}^sg_{sj}
-\Gamma_{kj}^sg_{is} = 0
\eqno(3.22a)$$
or, equivalently,
$$\nabla_kg^{ij} := \partial_k g^{ij} +\Gamma_{ks}^i g^{sj}
+\Gamma_{ks}^j g^{is} =0
\eqno(3.22b)$$
and
$$\Gamma_{ij}^k = \Gamma_{ji}^k.
\eqno(3.23)$$
(I recall that summation over twice repeated
indices here and below is assumed.
We will keep the symbol of summation over more than twice repeated
indices.)
Here the coefficients $\Gamma_{ij}^k$ of the connection (the
Christoffel symbols) can be expressed via the metric and its
derivatives as
$$\Gamma_{ij}^k = {1\over 2}g^{ks}\left( \partial_ig_{sj}
+\partial_j g_{is} - \partial_s g_{ij}\right) .
\eqno(3.24)$$
For us it will be more convenient to work with the {\it contravariant
components} of the connection
$$\Gamma^{ij}_k := (dx^i, \nabla_k dx^j)^{*} = -g^{is}\Gamma_{sk}^j.
\eqno(3.25)$$
The equations (3.22) and (3.23) for the contravariant components read
$$\partial_k g^{ij} = \Gamma^{ij}_k +\Gamma^{ji}_k
\eqno(3.26)$$
$$g^{is}\Gamma_s^{jk} = g^{js}\Gamma_s^{ik}.
\eqno(3.27)$$
It is also convenient to introduce operators
$$\nabla^i = g^{is}\nabla_s
\eqno(3.28a)$$
$$\nabla^i\xi_k = g^{is}\partial_s\xi_k + \Gamma^{is}_k\xi_s.
\eqno(3.28b)$$
For brevity we will call the operators $\nabla^i$ and the correspondent
coefficients $\Gamma_k^{ij}$ {\it contravariant connection}.

The {\it curvature tensor} $R_{slt}^k$ of the metric measures noncommutativity
of the operators $\nabla_i$ or, equivalently $\nabla^i$
$$(\nabla_s\nabla_l -\nabla_l\nabla_s)\xi_t = -R_{slt}^k\xi_k
\eqno(3.29a)$$
where
$$R_{slt}^k = \partial_s\Gamma_{lt}^k -\partial_l\Gamma_{st}^k
+\Gamma_{sr}^k \Gamma_{lt}^r -\Gamma_{lr}^k \Gamma_{st}^r.
\eqno(3.29b)$$
We say that the metric is {\it flat} if the curvature of it vanishes.
For a flat metric local {\it flat coordinates} $p^1$, ..., $p^n$ exist
such that in these coordinates the metric is constant and the components
of the Levi-Civit\`a connection vanish. Conversely, if a system of flat
coordinates for a metric exists then the metric is flat. The flat coordinates
are determined uniquely up to an affine  transformation with constant
coefficients. They
can be found from the following system
$$\nabla^i\delj p
 = g^{is}\partial_s\partial_jp + \Gamma^{is}_j\partial_sp = 0,
{}~i,j = 1,...,n.\eqno(3.30)$$
If we choose the flat coordinates orthonormalized
$$(dp^a,dp^b)^{*} =\delta^{ab}
\eqno(3.31)$$
then for the components of the metric and of the Levi-Civit\`a connection
the following formulae hold
$$g^{ij} = {\partial x^i\over \partial p^a}{\partial x^j\over \partial p^a}
\eqno(3.32a)$$
$$\Gamma^{ij}_kdx^k = {\partial x^i\over \partial p^a}
{\partial^2 x^j\over \partial p^a \partial p^b}dp^b.
\eqno(3.32b)$$

All these facts are standard in geometry (see, e.g., [55]). We need
to represent the formula (3.29b) for the curvature tensor in a slightly
modified form (cf. [53, formula (2.18)]).
\medskip
{\bf Lemma 3.3.} {\it For the curvature of a metric the following formula holds
$$R^{ijk}_l := g^{is}g^{jt} R_{slt}^k =
g^{is}\left( \partial_s\Gamma_l^{jk} - \partial_l\Gamma_s^{jk}\right)
+\Gamma_s^{ij}\Gamma_l^{sk} - \Gamma_s^{ik}\Gamma_l^{sj}.
\eqno(3.33)$$
}

Proof. Multiplying the formula (3.29b) by  $g^{is}g^{jt}$ and using (3.25) and
(3.26) we obtain (3.33). The lemma is proved.
\medskip
Let us consider now a manifold supplied with two nonproportional metrics
$(~,~)^{*}_1$ and $(~,~)^{*}_2$. In a coordinate system they are given by
their components $g^{ij}_1$ and $g^{ij}_2$ resp. I will denote by
$\Gamma_{1k}^{ij}$ and $\Gamma_{2k}^{ij}$ the correspondent Levi-Civit\`a
connections $\nabla_1^i$ and $\nabla_2^i$. Note that the difference
$$\Delta^{ijk} =
g_2^{is} \Gamma_{1s}^{jk} - g_1^{is} \Gamma_{2s}^{jk}
\eqno (3.34)$$
is a tensor on the manifold.
\medskip
{\bf Definition 3.1.} We say that the two metrics form a {\it flat pencil}
if:

1. The metric
$$g^{ij} = g_1^{ij} + \lambda g_2^{ij}
\eqno(3.35a)$$
is flat for arbitrary $\lambda$ and

2. The Levi-Civit\`a connection for the metric (3.35a) has the form
$$\Gamma_k^{ij} = \Gamma_{1k}^{ij} + \lambda \Gamma_{2k}^{ij}.
\eqno(3.35b)$$
\medskip
I will describe in more details the conditions for two metrics
to form a flat pencil in Appendix D below (it turns out that these conditions
are very close to the axioms of Frobenius manifolds).

Let us consider the metrics $(~,~)^*$ and $<~,~>^*$ on a Frobenius
manifold $M$
(the second metric
is induced on $T^*M$ by the invariant metric $<~,~>$). We will assume
further that the Euler vector field $E$ is linear (may be, linear
nonhomogeneous) in the flat coordinates.
\smallskip
{\bf Proposition 3.2.} {\it The metrics $(~,~)^*$ and $<~,~>^*$ on a
Frobenius manifold form a flat pencil.}
\smallskip
{\bf Lemma 3.4.} {\it In the flat coordinates the contravariant components
of the Levi-Civit\`a connection for the metric $(~,~)^*$ have the form
$$\Gamma_\gamma^{\alpha\beta} = (1+ d_\beta -{d_F\over 2})
c_\gamma^{\alpha\beta}.
\eqno(3.36)$$}

Proof. Substituting (3.36) to (3.26), (3.27) and (3.33)
we obtain identities. Lemma
is proved.
\smallskip
Proof of proposition. We repeat the calculation of the lemma for the same
connection and for the metric $g^{\alpha\beta} + \lambda\eta^{\alpha\beta}$.
The equations (3.26) and (3.27) hold true identically in $\lambda$. Now
substitute
the connection into the formula (3.33) for the curvature of the metric
$g^{\alpha\beta} + \lambda\eta^{\alpha\beta}$. We again obtain identity.
Proposition is proved.
\medskip
{\bf Definition 3.2.} The metric $(~,~)^*$ of the form (3.13) will be called
{\it intersection form of the Frobenius manifold}.
\medskip
We borrow this name from the singularity theory [5, 6]. The motivation
becomes clear from the consideration of the Example 1.7.
In this example the Frobenius manifold coincides with the universal
unfolding of the simple singularity of the $A_n$-type [5]. The metric (3.13)
for the example coincides with the intersection form of
(even-dimensional) vanishing cycles
of the singularity [3, 6, 66] as an inner product on the cotangent
bundle to the universal unfolding space (we identify [6] the tangent
bundle to the universal unfolding with the middle homology fibering
using the differential of the period mapping). It turns out that
the intersection form of odd-dimensional vanishing cycles (assuming
that the base of the bundle is even-dimensional) coincides
with the skew-symmetric form $<\hat V\, .\, ,\, .\, >^*$ where the operator
$\hat V$ was defined in (1.51). This can be derived from the
results of Givental [65] (see below (3.46)).
\smallskip
{\bf Example 3.1.} For the trivial Frobenius manifold corresponding
to a graded Frobenius algebra $A = \{ c_{\alpha\beta}^\gamma, ~
\eta_{\alpha\beta}, ~q_\alpha\} , ~d$ (see Example 1 of
Lecture 1) the intersection form is
a linear metric on the dual space $A^*$
$$g^{\alpha\beta} = \sum (1-q_\epsilon)t^\epsilon c_\epsilon^{
\alpha\beta},
\eqno(3.37)$$
for
$$c_\epsilon^{\alpha\beta} = \eta^{\alpha\sigma}c_{\sigma\epsilon}
^\beta ,~~ (\eta^{\alpha\beta}) := (\eta_{\alpha\beta})^{-1}.
$$
{}From the above considerations it follows that the Christoffel
coefficients for this flat metric are
$$\Gamma_\gamma^{\alpha\beta} = \left( {d+1\over 2} - q_\beta
\right) c_\gamma^{\alpha\beta}.
\eqno(3.38)$$
Flat linear metrics with constant Christoffel coefficients
were first studied by S.Novikov and A.Balinsky [12] due
to
their close relations to vector analogues of the Virasoro
algebra. We will come back to this example in Lecture 6
(see also Appendix G below).
\medskip
{\bf Remark 3.2.} Knowing the intersection form of a Frobenius manifold
and the Euler and the unity vector fields $E$ and $e$ resp. we can
uniquely reconstruct the Frobenius structure if $d_\alpha + d_\beta
-d_F + 2 \neq 0$ for any $1\leq \alpha,\, \beta \leq n$. Indeed,
we can put
$$<~,~>^*:= {\cal L}_e (~,~)^*
\eqno(3.39)$$
(the Lie derivative along $e$). Then we can choose the coordinates
$t^\alpha$ taking the flat coordinates for the metric $<~,~>^*$
and choosing them homogeneous for $E$.
Putting
$$g^{\alpha\beta}: = (dt^\alpha, dt^\beta )^*
\eqno(3.40)$$
$$\deg g^{\alpha\beta} := {{\cal L}_E g^{\alpha\beta} \over
g^{\alpha\beta} }
\eqno(3.41)$$
we can find the function $F$ from the
equations
$$g^{\alpha\beta} = \deg g^{\alpha\beta}F^{\alpha\beta}
\eqno(3.42)$$
($F^{\alpha\beta}$ are the contravariant components of the Hessian of
$F$, see formula (3.17)). This observation will be very important
in the constructions of the next lecture.
\medskip
{\bf Exercise 3.4.} Let $\omega$ will be the 1-form on a Frobenius
manifold defined by
$$\omega (~.~) = <e, ~.~>.
\eqno(3.43)$$
Show that the formula
$$\left\{ x^k, x^l\right\} : = \half {\cal L}_e
\left[ \left( dx^l, dx^i\right) \deli \left( dx^k, \omega\right)
- \left( dx^k, dx^i\right) \deli \left( dx^l, \omega\right)\right]
\eqno(3.44)$$
defines a Poisson bracket on the Frobenius manifold.
[Hint: prove that the bracket is constant in the flat coordinates
for the metric $<~,~>$,
$$\left\{ t^\alpha, t^\beta\right\} = -\half (q_\alpha-q_\beta)
\eta^{\alpha\beta}.]
\eqno(3.45)$$
Observe that the tensor of the Poisson bracket has the form
$$\{ \, .\, , \, .\,\} = -\half <\hat V . \, , \, . \, >^*.
\eqno(3.46)$$

For the case of the Frobenius manifold of Example 1.7
the Poisson structure coincides with the skew-symmetric
intersection form on the universal unfolding of the $A_n$ singularity
(see [65]). The formula (3.44) for this case was obtained by Givental
[65, Corollary 3].

\medskip
We add now the following {\it assumption of semisimplicity} on
the Frobenius manifold $M$. We say that a point $t\in M$ is
{\it semisimple} if the Frobenius algebra $T_tM$ is semisimple
(i.e. it has no nilpotents). It is clear that semisimplicity
is an open property of the point.
The assumption of semisimplicity for a Frobenius manifold $M$
means that a generic point of $M$ is semisimple. In physical
context this corresponds to massive perturbations of TCFT [27].
So we will also call $M$ satisfying the semisimplicity assumption
{\it massive} Frobenius manifold.
\smallskip
{\bf Main lemma.} {\it In a neighborhood of a semisimple point
local coordinates $u^1$, ..., $u^n$ exist such that
$$\deli\cdot\delj = \delta_{ij}\, \deli ,~~\deli = {\partial
\over \partial u^i}.
\eqno(3.47)$$}

Proof. In a neighborhood of a semisimple point $t$ {\it vector fields}
$\partial_1$, ..., $\partial_n$ exist such that
$\deli\cdot\delj = \delta_{ij}\, \deli $ (idempotents of the algebra
$T_tM$). We need to prove that these vector fields commute pairwise.
Let
$$[\deli , \delj ] =: f_{ij}^k \, \delk .
\eqno(3.48)$$
We rewrite the condition of flatness of the deformed connection
$\tilde\nabla (z)$ in the basis $\partial_1$, ..., $\partial_n$.
I recall that the curvature operator for a connection $\nabla$ is defined
by
$$R(X,Y)Z := [\nabla_X,\nabla_Y]Z - \nabla_{[X,Y]}Z.
\eqno(3.49)$$
We define the coefficients of the Euclidean connection on the
Frobenius manifold in the basis $\partial_1$, ..., $\partial_n$
by the formula (see [55], sect. 30.1)
$$\nabla_{\deli}\delj =: \Gamma_{ij}^k\delk .
\eqno(3.50)$$
Vanishing of the curvature of $\tilde\nabla (z)$ in the terms linear
in $z$ reads
$$\Gamma_{kj}^l \delta_i^l + \Gamma_{ki}^l \delta_{kj}
-\Gamma_{ki}^l \delta_j^l - \Gamma_{kj}^l \delta_{ki}
= f_{ij}^l \delta_k^l
\eqno(3.51)$$
(no summation over the repeated indices in this formula!).
For $l=k$ this gives $f_{ij}^k = 0$. Lemma is proved.
\medskip
{\bf Remark 3.3.} The main lemma can be reformulated in terms of
 {\it algebraic
symmetries} of a  massive Frobenius manifold. We say that
a diffeomorphism $f:M\rightarrow M$ of a Frobenius
manifold is algebraic symmetry if it preserves the multiplication
law of vector fields:
$$f_*(u\cdot v)=f_*(u)\cdot f_*(v)
\eqno(3.52)$$
(here $f_*$ is the induced linear map $f_*:T_xM\rightarrow T_{f(x)}M$).

It is easy to see that algebraic symmetries of a Frobenius manifold
form a finite-dimen\-si\-o\-nal Lie group $G(M)$.
The generators of action of $G(M)$ on $M$ (i.e. the representation of
the Lie algebra of $G(M)$ in the Lie algebra of vector fields on
$M$) are the vector fields $w$ such that
$$[w,u\cdot v]=[w,u]\cdot v+[w,v]\cdot u\eqno(3.53)$$
for any vector fields $u$, $v$.

Note that the group $G(M)$ always is nontrivial: it contains the one-parameter
subgroup of shifts along the coordinate $t^1$. The generator of this subgroup
coincides with the unity vector field $e$.
 \smallskip
{\bf Main lemma$'$.} {\it The connect component of the identity
in the group $G(M)$
of algebraic symmetries of a $n$-dimensional massive Frobenius manifold is
a $n$-dimensional commutative Lie group that acts localy transitively on $M$.}
\medskip
I will call the local coordinates $u^1$, ..., $u^n$ on a massive Frobenius
manifold {\it canonical coordinates}. They can be found as independent
solutions
of the system of PDE
$$\dgamma u \, c_{\alpha\beta}^\gamma (t) = \dalpha u \, \dbeta u
\eqno(3.54)$$
or, equivalently, the 1-form $du$ must be a homomorphism of the
algebras
$$du : T_tM \to \Cc .
\eqno(3.55)$$
The canonical coordinates are determined uniquely up to shifts and
permutations.

We solve now explicitly this system of PDE.
\smallskip
{\bf Proposition 3.3.} {\it In a neighborhood of a semisimple point
all the roots $u^1(t)$, ..., $u^n(t)$ of the characteristic equation
$$\det (g^{\alpha\beta}(t) - u\eta^{\alpha\beta}) = 0
\eqno(3.56)$$
are simple. They are canonical coordinates in this neighborhood.
Conversely, if the roots of the characteristic equation are simple
in a point $t$ then $t$ is a semisimple point on the Frobenius manifold
and $u^1(t)$, ..., $u^n(t)$ are canonical coordinates in the neighbourhood
of the point.}
\smallskip
{\bf Lemma 3.5.} {\it Canonical coordinates in a neighborhood
of a semisimple point can be chosen in such a way that
the Euler vector field $E$ have the form
$$E = \sum_i u^i\deli .
\eqno(3.57)$$}

Proof. Rescalings generated by $E$ act on the idempotents $\deli$
as $\deli \mapsto k^{-1}\deli$. So an appropriate shift of
$u^i$ provides $u^i\mapsto ku^i$. Lemma is proved.
\smallskip
{\bf Lemma 3.6.} {\it The invariant inner product $<~,~>$ is diagonal in
the canonical coordinates
$$<\deli , \delj > = \eta_{ii}(u) \, \delta_{ij}
\eqno(3.58)$$
for some nonzero functions $\eta_{11}(u)$, ..., $\eta_{nn}(u)$.
The unity vector field $e$ in the canonical coordinates has the form
$$e = \sum_i \deli .
\eqno(3.59)$$
}

The proof is obvious (cf. (1.41), (1.42)).

Proof of proposition. In the canonical coordinates of Main Lemma
we have
$$du^i\cdot du^j = \eta_{ii}^{-1} du^i\, \delta_{ij}.
\eqno(3.60)$$
So the intersection form reads
$$g^{ij}(u) = u^i \eta_{ii}^{-1}\delta_{ij}.
\eqno(3.61)$$
The characteristic equation (3.56) reads
$$\prod_i (u-u^i) = 0.
$$
This proves the first part of the proposition.

To prove the second part we consider the linear operators $U=(U_\beta^\alpha
(t))$ on $T_tM$ where
$$U_\beta^\alpha(t) := g^{\alpha\epsilon}(t)\eta_{\epsilon\beta}.
\eqno(3.62)$$
{}From (3.14) it folows that $U$ is the operator of multiplication by the
Euler vector field $E$. The characteristic equation for this operator
coincides with (3.56). So under the assumptions of the proposition
the operator of multiplication by $E$ in the point $t$ is a semisimple one.
This implies the semisimplicity of all the algebra $T_tM$ because of the
commutativity of the algebra.

Proposition is proved.
\medskip
Using canonical coordinates we reduce the problem of local classification
of massive Frobenius manifolds to an integrable system of ODE. To obtain
such a system we scrutinize the properties of the invariant metric in the
canonical coordinates. I recall that this metric has diagonal form in the
canonical coordinates. In other words, $u^1$, ..., $u^n$ are curvilinear
orthogonal coordinates in the (locally) Euclidean space with the (complex)
Euclidean metric $<~,~>$. The familiar object in the geometry of curvilinear
orthogonal coordinates is the {\it rotation coefficients}
$$\gamma_{ij}(u) := {\partial_j \sqrt{\eta_{ii}(u)}\over \sqrt{\eta_{jj}(u)}},
{}~~i\neq j
\eqno(3.63)$$
(locally we can fix some branches of $\sqrt{\eta_{ii}(u)}$).
They determine the law of rotation with transport
along the $u$-axes
of the natural orthonormal frame
related to the orthogonal system of coordinates (see [33]).
\smallskip
{\bf Lemma 3.7.} {\it The coefficients $\eta_{ii}(u)$ of the invariant metric
have the form
$$\eta_{ii}(u) = \deli t_1(u),~~i=1,\dots, n.
\eqno(3.64)$$}

Proof. According to (1.39) the invariant inner product $<~,~>$ has
the form
$$<a,b> = <e,a\cdot b> \equiv \omega(a\cdot b)
\eqno(3.65)$$
for any two vector fields $a$, $b$ where the 1-form $\omega$ is
$$\omega(\,.\,) := <e,.>.
\eqno(3.66)$$
Hence $\omega = dt_1$. Lemma is proved.
\medskip
We summarize the properties of the invariant metric in the canonical
coordinates in the following
\smallskip
{\bf Proposition 3.4.} {\it The rotation coefficients (3.63) of the invariant
metric
are symmetric
$$\gamma_{ij}(u) = \gamma_{ji}(u).
\eqno(3.67)$$
The metric is invariant w.r.t. the diagonal translations
$$\sum_k \delk \eta_{ii}(u) = 0,~~i=1,\dots, n.
\eqno(3.68)$$
The functions $\eta_{ii}(u)$ and $\gamma_{ij}(u)$ are homogeneous
functions of the canonical coordinates of the degrees $-d$
and $-1$ resp.}

Proof. The symmetry (3.67) follows from (3.64):
$$\gamma_{ij}(u) = \half{\deli\delj t_1(u)\over \sqrt{\deli t_1(u)
\delj t_1(u)}}.
\eqno(3.69)$$
To prove (3.68) we use (3.64) and covariant constancy of the vector field
$$e= \sum_i\deli .
$$
This reads
$$\sum_{k=1}^n \Gamma_{ik}^j = 0.
$$
For $i=j$ using the Christoffel formulae (3.24) we obtain (3.68).
The homogeneity follows from (1.50). Proposition is proved.
\medskip
{\bf Corollary 3.1.} {\it The rotation coefficients (3.63) satisfy the
following system of equations
$$\delk \gamma_{ij} = \gamma_{ik}\gamma_{kj},~~i,j,k~
{\rm are~distinct}
\eqno(3.70a)$$
$$\sum_{k=1}^n \delk\gamma_{ij} = 0
\eqno(3.70b)$$
$$\sum_{k=1}^n u^k\delk \gamma_{ij} = -\gamma_{ij}.
\eqno(3.70c)$$}

Proof. The equations (3.70a) and (3.70b) coincide with the
equations of flatness of the diagonal metric
obtained for the metrics of the form (3.64) by Darboux
and Egoroff [33]. The equation (3.70c) follows from homogeneity.
\medskip
We have shown that any massive Frobenius manifold determines
a scaling invariant (3.70c)
solution of the {\it Darboux - Egoroff system} (3.70a,b). We show now
that, conversely, any solution of the system (3.70)
under some genericity assumptions determines
locally a massive Frobenius manifold.

Let
$$\Gamma(u) := (\gamma_{ij}(u))$$
be a solution of (3.70).
\smallskip
{\bf Lemma 3.8.} {\it The linear system
$$\delk \psi_i = \gamma_{ik}\psi_k,~~i\neq k
\eqno(3.71a)$$
$$\sum_{k=1}^n \delk \psi_i = 0,~~i=1,\dots, n
\eqno(3.71b)$$
for an auxiliary vector-function $\psi = (\psi_1(u),\dots, \psi_n(u))^T$
has $n$-dimensional space of solutions.}

Proof. Compatibility of the system (3.71) follows from the Darboux - Egoroff
system (3.70). Lemma is proved.
\medskip
Let us show that, under certain genericity assumptions a basis of
homogeneous in $u$ solutions of (3.71) can be chosen. We introduce the
$n\times n$-matrix
$$V (u) := [\Gamma(u),U]
\eqno(3.72)$$
where
$$U := {\rm diag}(u^1,\dots, u^n),
\eqno(3.73)$$
$[~,~]$ stands for matrix commutator.
\smallskip
{\bf Lemma 3.9.} {\it The matrix $V(u)$ satisfies the following system
of equations
$$\delk V(u) = [V(u),[E_k,\Gamma ]],~ k=1, \dots, n
\eqno(3.74)$$
where $E_k$ are the matrix unities
$$(E_k)_{ij} = \delta_{ik}\delta_{kj}.
\eqno(3.75)$$
Conversely, all the differential equations (3.70) follow from (3.74).}

Proof. From (3.70) we obtain
$$\deli\gamma_{ij} ={1\over u^i-u^j}\left(\sum_{k\neq i,j}
(u^j-u^k)\gamma_{ik}\gamma_{kj} -\gamma_{ij}\right) .
\eqno(3.76)$$
The equation (3.74) follows from (3.70a) and (3.76). Lemma is proved.
\smallskip
{\bf Corollary 3.2.}
{\it \item{1).} The matrix $V(u)$
acts on the space of solutions
of the linear system (3.71). \item{2).} Eigenvalues of $V(u)$ do not depend
on $u$. \item{3).} A solution $\psi(u)$ of the system (3.71) is
a homogeneous function of $u$
$$\psi(cu) = c^\mu\psi(u)
\eqno(3.77)$$
{\rm iff} $\psi(u)$ is an eigenvector of the matrix $V (u)$
$$V(u)\psi(u) = \mu\psi(u).
\eqno(3.78)$$}

Proof. We rewrite first the linear system (3.71) in the matrix
form. This reads
$$\delk \psi = -[E_k , \Gamma ]\psi.
\eqno(3.79)$$
{}From (3.74) and (3.79) it follows immediately the first statement of the
lemma. Indeed, if $\psi$ is a solution of (3.79) then
$$\delk (V\psi) =(V [E_k,\Gamma ] - [E_k,\Gamma ]V )\psi
- V [E_k,\Gamma ]\psi = -[E_k,\Gamma ]V\psi .
$$
The second statement of the lemma also follows from (3.74).
The third statement is obvious since
$$\sum u^i\deli\psi = V\psi
$$
(this follows from (3.79)). Lemma is proved.
\medskip
{\bf Remark 3.4.} We will show below that a spectral parameter can be
inserted in the linear system (3.71). This will give a way
to integrate the system (3.70) using the isomonodromy deformations
technique.
\medskip
We denote by $\mu_1$, ..., $\mu_n$ the eigenvalues of the matrix
$V (u)$. Due to skew-symmetry of $V(u)$
these can be ordered in such a way
that
$$\mu_\alpha + \mu_{n-\alpha +1} = 0.
\eqno(3.80)$$
\smallskip
{\bf Proposition 3.5.} {\it For a massive Frobenius manifold corresponding
to a scaling invariant
solution of WDVV the matrix $V (u)$ is diagonalizable.
Its eigenvectors $\psi_\alpha$ $ = (\psi_{1\alpha}(u),\dots,
\psi_{n\alpha}(u))^T$
are
$$\psi_{i\alpha}(u) = {\deli t_\alpha (u)\over \sqrt{\eta_{ii}(u)}},
{}~~i,\, \alpha = 1,\dots, n.
\eqno(3.81)$$
The correspondent eigenvalues are
$$\mu_\alpha = q_\alpha - {d\over 2}
\eqno(3.82)$$
(the spectrum of the Frobenius manifold in the sense of Appendix C).
Conversely, let $V(u)$ be any diagonalizable solution of the system (3.74)
and $\psi_\alpha=(\psi_{i\alpha}(u))$ be the solutions of (3.71)
satisfying
$$V\psi_\alpha = \mu_\alpha \psi_\alpha.
\eqno(3.83)$$
Then the formulae
$$\eta_{\alpha\beta}= \sum_i \psi_{i\alpha}\psi_{i\beta}
\eqno(3.84a)$$
$$\deli t_\alpha = \psi_{i1}\psi_{i\alpha}
\eqno(3.84b)$$
$$c_{\alpha\beta\gamma} = \sum_i{\psi_{i\alpha}\psi_{i\beta}
\psi_{\gamma}\over \psi_{i1}}
\eqno(3.84c)$$
determine locally a massive Frobenius manifold with the scalingh dimensions
$$q_\alpha = \mu_\alpha -2 \mu_1, ~~d = -2\mu_1.
\eqno(3.84d)$$}

Proof is straightforward.

Note that we obtain a Frobenius manifold of the second type (1.22) if the
marked vector $\psi_{i1}$ belongs to the kernel of $V$.
\smallskip
{\bf Remark 3.5.} The construction of Proposition works also for Frobenius
manifolds with nondiagonalizable matrices $\nabla_\alpha E^\beta$. They
correspond to nondiagonalizable matrices $V(u)$. The reason of appearing
of linear nonhomogeneous terms in the Euler vector field (when some
of $q_\alpha$ is equal to 1) is more subtle. We will discuss it in terms
of monodromy data below.
\smallskip
{\bf Remark 3.6.} The change of the coordinates $(u^1, \dots, u^n)
\mapsto (t^1, \dots , t^n)$ is not invertible in the points where one of the
components
of the vector-function $\psi_{i1}(u)$ vanishes.
\medskip
{\bf Exercise 3.5.} Prove the formula
$$V_{ij}(u) = \sum_{\alpha,\,\beta}\eta^{\alpha\beta}
\mu_\alpha\psi_{i\alpha}(u)\psi_{j\beta}(u)
\eqno(3.85)$$
for the matrix $V(u)$ where $\psi_{i\alpha}(u)$ are given by the formula
(3.81).
\medskip
{\bf Remark 3.7.} From the construction it follows that a solution
$\gamma_{ij}(u)$ of the Darboux - Egoroff system (3.70) determines
$n$ essentially different (up to an equivalence)
solutions of WDVV. This comes from the freedom in the
choice of the solution $\psi_{i1}$ in the formulae (3.84). We will see now that
these ambiguity is described by the transformations (B.2) (or (B.11),
in the case
of coincidences between the eigenvalues of $V$).
\smallskip
{\bf Definition 3.3.} A 1-form $\sigma$ on a massive Frobenius manifold $M$ is
called {\it admissible} if the new invariant metric
$$<a,b>_\sigma := \sigma(a\cdot b)
\eqno(3.86)$$
together with the old multiplication law of tangent vectors and
with the old unity $e$ and the old Euler vector field $E$
determines on $M$ a structure of Frobenius manifold with the same
rotation coefficients $\gamma_{ij}(u)$.
\medskip
For example, the 1-form
$$\sigma = dt_1
$$
is an admissible one: it determines on $M$ the given Frobenius
structure.
\smallskip
{\bf Proposition 3.6.} {\it All the admissible forms on a massive Frobenius
manifold are
$$\sigma_c (~\cdot ~) := \big\langle \left(\sum_kc^k\partial_{\kappa_k}
\right)^2, ~\cdot ~\big\rangle
\eqno(3.87a)$$
for arbitary constants $c^k$ and
$$\deg t^{\kappa_1} = \deg t^{\kappa_2} = \dots .
\eqno(3.87b)$$
}

The form $\sigma_c$ can be written also as follows
$$\sigma_c = \sum_{i, j} c^ic^j F_{\kappa_i\kappa_j\alpha}dt^\alpha.
\eqno(3.87c)$$

Proof. Flat coordinates $t^{\alpha'}$ for a Egoroff metric
$$<~,~>' = \sum_i \eta_{ii}'(u) du^i
\eqno(3.88)$$
with the given rotation coefficients $\gamma_{ij}(u)$ are determined by
the system
$$\eqalign{\deli \psi'_{j\alpha} &= \gamma_{ij}(u) \psi'_{i\alpha}, ~~i\neq j
\cr
\sum_{i=1}^n \deli \psi'_{j\alpha} &= 0 \cr
\deli t'_\alpha &= \sqrt{\eta'_{ii}(u)} \psi'_{i\alpha}.\cr
}
\eqno(3.89)$$
Particularly,
$$\psi'_{i1}(u) = \sqrt{\eta'_{ii}(u)}.
$$
Also $\psi'_{i\alpha}(u)$ must be homogeneous functions of $u$.
{}From (3.89) we conclude, as in Corollary 3.2
that they must be eigenvectors
of the matrix $V(u)$. So we must have
$$\sqrt{\eta'_{ii}(u)} = \sum_k c^k \psi_{i\,\kappa_k}(u).
$$
This gives (3.87). Reversing the calculations we obtain that the
metric (3.87) is admissible. Proposition is proved.
\medskip
{}From Propositions 3.5 and 3.6 we obtain
\smallskip
{\bf Corollary 3.3.} {\it There exists a one-to-one correspondence
$$\left\{\matrix{{\rm Massive ~Frobenius ~manifolds} \cr
{\rm modulo ~ transformations ~ (B.11)}\cr}\right\}
\leftrightarrow \left\{\matrix{{\rm solutions ~ of ~the ~system ~(3.70)}
\cr {\rm with ~diagonalizable ~}V(u) \cr}\right\}
$$
}
\medskip

{\bf Remark 3.8.} Solutions of WDVV equations without semisimplicity
assumption depend on functional parameters. Indeed, for nilpotent
algebras the associativity conditions often are empty.
However, it is possible to describe a closure of the class of massive
Frobenius manifolds as the set of all Frobenius manifolds with
$n$-dimensional commutative group of algebraic symmetries. Let $A$ be a fixed
$n$-dimensional Frobenius
algebra with structure constants $c_{ij}^k$ and an invariant
inner nondegenerate inner product $\epsilon = (\epsilon_{ij})$. Let us
introduce matrices
$$C_i = (c_{ij}^k).\eqno(3.91)$$
An analogue of the Darboux -- Egoroff system (3.70) for an operator-valued
function
$$\gamma (u):A\to A,~\gamma =(\gamma_i^j(u)),~u=(u^1,\dots ,u^n)\eqno(3.92)$$
(an analogue of the rotation coefficients) where the operator $\gamma$ is
symmetric with respect to $\epsilon$,
$$\epsilon\gamma = \gamma^{{\rm T}}\epsilon\eqno(3.93)$$
has the form
$$[C_i,\partial_j\gamma ]-[C_j,\partial_i\gamma ]+
[[C_i,\gamma ],[C_j,\gamma ]] = 0,~i,j=1,\dots ,n,\eqno(3.94)$$
$\partial_i=\partial /\partial u^i$. This is an integrable system with the
Lax representation
$$\partial_i\Psi = \Psi (zC_i +[C_i,\gamma ]),~i=1,\dots ,n.\eqno(3.95)$$
It is convenient to consider $\Psi = (\psi_1(u),\dots ,\psi_n(u))$ as a
function with values in the dual space $A^*$. Note that $A^*$ also is a
Frobenius algebra with the structure constants $c^{ij}_k =
c_{ks}^i\epsilon^{sj}$ and the invariant inner product $<~,~>_*$ determined
by $(\epsilon^{ij}) = (\epsilon_{ij})^{-1}$.

Let $\Psi_\alpha (u)$, $\alpha = 1$, ... ,$n$ be a basis of solutions of
(3.95) for $z=0$
$$\partial \Psi_\alpha = \Psi_\alpha [C_i,\gamma ],~
\alpha =1,... ,n\eqno (3.96a)$$
such that the vector $\Psi_1(u)$ is invertible in $A^*$. We put
$$\eta_{\alpha\beta} = <\Psi_\alpha (u),\Psi_\beta (u)>_*\eqno(3.96b)$$
$${\rm grad}_u t_\alpha  = \Psi_\alpha (u)\cdot
\Psi_1 (u)\eqno(3.96c)$$
$$c_{\alpha\beta\gamma}(t(u)) =
{\Psi_\alpha (u)\cdot\Psi_\beta (u)\cdot\Psi_\gamma (u)
\over \Psi_1(u)}.\eqno(3.96d)$$

{\bf Theorem 3.1.} {\it Formulae (3.96) for arbitrary Frobenius
algebra $A$ localy parametrize all Frobenius manifolds with
$n$-dimensional commutative group of algebraic symmetries.}

Considering $u$ as a vector in $A$ and $\Psi_1^2=\Psi_1\cdot\Psi_1$ as
a linear function on $A$ one obtains the following analogue of Egoroff
metrics (on $A$)
$$ds^2 = \Psi_1^2(du\cdot du).\eqno(3.97)$$
\medskip
{\bf Examples.} We start with the simplest example $n=2$. The
equations (3.70) are linear in this case. They can be solved
as
$$\gamma_{12}(u) = \gamma_{21}(u) = {i\mu\over u^1 - u^2}
\eqno(3.98)$$
where $\pm\mu$ are the eigenvalues of the matrix $V(u)$ being
constant in this case. The basis $\psi_{i\alpha}(u)$ of solutions
of the system (3.71) has the form
$$\psi_1 = {1\over \sqrt{2}}\left(\matrix{r^\mu \cr ir^\mu \cr}\right),
{}~~\psi_2 = {1\over \sqrt{2}}\left(\matrix{r^{-\mu} \cr
-ir^{-\mu} \cr}\right),~~ r = u^1 - u^2
\eqno(3.99)$$
(we omit the inessential normalization constant). For $\mu\neq{-1/2}$
the flat coordinates are
$$t^1 = {u^1 +u^2\over 2}, ~~t^2 = {r^{2\mu +1}\over 2(2\mu +1)}.
\eqno(3.100)$$
We have
$$q_1 = 0, ~q_2 = -2\mu =d.
$$
For $\mu \neq \pm 1/2,~-3/2$ the function $F$ has the form
$$F = {1\over 2} (t^1)^2 t^2 + c (t^2)^{k}
$$
for
$$k = (2\mu+3)/(2\mu +1),~~
c = {(1+2\mu)^3\over 2(1-2\mu)(2\mu +3)}[2(2\mu + 1)]^{-4\mu /(2\mu + 1)}.
$$
For $\mu = 1/2$ the function $F$ has the form
$$F = {1\over 2} (t^1)^2 t^2 + {(t^2)^2\over 8}\left(\log{t^2} -{3\over 2}
\right).
$$
For $\mu = -3/2$ we have
$$F = {1\over 2} (t^1)^2 t^2 + {1\over 3\cdot 2^7} \log{t^2}.
$$
For $\mu = -1/2$ the flat coordinates are
$$t^1 = {u^1 +u^2\over 2}, ~~t^2 = {1\over 2} \log{r}.
\eqno(3.101)$$
The function $F$ is
$$F = {1\over 2} (t^1)^2 t^2 + 2^{-6} e^{4 t^2}.
$$
\medskip

For $n\geq 3$ the system (3.70) is non-linear. I will rewrite the first part
of it, i.e. the Darboux - Egoroff equations (3.70a,b) in a more recognizable
(for the theory of integrable systems) shape. Let us restrict the
functions $\gamma_{ij}(u)$ onto the plane
$$u^i = a_ix+b_it,~~i=1,\dots, n
$$
where the vectors $a=(a_1,\dots, a_n)$, $b=(b_1,\dots, b_n)$, and
$(1, 1, \dots, 1)$ are linearly independent. After the substitution we obtain
the following matrix form of the system (3.70a,b)
$$\partial_t [A,\Gamma ] -\partial_x [B,\Gamma ] +\left[ [A,\Gamma ],
[B,\Gamma]\right] = 0
\eqno(3.102a)$$
where
$$A = {\rm diag} (a_1, \dots, a_n),~~ B = {\rm diag} (b_1, \dots, b_n),~~
\Gamma = (\gamma_{ij})
\eqno(3.102b)$$
$[~,~]$ stands for the commutator of $n\times n$ matrices. This is a particular
reduction of the wellknown $n$-wave system [112] (let us forget at the moment
that all the matrices in (3.102) are complex but not real or hermitean or
etc.).
`Reduction' means that the matrices $[A,\Gamma ]$, $[B,\Gamma ]$ involved
in (3.102) are skew-symmetric but not generic. I recall [52] that particularly,
when the $x$-dependence drops down from (3.102), the system reduces to
the equations of free rotations of a $n$-dimensional solid (the so-called
{\it Euler - Arnold top} on the Lie algebra $so(n)$)
$$V_t = [\Omega,V ],
\eqno(3.103a)$$
$$V = [A,\Gamma ],~~\Omega = [B,\Gamma ],
\eqno(3.103b)$$
$$\Omega,~V\in so(n), ~~\Omega =ad_Bad_A^{-1}V.
\eqno(3.104)$$
This is a hamiltonian system on $so(n)$ with the standard Lie - Poisson
bracket [52] and with the quadratic Hamiltonian
$$H = - \half \,{\rm tr}\, \Omega V = \half \sum_{i<j}{b_i-b_j\over
a_i-a_j}V_{ij}^2
\eqno(3.105)$$
(we identify $so(n)$ with the dual space using the Killing foorm).

The equation (3.70c) determines a scaling reduction of the $n$-wave
system (3.102). It turns out that this has still the form of the Euler
-- Arnold top on $so(n)$ but the Hamiltonian depends
explicitly on time.
\smallskip
{\bf Proposition 3.7.} {\it The dependence(3.74)  of the matrix $V(u)$ on
$u^i$ is determined by a hamiltonian system on $so(n)$ with
a time-dependent quadratic Hamiltonian
$$H_i = \half \sum_{j\neq i} {V_{ij}^2\over u^i-u^j},~
\eqno(3.106)$$
$$\deli V = \left\{ V, H_i\right\} \equiv \left[ V, ad_{E_i}ad_U^{-1}
V\right]
\eqno(3.107)$$
for any $i=1, \dots, n$.}

Proof follows from (3.74), (3.103).
\medskip
The variables $u^1, \dots, u^n$ play the role of the times for the pairwise
commuting hamiltonian systems (3.106). From (3.107) one obtains
$$\eqalign{\sum_{i=1}^n \deli V & = 0\cr
\sum_{i=1}^n u^i \deli V & = 0.\cr}
\eqno(3.108)$$
So there are only $n-2$ independent parameters among the
\lq\lq times" $u^1, \dots, u^n$.
\medskip
It turns out that the non-autonomic tops (3.106) are still integrable.
But the integrability of them is more complicated as those for the
equations (3.103): they can be integrated by the method of isomonodromic
deformations.

The systems (3.74) have \lq\lq geometrical integrals"
(i.e. the Casimirs of the Poisson bracket on $so(n)$). These
are the $Ad$-invariant polynomials on $so(n)$. They
are the symmetric combinations
of the eigenvalues of the matrix $V(u)$. I recall that, due
to Corollary 3.2 the values of these geometrical integrals are
expressed via the scaling dimensions of the Frobenius manifold.
\smallskip

{\bf Example 3.2.} $n=3$. Take
the Hamiltonian
$$H: = (u_2 - u_1)H_3 =\half \left[ {\Omega_1^2\over s-1} + {\Omega_2^2\over s}
\right]
\eqno(3.109)$$
where we put
$$\Omega_k(s):= -V_{ij}(u)
\eqno(3.110)$$
$$s = {u^3-u^1\over u^2 - u^1},
\eqno(3.111)$$
and $(i,\,j,\,k)$ is an even permutation of $(1,\,2,\,3)$.
The Poisson bracket on $so(3)$ has the form
$$\{\Omega_1,\Omega_2\} = \Omega_3,~\{\Omega_2,\Omega_3\} = \Omega_1,
{}~\{\Omega_3,\Omega_1\} = \Omega_2.
\eqno(3.112)$$
The
correspondent hamiltonian system reads
$$\eqalign{{d \Omega_1\over ds} &= {1\over s}\Omega_2 \Omega_3 \cr
{d \Omega_2\over ds} &= -{1\over s-1} \Omega_1 \Omega_3 \cr
{d \Omega_3\over ds} &= {1\over s(s-1)} \Omega_1 \Omega_2. \cr}
\eqno(3.113)$$
The system has an obvious first integral (the Casimir of (3.112))
$$\Omega_1^2 + \Omega_2^2 + \Omega_3^2 = - R^2.
\eqno(3.114)$$
The value of the Casimir can be expressed via the scaling dimension
$d$. Indeed, the matrix $\Omega(u)$ (3.110) has the form
$$\Omega (u) = \left( \matrix{0 & \Omega_3 & -\Omega_2 \cr
                    -\Omega_3 & 0 & \Omega_1 \cr
                     \Omega_2 & -\Omega_1 & 0 \cr}
\right).
\eqno(3.115)$$
The eigenvalues of this matrix are $0$ and $\pm R$ where $R$ is defined
in (3.114). From (3.84) we know that the eigenvalues are related to the scaling
dimensions. For $n=3$ the scaling dimensions in the problem can be
expressed via one parameter $d$ as
$$q_1 = 0, ~q_2 = {d\over 2}, ~ q_3 = d.
\eqno(3.116)$$
{}From this we obtain that
$$R =- {d\over 2}
$$
(minus is chosen for convenience).
\smallskip
{\bf Proposition 3.8.} {\it WDVV equations for $n=3$ with the scaling
dimensions
(3.116) are equivalent to the following Painlev\'e-VI equation
$$y'' = {1\over 2}\left( {1\over y} +{1\over y-1}+{1\over
y-z}\right){(y')^2}
-\left({1\over z}+{1\over z-1}+{1\over y-z}\right)y'$$
$$+{y(y-1)(y-z)\over z^2(z-1)^2}\left({(1-d)^2\over 8}
-{(1-d)^2 z\over 8 y^2} +{z(z-1)\over 2(y-z)^2}\right).
\eqno(3.117)$$
}

Proof is given in Appendix E.
\medskip
For $n>3$ the system (3.74) can be considered as a high-order
analogue of the Painlev\'e-VI.
To show this we introduce a commutation representation [43]
of the Darboux - Egoroff system (3.70).
\smallskip
{\bf Lemma 3.10.} {\it The equations (3.70a,b) are equivalent to
the compatibility conditions of the following linear system
of differential equations depending on the
spectral parameter $z$ for an auxiliary function
$\psi = (\psi_1(u,z),\dots, \psi_n(u,z))$
$$\delk\psi_i = \gamma_{ik} \psi_k,~~i\neq k
\eqno(3.118a)$$
$$\sum_{k=1}^n \delk \psi_i = z \psi_i,~~i=1,\dots, n.
\eqno(3.118b)$$}

Proof is in a straightforward calculation.
\medskip
{\bf Exercise.} Show that the formula
$${\deli \tilde t(t(u),z)\over \sqrt{\eta_{ii}(u)}} = \psi_i(u,z)
\eqno(3.119)$$
establishes one-to-one correspondence between solutions of the linear
systems (3.118) and (3.5).

We can say thus that (3.118) gives a gauge transformation of the
\lq\lq Lax pair" of WDVV to the \lq\lq Lax pair" of the
Darboux - Egoroff system (3.70a,b).
\medskip
The equation (3.70c) specifies the similarity reduction of the system
(3.70a,b). It is clear that this is equivalent to a system of ODE
of the order $n(n-1)/2$. Indeed, for a given Cauchy data
$\gamma_{ij}(u_0)$ we can uniquely find the solution $\gamma_{ij}(u)$
of the system (3.70). We will consider now
this ODE system for arbitrary $n$ in more details.

We introduce first a very useful differential operator in $z$ with rational
coefficients
$$\Lambda = \partial_z - U - {1\over z}V(u).
\eqno(3.120)$$
Here
$$U={\rm diag} (u^1,\dots, u^n)
\eqno(3.121)$$
the matrix $V(u)$ is defined by (3.72).
\smallskip
{\bf Proposition 3.9.} {\it Equations (3.70) (or, equivalently, equations
(3.74))
are equivalent to the compatibility
conditions of the linear problem (3.118) with the differential equation in $z$
$$\Lambda \psi = 0.
\eqno(3.122)$$
}

Proof is straightforward (cf. the proof of Corollary 3.2).
\medskip
Solutions $\psi(u,z)
=(\psi_1(u,z), \dots, \psi_n(u,z))^T$ of
the equation (3.122) for a fixed $u$
are multivalued analytic functions in $\Cc\setminus{z=0\cup z=\infty}$.
The multivaluedness is described by {\it monodromy} of the operator
$\Lambda$. It turns out the parameters of monodromy of the operator
$\Lambda$ with the coefficients depending on $u$ as on the parameters
{\it do not depend on} $u$. So they are first integrals of the system
(3.74). We will see that a part of the first integrals are
the eigenvalues of the matrix $V(u)$ (I recall that these are expressed
via the degrees of the variables $t^\alpha$). But for $n \geq 3$ other
integrals are not algebraic in $V(u)$.

Let us describe the monodromy of the operator in more details. We will fix
some vector $u=(u^1,\dots, u^n)$ with the only condition $u^i-u^j\neq 0$
for $i\neq j$. For the moment we will concentrate ourselves on the
$z$-dependence of the solution of the differential equation (3.122)
taking aside the dependence of it on $u$.

The equation (3.122) has two singularities in the $z$-sphere $\Cc \cup\infty$:
the regular singularity at $z=0$ and the irregular one at $z=\infty$.
The matrix solution at the point $z=0$ (the regular singularity
of the operator $\Lambda$) has the form
$$\psi (z) \simeq z^{V}\,\psi_0
\eqno(3.123)$$
for arbitrary vector $\psi_0$.
If the matrix $V$ is diagonalizable and none of the differences
$\mu_\alpha - \mu_\beta$ for $\alpha\neq\beta$ is an integer
then a fundamental system of
solutions of (3.122) can be constructed such that
$$\psi^0_{j\alpha}(z) = z^{\mu_\alpha}\psi_{j\alpha}(1+ o(z)),
{}~~z\to 0.
\eqno(3.124)$$
Here $\alpha$ is the number of the solution of (3.122); the vector-functions
$\psi_{j\alpha}$ for any $\alpha = 1,\dots, n$ are the eigenvectors
of the matrix $V$ with the eigenvalues $\mu_\alpha$ resp.
It is convenient to consider the matrix-valued
solution $\Psi^0=\Psi^0(z)$ of (3.122) forming it of the vectors
(3.124) as of columns. The monodromy of the matrix around $z=0$
has the form
$$\Psi^0(z e^{2\pi i}) = \Psi^0(z)\diag (e^{2\pi i \mu_1},
\dots, e^{2\pi i \mu_n}).
\eqno(3.125)$$

Monodromy at $z=\infty$ (irregular singularity of the operator
$\Lambda$) is specified by a $n\times n$ {\it Stokes matrix}. I recall
here the definition of Stokes matrices adapted to the operators of
the form (3.120).

One of the definitions of the Stokes matrix of the operator
(3.120) is based on the theory of reduction of $\Lambda$ to a canonical form
by gauge transformations. The gauge transformations of $\Lambda$ have
the form
$$\Lambda \mapsto g^{-1}(z) \Lambda g(z)
\eqno(3.126a)$$
where the matrix valued function $g(z)$ is analytic near $z=\infty$
with
$$g(z) = 1 +O(z^{-1})
\eqno(3.126b)$$
satisfying
$$g(z)g^T(-z) = 1.
\eqno(3.126c)$$
According to the idea of Birkhoff [18] the orbits of the gauge
transformation (3.126) form a finite-dimensional family. The
local coordinates in this family are determined by the Stokes
matrix of the operator $\Lambda$.

To give a constructive definition of the Stokes
matrix one should use the asymptotic analysis of solutions of
the equation (3.122) near $z=\infty$.

Let us fix a vector of the parameters $(u^1, \dots, u^n)$
with
$u^i\neq u^j$ for $i\neq j$.
We define first {\it Stokes rays} in the complex $z$-plane. These
are the rays $R_{ij}$ defined for $i\neq j$ of the form
$$R_{ij} := \left\{ z\, | \, Re[z(u^i-u^j)]=0, ~Re[e^{i\epsilon}
z(u^i-u^j)]>0 ~{\rm for ~a ~small}~ \epsilon >0.\right\}.
\eqno(3.127)$$
The ray $R_{ji}$ is opposite to $R_{ij}$.

Let $l$ be an arbitrary oriented
line in the complex $z$-plane passing through
the origin containing no Stokes rays. It divides $\Cc$ into two half-planes
$\Cc_\rr$ and $\Cc_\ll$. There exist [13, 144]
two matrix-valued solutions $\Psi^\rr(z)$ and $\Psi^\ll(z)$ of (3.122) analytic
in the half-planes $\Cc_{\rr/\ll}$ resp. with the asymptotic
$$\Psi^{\rr/\ll}(z) = \left( 1+O\left( {1\over z}\right)\right) e^{zU}
{}~{\rm for }~ z\to\infty, ~z\in \Cc_{\rr/\ll}.
\eqno(3.128)$$

These functions can be analytically continued into some sectorial
neighbourhoods
of the half-planes.
On the intersection of the neighbourhoods of $\Cc_{\rr/\ll}$
the solutions $\Psi^\rr (z)$
and $\Psi^\ll (z)$ must be related by a linear transformation. To be more
specific let the line $l$ have the form
$$l = \{ z= \rho e^{i\phi_0}|\rho\in {\bf R}, ~\phi_0 ~{\rm is ~fixed}\}
\eqno(3.129)$$
with the natural orientation of the real $\rho$-line.
The half-planes $\Cc_{\rr/\ll}$ will be labelled in such a way that the
vectors
$$\pm ie^{i\phi_0}
\eqno(3.130)$$
belong to $\Cc_{\ll/\rr}$ resp. For the matrices $\Psi^{\rr/\ll}(z)$ we obtain
$$\eqalign{\Psi^\ll (\rho e^{i\phi_0}) &= \Psi^\rr(\rho e^{i\phi_0})S_+
\cr
\Psi^\ll(-\rho e^{i\phi_0}) &= \Psi^\rr(-\rho e^{i\phi_0})S_-
\cr}
\eqno(3.131)$$
$\rho > 0$ for some constant nondegenerate matrices $S_+$ and $S_-$.
The boundary-value problem (3.131) together with the asymptotic (3.128)
is a particular case of {\it Riemann -- Hilbert} b.v.p.
\smallskip
{\bf Proposition 3.10.} {\it \item{1.} The matrices $S_\pm$ must have the form
$$S_+ = S, ~~S_- = S^T
\eqno(3.132a)$$
$$S\equiv (s_{ij}), ~~s_{ii}=1, ~~s_{ij} = 0 ~{\rm if}~ i\neq j
{}~{\rm and}~ R_{ij} \subset \Cc_\rr .
\eqno(3.132b)$$
\item{2.} The eigenvalues of the matrix
$$M := {S^T}S^{-1}
\eqno(3.133)$$
are
$$(e^{2\pi i \mu_1},
\dots, e^{2\pi i \mu_n})\eqno(3.134)$$
where $\mu_1, \dots, \mu_n$ is the spectrum of the Frobenius variety.
The solution $\Psi^0(z)$ (3.125) has the form
$$\Psi^0(z) = \Psi^\rr (z)\,C
\eqno(3.135)$$
where the columns of the matrix $C$ are the correspondent eigenvectors
of the matrix $M$.}

Proof.
We will show first that the restriction for the matrices
$S_+$ and $S_-$ follows from the skew-symmetry of the matrix $V$.
Indeed, the skew-symmetry is equivalent to constancy of the natural
inner product in the space of solutions of (3.122)
$$\Psi^T(-z)\Psi(z) = const.
\eqno(3.136)$$
Let us proof that for the piecewise-analytic function
$$\Psi(z) := \Psi^{\rr/\ll}(z)
\eqno(3.137)$$
the identity
$$\Psi(z)\Psi^T(-z) = 1
\eqno(3.138)$$
follows from the restriction (3.132a). Indeed, for $\rho >0$,
$z = \rho e^{i(\phi_0 + 0)}$,  the l.h.s. of (3.138) reads
$$\Psi^\rr (z) {\Psi^\ll}^T(-z) = \Psi^\rr (z) S
{\Psi^\rr}^T(-z).
\eqno(3.139)$$
For $z = \rho^{i(\phi_0 - 0)}$ we obtain the same expression for the l.h.s.
So the piecewise-analytic function $\Psi(z)\Psi^T(-z) $ has no jump
on the semiaxis $z = \rho^{i\phi_0}$, $\rho > 0$. Neither it has a jump on the
opposite semiaxis
(it can be verified similarly). So the matrix-valued function is analytic
in the whole complex $z$-plane. As $z\to\infty$ we have from the asymptotic
(3.128) that this function tends to the unity matrix. The Liouville theorem
implies (3.138). So the condition (3.132a) is sufficient for the skew-symmetry
of
the matrix $V$. Inverting the considerations we obtain also the
necessity of the condition.

To prove the restrictions (3.132b) for the Stokes matrix we put
$$\Psi^{\rr/\ll}(z) =: \Phi^{\rr/\ll}(z) \exp{zU}.
\eqno(3.140)$$
For the matrix-valued functions $\Phi^{\rr/\ll}(z)$ we have the following
Riemann - Hilbert problem
$$\Phi^\ll (\rho e^{i\phi_0}) = \Phi^\rr (\rho e^{i\phi_0})\,\tilde S
\eqno(3.141a)$$
$$\Phi^\ll (-\rho e^{i\phi_0}) = \Phi^\rr (-\rho e^{i\phi_0})\,\tilde S^T
\eqno(3.141b)$$
$$\Phi^{\rr/\ll}(z) = 1 + O\left({1\over z}\right),~~z\to\infty .
\eqno(3.141c)$$
Here
$$\tilde S := e^{zU} S e^{-zU}.
\eqno(3.141d)$$
{}From the asymptotic (3.128) we conclude that the matrix $\tilde S$ must tend
to 1 when $z= \rho e^{i\phi_0}$,
$\rho\to +\infty$. This gives (3.132b).

To prove the second statement of the proposition it is enough to compare
the monodromy around $z=0$ of the solution of the Riemann - Hilbert
problem and of $\Psi^0$. Proposition is proved.
\medskip
{\bf Definition 3.4.} The matrix $S$ in (3.132) is called {\it Stokes matrix}
of the operator (3.120).
\medskip
Observe that precisely $n(n-1)/2$ off-diagonal entries of the Stokes matrix
can be nonzero due to (3.132).

The Stokes matrix changes when the line $l$ passes through a separating
ray $R_{ij}$. We will describe these changes in Appendix F.

Note that if the matrix $M$ has simple spectrum then the
solution $\Psi^0$ is completely determined by the Stokes  matrix $S$.
Non-semisimplicity of the matrix $M$ can happen when
some of the differences between the eigenvalues of $V$ is equal
to an integer.

The eigenvalues of the skew-symmetric complex matrix $V$ in general
are complex numbers. We will obtain below sufficient conditions for
them to be real.
\smallskip
{\bf Exercise 3.7.} Prove that the expansion of the solution of the Riemann
- Hilbert problem (3.128), (3.131)  for $z\to\infty$ has the form
$$\Psi^{\rr/\ll} (z) = \left( 1+ {\Gamma\over z} +O\left({1\over z^2}\right)
\right) e^{zU}
\eqno(3.142a)$$
where
$$V = [\Gamma ,U].
\eqno(3.142b)$$
\medskip
Let us assume that the Riemann - Hilbert problem (3.128), (3.131), (3.132)
 for
a given Stokes matrix $S$ has
a unique solution for a given  $u$. Since the solvability of the problem is
an open property, we obtain for the given $S$ locally a well-defined
skew-symmetric matrix-valued function $V(u)$. We show now that this
is a solution of the system (3.74). More precisely,
\smallskip
{\bf Proposition 3.11.} {\it If the dependence on $u$ of the matrix $V(u)$
of the coefficients of the
system (3.122) is specified by the system (3.74) then the Stokes matrix $S$
does not depend on $u$. Conversely, if the $u$-dependence
of $V(u)$ preserves the matrix $S$ unchanged then $V(u)$
satisfies the system (3.74).}

Proof. We prove first the second part of the proposition. For the
piecewise-analytic function $\Psi(u,z) = \Psi^{\rr/\ll}(u,z)$ determined
by the Riemann - Hilbert problem (3.128), (3.131), (3.132) the combination
$$\deli \Psi(u,z)\cdot \Psi^{-1}(u,z)
\eqno(3.143)$$
for any $i=1, \dots, n$
has no jumps on the line $l$. So it is analytic in the whole
complex $z$-plane. From (3.142) we obtain that for $z\to\infty$
$$\deli \Psi(u,z)\cdot \Psi^{-1}(u,z)= zE_i - [E_i, \Gamma ]
+ O(1/z).
$$
Applying the Liouville theorem we conclude that the solution of
the Riemann - Hilbert problem satisfies the linear system
$$\deli \Psi (u,z) = (zE_i - [E_i,\Gamma(u)])\Psi (u,z),
{}~~i=1,\dots, n.
\eqno(3.144a)$$
Furthermore, due to $z$-independence of the Stokes matrix $S$
the function $\Psi (cu, c^{-1}z)$ is a solution of the same
Riemann - Hilbert problem. From the uniqueness we obtain that
the solution satisfies also the condition
$$\left(z{d\over dz} - \sum u^i\deli \right) \Psi (u,z) =0.
\eqno(3.144b)$$
Compatibility of the equations (3.144a) reads
$$0 = (\deli\delj - \delj\deli )\Psi(u,z) \equiv
\left( [E_j,\deli \Gamma ]-[E_i,\delj \Gamma ] +
\left[ [E_i,\Gamma ],[E_j,\Gamma ]\right] \right) \Psi (u,z).
$$
Since the matrix
$$\left( [E_j,\deli \Gamma ]-[E_i,\delj \Gamma ] +
\left[ [E_i,\Gamma ],[E_j,\Gamma ]\right] \right)$$
does not depend on $z$, we conclude that
$$\left( [E_j,\deli \Gamma ]-[E_i,\delj \Gamma ] +
\left[ [E_i,\Gamma ],[E_j,\Gamma ]\right] \right)=0.
$$
This coincides with the equations (3.70a,b). From (3.144b) we obtain the
scaling
condition
$$\Gamma(cu) = c^{-1} \Gamma(u).
$$
This gives the last equation (3.70c).

Conversely, if the matrix $\Gamma(u)$ satisfies the system (3.70)
(or, equivalently, $V=[\Gamma, U]$ satisfies the system (3.74))
then the equations (3.144a) are compatible with the equation (3.144b).
Hence for a solution of the Riemann - Hilbert problem the matrices
$$(\deli -zE_i+ [E_i,\Gamma ])\Psi$$
and
$$\left(z{d\over dz} - \sum u^i\deli \right) \Psi$$
satisfy the same differential equation (3.122). Hence
$$(\deli -zE_i+ [E_i,\Gamma ])\Psi (u,z)= \Psi (u,z)\, T_i
\eqno(3.145a)$$
and
$$\left(z{d\over dz} - \sum u^i\deli \right) \Psi(u,z) = \Psi(u,z)\,
T\eqno(3.145b)$$
for some matrices $T_i = T_i(u)$, $T= T(u)$. Comparing the expansions
of the both sides of (3.145) at $z\to\infty$ we obtain that $T_i = T = 0$.
So the solution of the Riemann - Hilbert problem satisfies the equations
(3.144). From this immediately follows that the Stokes matrix does not depend
on $u$. Proposition is proved.
\medskip
According to the proposition the Stokes matrices of the operators (3.120)
locally
parame\-trize the solutions of the system (3.74). Note that, due to the
conditions (3.132) there are precisely $n(n-1)/2$ independent complex
parameters in
the Stokes matrix of the operator (3.120).

{}From Miwa's results [106] on the Painlev\'e property of the equations
of isomonodromy deformations we obtain
\smallskip
{\bf Corollary 3.4.} {\it Any solution $V(u)$ of the system (3.74)
is a single-valued meromorphic function on the universal covering of
domain $u^i\neq u^j$ for all $i\neq j$ i.e., on $CP^{n-1}\setminus
{\rm diagonals}$.
}
\medskip
{\bf Remark 3.9.} One can obtain the equations (3.74) also as the equations
of isomonodromy deformations of an operator with regular singularities
$${d\phi\over d\lambda} +\sum_{i=1}^n{A_i\over \lambda -u_i}\phi = 0.
\eqno(3.146)$$
The points $\lambda = u_1, \dots, \lambda =u_n$ are the regular singularities
of the coefficients. If $A_1 + \dots +A_n \neq 0$ then
$\lambda =\infty$ is also a regular singularity. The monodromy
preserving deformations of (3.146) were described by Schlesinger [130].
They can be represented in the form of compatibility conditions
of (3.146) with linear system
$$\deli \phi = \left( {A_i\over \lambda - u_i} + B_i\right)\phi, ~~
i=1, \dots, n
\eqno(3.147)$$
for some matrices $B_i$. To represent (3.74) as a reduction of the Schlesinger
equations one put
$$A_i = E_iV, ~~B_i = - ad_{E_i}ad_U^{-1}V.
\eqno(3.148)$$
Observe that the hamiltonian structure (3.107) of the equations (3.74) is
obtained
by the reduction
(3.148) of the hamiltonian structure of general Schlesinger equations found
in [127]. The Painlev\'e property of the equations (3.74) follows thus also
from the results of Malgrange (see in [102].

To obtain (3.146) from (3.120) and (3.122) we apply the following trick
(essentially due to Poincar\'e and Birkhoff, see the
textbook [75]). Do (just formally)
the inverse Laplace transform
$$\psi(z) = z \oint e^{\lambda z} \phi(\lambda)\, d\lambda.
\eqno(3.149)$$
Substituting to (3.122), (3.144a) and integrating by parts we obtain
(3.146) and (3.147).
\medskip
We will show now that non-semisimplicity of the matrix $M=S^TS^{-1}$
is just the reason of appearing of linear nonhomogeneous terms
in the Euler vector field $E(t)$. We consider here only the simplest
case of Frobenius manifolds with the pairwise distinct scaling
dimensions satisfying the inequalities
$$0\leq q_\alpha < q_n = d \leq 1.
\eqno(3.150)$$
Such Frobenius manifolds were called reduced in Appendix B. We showed
that any massive Frobenius manifold can be reduced to the form
(3.150) by the transformations of Appendix B. Later we will show that these
transformations essentially do not change  the Stokes matrix.
\smallskip
{\bf Proposition 3.12.} {\it 1). For the case $d<1$ the matrix $M$ is
diagonalizable.

2). For $d=1$ the matrix $M$ is diaagonalizable {\rm iff} $E(t)$ is a linear
homogeneous vector field.}

Proof. If $d<1$ then all the numbers $e^{2\pi i \mu_1}, \dots,
e^{2\pi i \mu_n}$ are pairwise distinct (I recall that the scaling
dimensions are assumed to be pairwise distinct). This gives diagonalizability
of the matrix $M$.

If $d=1$ then $\mu_1 = -\half$, $\mu_n = \half$. So $e^{2\pi i \mu_1}=
e^{2\pi i \mu_n}=-1$, and the characteristic roots of the matrix
$M$ are not simple. To prove diagonalizability of the matrix $M$ we
are to construct a fundamental system of
solutions of the equation $\Lambda \psi = 0$
of the form
$$\psi(u,z) = (\psi(u) + O(z))z^\mu .
\eqno(3.151)$$
We will show that the linear nonhomogeneous terms in the Euler vector
field give just the obstruction to construct such a fundamental system.
\smallskip
{\bf Lemma 3.11.} {\it If the Euler vector field of a Frobenius manifold
with $d=1$ is
$$E = \sum_{\alpha =1}^{n-1} (1-q_\alpha)t^\alpha\dalpha + r\partial_n
\eqno(3.152)$$
then a fundamental system of solutions $\psi_{i\alpha}(u,z)$
of the equation $\Lambda \psi = 0$ exists such that
$$\eqalignno{\psi_{i\alpha}(u,z) &= \left( \psi_{i\alpha}(u)+O(z)\right)
 z^{\mu_\alpha}, ~~\alpha\neq 1
&(3.153a)\cr
\psi_{i1}(u,z) &= {1\over \sqrt{z}} \left[\psi_{i1}(u)
+ r z\log z \psi_{in}(u) +O(z)\right]
&(3.153b)\cr}
$$
where $\psi_{i\alpha}(u)$ are defined in (3.81).}

Proof is obtained by substitution of the expansions (3.153) to the equation
(3.122).
\medskip
We conclude that the basis of solutions of (3.122) being also eigenvectors
of the monodromy around $z=0$ can be constructed {\it iff} $r=0$.
Proposition is proved.
\medskip
General theory of solvability of the Riemann - Hilbert problem (3.128),
(3.131), (3.132)
is far from being completed. If the Stokes matrix is sufficiently close
to the identity then existence and uniqueness of the solution of
the problem can be proved using the following reduction of the
Riemann - Hilbert problem to a system of singular linear integral
equations.

To obtain these equations we assume for the sake of simplicity that
$$Re(u^i - u^j) \neq 0 ~{\rm for }~i\neq j.
\eqno(3.154)$$
We
introduce the matrix $R(u,\rho )$ putting
$$R(u,\rho ) = \cases{ e^{\rho U}Se^{-\rho U} & if $ \rho > 0$ \cr
e^{\rho U}S^Te^{-\rho U} & if $ \rho < 0$ \cr}.
\eqno(3.155)$$
\smallskip
{\bf Proposition 3.13.} {\it For $S$ sufficiently close to the identity
the Riemann - Hilbert b.v.p. (3.128), (3.131), (3.132)
 is equivalent to the integral
equation
$$\Phi(u,\rho ) = 1 -{1\over 2\pi i}\int_{-\infty}^{\infty}
{\Phi(u,\rho' )(R(u,\rho')-1)\over \rho - \rho'-i0} d\rho'.
\eqno(3.156)$$}

Proof. For the piecewise-analytic function $\Phi(u,z) = \left(\Phi^\rr (u,z),
\Phi^\ll(u,z)\right)$
$$\Phi^{\rr/\ll}(u,z) = \Psi^{\rr/\ll}(u,z) e^{-zU}
$$
we have the following Riemann - Hilbert problem
$$\Phi^\ll (u,\rho ) = \Phi^\rr (u,\rho ) R(u,\rho )
\eqno(3.157a)$$
for any real $\rho$,
$$\Phi(u,z) = 1+O\left({1\over z}\right) ~{\rm for}~ z\to\infty .
\eqno(3.157b)$$
If we knew the jump
$$\Delta \Phi (u,\rho ):= \Phi^\rr (u,\rho ) - \Phi^\ll (u,\rho)=
\Phi(u, \rho -i0) - \Phi(u,\rho+i0)
\eqno(3.158)$$
then we could reconstruct the functions $\Phi^{\rr/\ll}$ using the Cauchy
integral
$$\Phi^{\rr/\ll} (u,z) = 1 \mp {1\over 2\pi i} \int_{-\infty}^{\infty}
{\Delta \Phi(u,\rho)\over z-\rho}d\rho.
\eqno(3.159)$$
We put now
$$\Phi(u,\rho ) := \Phi^\rr (u,\rho).\eqno(3.160)$$
Then
$$\Delta \Phi (u, \rho ) = \Phi(u,\rho)(R(u,\rho) -1).
\eqno(3.161)$$
Substituting to the Cauchy integral (3.159) for $\Phi^\rr$ we obtain
the integral equation (3.156). If $S$ is sufficiently close to the identity
matrix then the kernel of the equation is small. This provides
existence and uniqueness of the solution of (3.156). Having known the function
$\Phi(u,\rho)$ we can find the jump (3.161) and then reconstruct the solution
of the Riemann - Hilbert b.v.p. using the Cauchy integrals (3.159).
Proposition is proved.
\medskip
We are close now to formulate the precise statement about parametrization
of solutions of WDVV by Stokes matrices of the operators (3.120).
\smallskip
{\bf Lemma 3.13.} {\it For two equivalent Frobenius manifolds satisfying
the semisimplicity condition the corresponding solutions
$V(u) = (V_{ij}(u))$ of the
system (3.74) are related by a permutation of coordinates
$$(u^1,\dots, u^n)^T\mapsto P(u^1,\dots, u^n)^T,
\eqno(3.162a)$$
$$V(u)\mapsto\epsilon P^{-1}V(u) P\epsilon,
\eqno(3.162b)$$
$P$ is the matrix of the permutation, $\epsilon$ is an arbitrary diagonal
matrix
with $\pm 1$ diagonal entries.}
\medskip
Observe that the permutations act on the differential operators
$\Lambda$ as
$$\Lambda \mapsto \epsilon P^{-1}\Lambda P\epsilon.
\eqno(3.163)$$
The Stokes matrix $S$ of the
operator $\Lambda$ changes as
$$S\mapsto P^{-1} \epsilon\, S\, \epsilon\, P.
\eqno(3.164)$$

Note that the Legendre-type transformations (B.2) change the operator
$\Lambda$ only as in (3.163).
So the correspondent transformations of the Stokes matrix have
the form (3.164).

Summarizing the considerations of this section, we obtain
\smallskip
{\bf Theorem 3.2.} {\it There exists a  local one-to-one correspondence
$$\left\{ \matrix{{\rm Massive ~Frobenius~ manifolds} \cr
{\rm modulo~ transformations ~(B.2)}\cr}
\right\} \leftrightarrow
\left\{ \matrix{{\rm Stokes~ matrices~ of~ differential}\cr
{\rm operators~} \Lambda {\rm ~modulo ~transformations ~(3.164)}\cr}\right\}.
$$
}
\medskip
{\bf Definition 3.5.} The Stokes matrix $S$ of the operator (3.120) considered
modulo the transformations (3.164) will be called {\it Stokes matrix of
the Frobenius manifold}.
\medskip
In the paper [27] Cecotti and Vafa found a physical interpretation
of the matrix entries $S_{ij}$ for a Landau - Ginsburg TFT as the algebraic
numbers of solitons propagating between classical vacua. In this interpretation
$S$ always is an integer-valued matrix.

It is interesting that {\it the same} Stokes matrix appears, according to
[27], in the Riemann - Hilbert problem of [47] specifying the Zamolodchikov
(or $t\, t^*$) hermitean metric on these Frobenius manifolds.
\medskip
At the end of this section we explain the sense of the transformations
(B.13) of WDVV from the point of view of the operators (3.120).
\smallskip
{\bf Proposition.} {\it The rotation coefficients $\gamma_{ij}(u)$
and $\hat\gamma_{ij}(u)$ of two Frobenius manifolds related by the inversion
(B.11) are related by the formula
$$\hat\gamma_{ij} = \gamma_{ij} - A_{ij}
\eqno(3.165a)$$
where
$$A_{ij} := {\sqrt{\deli t_1 \delj t_1}\over t_1}.
\eqno(3.165b)$$
The solutions $\psi(u,z)$ and $\hat\psi(u,z)$ of the correspondent systems
(3.122) are related by the gauge transformation
$$\psi = \left( 1 + {A\over z}\right) \hat\psi
\eqno(3.166)$$
for $A = (A_{ij})$.
}

We leave the proof of this statement as an exercise for the reader.
\medskip
The gauge transformations of the form
$$\psi (z) \mapsto g(z)\psi (z)
$$
with rational invertible matrix valued function $g(z)$ preserving the form
of the operator $\Lambda$ are called {\it Schlesinger transformations} of
the operator [127]. They preserve unchanged the monodromy property of the
operator. However, they change some of the eigenvalues of the matrix $V$
by an integer (see (B.16)).

It can be proved that all {\it elementary}
Schlesinger transformations of the system
(3.122) where $g(z) = (1 + A\, z^{-1})$
are superpositions of the transformation (B.13) and of the Legendre-type
transformations (B.2). These generate all the group of
Schlesinger transformations of (3.120). This group is a group of symmetries
of WDVV according to Appendix B.
\medskip
To conclude this long lecture we will discuss briefly the reality
conditions of the solutions of WDVV. We say that the Frobenius manifold is
real if it admits an antiholomorphic automorphism $\tau: M \to M$.
This means that
in some coordinates on $M$ the structure functions $c_{\alpha\beta}^\gamma(t)$
all are real. The scaling dimensions $q_\alpha$ also are to be real.

The antiinvolution $\tau$ could either preserve or permute
the canonical coordinates $u^1(t)$, \dots, $u^n(t)$. We consider here
only the case when the canonical coordinates are $\tau$-invariant
near some real point $t\in M$.
\smallskip
{\bf Exercise 3.8.} Prove that the canonical coordinates are real near
a point $t\in M$ where the intersection form is definite positive.
Prove that in this case for even $n$ half of the canonical coordinates
are positive and half of them are negative, while for odd $n$ one
obtains $(n+1)/2$ negative and $(n-1)/2$ positive canonical coordinates.
\medskip
For real canonical coordinates the diagonal metric $\eta_{ii}(u)$ is
real as well. We put
$$J_i := sign\, \eta_{ii}(u), ~~i=1, \dots, n
\eqno(3.167)$$
near the point $u$ under consideration. The matrix $\Gamma(u)$ of the rotation
coefficients and, hence, the matrix $V(u)$ obeys the symmetry
$$\Gamma^\dagger = J \Gamma J, ~~V^\dagger = - J V J
\eqno(3.168a)$$
where
$$J = \diag (J_1, \dots, J_n).
\eqno(3.168b)$$
Here dagger stands for the hermitean conjugation.
\smallskip
{\bf Proposition 3.14.} {\it If the coefficients of the operator $\Lambda$
for real $u$ satisfy the symmetry (3.168) then the Stokes matrix satisfies
the equation
$$\bar S J S J = 1.
\eqno(3.169)$$
Conversely, if the Stokes matrix satisfies the equation (3.169) and
the Riemann - Hilbert problem (3.128), (3.131), (3.132)
 has a unique solution
for a given real $u$ then the corresponding solution of the system
(3.74) satisfies (3.168).}

Here the bar denotes the complex conjugation of all the entries of
$S$.

Proof. Let $l$ be the real line on the $z$-plane.
As in the proof of Proposition 3.10 we obtain that the equation
(3.169) is equivalent to the equation
$$\Psi_{\rr/\ll}(u, z) =\left(\Psi_{\ll/\rr}(u,\bar z)\right)^\dagger.
\eqno(3.170)$$
Proposition is proved.
\medskip
To derive from (3.168) the reality of the Frobenius manifold we are to
provide also reality of the Euler vector fields. For this we
need the eigenvalues of the matrix $M= S^TS^{-1}$ to be unimodular.
\smallskip
{\bf Lemma 3.14.} {\it The eigenvalues $\lambda$ of a matrix $M=S^TS^{-1}$
with the matrix $S$ satisfying (3.169) are invariant w.r.t. the transformations
$$\lambda \mapsto \lambda^{-1}, ~~\lambda \mapsto \bar \lambda.
\eqno(3.171)$$}

Proof is obvious.

We conclude that for a generic matrix $S$ satisfying (3.169) the collection
of the eigenvalues must consist of:

1). Quadruples $\lambda, \lambda^{-1}, \bar\lambda, \bar\lambda^{-1}$
for a nonreal $\lambda$ with $|\lambda|\neq 1$.

2). Pairs $\lambda, \bar\lambda$ for a nonreal $\lambda$ with
$|\lambda|=1$.

3). Pairs $\lambda, \lambda^{-1}$ for a real $\lambda$ distinct from
$\pm 1$.

4). The point $\lambda =1$ for the matrices of odd dimension.

All these types of configurations of eigenvalues are stable
under small perturbations of $S$. Absence of the eigenvalues
of the types 1 and 3 specifies an open domain in the space of
all complex $S$-matrices.
\smallskip
{\bf Example 3.3.} For $n=3$ and $J= \diag (-1, -1, 1)$
the matrices satisfying (3.169) are parametrized by
3 real numbers $a, b, c$ as
$$S = \left(\matrix{1 & ia & b+{i\over 2}ac\cr
0 & 1 & c \cr
0 & 0 & 1\cr}\right).
\eqno(3.172)$$
The eigenvalues of $S^TS^{-1}$ are unimodular {\it iff}
$$0\leq b^2+c^2 + {1\over 4} a^2c^2 - a^2 \leq 4.
\eqno(3.173)$$
\vfill\eject
\bigskip
\centerline{\bf Appendix D.}
\smallskip
\centerline{\bf Geometry of flat pencils of metrics.}
\medskip
{\bf Proposition D.1.} {\it For a flat pencil of metrics a vector field
$f = f^i\partial_i$ exists
such that the difference tensor (3.34) and the metric $g_1^{ij}$ have the
form
$$\Delta^{ijk} = \nabla_{2}^i\nabla_{2}^jf^k
\eqno(D.1a)$$
$$g_1^{ij} = \nabla_{2}^if^j + \nabla_{2}^jf^i + c g_2^{ij}
\eqno(D.1b)$$
for a constant $c$.
The vector field satisfies the equations
$$\Delta_s^{ij} \Delta_l^{sk} = \Delta_s^{ik}\Delta_l^{sj}
\eqno(D.2)$$
where
$$\Delta_k^{ij} := g_{2ks} \Delta^{sij} = \nabla_{2k} \nabla_2^i f^j,$$
and
$$(g_1^{is}g_2^{jt} - g_2^{is}g_1^{jt} ) \nabla_{2s}\nabla_{2t}f^k = 0.
\eqno(D.3)$$
Conversely, for a flat metric $g_2^{ij}$ and for a solution $f$ of the system
(D.2), (D.3) the metrics $g_2^{ij}$ and (D.1b) form a flat pencil.}

Proof. Let us assume that $x^1$, ..., $x^n$ is the flat coordinate
system for the metric $g_2^{ij}$. In these coordinates we have
$$\Gamma_{2k}^{ij} = 0, ~ \Delta_k^{ij}
:= g_{2ks} \Delta^{sij} = \Gamma_{1k}^{ij} .
\eqno(D.4)$$
The equation $R^{ijk}_l = 0$ in these coordinates reads
$$\left( g_1^{is} + \lambda g_2^{is}\right)
\left( \partial_s\Delta_l^{jk} - \partial_l\Delta_s^{jk}\right)
+ \Delta_s^{ij} \Delta_l^{sk} - \Delta_s^{ik} \Delta_l^{sj} =0.
\eqno(D.5)$$
Vanishing of the linear in $\lambda$ term provides existence of a
tensor $f^{ij}$ such that
$$\Delta_k^{ij} = \partial_kf^{ij}.$$
The rest part of (D.5) gives (D.2). Let us use now the condition
of symmetry (3.27) of the connection $\Gamma{1\, k}^{ij} +\lambda \Gamma_{2\,
k}
^{ij}$.
In the coordinate system this reads
$$\left( g_1^{is} +\lambda g_2^{is}\right) \partial_s f^{jk} =
\left( g_1^{js} +\lambda g_2^{js}\right) \partial_s f^{ik}.
\eqno(D.6)$$
Vanishing of the terms in (D.6) linear in $\lambda$ provides existence
of a vector field $f$ such that
$$f^{ij} = g_2^{is}\partial_s f^j.$$
This implies (D.1a).
The rest part of the equation (D.6) gives (D.3). The last equation (3.26)
gives (D.1b). The first part of the proposition is proved. The converse
statement follows from the same equations.
\medskip
{\bf Remark D.1.} The theory of S.P.Novikov and the author establishes a
one-to-one correspondence between flat contravariant metrics on a manifold
$M$ and Poisson brackets of hydrodynamic type on the loop space
$$L(M) := \{ {\rm smooth ~maps} ~ S^1 \to M\} $$
with certain nondegeneracy conditions [53, 54]. For a flat metric $g^{ij}(x)$
and the correspondent contravariant connection $\nabla^i$ the Poisson bracket
of two functionals of the form
$$I = I[x] = {1\over 2\pi} \int_0^{2\pi} P(s,x(s))\, ds, ~~
J = J[x] = {1\over 2\pi} \int_0^{2\pi} Q(s,x(s))\, ds,
$$
$x = (x^i(s))$, $x(s+2\pi )
= x(s)$ is defined by the formula
$$\{ I,J\} := {1\over 2\pi }\int_0^{2\pi} {\delta I\over \delta x^i(s)}
\nabla^i{\delta J\over \delta x^j(s)} \, dx^j(s) +
{1\over 2\pi }\int_0^{2\pi} {\delta I\over \delta x^i(s)}
g^{ij}(x) d_s{\delta J\over \delta x^j(s)}.
\eqno(D.7)$$
Here the variational derivative $\delta I / \delta x^i(s) \in
T_*M|_{x=x(s)}$ is defined by the equality
$$I[x+\delta x] - I[x] =
{1\over 2\pi}\int_0^{2\pi} {\delta I\over \delta x^i(s)} \delta
x^i(s)\, ds + o(|\delta x|);
\eqno(D.8)$$
$\delta J / \delta x^j(s) $ is defined by the same formula,
$d_s := ds {\partial\over \partial s}$. The Poisson bracket can be uniquely
extended
to all \lq\lq good" functionals on the loop space by Leibnitz rule [53, 54].
Flat pencils of metrics correspond to compatible pairs of Poisson
brackets of hydrodynamic type.
By the definition, Poisson brackets $\{ ~,~\}_1$ and
$\{ ~,~\}_2$ are called compatible if an arbitrary linear combination with
constant coefficients
$$a\{ ~,~\}_1 + b\{ ~,~\}_2$$
again is a Poisson bracket. Compatible pairs of Poisson brackets are
important in the theory of integrable systems [101].
\medskip

The main source of flat pencils is provided by the following statement.
\medskip
{\bf Lemma D.1.} {\it If for a flat metric  in some coordinate
system $x^1$, ..., $x^n$ both the components $g^{ij}(x)$ of the metric and
$\Gamma_k^{ij}(x)$ of the correspondent Levi-Civit\`a connection depend
linearly
on the coordinate $x^1$ then the metrics
$$g_1^{ij} := g^{ij} ~ {\rm and }~ g_2^{ij} := \partial_1 g^{ij}
\eqno(D.9)$$
form a flat pencil if ${\rm det} (g_2^{ij}) \neq 0$. The correspondent
Levi-Civit\`a connections have the form
$$\Gamma_{1k}^{ij} := \Gamma_k^{ij}, ~\Gamma_{2k}^{ij} :=
\partial_1 \Gamma_k^{ij}.
\eqno(D.10)$$
}

Proof. The equations (3.26), (3.27) and the equation of vanishing of the
curvature have constant coefficients. Hence the transformation
$$g^{ij}(x^1, ..., x^n)\mapsto g^{ij}(x^1 +\lambda , ...,
x^n), ~\Gamma_k^{ij}(x^1, ..., x^n)\mapsto
\Gamma_k^{ij}(x^1 +\lambda , ..., x^n)$$
for an arbitrary $\lambda$ maps the solutions of these equations to
the solutions. By the assumption we have
$$g^{ij}(x^1+\lambda , ..., x^n) = g_1^{ij}(x) +\lambda g_2^{ij}(x),
{}~\Gamma_k^{ij}(x^1+\lambda , ..., x^n)
=\Gamma_{1k}^{ij}(x) +\lambda \Gamma_{2k}^{ij}(x).$$
The lemma is proved.
\medskip
All the above considerations can be applied also to complex (analytic)
manifolds where the metrics are nondegenerate
quadratic forms analyticaly depending
on the point of $M$.
\bigskip
\vfill\eject
\bigskip
\centerline{\bf Appendix E.}
\smallskip
\centerline{\bf WDVV and Painlev\'e-VI.}
\medskip
This Appendix is based on the papers [61, 62] (the case R = -1/2
in Lemma E.2 is missed in [61]).
\smallskip
{\bf Proposition E.1.} {\it On the symplectic leaves (3.114) of $so(3)$ the
time-dependent Euler top (3.113) is reduced to the particular case of
Painlev\'e-VI equation
$$y'' = {1\over 2}\left( {1\over y} +{1\over y-1}+{1\over
y-z}\right){(y')^2}
-\left({1\over z}+{1\over z-1}+{1\over y-z}\right)y'$$
$$+{y(y-1)(y-z)\over z^2(z-1)^2}\left({(2R+1)^2\over 8}
-{(2R+1)^2 z\over 8 y^2} +{z(z-1)\over 2(y-z)^2}\right).
\eqno(E.1)$$
}

{\bf Remark E.1.} The equation is a time-dependent
Hamiltonian system with one degree of freedom
$${dy\over dz} = {\partial H_R\over\partial p},~~
{dp\over dz} = -{\partial H_R\over \partial y}
\eqno(E.2a)$$
and with the Hamiltonian
$$H_R = {1\over z(z-1)}\left\{ y(y-1)(y-z)p^2 +
(y-1)\left[ y -(R+1/2)(y-z)\right] p -\half R (y-z)\right\}.
\eqno(E.2.b)$$
It can be proved that this Hamiltonian structure is inherited from
the $so(3)$ Hamiltonian structure of the equations (3.113).

Proof follows from the following two lemmata.
\smallskip
{\bf Lemma E.1.} {\it After the substitution
$$\Omega_3(s) = \sqrt{-1} \,\phi(z)
\eqno(E.3a)$$
where
$$s = \left({\sqrt{z} -1\over \sqrt{z} +1}\right)^2
\eqno(E.3b)$$
the system (3.113) on the level surface (3.114) reads
$$(z-1)^2\left[ \phi'' +{3z-1\over 2z(z-1)}\phi' +
{2\phi(\phi^2 - R^2)\over z(z-1)^2}\right]^2
= {(z+1)^2\over z^2}\phi^2\left[ (\phi')^2 + {(\phi^2 - R^2)^2
\over z(z-1)^2}\right].
\eqno(E.4)$$}

Proof. Differentiating the third of the equations (3.113) w.r.t. $s$
we obtain
$${d^2\Omega_3\over ds^2} = -{1\over s} {d \Omega_3\over ds}
+ {\Omega_3\over s(s-1)} \left( {1\over s} \Omega_2^2 - \Omega_1^2\right).
\eqno(E.5)$$
We can express $\Omega_1^2$ and $\Omega_2^2$ via $\Omega_3$
and $d\Omega_3/ds$ from (3.113)
and
$$ \Omega_1^2 \Omega_2^2 = s^2 \left( d\Omega_3/ds\right)^2
$$
and substitute the results to (E.5).
Changing the independent variable $s\to z$ after simple calculations
we obtain (E.4). Lemma is proved.

\smallskip
{\bf Lemma E.2.} {\it The transformation $y\leftrightarrow \phi$ of the
form
$$\phi = z{y'\over y} - {(2R+1)y\over 2(z-1)} + {(2R+1)z
\over 2(z-1)y} - {1\over 2}
\eqno(E.6)$$
and, conversely,
$$y = -B/A
\eqno(E.7a)$$
where
$$A= - 2\phi' + {1\over z}\phi^2 + 2 \phi R {z+1\over z(z-1)} +
{R^2\over z}
\eqno(E.7b)$$
$$B=(z+1)\phi' -{4 R\over z-1}\phi +{z(z-1)^2\over \phi (z+1)}
\left[ \phi'' +{3z-1\over 2z(z-1)}\phi' +
{2\phi(\phi^2 - R^2)\over z(z-1)^2}\right]
\eqno(E.7c)$$
for $R\neq 0,~-1/2$ establishes
one-to-one-correspondence between the solutions of the
equation (E.4) and of the particular case (E.1) of the
Painlev\'e-VI equation.
For $R=0$ the Painlev\'e-VI equation has a 1-parameter family
solutions of the form
$$y = {\sqrt{z} + c z\over c + \sqrt{z}}
\eqno(E.8)$$
($c$ is an arbitrary constant).
Any of these solutions corresponds to the trivial solution $\phi =0$
of the equation (E.4).

For $R = -1/2$ the equation (E.4) has a one-parameter family of solutions
of the form
$$\phi = -{1\over 2} - 2z {d\over dz} \log{}f(\half , \half , 1, z)
\eqno(E.9)$$
where $f(\half , \half , 1, z)$ stands for the general solution of
the hypergeometric equation
$$z(z-1)f'' + (2z-1) f' + {1\over 4} f = 0.
\eqno(E.10)$$
The transformation (E.7) is not defined for the solutions of (E.9).}

Proof can be obtained by a long but straightforward calculation.
\medskip
{\bf Remark E.2.} For a particular values of the parameter $R$ some
solutions of the Painlev\'e-VI equation (E.1) can be expressed via
classical transcendental functions [114]. The symplest of these
is the case $R = -1/2$ where the general solution of (E.1) was
obtained by E.Picard (see in [114]). This corresponds to the general
solution of Chazy equation (C.5).
K.Okamoto [114] related the theory
of classical solutions of the Painlev\'e-VI as of the Hamiltonian system (E.2)
to an action of the affine
Weyl group $W_a(D_4)$ of the root system of the type $D_4$ in the space
of parameters of the Painlev\'e-VI. According to this theory the Painlev\'e-VI
admits a classical solution if the vector of parameters of the equation
belongs to the boundary of the Weyl chamber $W_a(D_4)$.

Strictly speaking, all the family (E.1) belongs to the boundary. However,
for $R\neq -1/2$ the classical solutions of (E.1) given by Okamoto
all are singular (they have the form $y\equiv 1$.). So the solutions
of the equation of Appendiix A that can be obtained from the above procedure
using the polynomial solutions (A.7), (A.8), (A.9) of WDVV look to be new
(for them $R = -1/4, ~-1/3, ~-2/5$ resp.). These solutions can be
expressed via algebraic functions.

I learned recently from N.Hitchin [72] that he also found a particular
solution of the Painlev\'e-VI equation in terms of algebraic functions.
The Coxeter group $A_3$ also played an important role in his
constructions. It would be interesting to compare the solutions
of [72] and those coming from Appendix A.
\smallskip
{\bf Exercise E.1}. 1). Show that the canonical coordinates $u_1$, $u_2$,
$u_3$ for a 3-dimensional Frobenius manifold with the free energy
of the form (C.2) with $\gamma = \gamma(t_3)$ being an arbitrary solution
of the Chazy equation (C.5) have the form
$$u_i = t_1 + \half t_2^2 \omega_i(t_3), ~~i = 1, 2, 3
\eqno(E.11)$$
where $\omega_i(t_3)$ are the roots of the cubic equation (C.7).

2). Show that the correspondent solutions of the Euler equations
(3.113) has the form
$$\Omega_i = {\omega_i\over 2\sqrt{(\omega_j - \omega_i)(
\omega_k - \omega_i)}}
\eqno(E.12)$$
(here $i,\, j,\, k$ are distinct indices) where the dependence
$t_3 = t_3(s)$ is determined by the equation
$${\omega_3(t_3) - \omega_1(t_3)\over \omega_2(t_3)-\omega_1(t_3)}
= s.
\eqno(E.13)$$
\vfill\eject
\medskip
\centerline{\bf Appendix F.}
\smallskip
\centerline{\bf Branching of solutions of the equations}
\smallskip
\centerline{\bf of isomonodromic deformations and braid group}
\medskip
We consider here only isomonodromy deformations of the operator
$$\Lambda = {d\over dz} - U -{1\over z} V
\eqno(F.1)$$
where, as above, $U = {\rm diag}\, (u^1, \dots, u^n)$ is a
constant diagonal matrix with $u^i \neq u^j$ for $i\neq j$ and
$V$ is a skew-symmetric matrix. In this Appendix we will explain how
to relate the {\it nonlinear monodromy} of the isomonodromy deformations
of the operator (F.1) around the fixed critical locus $u^i=u^j$ for some
$i\neq j$ with the linear monodromy of the operator.

We define first {\it Stokes factors}
of the operator (see [13]).

For any ordered pair $i\, j$
with $i\neq j$ we define the {\it Stokes ray} $R_{ij}$
$$R_{ij} = \left \{ z = - i r \left( \bar u^i - \bar u^j\right)
| r\geq 0\right \} .
\eqno(F.2)$$
Note that the ray $R_{ji}$ is the opposite to $R_{ij}$. The line
$R_{ij}\cup R_{ji}$ divides ${\bf C}$ into two half-planes
$P_{ij}$ and $P_{ji}$ where the half-plan $P_{ij}$ is on the left
of the ray $R_{ij}$. We have
$$|e^{z u^i}| > |e^{zu^j}| ~{\rm for} ~ z\in P_{ij}.
\eqno(F.3)$$
{\it Separating rays} are those who coincide with some of the Stokes
rays.

Let $l$ be an oriented line going through the origin not containing
Stokes rays. It divides ${\bf C}$ into two half-planes $P_\ll$ and $P_\rr$.
We order the separating rays $R_1, \dots, R_{2m}$ starting with the first
one in $P_\rr$. In the formulae below the numbers of the separating rays
will be considered modulo $2m$.

Let $\Psi_j$ be the matrix solution of the equation
$$\Lambda \Psi = 0
\eqno(F.4)$$
uniquely determined by the asymptotic
$$\Psi = \left( 1 + O\left( {1\over z}\right)\right)e^{zU}
\eqno(F.5)$$
in the sector from $R_j e^{-i\epsilon/2}$ to
$R_{m+j} e^{- i \epsilon}$. This can be extended analytically into the open
sector from $R_{j-1}$ to $R_{m+j}$. On the intersection of two
such subsequent sectors we have
$$\Psi_{j+1} = \Psi_j K_{R_j}
\eqno(F.6)$$
for some nondegenerate matrix $K_{R_j}$.

For a given choice of the oriented line $l$ we obtain thus a matrix
$K_R$ for any separating ray $R$. These matrices will be called
{\it Stokes factors} of the operator $\Lambda$.
\smallskip
{\bf Lemma F.1}. {\it The matrices $K = K_R$ satisfy the conditions
$$K_{ii} = 1, ~i = 1, \dots, n, ~ K_{ij}\neq 0 ~{\rm only ~for }~
R_{ij}\subset R
\eqno(F.7a)$$
$$K_{-R} = K^{-T}_R.
\eqno(F.7b)$$
The solutions $\Psi_\rr$ and $\Psi_\ll$ of Lecture 3 have the form
$$\Psi_\rr = \Psi_1, ~~\Psi_\ll = \Psi_{m+1}.
\eqno(F.8)$$
The Stokes matrix $S$ is expressed via the Stokes factors in the
form
$$S = K_{R_m}K_{R_{m-1}}\dots K_{R_1}.
\eqno(F.9)$$
Conversely, for a given configuration of the line $l$ and of the
Stokes rays
the Stokes factors of the form (F.7) are uniquelly
determined from the equation (F.9).}

Proof of all of the statement of the lemma but (F.7b) can be found in [13].
The relation (F.7b) follows from the skew-symmetry of $V$ as in
Proposition 3.10.
\medskip
{\bf Example F.1.} For generic $u_1, \dots, u_n$ all the Stokes rays
are pairwise distinct. Then the Stokes factors have only one nonzero
off-diagonal element, namely
$$\left(K_{R_{ij}}\right)_{ij}\neq 0.
\eqno(F.10)$$
\medskip
Let the matrix $V = V(u)$ depend now on $u = (u^1, \dots ,u^n)$ in such
a way that
small deformations of $u$ are isomonodromic. After a large deformation
the separating rays could pass through the line $l$. The correspondent
change of the Stokes
matrix is described by the following
\smallskip
{\bf Corollary F.1.} {\it If a separating ray $R$ passes through the positive
half-line $l_+$ moving clockwise then the solutions $\Psi_\rr$,
$\Psi_\ll$ and the Stokes matrix $S$ are transformed as follows
$$\Psi_\rr = \Psi_\rr ' K_R, ~\Psi_\ll '= \Psi_\ll K_R^T, ~
S' = K_R S K_R^T.
\eqno(F.11)$$
}
\medskip
{\bf Remark F.1.} A similar statement holds as well without skew-symmetry
of the matrix $V$. Instead of the matrices $K_R$ and $K_R^T$ in the formulae
there will be two independent matrices $K_R$ and $K_{-R}^{-1}$.
\medskip
Particularly, let us assume that the real parts of $u^i$ are pairwise distinct.
We order them such that
$${\rm Re}\, u^1 < \dots < {\rm Re}\, u^n.
\eqno(F.12)$$
The real line with the natural orientation will be chosen
as the line $l$. The correspondent Stokes matrix will be upper triangular
for such a choice. Any closed path  in the space of pairwise distinct
ordered
parameters $u$ determines a transformation of the Stokes matrix $S$
that can be read of (F.11) (just permutation of the Stokes factors).
We obtain an action of the pure braid group on the space of the Stokes
matrices. We can extend it onto all the braid group adding permutations
of $u_1, \dots, u_n$.
Note that the eigenvalues of the matrix $S^T S^{-1}$ (the monodromy
of (F.1) in the origin) are preserved by the action (F.11).
\smallskip
{\bf Proposition F.1.} {\it The solution $V(u)$ of the equations of the
isomonodromy
deformations of the operator $\Lambda$ with the given Stokes matrix $S$
is an algebraic function with branching along the diagonals  $u^i = u^j$
only if $S$ belongs to a finite orbit of the action of the pure
braid group.}

Proof. We know from Miwa's theorem [106] that the matrix function
$V(u)$ is meromorphic on the universal covering of
$CP^{n-1}\setminus \cup \{ u^i= u^j\}$.
Closed paths in the deformation space will interchange the branches
of this function. Due to the assumptions we will have only finite
number of branches. Proposition is proved.
\medskip
In the theory of Frobenius manifolds the parameters $u^i$ (i.e. the
canonical coordinates) are determined only up to a permutation.
So we obtain the action of the braid group $B_n$ on the space of
Stokes matrices. Explicitly, the standard generator $\sigma_i$ of $B_n$
($1 \leq i \leq n-1$) interchanging $u^i$ and $u^{i+1}$ moving $u^i$
clockwise around $u^{i+1}$ acts as follows
$$\Psi'_\rr = \Psi_\rr K^{-1}, ~\Psi'_\ll = \Psi_\ll K^T,~~ S' = K S K^T
\eqno(F.13)$$
where the matrix $K = K_i(S)$ has the form
$$K_{ss} = 1, s=1, \dots, ~s\neq i,~ K_{ii}
= 0, ~ K_{i\, i+1} = K_{i+1 \, i} = 1, ~ K_{i+1\, i+1} = - s_{i\, i+1}
\eqno(F.14)$$
other matrix entries of $K$ vanish.
\smallskip
{\bf Exercise F.1.} Verify that the braid
$$\left( \sigma_1 \dots \sigma_{n-1}\right)^{n}
\eqno(F.15)$$
acts trivially on the space of Stokes matrices.
\medskip
The braid (F.15) is the generator of the center of the braid group
$B_n$ for $n\geq 3$ [19]. We obtain thus an action of the quotient
$$A_n^* = B_n/{\rm center}
\eqno(F.16)$$
on the space of the Stokes matrices. Note that $A_n^*$ coincides
with the mapping class group of the plane with $n$ marked points [19].
For $n=3$ the group
$A_n^*$ is isomorphic to the modular group $PSL(2,{\bf Z})$.
\medskip
{\bf Example F.2.} For $n=3$ we put $s_{12} = x$, $s_{13} = y$,
$s_{23} = z$. The transformations of the braid group act as follows:
$$\sigma_1: ~(x,y,z) \mapsto (-x, z, y -xz),
\eqno(F.17)$$
$$\sigma_2: ~(x,y,z) \mapsto (y, x-yz, -z).
\eqno(F.18)$$
These preserve the polynomial
$$x^2+y^2+z^2 - x y z.
\eqno(F.19)$$
Indeed, the characteristic equation of the matrix $S^TS^{-1}$
has the form
$$(\lambda -1)[\lambda^2+(x^2+y^2+z^2-xyz-2)\lambda +1]=0.
\eqno(F.20)$$
For integer $x, y, z$ this action was discussed first by Markoff
in 1876 in the theory of Diophantine approximations [25]. The general action
(F.13), (F.14) (still on integer valued matrices)
 appeared also in the theory of exceptional vector bundles
over projective spaces [121]. Essentially it was also found from
physical considerations in [27] (again for integer matrices $S$)
describing \lq\lq braiding of Landau - Ginsburg superpotential".

It is clear that finite orbits of the full braid group $B_n$
must consist of finite orbits of the subgroup of pure braids.
So it is sufficient to find finite orbits of the action (F.13).

For $n=3$ I found 5 finite orbits of the braid group (F.17), (F.18).
These are the orbits of the points
$$(0,-1,-1), ~(0,-1 ,-\sqrt{2}), ~ (0, -1, -{\sqrt{5} + 1\over 2}),
{}~(0,{1-\sqrt{5}\over 2},-{1+\sqrt{5}\over 2}), ~(0,-1,{1-\sqrt{5}\over
2}).
$$
The orbits consist of 16, 36, 40, 72, and 40 points resp. The first
three orbits correspond to the Stokes matrices of the solutions
(A.7), (A.8), (A.9)
(i.e., to the tetrahedron, cube and icosahedron resp.). The nature of the
last two orbits is not clear (may be, they correspond to the two
pairs of the Kepler - Poinsot regular star-polyhedra in the
three-dimensional space).

It would be an interesting problem to classify periodic orbits of
the action (F.13), (F.14) of the braid group, and to figure out
what of them correspond to algebraic solutions of the
Painlev\'e-type equations (3.74).
\vfill\eject
\medskip
\centerline{\bf Appendix G.}
\smallskip
\centerline{\bf Monodromy group of a Frobenius manifold.}
\medskip
In Lecture 3 I have defined the intersection form of arbitrary Frobenius
manifold $M$. This is another flat (contravariant) metric $(~,~)^*$
on $M$ determined by the formula (3.13). In this Appendix we will study the
Euclidean structure on $M$ determined by the new metric.
Let us assume that the Frobenius manifold $M$ is {\it analytic}.
This means that the structure functions $c_{\alpha\beta}^\gamma(t))$
are analytic in $t$. As it follows from the resullts of Lecture 3
the assumption is not restrictive in the semisimple case.

The contravariant metric $(~,~)^*$ degenerates on the sublocus
where
the determinant
$$\Delta(t) := \det\left( g^{\alpha\beta}(t)\right).
\eqno(G.1)$$
vanishes.
Let $\Sigma \subset M$ be specified
by the equation
$$\Sigma := \left\{ t |\Delta(t):= \det (g^{\alpha\beta}(t) = 0\right\}.
\eqno(G.2)$$
This is a proper analytic subset in $M$. We will call it {\it
the discriminant locus} of the Frobenius manifold. The analytic function
$\Delta(t)$ will be called {\it the discriminant} of the manifold.

On $M\setminus \Sigma$ we have a locally Euclidean metric determined by the
inverse of the intersection form. This specifies an isometry
$$\Phi : \Omega\to \widehat{M\setminus\Sigma}
\eqno(G.3)$$
of a domain $\Omega$ in the standard $n$-dimensional (complex) Euclidean space
$E^n$ to the universal covering of $M\setminus\Sigma$. Action of the
fundamental
group $\pi_1(M\setminus\Sigma)$ on the universal covering can be lifted
to an action by isometries on $E^n$. We obtain a representation
$$\mu :\pi_1(M\setminus\Sigma)\to Isometries (E^n).
\eqno(G.4)$$
\smallskip
{\bf Definition G.1. } The group
$$W(M) := \mu\left(\pi_1(M\setminus\Sigma)\right)\subset Isometries (E^n)
\eqno(G.5)$$
is called {\it the monodromy group of the Frobenius manifold}.
\medskip
To construct explicitly the isometry (G.3) we are to fix a point $t\in
M\setminus\Sigma$ and to find the flat coordinates of the intersection
form in a neibourghood of the point.
The flat coordinates $x = x(t^1, \dots, t^n)$ are to be found from the system
of differential equations
$$\hat\nabla^\alpha\hat\nabla_\beta x := g^{\alpha\epsilon}(t)
\partial_\epsilon\partial_\beta x + \Gamma_\beta^{\alpha\epsilon}(t)\partial_
\epsilon x = 0, ~~\alpha, \beta = 1, \dots, n.
\eqno(G.6)$$
(Here $\hat\nabla$ is the Levi-Civit\`a connection for $(~,~)^*$; the
components of the metric and of the connection are given by the formulae
(3.15), (3.36).) This is an overdetermined holonomic system.
Indeed, vanishing of the curvature
of the intersection form (Proposition 3.2) provides compatibility of
the system. More precisely,
\smallskip
{\bf Proposition G.1.} {\it Near a point $t_0\in M$ where
$$\det \left( g^{\alpha\beta}(t_0)\right) \neq 0
$$
the space of solutions of the system (G.6) modulo constants has dimension
$n$. Any linearly independent (modulo constants) solutions
$x^1(t), \dots, x^n(t)$ of (G.6) can serve as local coordinates near
$t_0$. The metric $g^{\alpha\beta}(t)$ in these coordinates is
constant
$$g^{ab} := {\partial x^a\over \partial t^\alpha}
{\partial x^b\over \partial t^\beta} g^{\alpha\beta}(t) = const.
$$
}

This is a reformulation of the standard statement about the flat
coordinates of a zero-curvature metric.
\medskip
{\bf Exercise G.1.} For $d\neq 1$ prove that: \item{1.} $x_a(t)$ are
quasihomogeneous functions of $t^1$, ..., $t^n$ of the degree
$$\deg x_a(t) = {1-d\over 2};\eqno(G.7)$$
\item{2.} that
$$t_1 \equiv \eta_{1\alpha} t^\alpha = {1-d\over 4} g_{ab}x^a x^b
\eqno(G.8)$$
where $\left( g_{ab}\right) = \left( g^{ab}\right)^{-1}$.
\medskip
{\bf Example G.1.} For the two-dimensional Frobenius manifold
with the polynomial free energy
$$F(t_1,t_2) = \half t_1^2t_2 + t_2^{k+1}, ~~k\geq 2
\eqno(G.9)$$
the system (G.6) can be easily solved in elementary functions.
The flat coordinates $x$ and $y$ can be introduced in such a way that
$$\eqalign{t_1 & = 4\sqrt{k(k^2-1)}{\rm Re}\, (x+iy)^k\cr
t_2 & =x^2+y^2.\cr}
\eqno(G.10)$$
Thus the monodromy group of the Frobenius
manifold (G.9) is the group $I_2(k)$ of symmetries of the regular $k$-gon.

For the polynomial solutions (A.7), (A.8), (A.9)
the calculation of the monodromy group
is more involved. In the next Lecture we will see that the monodromy
groups of these three polynomial solutions of WDVV coincide with
the groups $A_3$, $B_3$,  $H_3$ of symmetries of the regular tetrahedron,
cube and icosahedron in the three-dimensional space.

More generally, for the Frobenius manifolds of Lecture 4
where $M = \Cc^n /W$ for a finite Coxeter group $W$,
the solutions of the system (G.6) are the Euclidean coordinates
in $\Cc^n$ as the functions on the space of orbits. If we
identify the space of orbits with the universal unfolding of
the correspondent simple singularity [5, 6, 23, 131] then the map
$$M\ni t\mapsto (x^1(t), \dots, x^n(t)) \in \Cc^n
\eqno(G.11)$$
coincides with the period mapping. The components $x^a(t)$
are sections of the bundle of vanishing cycles being locally
horizontal w.r.t. the Gauss -- Manin connection. Note that
globally (G.11) is a multivalued mapping. The multivaluedness
is just described by the action of the Coxeter group $W$ coinciding
with the monodromy group of the Frobenius manifolds.
\medskip
Basing on this example we introduce
\smallskip
{\bf Definition G.2.} The system (G.6) is called {\it Gauss -- Manin
equation} of the Frobenius manifold.
\medskip

Note that the coefficients of Gauss -- Manin
equation on an analytic Frobenius manifold also are
analytic in $t$. This follows
from (3.15), (3.36). However, the solutions may not be analytic
everywhere. Indeed, if we rewrite Gauss -- Manin equations
in the form solved for the second order derivatives
$$\partial_\alpha \partial_\beta x - \Gamma_{\alpha\beta}^\gamma
(t) \partial_\gamma x = 0
\eqno(G.12)$$
$$\Gamma_{\alpha\beta}^\gamma(t) := - g_{\alpha\epsilon}\Gamma^{\epsilon
\beta}_\gamma(t), ~~ \left( g_{\alpha\epsilon}\right) := \left(
g^{\alpha\epsilon}\right)^{-1}
$$
then the coefficients will have poles on the discriminant $\Sigma$.

So the solutions of Gauss -- Manin equation (G.6) are analytic in $t$
on $M\setminus \Sigma$. Continuation of some basic solution $x^1(t), \dots,
x^n(t)$ along a closed path $\gamma$ on $M\setminus \Sigma$ can give
new basis of solutions $\tilde x^1(t), \dots,
\tilde x^n(t)$. Due to Proposition G.1 it must have the form
$$\tilde x^a(t) = A_b^a(\gamma) x^b(t) + B^a(\gamma)
\eqno(G.13)$$
for some constants $A_b^a(\gamma), B^a(\gamma)$. The matrix $A_b^a
(\gamma)$ must be orthogonal w.r.t. the intersection form
$$A_b^a(\gamma) g^{bc} A_c^d(\gamma) = g^{ad}.
\eqno(G.14)$$
The formula (G.13) determines the representation (G.4) of the fundamental group
$\pi_1(M\setminus \Sigma, t)\ni\gamma$ to the group of isometries
of the $n$-dimensional
complex Euclidean space $E^n$. This is just the monodromy representation
(G.5).

By the construction
$$M \setminus\Sigma= \Omega/W(M).
$$
\smallskip
{\bf Proposition G.2.} {\it For $d\neq 1$ the monodromy group is
a subgroup in $O(n)$ (linear orthogonal transformations).}

Proof. Due to Exercise G.1 for $d\neq 1$ one can choose the
coordinates $x^a(t)$ to be invariant w.r.t. the scaling transformations
$$x^a(c^{\deg t^1} t^1, \dots, c^{\deg t^n} t^n) =
c^{1-d\over 2} x^a(t^1, \dots, t^n).
\eqno(G.15)$$
The monodromy preserves such an invariance. Proposition is proved.
\medskip
{\bf Example G.2.} If some of the scaling dimensions $q_\alpha = 1$
then the Frobenius structure may admit a discrete group of translations
along these variables. The Gauss - Manin equations then will be a system
with periodic coefficients. The correspondent monodromy transformation
(i.e. the shift of solutions of (G.6) along the periods) will contribute to
the monodromy group of the Frobenius manifold.

To see what happens for $d=1$ let us find the monodromy
group for the two-dimensional Frobenius manifold with
$$F = \half {t^1}^2 t^2 + e^{t^2}.
\eqno(G.16)$$
I recall that this describes the quantum cohomology of $CP^1$.
The manifold $M$ is the cylinder $\left( t^1, t^2 ({\rm mod} \, 2\pi
i)\right)$.
The Euler vector field is
$$E= t^1 \dela + 2 \delb .
\eqno(G.17)$$
The intersection form has the matrix
$$\left( g^{\alpha\beta}\right)
= \left( \matrix{ 2e^{t^2} & t^1 \cr t^1 & 2 \cr}\right) .
\eqno(G.18)$$
The Gauss -- Manin system reads
$$\eqalign{ 2 e^{t^2} \dela^2 x + t^1 \dela\delb x &= 0\cr
2 e^{t^2}\dela\delb x + t^1 \delb^2 x + e^{t^2} \dela x &= 0\cr
t^1 \dela^2 x + 2 \dela\delb x + \dela x &= 0\cr
t^1 \dela\delb x + 2 \delb^2 x &= 0.\cr}
\eqno(G.19)$$
The basic solutions are
$$\eqalign{ x^1 &= -i t^2\cr
x^2 &= 2\arccos \half t^1 e^{-{t^2\over 2}}.\cr}
\eqno(G.20)$$
The intersection form in these coordinates is
$$\half \left(- {dx^1}^2 + {dx^2}^2\right) .
\eqno(G.21)$$
The Euler vector field reads
$$E = {\partial\over\partial x^1}.
\eqno(G.22)$$
The discriminant locus is specified by the equation
$$t^1 = \pm 2 e^{t^2\over 2}
\eqno(G.23a)$$
or, equivalently
$$x^2 = 0, ~~x^2 = 2\pi .
\eqno(G.23b)$$
The monodromy group is generated by the transformations of the following
two types.

The transformations of the first type are obtained by continuation
of the solutions (G.20) along the loops around the discriminant locus.
This gives the transformations
$$\left( x^1, x^2 \right) \mapsto \left( x^1,
\pm x^2 + 2\pi n\right), ~~n \in {\bf Z}.
\eqno(G.24a)$$
This is nothing but the action of the simplest affine Weyl group
of the type $A_1^{(1)}$.
The transformations of the second type
$$\left( x^1, x^2\right) \mapsto \left( x^1 + 2\pi n, (-1)^n x^2\right),
{}~~n\in{\bf Z}
\eqno(G.24b)$$
are generated by the closed loops $t^2 \mapsto t^2 \exp 2\pi i n$ on $M$.
This gives an extension of the affine Weyl group (G.24a).
So $M$ is the quotient of $\Cc^2$ over the extended affine Weyl group
(G.24) (although $t^2$ is not a globally
single-valued function). The coordinate
$$t^1 = 2 e^{ix^1/2}\cos{x^2\over 2}
\eqno(G.25)$$
is the basic invariant of the group (G.24) homogeneous w.r.t.
the Euler vector field (G.22). That means that any other invariant
is a polynomial (or a power series) in $t^1$ with the coefficients
being arbitrary $2\pi i$-periodic functions in $t^2$.
 Another flat coordinate
$$t^2 =ix^1
$$ is invariant w.r.t. the affine Weyl group (G.24a) and it gets a shift
w.r.t. the transformations (G.24b).
\medskip
{\bf Example G.3.} The monodromy group of a Frobenius manifold can be
in principle
computed even if we do not know the structure of it using the isomonodromicity
property of the solutions of the Gauss - Manin system. For the example
of the $CP^2$ model (see above Lecture 2) it is enough to compute
the monodromy group on the sublocus $t^1 = t^3 = 0$. We obtain that the
monodromy of the $CP^2$-model is generated by the monodromy group of the
operator with rational coefficients
$$\left(\matrix{3q & 0 & \lambda\cr
0 & \lambda & 3\cr
\lambda & 3 & 0}\right) {\partial \xi\over \partial\lambda}
+ \left( \matrix{0 & 0 & -1/2\cr
0 & 1/2 & 0\cr
3/2 & 0 & 0\cr}\right) \xi = 0
\eqno(G.26a)$$
for
$$q = e^t, ~~ t = t_2
$$
and of the operator with $2\pi i$-periodic coefficients
$$\left(\matrix{3q & 0 & \lambda\cr
0 & \lambda & 3\cr
\lambda & 3 & 0}\right) {\partial \xi\over\partial t}
+\left( \matrix{ {3\over 2} q & 0  & 0\cr
0 & 0 & -1/2\cr
0 & 1/2 & 0\cr}\right) \xi = 0.
\eqno(G.26b)$$
The last one can be also reduced to an operator with rational coefficients
by the substitution $t\to q$.
It is still an open problem to solve these equations
and to compute the monodromy.
\medskip
The monodromy group can be defined also for twisted Frobenius manifolds.
Particularly, if the inversion (B.11) acts as a conformal
transformation of the intersection form.
We will see in Appendix J an example of such a situation.
\medskip

Another important sublocus in a Frobenius manifold satisfying the
semisimplicity condition is {\it the nilpotent locus}
$\Sigma_{\rm nil}$ consisting of all the points $t$ of $M$ where
the algebra on $T_tM$ is not semisimple. According to Proposition 3.3
the nilpotent discriminant is contained in the discriminant locus
of the polynomial
$$\det \left( g^{\alpha\beta}(t) - \lambda \eta^{\alpha\beta}\right)
= \Delta (t^1-\lambda, t^2,\dots, t^n).
\eqno(G.27)$$
The discriminant $\Delta_{\rm nil}(t)$
of (G.27) as of the polynomial on $\lambda$ will be called
{\it nilpotent discriminant}.
\smallskip
{\bf Example G.4.} For the Frobenius manifold of Example 1.7
the discriminant locus
$$\Sigma = \left\{ {\rm set~ of~ polynomials~} \lambda(p) ~{\rm having~
a~ critical~ value~} \lambda = 0\right\} .
\eqno(G.28)$$
The discriminant $\Delta(t)$ coincides with
the discriminant
of the polynomial $\lambda(p)$ (I recall that the coefficients of the
polynomial $\lambda(p)$ are certain functions on $t$). The nilpotent
locus of $M$  is the {\it caustic} (see [6])
$$\Sigma_{\rm nil} = \left\{ {\rm set~ of~ polynomials~} \lambda(p)
{}~{\rm with~ multiple~ critical~ points~} p_0,~\right.
$$
$$\left.\lambda^{(k)}(p_0)=0 ~{\rm for ~} k= 1, 2, \dots, k_0\geq 2\right\} .
\eqno(G.29)$$
The nilpotent discriminant is a divisor of
the discriminant of the
polynomial ${\rm discr}_p(\lambda(p) - \lambda)$ as of the polynomial
in $\lambda$.
\medskip
On the complement $M\setminus \Sigma$ a metric
$$ds^2 := g_{\alpha\beta}(t)dt^\alpha dt^\beta
\eqno(G.30)$$
is well-defined. Here the matrix $g_{\alpha\beta}(t)$ is the inverse
to the matrix $g^{\alpha\beta}(t)$ (3.15). The metric
has a pole on the discriminant locus. We will show that the singularity
of the metric
can be eliminated after lifting to a covering of $M$.

Let $\hat M$ be the two-sheet covering of the Frobenius manifold $M$
ramifying along the discriminant locus
$$\hat M := \left\{ (w,t),~ w\in \Cc,~ t\in M | w^2 = \Delta(t)
\right\} .
\eqno(G.31)$$
We have a natural projection
$$\pi : \hat M \to M.
\eqno(G.32)$$
\smallskip
{\bf Lemma G.1.} {\it The pullback $\pi^* \, ds^2$ of the metric (G.30)
onto $\hat M$ is analytic on $\hat M \setminus (\Sigma\cap
\Sigma_{\rm nil})$.
}

Proof. Outside of $\Sigma$
on $M$ we can use the canonical coordinates.
In these the metric has the form
$$ds^2 = \sum_{i=1}^n {\eta_{ii}(u)\over u^i} (du^i)^2.
\eqno(G.33)$$
The canonical coordinates serve near a point $t_0\in M\setminus (\Sigma\cap
\Sigma_{\rm nil})$ either, but some of them vanish. The vanishing is
determined by a splitting $(i_1,\dots, i_p)\cup (j_1,\dots, j_q)
= (1, 2, \dots, n)$, $p+q = n$ such that
$$u^{i_1}(t_0)=0, \dots, u^{i_p} (t_0) = 0, ~~
u^{j_i}(t_0)\neq 0, \dots, u^{j_q}\neq 0.
\eqno(G.34)$$
The local coordinates near the correspondent point $\pi^{-1}(t_0)
\in \hat M$ are
$$\sqrt{u^{i_1}}, \dots, \sqrt{u^{i_p}}, u^{j_1}, \dots, u^{i_q}.
\eqno(G.35)$$
Rewriting (G.33) near the point $t_0$ as
$$ds^2 = 4 \sum_{s=1}^p \eta_{i_si_s}(u)(d\sqrt{u^{i_s}})^2
+ \sum_{s=1}^q \eta_{j_sj_s}(u) (du^{j_s})^2
\eqno(G.36)$$
we obtain analyticity of $ds^2$ on $\hat M$. Lemma is proved.
\medskip
Due to Lemma G.1 the flat coordinates $x^a(t)$ as the functions
on $\hat M$ can be extended to any component of $\Sigma\setminus
\Sigma_{\rm nil}$. The image
$$\left( x^1(t), \dots, x^n(t)\right)_{t\in \Sigma\setminus
\Sigma_{\rm nil}}
\eqno(G.37)$$
is the discriminant locus written in the flat coordinates
$x^1, \dots, x^n$.
\smallskip
{\bf Lemma G.2.} {\it Any component of $\Sigma\setminus
\Sigma_{\rm nil}$ in the coordinates $x^1(t), \dots, x^n(t)$
is a hyperplane.}

Proof. We will show that the second fundamental form of the hypersurface
$\Delta(t) = 0$ in $\hat M$ w.r.t. the metric (G.30) vanishes. Near
$\Sigma\setminus
\Sigma_{\rm nil}$ we can use the canonical coordinates $u^1, \dots, u^n$.
Let $\Sigma\setminus
\Sigma_{\rm nil}$ be specified locally by the equation $u^n = 0$.
Let us first calculate the second fundamental form of the
hypersurface
$$u^n = u^n_0 \neq 0.
\eqno(G.38)$$
The unit normal vector to the hypersurface is
$$N = \sqrt{u^n_0\over \eta_{nn}} \partial_n.
\eqno(G.39)$$
The vectors $\deli, ~i\neq n$ span the tangent plane to the hypersurface.
The second fundamental form is
$$b_{ij} := \left( \hat\nabla_i\delj , N\right) =
\sqrt{\eta_{nn}\over u^n_0} \Gamma_{ij}^n, ~~1\leq i, j \leq n-1
$$
$$= - {\delta_{ij}\over 2} \sqrt{u^n_0\over \eta_{nn}} \partial_{n}
\left({\eta_{ii}\over u^i}\right) = -\delta_{ij}
\sqrt{u^n_0\over \eta_{nn}}{\psi_{i1}\psi_{n1}\gamma_{in}
\over u^i}
\eqno(G.40)$$
in the notations of Lecture 3 (but $\Gamma_{ij}^k$ here are the Christoffel
coefficients for the intersection form). This vanishes when
$u^n_0\to 0$. Lemma is proved.
\medskip
I recall that a linear
orthogonal transformation $A: \Cc^n \to \Cc^n$ of a complex Euclidean space
is called {\it reflection} if
$A^2 = 1$ and $A$ preserves the points of a hyperplane in $\Cc^n$.
\smallskip
{\bf Theorem G.1.} {\it For $d\neq 1$ the monodromy along a small loop
around the discriminant on an
analytic Frobenius manifold satisfying
semisimplicity condition is
a reflection.}

Proof. Since the flat coordinates are analytic and single valued
on the two-sheet covering $\hat M$ the monodromy transformations $A$
along loops around $\Sigma$ are involutions, $A^2=1$.They preserves
the hyperplanes (G.28). Note that the hyperplanes necessary pass
through the origin for $d\neq 1$. Theorem is proved.
\medskip
In Lecture 4 we will show that any finite reflection group arises
as the monodromy group of a Frobenius manifold. This gives a very
simple constructionn of polynomial Frobenius manifolds with $d<1$.
Similarly,
the construction of Example G.2
can be generalized to arbitrary affine Weyl groups (properly extended).
This gives a Frobenius structure with
$d=1$ and linear nonhomogeneous Euler vector field $E$ on their  orbit spaces
(i.e. ${\cal L}_E t^n \neq 0$).
We will consider the details of this construction in a separate
publication. Finally, in Appendix  J we will construct a twisted
Frobenius manifold whose monodromy is a simplest extended complex
crystallographic
group. This will give Froobenius manifolds again with $d=1$ but with
linear homogeneous Euler vector field (i.e. ${\cal L}_E t^n = 0$).
\vfill\eject
{\bf Appendix H.}
\smallskip
\centerline{\bf Generalized hypergeometric equation}
\smallskip
\centerline{\bf associated to a Frobenius manifold and its monodromy.}
\medskip
The main instrument to calculate the monodromy of a Frobenius
manifold coming from the loops winding around the discriminant
is a differential equation with rational coefficients wich we are going to
define now.  This will be the equation for the flat coordinates
of the linear pencil of the metrics
$$(~,~)_\lambda^* := (~,~)^* - \lambda <~,~>^* = \left(g^{\alpha\beta}
-\lambda\eta^{\alpha\beta}\right)
\eqno(H.1)$$
as the functions of the
parameter $\lambda$. We will obtain also an integral transform
relating the flat coordinates of the deformed connection (3.3) and
the flat coordinnates of the deformed metric (H.1).

Let $x_1 = x_1(t)$, ..., $x_n = x_n(t)$ be the orthonormal flat
coordinates for the metric $g^{\alpha\beta}(t)$.
The flat coordinates for the flat pencil H.1 can be construct easily.
\smallskip
{\bf Lemma H.1.} {\it The functions
$$\tilde x_a(t,\lambda) := x_a(t^1-\lambda ,t^2,\dots, t^n),
{}~~a=1,\dots ,n
\eqno(H.2)$$
are the flat coordinates for the metric $(~,~)^* - \lambda <~,~>^*$.
}

Proof. The linear combination (H.1) can be written in the form
$$g^{\alpha\beta}(t) - \lambda \eta^{\alpha\beta} =
g^{\alpha\beta}(t^1 - \lambda, t^2, \dots, t^n).
\eqno(H.3)$$
Lemma is proved.
\medskip
\smallskip
{\bf Corollary H.1.} {\it The gradient $\xi_\epsilon := \partial_\epsilon
x(\lambda , t)$ of the flat coordinates of the pencil
$g^{\alpha\beta}(t) - \lambda \eta^{\alpha\beta}$ satisfies the following
system of linear differential equations in $\lambda$
$$\left(\lambda\eta^{\alpha\epsilon} - g^{\alpha\epsilon}(t)\right)
{d\over d\lambda} \xi_\epsilon = \eta^{\alpha\epsilon}\left(-\half
+ \mu_\epsilon
\right) \xi_\epsilon .
\eqno(H.4)$$
}

Proof. We have from (3.38)
$$\Gamma_1^{\alpha\epsilon} = \left({d+1\over 2}-q_\epsilon\right)
c_1^{\alpha\epsilon} = \left(\half - \mu_\epsilon\right) \eta^{\alpha
\epsilon}.
\eqno(H.5)$$
So the equation (G.6) for $\beta = 1$ reads
$$\left( g^{\alpha\epsilon}(t) -\lambda \eta^{\alpha\epsilon}\right)
\partial_1\partial_\epsilon x + \eta^{\alpha\epsilon}
\left( \half - \mu_\epsilon\right) \partial_\epsilon x = 0.
\eqno(H.6)$$
Due to Lemma H.1 we have
$$\partial_1 = - {d\over d\lambda}.
\eqno(H.7)$$
Corollary is proved.
\medskip
The equation (H.4) is a system of linear ordinary differential equations
with rational coefficients depending on the parammeters
$t^1, \dots, t^n$. The coefficients have poles on the shifted
discriminant locus
$$\Sigma_\lambda := \left\{ t~|\Delta(t^1-\lambda, t^2,\dots, t^n) =0
\right\} .
\eqno(H.8)$$
\smallskip
{\bf Lemma H.2.} {\it Monodromy of the system (H.4) of differential
equations with rational coefficients around $\Sigma_\lambda$ coincides
with the monodromy of the Frobenius manifold around the discriminant
$\Sigma$. The monodromy does not depend on the parameters $t$.}

Proof. The first statement is obvious. The second one follows from the
compatibility of the equations (G.6) with the equations in $\lambda$ (H.4).
\medskip
{\bf Definition H.1.} The differential equation (H.4) with rational
coefficients
will be called {\it generalized hypergeometric equation
associated with the Frobenius manifold}.
\medskip
In this definition we are motivated also by [115] wher it was shown
that the differential equations for the functions $_nF_{n-1}$ are
particular cases of (H.4) (however, in general without the skew-symmetry
of the matrix $V$).

We construct now an integral transform relating the
flat coordinates $\tilde t(t,z)$ of
the deformed connection (3.3) and the flat coordinates $x(t,\lambda)$ of the
pencil of metrics (H.1).
We recall that the coordinates $x(t,\lambda)$ are the solutions of the
differential
equations
$$\left(g^{\alpha\epsilon}-\lambda\eta^{\alpha\epsilon}\right)
\partial_\epsilon\dbeta x +\Gamma_\beta^{\alpha\epsilon}
\partial_\epsilon x = 0.
\eqno(H.9)$$
\smallskip
{\bf Proposition H.1.} {\it Let $\tilde t(t,z)$ be a flat coordinate of the
deformed connection (3.3) normalized by the condition
$$z\partial_z \tilde t = {\cal L}_E \tilde t.
\eqno(H.10)$$
Then the function
$$x(t,\lambda) := \oint z^{d-3\over 2} e^{-\lambda z}
\tilde t(t,z)\, dz
\eqno(H.11)$$
is a flat coordinate for the pencil (H.1).
}

Here the integral is considered along any closed loop in the extended complex
plane $z\in\Cc\cup \infty$. We will specify later how to choose the contour
of the integration to obtain a well-defined integral.

Proof is based on the following
\smallskip
{\bf Lemma H.3.} {\it The following identity holds true
$$\left( dt^\alpha , \hat\nabla_\gamma d\tilde t\right)^* dt^\gamma
= z^{-1} d\left[ <dt^\alpha, \left( z\partial_z +{d-3\over 2}\right) d\tilde
t>^*
\right].
\eqno(H.12)$$
Here $\tilde t = \tilde t(t,z)$ is a flat coordinate of the deformed
affine connection (3.3), $\hat\nabla$ is the Levi-Civit\`a
connection for the intersection form.
}

Proof. The l.h.s. of (H.12) reads
$$\left( g^{\alpha\sigma}\partial_\sigma\dgamma\tilde t
+ \Gamma_\gamma^{\alpha\sigma}\partial_\sigma\tilde t\right) dt^\gamma =
$$
$$= \left( z\sum_\rho E^\rho c_\rho^{\alpha\sigma}c_{\sigma\gamma}^\nu
\partial_\nu\tilde t + \sum_\sigma \left({d+1\over 2}-q_\sigma\right)
c^{\alpha\sigma}_\gamma \partial_\sigma\tilde t\right) dt^\gamma.
$$
On the other side, using the equation (H.10) we obtain
$$\partial_\gamma\left( z\partial_z + {d-3\over 2}\right)\partial_\epsilon
\tilde t = z^2c_{\epsilon\gamma}^\sigma U_\sigma^\nu \partial_\nu\tilde t
+ z c_{\epsilon\gamma}^\sigma \partial_\sigma \tilde t
+ z\sum_\sigma c_{\epsilon\gamma}^\sigma
(1-q_\sigma)\partial_\sigma\tilde t +
z {d-3\over 2} c_{\epsilon\gamma}^\sigma\partial_\sigma \tilde t =
$$
$$= z^2 E^\rho c_{\rho\sigma}^\nu c_{\epsilon\gamma}^\sigma \partial_\nu
\tilde t + z \sum_\sigma c_{\epsilon\gamma}^\sigma \left({d+1
\over 2} - q_\sigma\right) \partial_\sigma \tilde t.
$$
Multiplying by $\eta^{\alpha\epsilon}$ and using the associativity condition
$$c_{\epsilon\gamma}^\sigma c_{\rho\sigma}^\nu =
c_{\rho\epsilon}^\sigma c_{\sigma\gamma}^\nu
$$
we obtain, after multiplication by $dt^\gamma $ and division over $z$,
the expression (H.12). Lemma is proved.
\medskip
Proof of Proposition. For the function $x = x(t,\lambda)$ of the form (H.11)
we obtain, using Lemma and integrating by parts
$$\left( dt^\alpha, \hat\nabla_\gamma dx\right)^* dt^\gamma
= dt^\gamma \oint z^{d-3\over 2} e^{-\lambda z} \left( dt^\alpha ,
\hat\nabla_\gamma d\tilde t\right)^* dz =
$$
$$=d_t\oint z^{d-3\over 2}e^{-\lambda z} <dt^\alpha , \partial_z d\tilde t
+ {d-3\over 2 z} d\tilde t>^* dz +
$$
$$= d_t\left\{ \oint\left( \lambda z^{d-3\over 2}e^{-\lambda z}
-{d-3\over 2} z^{d-5\over 2} e^{-\lambda z}\right) <dt^\alpha,
d\tilde t>^* dz+ {d-3\over 2}\oint z^{d-5\over 2} e^{-\lambda z}
<dt^\alpha, d\tilde t>^*dz\right\} =
$$
$$= \lambda d_t \oint z^{d-3\over 2} e^{-\lambda z} <dt^\alpha, d\tilde t
>^*dz = \lambda \eta^{\alpha\epsilon}\partial_\epsilon\dgamma x\, dt^\gamma.
$$
So $x$ satisfies the differential equation (H.9).
Proposition is proved.
\medskip
We study now the monodromy of our  generalized hypergeometric equation (G.6)
in a neiborghood of a semisimple point $t\in M$. First we rewrite the
differential
equations (3.5) and (H.9) and the integral transform (H.11) in the canonical
coordinates $u^i$.
\smallskip
{\bf Proposition H.2.} {\it Let $x = x(t,\lambda)$ be a flat coordinate
of the metric (H.1). Put
$$\phi_i(u,\lambda) :=\deli x(t,\lambda) /\sqrt{\eta_{ii}(u)}, ~~t = t(u).
\eqno(H.13)$$
The vector-function $\phi = (\phi_1, \dots, \phi_n)^T$,
$\phi_i = \phi_i(u,\lambda)$ satisfies the system
$$\eqalignno{\delj \phi_i & = \gamma_{ij}\phi_j, ~i\neq j
&(H.14a)\cr
\sum_{{k=1}}^n (u^k - \lambda ) \delk \phi_i & = -\half \phi_i
&(H.14b)\cr}
$$
and also
the following differential equation in $\lambda$
$$\left(\lambda \cdot 1 - U\right) {d\over d\lambda}\phi
= -\left( \half \cdot 1 + V(u)\right) \phi
\eqno(H.15)$$
where $U = {\rm diag}\, (u^1, \dots, u^n)$ and $V(u) = \left(
(u^j - u^i) \gamma_{ij}(u)\right)$ (cf. (3.146) above).

If $\psi = (\psi_1, \dots, \psi_n)^T$ is a solution to the linear system
(3.118), (3.122) then
$$\phi(u,\lambda) = \oint e^{-\lambda z} \psi(u,z) {dz\over \sqrt{z}}
$$
satisfies the system (H.9).
}

The proof is omitted.

We consider now the case where $0\leq q_\alpha \leq d <1$. In this case
instead of
the loop integrals (H.11) (or (H.16)) it's better to use more convenient
Laplace
integrals. We will use these Laplace integrals to express the monodromy
of our generalized hypergeometric equation in terms of the Stokes
matrix of the Frobenius manifold.

Let $\Psi (u,z) = \left( \psi_{i a}(u,z)\right)$, $i, a = 1, \dots , n$
be a solution of the equation (3.122) analytic in a half-plane. Let us assume
that
$$0\leq q_\alpha \leq d < 1.
\eqno(H.17)$$
We construct
the functions $x_a(u,\lambda) $ taking the Laplace transform of these
solutions:
$$\deli x_a(u,\lambda) =
\sqrt{\eta_{ii}(u)}\hat\psi_{ia}(u,\lambda)
\eqno(H.18a)$$
where
$$\hat\psi_{i a}(u,\lambda):= {1\over\sqrt{-2\pi}}
\int_0^\infty e^{-\lambda z}\psi_{i a}(u,z)
{dz\over\sqrt z}.
\eqno(H.18b)$$
We can normalize them uniquelly by the homogeneity requirement
$$\left(\lambda {d\over d\lambda} - {\cal L}_E\right)
x_a(u,\lambda) = {1-d\over 2} x_a(u,\lambda).
\eqno(H.19)$$
\smallskip
{\bf Theorem H.1.} {\it Functions $x_a(u,\lambda)$ are flat coordinates
of the flat metric $(~,~)^* - \lambda <~,~>^*$.}

Proof coincides with the proof of Proposition H.2
(due to the inequalities (H.17)
the boundary terms at $z=0$ vanish).
\medskip
{\bf Corollary H.2.} {\it The intersection form
$$g^{ab} := (dx_a(u,\lambda), dx_b(u,\lambda))_\lambda ^*
= \sum_{i=1}^n(u^i-\lambda ) \hat\psi_{ia}(u,\lambda)\hat\psi_{ib}(u,\lambda)
\eqno(H.20)$$
does not depend on $\lambda$ neither on $u$.
}
\medskip
The coordinates are multivalued analytic functions of $\lambda$. They have
also singularities at the points $\lambda = u^i$. The monodromy of these
functions coincide with the monodromy of the differential operator
(H.15) with regular singular points at $\lambda = u^i$, $i=1, \dots, n$
and $\lambda = \infty$. We will calculate now this monodromy in terms
of the Stokes matrix of the original operator.

To calculate the monodromy I will use the following elementary way
of analytic continuation of Laplace transforms of a function
analytic in a halfplane.
\smallskip
{\bf Lemma H.4.} {\it Let the function $\psi(z)$ be analytic in the
right halfplane and
$$\eqalign{|\psi(z)| & \to 1~ {\rm for}~ z \to \infty\cr
z|\psi(z)| & \to 0~{\rm for}~ |z|\to 0\cr}
\eqno(H.21)$$
uniformly in the sector $-{\pi\over 2} +\epsilon
\leq \arg z \leq {\pi\over 2} - \epsilon$ for arbitrary small $\epsilon > 0$.
Then the Laplace transform
$$\hat\psi(\lambda) := \int_0^\infty e^{-\lambda z}\psi(z)\, dz
\eqno(H.22)$$
can be analytically continued in the complex $\lambda$-plane
with a cut along the negative real half-line.}

Proof. $\hat\psi(\lambda)$ is an analytic function in the right half-plane
${\rm Re} \lambda >0$. Let us show that for these $\lambda$ the equality
$$\hat\psi(\lambda) = \int_0^\infty e^{-\lambda z e^{i\alpha}}
\psi(z e^{i\alpha})\, d(z e^{i\alpha})
\eqno(H.23)$$
holds true for any $\alpha$ such that
$$-{\pi\over 2} - \arg \lambda < \alpha < {\pi\over 2} - \arg \lambda .
$$
Indeed, let us consider the contour integral
$$\oint_C e^{-\lambda z} \psi(z) \, dz.
$$
Integrals along the arcs tend to zero when $r\to 0$, $R\to\infty$
(see Fig.15). In the limit we obtain (H.23).
\vskip 2cm
\centerline{Fig.15}
\vskip 2cm
Now observe that the r.h.s. of (H.23) is analytic in the halfplane
$$-{\pi\over 2}-\alpha < \arg \lambda <{\pi\over 2}-\alpha .
$$
Varying $\alpha$ from $-{\pi\over 2} + 0$ to ${\pi\over 2}- 0$ we obtain
the needed analytic continuation.
\medskip
Let us fix some oriented line $l = l_+ \cup l_-$ not containing the separating
rays of the operator (3.120). Let $\Psi^{\rr\,/ \ll}(u,z) = \left( \psi_{i a}^
{\rr\, /\ll}(u,z)\right)$ be the canonical solutions (3.128) of (3.122)
in the correspondent half-planes. Their Laplace transforms will be
defined by the integrals
$$\hat\psi_{i a}^\rr (u,\lambda) = {1\over\sqrt{-2\pi}}
\int_{i l_-}e^{-\lambda z}
\psi _{i a}^\rr(u,z){dz\over \sqrt z}
\eqno(H.24)$$
(analytic function outside the cut $u_a+i \bar l_+$) where we chose the branch
of $\sqrt z$ with the cut along $l_-$, and
$$\hat\psi_{i a}^\ll (u,\lambda) = {1\over\sqrt{-2\pi}}
\int_{i l_+} e^{-\lambda z}\psi_{ia}^\ll
(u,z) {dz\over \sqrt{z}}
\eqno(H.25)$$
(analytic in $\lambda$ outside the cut $u_a+i \bar l_+$). By
$x_a^\rr (u,\lambda)$
and $x_a^\ll (u,\lambda)$ we denote the correspondent coordinates (H.18).
By
$$A := S + S^T
\eqno(H.26)$$
we denote the symmetrized Stokes matrix. Note that the rays $u_1+i\bar l_+$,
\dots, $u_n+i\bar l_+$ are pairwise distinct (this is equivalent to
nonintersecting
of the Stokes rays (F.2) with the line $l$). We chose generators $g_a$ in the
fundamental
group $\pi_1(\Cc\setminus (u^1\cup \dots \cup u^n)) $ taking the loops going
from
$\infty$
along these rays to $u^a$ then around $\lambda = u^a$ and then back to infinity
along
the same ray. The monodromy group of the differential equation (H.15) w.r.t.
the chosen basis
of the fundamental group is described by
\smallskip
{\bf Theorem H.2.} {\it Monodromy of the functions $x^\rr_1(u,\lambda)$, \dots,
$x_n^\rr (u,\lambda)$ around the point $\lambda = u_b$ is the reflection
$$\eqalign{x_a^\rr (u,\lambda) & \mapsto x_a^\rr (u,\lambda) ~{\rm for}~
a\neq b\cr
x_b^\rr (u,\lambda) &\mapsto x_b^\rr (u,\lambda) - \sum_{a=1}^n A_{ba}
x_a^\rr (u,\lambda).\cr}
\eqno(H.27)$$
}
{\bf Remark.} Monodromy at infinity is specified by the matrix
$$- T = -S^T S^{-1}.
\eqno(H.28)$$
This is in agreement with the Coxeter identity [32]
for the product of the reflections (H.27):
$$T_1 \dots T_n = - S^T S^{-1}
\eqno(H.29)$$
where $T_b$ is the matrix of the reflection (H.27).
\smallskip
Proof. When $\lambda$ comes clockwise/counter-clockwise to the cut
$u_a+i\bar l_-$ the ray of integration in the Laplace integral
(H.24) for $b=a$ (only!) rotates counter-clockwise/clockwise to
$l_+/l_-$. To continue the integrals through the cut we express them
via $x_b^\ll (u,\lambda)$ using the formula (3.131). Since the functions
$x_b^\ll (u,\lambda)$ and the functions $x_a^\rr (u,\lambda)$ for $a\neq b$
have no jump on the cut $u_a + i\bar l_-$ we obtain the monodromy
transformation
$$x_b^\rr + \sum_{R_{ab}\subset P_\rr} s_{ab} x_a^\rr
\mapsto -\left( x_b^\rr + \sum_{R_{ab}\subset P_\ll}
s_{ba} x_a^\rr\right)
$$
the sign \lq\lq $-$" is due to the change of the branch of $\sqrt{z}$ when
the ray of the integration is moving through $l_-$. This coincides with (H.27).
Theorem is proved.
\medskip
We can construct another system of flat coordinates using the Laplace
transform of the columns of the matrix $\Psi_0(u,z) =\Psi_\rr (u,z) C$.
I recall that the matrix $C$ consists of the eigenvectors of $S^TS^{-1}$
$$S^TS^{-1}C = C e^{2\pi i \mu}.
\eqno(H.30)$$
Here we introduce a diagonal matrix
$$\mu = {\rm diag}\, (\mu_1, \dots, \mu_n).
\eqno(H.31)$$
The numbers $\mu_\alpha = q_\alpha - {d\over 2}$ are ordered
in such a way that
$$\mu_\alpha + \mu_{n-\alpha +1} = 0.
\eqno(H.32)$$
This gives a useful identity
$$\mu\eta + \eta\mu = 0.
\eqno(H.33)$$
Note that the case $0\leq {\rm Re} \, q_\alpha \leq{\rm Re} \,
d<1$ corresponds to
$$-{1\over 2} < \mu_\alpha < {1\over 2}.
\eqno(H.34)$$

We normalize the eigenvectors (H.30) in such a way that the entries
of the matrix $\Psi_0(u,z) = \left( {\psi_0}_{i\alpha}(u,z)\right)$
have the following expansions near the origin
$${\psi_0}_{i\alpha}(u,z) = z^{\mu_\alpha} \left( \psi_{i\alpha} (u) + O(z)
\right) ~{\rm when} ~ z\to 0, ~ z\in P_\rr .
\eqno(H.35)$$
We continue analytically the matrix $\Psi_0(u,z)$ in the left half-plane
$z\in P_\ll$ with a cut along the ray $il_+$.
\smallskip
{\bf Lemma H.5.} {\it Under the assumption (H.34) and the normalization (H.35)
the
matrix $C$ satisfies the relations
$$C\eta e^{\pi i \mu} C^T = S, ~~C \eta e^{-\pi i \mu} C^T = S^T.
\eqno(H.36)$$
}

Proof. We use the identity
$$\Psi_\rr (u,z) \Psi_\ll^T (u,-z) =1
\eqno(H.37)$$
(see above). Let $z$ belong to the sector from $il_-$ to $l_+$. Then
$-z = z e^{-\pi i}$ belongs to the sector from $i l_+$ to $l_-$. For such $z$
we have
$$\Psi_0(u,ze^{-\pi i}) =\Psi_\ll (u,-z)S^{-T} C.
\eqno(H.38)$$
Substituting (H.38) and (H.33) to (H.37), we obtain
$$\Psi_0(u,z) C^{-1} S C^{-T} \Psi_0^T(u,ze^{-\pi i}) = 1,
$$
or
$$C^{-1}SC^{-T} = \Psi_0^{-1}(u,z)\Psi_0^{-T}(u,ze^{-\pi i}).
\eqno(H.39)$$
Let $z$ tend to zero keeping it within the sector from $i l_-$ to $l_+$.
Using the identity
$$\Psi^{-1}(u)\Psi^{-T}(u) \equiv \eta ,
$$
where
$$\Psi(u) := \left( \psi_{i\alpha}(u)\right)
$$
in the leading term in $z$ we obtain from (H.39) and (H.35)
$$C^{-1}SC^{-T} = z^{-\mu}\eta z^{-\mu}e^{\pi i \mu} = \eta e^{\pi i\mu}
$$
and we can truncate the terms $O(z)$ off the expansion due to (H.34). This
proves the first of the equations (H.36). Transposing this equation and
applying
again the identity (H.33) we obtain the second equation (H.36). Lemma is
proved.
\medskip
We introduce the coordinates $y_\alpha(u,\lambda)$ such that
$$\deli y_\alpha(u,\lambda) = \sqrt{\eta_{ii}(u)}
\int_{il_-}e^{-\lambda z} {\psi_0}_{i\alpha}(u,z){dz\over \sqrt z}
\eqno(H.40)$$
normalizing them as in (H.19). They are still flat coordinates of the
metric $(~,~)^* - \lambda <~,~>^*$. We calculate now the matrix
of the correspondent covariant metric in these coordinates.
\smallskip
{\bf Lemma H.6.} {\it In the coordinates $y_\alpha(u,\lambda)$ the metric
$g_{\alpha\beta}(u,\lambda ) := \left( g^{\alpha\beta}(u) - \lambda
\eta^{\alpha\beta}\right)^{-1}$ is a constant matrix of the form
$$-{1\over \pi} \eta \cos{\pi\mu}.
\eqno(H.41)$$
}

Proof. The contravariant metric $g^{\alpha\beta} - \lambda \eta^{\alpha\beta}$
in the coordinates (H.40) has the matrix
$$\hat g^{\alpha\beta} = \sum_{i=1}^n (u^i-\lambda)
\int_{il_-}e^{-\lambda z}{\psi_0}_{i\alpha}(u,z){dz\over \sqrt z}
\int_{il_-} e^{-\lambda z} {\psi_0}_{i\beta}(u,z){dz\over\sqrt z}.
\eqno(H.42)$$
The matrix is $\lambda$-independent. So we can calculate it
taking the asymptotic with $\lambda\to\infty$. From (H.35) we obtain
asymptotically
$$\int_{il_-}e^{-\lambda z}{\psi_0}_{i\alpha}(u,z){dz\over \sqrt z}
\simeq \Gamma\left({1\over 2} +\mu_\alpha\right)\lambda^{
-\left({1\over 2} +\mu_\alpha\right)}\psi_{i\alpha}(u).
\eqno(H.43)$$
So
$$\hat g^{\alpha\beta}= -\lambda \Gamma\left({1\over 2}+\mu_\alpha\right)
\Gamma\left({1\over 2} + \mu_\beta\right) \lambda^{-(1+\mu_\alpha +\mu_\beta)}
\eta_{\alpha\beta} =
$$
$$ -\delta_{\alpha +\beta, n+1}\Gamma\left(
{1\over 2}+\mu_\alpha\right)\Gamma\left({1\over 2}-\mu_\alpha\right)
=-{\pi \delta_{\alpha+\beta, n+1}\over \cos{\pi \mu_\alpha}}.
\eqno(H.44)$$
Taking the inverse matrix, we obtain (H.41). Lemma is proved.
\medskip
{\bf Corollary H.3.} {\it In the coordinates $x_a^\rr$
the intersection form has the
matrix
$$G=A = S + S^T.
\eqno(H.45)$$
}

Proof. From (H.30) we obtain the transformation of the coordinates
$$(y_1, \dots, y_n) =\sqrt{-2\pi}(x_1, \dots,x_n)C.
\eqno(H.46)$$
Here we denote $x_a = x_a^\rr$.
So the matrix $G$ has the form, due to Lemma H.6
$$G = 2 C \eta \cos{\pi \mu}C^T = S + S^T.
$$
Corollary is proved.
\medskip
The corollary shows that $x_1^\rr$, \dots, $x_n^\rr$ are the coordinates
w.r.t. the basis of the root vectors
of the system of generating
reflections (H.27).
This establishes a relation of the monodromy of the
differential operator (3.120) to the monodromy group of the Frobenius
manifold.

I recall that the root vectors $e_1$, \dots, $e_n$
of the system of reflections $T_1$, \dots, $T_n$
are defined by the following two
conditions
$$T_i e_i = -e_i, ~i=1, \dots, n
$$
$$(e_i, e_i) = 2.
$$
The reflection
$T_i$ in the basis of the root vectors acts as
$$T_i(e_j) = e_j - (e_i,e_j)e_i.
$$
The matrix
$$A_{ij} := (e_i, e_j)
$$
is called {\it Coxeter matrix} of the system of reflections. For the
monodromy group of our generalized hypergeometric equation (H.15)
the Coxeter matrix coincides with the symmetrized Stokes matrix (H.45).
The assumption $d<1$ is equivalent to nondegenerateness of the
symmetrized Stokes matrix.

For the finite
Coxeter groups (see Lecture 4 below) the system of generating reflections
can be chosen in such a way that all $A_{ij}$ are nonpositive. For the
particular subclass of Weyl groups of simple Lie algebras the Coxeter
matrix coincides with the symmetrized Cartan matrix of the Lie algebra.

The coordinates $y_1$, \dots, $y_n$ are dual to the basis of eigenvectors
of the {\it Coxeter transform} $T_1 \dots T_n$ due to the formula
(H.29). The basis of the eigenvectors $f_1$, \dots, $f_n$
of the Coxeter transform due to (H.41) is
normalized as
$$(f_\alpha,f_\beta) = -{1\over \pi}\cos \pi \mu_\beta \delta_{\alpha
+\beta, n+1}.
\eqno(H.47)$$

At the end of this Appendix we consider an application of the integral formula
(H.11) to computation of the flat coordinates of the intersection form
on a trivial Frobenius manifold. In this case we have a linear contravariant
metric (3.37) parametrized by a graded Frobenius algebra $A$. The gradings
$q_\alpha$ of the basic vectors $e_\alpha$ of the algebra are determined
up to a common nonzero factor
$$q_\alpha \mapsto \kappa q_\alpha, ~~d\mapsto \kappa d.
$$
The normalized flat coordinates $\tilde t_\alpha(t,z)$ of
the deformed connection
can be found easily ((3.5) is an equation with constant coefficients)
$$\tilde t_\alpha(t,z) = z^{q_\alpha-d} <e_\alpha , e^{z{\bf t}} -1>, ~~
\alpha = 1, \dots, n
\eqno(H.48)$$
for
$${\bf t} =t^\alpha e_\alpha \in A.
$$
So the integral formula (H.11) for the flat coordinates of the pencil (H.1)
reads
$$x_\alpha(t,\lambda) = <e_\alpha, \int z^{-{3\over 2}+\mu_\alpha}
\left(e^{z({\bf t}-\lambda)}-e^{-\lambda z}\right)\, dz>=
$$
$$= \Gamma\left(-\half + \mu_\alpha\right) <e_\alpha,
(\lambda - {\bf t}^{\half - \mu_\alpha}>
\eqno(H.49)$$
for $\mu_\alpha = q_\alpha - {d\over 2}$ if $\mu_\alpha \neq 1\over 2$.
For $\lambda = 0$ renormalizing (H.49) we obtain the flat coordinates
$x_\alpha(t)$ of the intersection form (3.37)
$$x_\alpha(t) = <e_\alpha, {\bf t}^{\half - \mu_\alpha}>, ~~\alpha = 1,
\dots, n.
\eqno(H.50)$$
The r.h.s. is a polynomial in $t^2, \dots, t^n$ but it ramifies as
a function of $t^1$.
\medskip
{\bf Remark H.1.} Inverting the Laplace-type integrals (H.11) and integrating
by parts we arrive to an integral representation of the deformed coordinates
$\tilde t(t,z)$ via the flat coordinates of the intersection form (therefore,
via solutions of our generalized hypergeometric equation). This gives
\lq\lq oscillating integrals" for the solutions of (3.5), (3.6):
$$\tilde t(t,z) = -z^{1-d\over 2}\oint e^{z\lambda(x,t)}dx
\eqno(H.51)$$
where $\lambda = \lambda(x,t)$ is a function inverse to $x = x(t,\lambda)
= x(t^1-\lambda, t^2, \dots, t^n)$ for a flat coordinate $x(t)$ of the
intersection form.
\vfill\eject
{\bf Appendix I.}
\smallskip
\centerline{\bf Determination of a superpotential of a Frobenius manifold.}
\medskip
In this Appendix we will show that {\it any} irreducible massive
Frobenius manifold with $d<1$ can be described by the formulae
(2.94) for some LG superpotential $\lambda(p;t)$. The superpotential
will always be a function of {\it one} variable $p$ (may be, a multivalued
one) depending on the parameters $t=(t^1, \dots, t^n)$.

We first construct a function $\lambda(p;t)$ for any massive Frobenius
manifold such that the critical values of it are precisely the canonical
coordinates on the Frobenius manifold
$$u^i(t) = \lambda(q^i(t);t), ~~{d\lambda\over dp}|_{p=q^i(t)} = 0, ~~
i=1, \dots, n.
\eqno(I.1)$$
For the construction we will use the flat coordinates of the flat
pencil (H.1) of metrics on $M$. I recall that these can be represented
as
$$x_a(\lambda; t^1, \dots, t^n) = x_a(t^1-\lambda, t^2, \dots, t^n)
\eqno(I.2)$$
where $x_a(t)$ are flat coordinates of the intersection form.

Due to Lemma G.1 the coordinates $x_a(t^1 -\lambda, t^2,\dots, t^n)$
are analytic in $t$ and $\lambda$ outside of the locus
$$\Delta(t^1-\lambda , t^2,\dots, t^n) = 0.
\eqno(I.3)$$
On the semisimple part of the locus we have
$$\lambda = u^i(t)
\eqno(I.4)$$
for some $i$. Near such a point $x_a(t^1 - \lambda, t^2,\dots, t^n)$
is analytic in $\sqrt{\lambda - u^i(t)}$.

Let us fix some $t_0\in M\setminus (\Sigma\cup \Sigma_{\rm nil}$
and some $a$ between $1$ and $n$ such that
$$\partial_1 x_a(t_0^1 - u^i(t_0, t^2_0,\dots, t^n_0)\neq 0
\eqno(I.5)$$
for any $i = 1,\dots, n$
(such $a$ exists since $x_1$, ..., $x_n$ are local coordinates)
and put
$$p = p(\lambda, t) := x_a( t^1 -\lambda, t^2,\dots, t^n).
\eqno(I.6)$$
By $\lambda = \lambda (p,t)$ we denote the inverse function.
\smallskip
{\bf Proposition I.1.} {\it For $t$ close to $t_0$
the critical points of the
function $\lambda(p,t)$ are
$$q^i = p(u^i(t),t),~~ i = 1,\dots, n.
\eqno(I.7)$$
The correspondent critical values equal $u^i(t)$.}

Proof. Near $\lambda = u^i(t_0)$ we have
$$x_a = q^i + x_a^1 \sqrt{\lambda - u^i} + O(\lambda - u^i)
$$
and $x_a^1\neq 0$ by the assumption. For the inverse function we have
locally
$$\lambda = u^i(t) + (x_a^1)^{-1} (p-q^i)^2 + o\left((p-q^i)^2\right).
\eqno(I.8)$$
This proves that $\lambda$ has the prescribed critical values.
\medskip
Let us assume now that $d<1$. In this case we will construct a particular
flat coordinate $p = p(t)$
of the intersection form such that the function inverse to $p = p(t,\lambda)
= p(t^1-\lambda, t^2, \dots, t^n)$
is the LG superpotential of the Frobenius manifold.

I will use the flat coordinates $x_a^\rr (u,\lambda)$ constructed in the
previous Appendix.
\smallskip
{\bf Lemma I.1.} {\it For $\lambda \to u_i$
$$x_a^\rr (u,\lambda) = q_{ai}(u) + \delta_{ai}\sqrt{2\eta_{ii}(u)}
\sqrt{u_i-\lambda} + O(u_i-\lambda)
\eqno(I.9)$$
for some functions $q_{ai}(u)$.}

Proof. Using the asymptotic
$$\psi_{ia}^\rr (u,z) = \left[ \delta_{ai} + O\left({1\over z}\right)\right]
e^{zu_i}
$$
for $z\to\infty$ we obtain (I.9).
Lemma is proved.
\medskip
We consider now the following particular flat coordinate
$$p = p(t;\lambda) := \sum_{a=1}^nx_a^\rr (u,\lambda).
\eqno(I.10)$$
By $\lambda = \lambda(p;t)$ we denote, as above, the inverse function.
\smallskip
{\bf Theorem I.1.} {\it For the metrics $<~,~>$, $(~,~)$ and for the
trilinear form
(1.46) the following formulae hold true
$$\eqalignno{<\partial',\partial''>_t &= -\sum_{i=1}^n
\res_{p = q^i} {\partial'(\lambda(p,t)dp)\,\partial''(\lambda
(p,t)dp)\over d\lambda (p,t)}
&(I.11)\cr
(\partial',\partial'')_t &= -\sum_{i=1}^n
\res_{p=q^i} {\partial'(\log\lambda(p,t)dp)\,\partial''(\log\lambda
(p,t)dp)\over d\log\lambda (p,t)}
&(I.12)\cr
c(\partial', \partial'', \partial''')_t &=
-\sum_{i=1}^n
\res_{p=q^i} {\partial'(\lambda(p,t)dp)\,\partial''(\lambda
(p,t)dp)\, \partial'''(\lambda(p,t)dp)
\over dp\, d\lambda (p,t)}.
&(I.13)\cr}
$$
}

In these formulae
$$d\lambda := {\partial\lambda(p,t)\over \partial p}dp, ~~
d\log\lambda := {\partial\log\lambda(p,t)\over\partial p}dp.
$$

Proof. From (I.9) we obtain
$$p(\lambda,t) = \sum_ax_a^\rr (u,\lambda) =
q_i(u) + \sqrt{2\eta_{ii}(u)}\sqrt{u_i-\lambda} +O(u_i-\lambda)
\eqno(I.14)$$
near $\lambda = u_i$ where we put
$$q_i(u) := \sum_a q_{ai}(u).
\eqno(I.15)$$
So $q_i(u)$ is a critical point of $\lambda$ with the critical value
$u_i$. Near this point
$$\lambda = u_i- {(p-q_i)^2\over 2\eta_{ii}(u)} + O(p-q_i)^3.
\eqno(I.16)$$
{}From this formula we immediately obtain that for $\partial' = \deli$,
$\partial'' = \delj$ the r.h.s. of the formula (I.11) is equal to
$$\eta_{ii}(u)\delta_{ij} = <\deli, \delj >_u.
$$
This proves the equality (I.11). The other equalities are proved in a similar
way. Theorem is proved.
\medskip
{\bf Example I.1.} Using the flat coordinates from Example G.2 we obtain
the LG superpotential of the $CP^1$-model (see Lecture 2)
$$\lambda(p; t_1,t_2) = t_1 - 2e^{t_2\over 2}\cos p.
\eqno(I.17)$$
The cosine is considered as an analytic function on the cylinder $p\simeq
p+2\pi$, so it has only 2 critical points $p=0$ and $p=\pi$.
\medskip
Other examples will be considered in the next Lecture.
\vfill\eject
\centerline{\bf Lecture 4.}
\medskip
{\bf Frobenius structure on the space of orbits of a Coxeter group.}
\medskip
Let $W$ be a {\it Coxeter group}, i.e. a finite group of linear transformations
of real $n$-dimensional space $V$ generated by reflections. In this Lecture
we construct Frobenius manifolds whose monodromy is a given Coxeter group
$W$. All of these Frobenius structures will be polynomial. The results of
Appendix A suggest that the construction of this Lecture gives all the
polynomial solution of WDVV with $d<1$ satisfying the semisimplicity
assumption, although this is still to be proved.

We always
can assume the transformations of the group $W$
to be orthogonal w.r.t. a Euclidean
structure on $V$. The complete classification of irreducible Coxeter
groups was obtained in [31]; see also [21]. The complete list consists
of the groups (dimension of the space $V$ equals the subscript in the name
of the group) $A_n$, $B_n$, $D_n$, $E_6$, $E_7$, $E_8$, $F_4$, $G_2$
(the Weyl groups of the correspondent simple Lie algebras), the groups
$H_3$ and $H_4$ of symmetries of the regular
icosahedron and of the regular 600-cell in the 4-dimensional space
resp. and the groups $I_2(k)$ of symmetries of
the regular $k$-gone on the plane. The group $W$ also acts on the symmetric
algebra $S(V)$ (polynomials of the coordinates of $V$) and on the $S(V)$-module
$\Omega (V)$ of differential forms on $V$ with polynomial coefficients.
The subring $R=S(V)^W$ of $W$-invariant polynomials is generated by $n$
algebraicaly independent homogeneous polynomials $y^1$, ..., $y^n$ [21].
The submodule $\Omega(V)^W$ of the $W$-invariant differential forms with
polynomial coefficients is a free $R$-module with the basis
$dy^{i_1}\wedge ...\wedge dy^{i_k}$ [21]. Degrees of the basic invariant
polynomials are uniquely determined by the Coxeter group. They can be expressed
via the {\it exponents} $m_1$, ..., $m_n$ of the group, i.e. via the
eigenvalues
of a Coxeter element $C$ in $W$ [21]
$$d_i := {\rm deg}\, y^i = m_{i} +1,
\eqno(4.1a)$$
$$\{ {\rm eigen}~ C \} = \{ \exp {2\pi i (d_1 -1)\over h}, ...,
\exp {2\pi i (d_n -1)\over h}\} .
\eqno(4.1b)$$
The maximal degree $h$ is called {\it Coxeter number} of $W$. I will use the
reversed ordering of the invariant polynomials
$$d_1 =h >d_2 \geq ...\geq d_{n-1} >d_2=2.
\eqno(4.2)$$
The degrees satisfy the {\it duality condition}
$$d_i +d_{n-i+1} = h+2, ~i=1, ..., n.
\eqno(4.3)$$
The list of the degrees for all the Coxeter groups is given in Table 1.
\medskip
$$\matrix{W & d_1, ~..., ~d_n \cr
{} & {} \cr
A_n & d_i =n+2-i \cr
B_n & d_i = 2(n-i+1) \cr
D_n,~ n=2k & d_i=2(n-i), ~i\leq k, \cr
{} & d_i=2(n-i+1), ~k+1\leq i \cr
D_n, ~n=2k+1 & d_i=2(n-i), ~i\leq k, \cr
{} & ~d_{k+1} = 2k+1, \cr
{} & ~ d_i=2(n-i+1),
{}~k+2\leq i \cr
E_6 & 12,~9,~8,~6,~5,~2 \cr
E_7 & 18,~14,~12,~10,~8,~6,~2 \cr
E_8 & 30,~24,~20,~18,~14,~12,~8,~2 \cr
F_4 & 12,~8,~6,~2 \cr
G_2 & 6,~2 \cr
H_3 & 10,~6,~2 \cr
H_4 & 30,~20,~12,~2 \cr
I_2(k) & k,~2}$$
\smallskip
\centerline{ Table 1.}
\medskip
\medskip
I will extend the action of the group $W$ to the complexified space $V\otimes
{\bf C}$. The space of orbits
$$M = V\otimes {\bf C}/W$$
has a natural structure of an affine algebraic variety: the coordinate ring
of $M$ is the (complexified) algebra $R$ of
invariant polynomials of the group $W$.
The coordinates $y^1$, ..., $y^n$ on $M$ are defined up to an invertible
transformation
$$y^i\mapsto {y^i}'(y^1,...,y^n),
\eqno(4.4)$$
where ${y^i}'(y^1,...,y^n)$ is a graded homogeneous polynomial of the same
degree
$d_i$ in the variables $y^1$, ..., $y^n$, ${\rm deg}\, y^k = d_k$. Note that
the
Jacobian ${\rm det}(\partial {y^i}'/\partial y^j)$ is a constant (it should not
be zero). The transformations (4.4) leave invariant the vector field
$\partial_1 := \partial /\partial y^1$ (up to a constant factor) due to the
strict inequality $d_1 > d_2$. The coordinate $y^n$ is determined uniquely
within a factor. Also the vector field
$$E ={1\over h}\left( d_1y^1\partial_1 + ... +d_ny^n\partial_n \right) =
{1\over h} x^a {\partial \over \partial x^a}
\eqno(4.5)$$
(the generator of scaling transformations) is well-defined on $M$. Here
we denote by $x^a$ the coordinates in the linear space $V$.

Let $(~,~)$ denotes the $W$-invariant Euclidean metric in the space $V$.
I will use the orthonormal coordinates $x^1$, ..., $x^n$ in $V$ with respect to
this metric. The invariant $y^n$ can be chosen as
$$y^n = {1\over 2h} ((x^1)^2 + ... + (x^n)^2).
\eqno(4.6)$$
We extend $(~,~)$ onto $V\otimes {\bf C}$ as a complex quadratic form.

The factorization map $V\otimes {\bf C} \to M$ is a local diffeomorphism
on an open subset of $V\otimes {\bf C}$. The image of this subset in $M$
consists of {\it regular orbits} (i.e. the number of points of the orbit
equals $\#\,  W$). The complement is the {\it discriminant locus}
${\rm Discr} \, W$.
By  definition it consists of all irregular orbits. Note that the linear
coordinates in $V$ can serve also as local coordinates in small
domains in $M\setminus {\rm Discr} \, W$. It defines a metric $(~,~)$
(and $(~,~)^*$) on $M\setminus {\rm Discr} \, W$. The contravariant metric
can be extended onto $M$ according to the following statement
(cf. [126, Sections 5 and 6]).
\medskip
{\bf Lemma 4.1.} {\it The Euclidean metric
of $V$ induces polynomial contravariant
metric $(~,~)^*$ on the space of orbits
$$g^{ij}(y) = (dy^i,dy^j)^* := {\partial y^i\over \partial x^a}
{\partial y^j\over \partial x^a}
\eqno(4.7)$$
and the correspondent contravariant Levi-Civit\`a connection
$$\Gamma_k^{ij}(y)dy^k = {\partial y^i\over \partial x^a}
{\partial^2y^j\over \partial x^a \partial x^b} dx^b
\eqno(4.8)$$
also is a polynomial one.}

Proof. The right-hand sides in (4.7)/(4.8)  are $W$-invariant
polynomials/differential forms with polynomial coefficients. Hence
$g^{ij}(y)$/$\Gamma_k^{ij}(y)$ are polynomials in $y^1$, ..., $y^n$.
Lemma is proved.
\medskip
{\bf Remark 4.1.} The matrix $g^{ij}(y)$ does not degenerate on
$M\setminus {\rm Discr} \, W$ where the factorization
$V\otimes {\bf C} \to M$ is a local diffeomorphism. So the polynomial
(also called {\it discriminant} of $W$)
$$D(y) := {\rm det} (g^{ij}(y))
\eqno(2.9)$$
vanishes precisely on the discriminant locus ${\rm Discr} \,
W$ where the variables
$x^1$, ..., $x^n$ fail to be local coordinates. Due to this fact the matrix
$g^{ij}(y)$ is called {\it discriminant matrix} of $W$.
The contravariant metric (4.7) was introduced by V.I.Arnold [3] in the form
of operation of convolution of invariants $f(x)$, $g(x)$ of a reflection
group
$$f, ~g \mapsto (df,\, dg)^* = \sum {\partial f\over \partial x^a}
{\partial g\over \partial x^a}.
$$
Note that the image of $V$ in the real
part of $M$ is specified by the condition of positive semidefiniteness
of the matrix $(g^{ij}(y))$ (cf. [118]). The Euclidean connection
(4.8) on the space of orbits is called {\it Gauss - Manin connection}.
\medskip
The main result of this lecture is
\smallskip
{\bf Theorem 4.1.} {\it There exists a unique, up to an equivalence, Frobenius
structure on the space of orbits of a finite Coxeter group with
the intersection form (4.7), the Euler vector field (4.5) and the unity vector
field $e := \partial /\partial y^1$.
}
\medskip
We start the proof of the theorem with the following statement.
\smallskip
{\bf Proposition 4.1.} {\it The functions $g^{ij}(y)$ and $\Gamma_k^{ij}(y)$
depend linearly on $y^1$.}

Proof. From the definition one has that $g^{ij}(y)$ and $\Gamma_k^{ij}(y)$
are graded homogeneous polynomials of the degrees
$${\rm deg}\, g^{ij}(y) = d_i + d_j -2
\eqno(4.10)$$
$${\rm deg}\,\Gamma_k^{ij}(y) = d_i + d_j - d_k - 2.
\eqno(4.11)$$
Since $d_i +d_j \leq 2h = 2d_1$ these polynomials can be at most linear
in $y^1$. Proposition is proved.
\medskip
{\bf Corollary 4.1} (K.Saito) {\it The matrix
$$\eta^{ij}(y) := \partial_1 g^{ij}(y)
\eqno(4.12)$$
has a triangular form
$$\eta^{ij}(y) = 0 ~{\rm for}~ i+j > n+1,
\eqno(4.13)$$
and the antidiagonal elements
$$\eta^{i(n-i+1)} =: c_i
\eqno(4.14)$$
are nonzero constants. Particularly,
$$c := {\rm det} (\eta^{ij}) = (-1)^{n(n-1)\over 2} c_1...c_n \neq 0.
\eqno(4.15)$$}

Proof. One has
$${\rm deg}\, \eta^{ij}(y) = d_i +d_j - 2 - h.$$
Hence deg $\eta^{i(n-i+1)} = 0$ (see (4.3)) and deg $\eta^{ij}<0$
for $i+j >n+1$. This proves triangularity of the matrix and
constancy of the antidiagonal entries $c_i$. To prove
nondegenerateness of $(\eta^{ij}(y))$ we consider, following
Saito, the discriminant (4.9) as a polynomial in $y^1$
$$D(y) = c(y^1)^n + a_1(y^1)^{n-1} + ... + a_n$$
where the coefficients $a_1$, ..., $a_n$ are quasihomogeneous
polynomials in
$y^2$, ..., $y^n$ of the degrees $h$, ..., $nh$ resp.
and the leading coefficient $c$ is given in (4.15).
Let $\gamma$ be the eigenvector of a Coxeter transformation $C$
with the eigenvalue $\exp (2\pi i/h)$. Then
$$y^k(\gamma ) = y^k(C\gamma) = y^k(\exp (2\pi i/h) \gamma ) =
\exp (2\pi id_k/h)y^k(\gamma).$$
For $k>1$ we obtain
$$y^k(\gamma) = 0,~ k=2,...,n.$$
But $D(\gamma)\neq 0$ [21]. Hence the leading coefficient $c\neq 0$.
Corollary is proved.
\medskip
{\bf Corollary 4.2.} {\it The space $M$ of orbits of a finite Coxeter
group carries a flat pencil of metrics $g^{ij}(y)$ (4.7) and $\eta^{ij}(y)$
(4.12) where the matrix $\eta^{ij}(y)$ is polynomialy invertible
globaly on $M$.}
\medskip
We will call (4.12) {\it Saito metric} on
the space of orbits. This was introduced
by Saito, Sekiguchi and Yano in [124] using the classification of Coxeter
groups for all of them but $E_7$ and $E_8$. The general proof of
flateness was obtained in [123].

This metric
will be denoted by $<~,~>^*$ (and by $<~,~>$ if considered on the tangent
bundle $TM$).
Let us denote by
$$\gamma^{ij}_k(y) := \partial_1\Gamma_k^{ij}(y)
\eqno(4.16)$$
the components of the Levi-Civit\`a connection for the metric $\eta^{ij}(y)$.
These are quasihomogeneous polynomials of the degrees
$${\rm deg}\,\gamma_k^{ij}(y) = d_i+d_j-d_k-h-2.
\eqno(4.17)$$
\medskip
{\bf Corollary 4.3} (K.Saito). {\it There exist homogeneous polynomials
$t^1(x)$, ..., $t^n(x)$ of degrees $d_1$, ..., $d_n$ resp. such that
the matrix
$$\eta^{\alpha\beta} := \partial_1 (dt^\alpha ,dt^\beta )^*
\eqno(4.18)$$
is constant.}

The coordinates $t^1$, ..., $t^n$ on the orbit space
will be called {\it Saito flat coordinates}. They can be chosen in such a
way that the matrix (4.18) is antidiagonal
$$\eta^{\alpha\beta} = \delta^{\alpha + \beta , n+1}.$$
Then the Saito flat coordinates are defined uniquely up to an $\eta$-orthogonal
transformation
$$t^\alpha \mapsto a^\alpha_\beta t^\beta ,$$
$$\sum_{\lambda + \mu = n+1} a^\alpha_\lambda a^\beta_\mu =
\delta^{\alpha + \beta , n+1}.$$

Proof. From flatness of the metric $\eta^{ij}(y)$ it follows that the flat
coordinates $t^\alpha (y)$, $\alpha = 1$, ..., $n$ exist at least localy.
They are to be determined from the following system
$$\eta^{is}\partial_s\partial_j t + \gamma_j^{is}\partial_st = 0
\eqno(4.19)$$
(see (3.30)). The inverse matrix $(\eta_{ij}(y)) = (\eta^{ij}(y))^{-1}$
also is polynomial in $y^1$, ..., $y^n$. So rewriting the system (4.19)
in the form
$$\partial_k\partial_l t + \eta_{il}\gamma_k^{is}\partial_st = 0
\eqno(4.20)$$
we again obtain a system with polynomial coefficients. It can be written as a
first-order system for the entries $\xi_l = \partial_lt$,
$$\partial_k\xi_l  + \eta_{il}\gamma_k^{is}\xi_s = 0, ~
k,l = 1,...,n
\eqno(4.21)$$
(the integrability condition $\partial_k\xi_l = \partial_l\xi_k$
follows from vanishing of the curvature).
This is an overdetermined holonomic system.
So the space of solutions has dimension $n$. We can choose a fundamental
system of solutions $\xi_l^\alpha (y)$ such that $\xi_l^\alpha (0) =
\delta_l^\alpha$. These functions are analytic in $y$ for sufficiently
small $y$.
We put $\xi_l^\alpha (y) =: \partial_lt^\alpha (y)$, $t^\alpha (0) = 0$.
The system of solutions is invariant w.r.t. the scaling transformations
$$y^i \mapsto c^{d_i} y^i,~ i=1, ..., n.$$
So the functions $t^\alpha (y)$ are quasihomogeneous in $y$ of the same degrees
$d_1$, ..., $d_n$. Since all the degrees are positive the power series
$t^\alpha (y)$ should be polynomials in $y^1$, ..., $y^n$. Because of the
invertibility of the transformation $y^i\mapsto t^\alpha$ we conclude that
$t^\alpha(y(x))$ are polynomials in $x^1$, ..., $x^n$. Corollary is proved.
\medskip
We need to calculate particular components of the metric $g^{\alpha\beta}$
and of the correspondent Levi-Civit\`a connection in the coordinates
$t^1$, ..., $t^n$ (in fact, in arbitrary homogeneous coordinates
$y^1$, ..., $y^n$).
\medskip
{\bf Lemma 4.2.} {\it Let the coordinate $t^n$ be normalized as in (4.6).
Then the following formulae hold:
$$g^{n\alpha} = {d_\alpha \over h}t^\alpha
\eqno(4.22)$$
$$\Gamma_\beta^{n\alpha} = {(d_\alpha -1) \over h}\delta_\beta^\alpha .
\eqno(4.23)$$}

(In the formulae there is no summation over the repeated indices!)

Proof. We have
$$g^{n\alpha} = {\partial t^n\over \partial x^a}
{\partial t^\alpha \over \partial x^a} = {1\over h}
x^a {\partial t^\alpha\over \partial x^a} = {d_\alpha \over h}t^\alpha$$
due to the Euler identity for the homogeneous functions $t^\alpha (x)$.
Furthermore,
$$\Gamma_\beta^{n\alpha} dt^\beta = {\partial t^n\over \partial x^a}
{\partial^2 t^\alpha \over \partial x^a\partial x^b} dx^b ={1\over h}
x^a {\partial^2t^\alpha\over \partial x^a\partial x^b} dx^b ={1\over h}
x^a d\left( {\partial t^\alpha \over \partial x^a}\right)
$$
$$={1\over h}d \left( x^a  {\partial t^\alpha \over x^a}\right) -
{1\over h}
{\partial t^\alpha \over \partial x^a} dx^a =
{(d_\alpha -1) \over h}dt^\alpha .$$
Lemma is proved.
\medskip
We can formulate now
\medskip
{\bf Main lemma.} {\it Let $t^1$, ..., $t^n$ be the Saito flat coordinates
on the space of orbits of a finite Coxeter group and
$$\eta^{\alpha\beta} = \partial_1 (dt^\alpha, dt^\beta )^*
\eqno(4.24)$$
be the correspondent constant Saito metric.
Then there exists a
quasihomogeneous polynomial $F(t)$ of the degree $2h+2$ such that
$$(dt^\alpha, dt^\beta )^* = {(d_\alpha +d_\beta -2)\over h}
\eta^{\alpha\lambda}
\eta^{\beta\mu}\partial_\lambda\partial\mu F(t).
\eqno(4.25)$$
The polynomial $F(t)$ determines on the space of orbits a polynomial Frobenius
structure with the structure constants
$$c_{\alpha\beta}^\gamma (t) = \eta^{\gamma\epsilon}
\partial_\alpha\partial_\beta\partial_\epsilon F(t)
\eqno(4.26a)$$
the unity
$$e = \partial_1
\eqno(4.26b)$$
the Euler vector field
$$E= \sum \left( 1-{\deg t^\alpha\over h}\right) t^\alpha \dalpha
$$
and the invariant inner product $\eta$.}

Proof. Because of Corollary 4.3 in the flat coordinates the tensor
$\Delta_\gamma^{\alpha\beta} = \Gamma_\gamma^{\alpha\beta}$ should
satisfy the equations (D.1) - (D.3) where $g_1^{\alpha\beta}
= g^{\alpha\beta}(t)$, $g_2^{\alpha\beta} = \eta^{\alpha\beta}$.
First of all according to (D.1a) we can represent the tensor
$\Gamma_\gamma^{\alpha\beta}(t)$ in the form
$$\Gamma_\gamma^{\alpha\beta}(t) = \eta^{\alpha\epsilon}
\partial_\epsilon\partial_\gamma f^\beta (t)
\eqno(4.27)$$
for a vector field $f^\beta (t)$. The equation (3.27) (or,
equivalently, (D.3)) for the metric $g^{\alpha\beta}(t)$ and the
connection (4.27) reads
$$g^{\alpha\sigma}\Gamma_\sigma^{\beta\gamma} = g^{\beta\sigma}
\Gamma_\sigma^{\alpha\gamma}.$$
For $\alpha = n$ because of Lemma 4.2 this gives
$$\sum_\sigma d_\sigma t^\sigma \eta^{\beta\epsilon}
\partial_\sigma\partial_\epsilon f^\gamma = (d_\gamma -1)g^{\beta\gamma}.
$$
Applying to the l.h.s. the Euler identity (here deg $\partial_\epsilon
f^\gamma = d_\gamma - d_\epsilon +h$) we obtain
$$(d_\gamma - 1) g^{\beta\gamma} =
\sum_\epsilon \eta^{\beta\epsilon} (d_\gamma - d_\epsilon +h)
\partial_\epsilon f^\gamma =
(d_\gamma + d_\beta - 2)\eta^{\beta\epsilon}\partial_\epsilon f^\gamma.
\eqno(4.28a)$$
{}From this one obtains the symmetry
$${\eta^{\beta\epsilon}\partial_\epsilon f^\gamma \over d_\gamma -1} =
{\eta^{\gamma\epsilon}\partial_\epsilon f^\beta \over d_\beta -1}.
$$
Let us denote
$${f^\gamma\over d_\gamma -1} =: {F^\gamma\over h} .
\eqno(4.28b)$$
We obtain
$$\eta^{\beta\epsilon}\partial_\epsilon F^\gamma =
\eta^{\gamma\epsilon}\partial_\epsilon F^\beta .$$
Hence a function $F(t)$ exists such that
$$F^\alpha = \eta^{\alpha\epsilon} \partial_\epsilon F.
\eqno(4.28c)$$
It is clear that $F(t)$ is a quasihomogeneous polynomial of
the degree $2h+2$. From the formula (4.28) one immediately obtains (4.25).

Let us prove now that the coefficients (4.26a) satisfy the associativity
condition. It is more convenient to work with the dual
structure constants
$$c_\gamma^{\alpha\beta}(t) = \eta^{\alpha\lambda}\eta^{\beta\mu}
\partial_\lambda\partial_\mu\partial_\gamma F.$$
Because of (4.27), (4.28) one has
$$\Gamma_\gamma^{\alpha\beta} = {d_\beta -1\over h} c_\gamma^{\alpha\beta}.$$
Substituting this in (D.2) we obtain associativity. Finaly,
for $\alpha = n$ the formulae (4.22), (4.23) imply
$$c_\beta^{n\alpha} =  \delta_\beta^\alpha .$$
Since $\eta^{1n} = 1$, the vector (4.26b) is the unity of the algebra.
Lemma is proved.
\medskip
Proof of Theorem.

Existence of a Frobenius structure on the space of orbits satisfying
the conditions of Theorem 4.1 follows from Main lemma. We are now to prove
uniqueness.
Let us consider a polynomial Frobenius structure
on $M$ with the Euler vector field (4.5)
and with the Saito invariant metric. In the Saito flat coordinates we have
$$dt^\alpha\cdot dt^\beta = \eta^{\alpha\lambda} \eta^{\beta\mu}
\partial_\lambda \partial_\mu \partial_\gamma F(t)dt^\gamma .$$
The r.h.s. of (3.13) reads
$$i_E (dt^\alpha \cdot dt^\beta ) = {1\over h}\sum_\gamma d_\gamma t^\gamma
\eta_{\alpha\lambda} \eta^{\beta\mu}
\partial_\lambda \partial_\mu \partial_\gamma F(t)
= {1\over h}(d_\alpha + d_\beta - 2) \eta_{\alpha\lambda} \eta^{\beta\mu}
\partial_\lambda \partial_\mu  F(t) .$$
This should be equal to $(dt^\alpha , dt^\beta )^*$. So the function
$F(t)$ must satisfy (4.25). It is determined uniquely by this equation
up to terms quadratic in $t^\alpha$. Such an ambiguity does not affect
the Frobenius structure. Theorem is proved.
\medskip
We will show know that the Frobenius manifolds we have constructed
satisfy the semisimplicity condition. This will follow from the
following construction.

Let $R = {\bf C}[y_1, ...,y_n]$ be the coordinate ring of the orbit space $M$.
The Frobenius algebra structure on the tangent planes
$T_yM$ for any $y\in M$ provides the $R$-module $Der\, R$ of invariant
vector fields with a structure of Frobenius algebra over $R$.
To describe this structure
let us  consider a homogeneous basis of invariant polynomials $y_1$, ..., $y_n$
of the Coxeter group.
Let $D(y_1, ..., y_n)$ be the discriminant of the group.
We introduce a polynomial of degree $n$ in an auxiliary variable $u$ putting
$$P(u; y_1, ..., y_n) := D(y_1 -u, y_2, ..., y_n).
\eqno(4.29)$$
Let $D_0(y_1, ..., y_n)$ be the discriminant of this polynomial in $u$.
It does not vanish identicaly on the space of orbits.
\medskip
{\bf Theorem 4.2.} {\it The map
$$1\mapsto e, ~~u\mapsto E
\eqno(4.30a)$$
can be extended uniquely to an isomorphism of $R$-algebras
$$R[u]/(P(u;y)) \to Der\, R.
\eqno(4.30b)$$}
\medskip
{\bf Corollary 4.4.} {\it The algebra on $T_yM$ has no nilpotents
outside the zeroes of the polynomial $D_0(y_1, ..., y_n)$.}
\medskip

We start the proof with an algebraic remark: let $T$
be a $n$-dimensional space and $U: T\to T$ an endomorphism (linear
operator). Let
$$P_U(u) := {\rm det}\, (U-u\cdot 1) $$
be the characteristic polynomial of $U$. We say that the endomorphism $U$
is semisimple if all the $n$ roots of the characteristic polynomial are
simple. For a semisimple endomorphism there exists a cyclic vector
$e\in T$ such that
$$T = {\rm span}\, (e, Ue, ..., U^{n-1}e).$$
The map
$${\bf C}[u]/(P_U(u)) \to T, ~~u^k\mapsto U^ke, ~k=0, 1, ..., n-1
\eqno(4.31)$$
is an isomorphism of linear spaces.
\medskip
Let us fix a point $y\in M$. We define a linear operator
$$U = (U_j^i(y)): T_yM \to T_yM
\eqno(4.32)$$
(being also an operator on the cotangent bundle of the space of orbits)
taking the ratio of the quadratic forms $g^{ij}$ and $\eta^{ij}$
$$<U\omega_1,\omega_2>^* = (\omega_1,\omega_2)^*
\eqno(4.33)$$
or, equivalently,
$$U_j^i(y) := \eta_{js}(y)g^{si}(y).
\eqno(4.34)$$
\medskip
{\bf Lemma 4.3.} {\it The characteristic polynomial of the operator
$U(y)$ is given up to a nonzero factor $c^{-1}$ (4.15) by the formula
(4.29).}

Proof. We have
$$P(u;y^1, \dots , y^n) :=
{\rm det}(U - u\cdot 1) = {\rm det}(\eta_{js}){\rm det}
(g^{si} - u\eta^{si}) = $$
$$c^{-1}{\rm det}(g^{si}(y^1-u, y^2,
\dots , y^n) = c^{-1}D(y^1 - u, y^2, \dots , y^n).$$
Lemma is proved.
\medskip
{\bf Corollary 4.5.} {\it The operator $U(y)$ is semisimple at a generic
point $y\in M$.}

Proof. Let us prove that the discriminant $D_0(y^1, \dots , y^n)$ of the
characteristic polynomial $P(u;y^1, \dots , y^n)$ does not vanish identicaly
on $M$. Let us fix a Weyl chamber $V_0 \subset V$ of the group $W$. On the
inner
part of $V_0$ the factorization map
$$V_0 \to M_{Re}$$
is a diffeomorphism. On the image of $V_0$ the discriminant $D(y)$ is
positive. It vanishes on the images of the $n$ walls of the Weyl chamber:
$$D(y)_{i-{\rm th ~ wall}} = 0,~~ i=1, \dots , n.
\eqno(4.35)$$
On the inner part of the $i$-th wall (where the surface
(4.35) is regular) the equation (4.35) can be solved for $y^1$:
$$y^1 = y^1_i(y^2, \dots , y^n).
\eqno(4.36)$$
Indeed, on the inner part
$$(\partial_1D(y))_{i-{\rm th ~ wall}} \neq 0.$$
This holds since the polynomial
$D(y)$ has simple zeroes at the generic point of the discriminant of $W$
(see, e.g., [4]) .

Note that the functions (4.36) are the roots of the equation $D(y) = 0$
as the equation in the unknown $y^1$. It follows from above that this equation
has simple roots for generic $y^2$, ..., $y^n$. The roots of the characteristic
equation
$$D(y^1-u, y^2, \dots , y^n) = 0
\eqno(4.37a)$$
are therefore
$$u_i = y^1 - y^1_i(y^2, \dots , y^n), ~~i=1, \dots , n.
\eqno(4.37b)$$
Genericaly these are distinct. Lemma is proved.
\medskip
{\bf Lemma 4.4.} {\it The operator $U$ on the tangent planes $T_yM$ coincides
with the operator of multiplication by the Euler
vector field $E$.}

Proof. We check the statement of the lemma in the Saito flat coordinates:
$$\sum_\sigma {d_\sigma \over h} t^\sigma c_{\sigma\beta}^\alpha
= {h-d_\beta +d_\alpha\over h}\eta^{\alpha\epsilon}
\partial_\epsilon \partial_\beta F
=$$
$$ \sum_\lambda {d_\lambda + d_\alpha - 2 \over h}
\eta_{\beta\lambda}\eta^{\alpha\epsilon} \eta^{\lambda\mu}
\partial_\epsilon \partial_\mu F = \eta_{\beta\lambda}g^{\alpha\lambda}
= U_\beta^\alpha .$$
Lemma is proved.
\medskip
Proof of Theorem 4.2.

Because of Lemmas 4.3, 4.4  the vector fields
$$e, ~E, ~E^2, \dots , E^{n-1}
\eqno(4.38)$$
genericaly are linear independent on $M$. It is easy to see
that these are polynomial vector fields on $M$. Hence $e$ is a cyclic vector
for the endomorphism $U$ acting on $Der \, R$.
So in generic
point $y\in M$ the map (4.30a) is an isomorphism of Frobenius
algebras
$${\bf C}[u]/(P(u;x)) \to T_xM.$$
This proves Theorem 2.
\medskip
{\bf Remark 4.2.} The Euclidean metric (4.7) also defines an invariant
inner product for the Frobenius algebras (on the cotangent planes $T_*M$).
It can be shown also that the trilinear form
$$(\omega_1\cdot \omega_2, \omega_3)^*$$
can be represented (localy, outside the discriminant locus $Discr\, W$)
in the form
$$(\hat\nabla^i \hat\nabla^j \hat\nabla^k \hat F(x))
\partial_i \otimes \partial_j \otimes \partial_k$$
for some function $\hat F(x)$. Here $\hat\nabla$ is the Gauss-Manin
connection (i.e. the Levi-Civit\`a connection for the metric (4.7)).
The unity $dt^n/h$ of the Frobenius algebra on $T_*M$ is not
covariantly constant w.r.t. the Gauss-Manin connection.
\medskip
{\bf Remark 4.3.} The vector fields
$$l^i := g^{is}(y)\partial_s,~~i = 1, \dots , n
\eqno(4.39)$$
form a basis of the $R$-module Der$_R(-\log (D(y))$ of the vector
fields on $M$ tangent to the discriminant locus [4]. By the definition,
a vector field $u\in$ Der$_R(-\log (D(y))$ {\it iff}
$$uD(y) = p(y)D(y)$$
for a polynomial $p(y)\in R$. The basis (4.39) of
Der$_R(-\log (D(y))$ depends on the choice of coordinates
on $M$. In the Saito flat coordinates commutators of the basic vector
fields can be calculated via the structure constants of the Frobenius
algebra on $T_*M$. The following formula holds:
$$[l^\alpha , l^\beta ] = {d_\beta - d_\alpha \over h}
c^{\alpha\beta}_\epsilon l^\epsilon .
\eqno(4.40)$$
This can be proved using (4.25).
\medskip
{\bf Example 4.1.} $W = I_2(k)$, $k\geq 0$. The action of the group on the
complex $z$-plane is generated by the transformations
$$z\mapsto e^{{2\pi i\over k}}z,~~z\mapsto \bar z.
$$
The invariant metric on ${\bf R}^2 = \Cc$ is
$$ds^2 = dz d\bar z,
$$
the basic invariant polynomials are
$$t^1 = z^k + \bar z^k, ~~\deg t^1 = k,
$$
$$t^2 = {1\over 2k}z \bar z,~~ \deg t^2 = 2.
$$
We have
$$\eqalign{g^{11}(t) &= (dt^1, dt^1) =4
{\partial t^1\over \partial z} {\partial t^1\over \partial\bar z}
= 4k^2 (z\bar z)^{k-1} = (2k)^{k+1}(t^2)^{k-1}\cr
g^{12}(t) &= (dt^1, dt^2) =2 \left(
{\partial t^1\over \partial z}{\partial t^2\over \partial\bar z}
+{\partial t^1\over \partial\bar z}{\partial t^2\over \partial z}\right)
=(z^k +\bar z^k) = t^1\cr
g^{22}(t) &= 4{\partial t^2\over \partial z}{\partial t^2\over \partial\bar z}
={2\over k} t^2.\cr}
$$
The Saito metric (4.12) is constant in these coordinates. The formula (4.25)
gives
$$F(t^1,t^2) = \half (t^1)^2 t^2 +{(2k)^{k+1}\over 2(k^2-1)}(t^2)^{k+1}.
$$
This coincides with (1.24a) (up to an equivalence) for $\mu =\half
(k-1)/(k+1)$.
Particularly, for $k=3$ this gives the Frobenius structure on
$\Cc^2 /A_2$, for $k=4$ on $\Cc^2 /B_2$, for $k=6$ on $\Cc^2 /G_2$.
\medskip
{\bf Example 4.2.} $W = A_n$. The group acts on the $(n+1)$-dimensional
space ${\bf R}^{n+1} = \{ (\xi_0,\xi_1,\dots, \xi_n)\}$ by the permutations
$$(\xi_0,\xi_1,\dots, \xi_n)\mapsto (\xi_{i_0},\xi_{i_1},\dots, \xi_{i_n}).
$$
Restricting the action onto the hyperplane
$$\xi_0 + \xi_1 + \dots, + \xi_n = 0
\eqno(4.41)$$
we obtain the desired action of $A_n$ on the $n$-dimensional space (4.41).
The invariant metric on (4.41) is obtained from the standard Euclidean
metric on ${\bf R}^{n+1}$ by the restriction.

The invariant polynomials on (4.41) are symmetric polynomials on
$\xi_0,\xi_1,\dots, \xi_n$. The elementary symmetric polynomials
$$a_k = (-1)^{n-k+1}(\xi_0\xi_1\dots \xi_k + \dots ),~~
k= 1,\dots, n
\eqno(4.42)$$
can be taken as a homogeneous basis in the graded ring of the
$W$-invariant polynomials on (4.41). So the complexified space
of orbits $M = \Cc^n/A_n$ can be identified with the space of
polynomials $\lambda(p)$ of an auxiliary variable $p$ of the form
(1.65).

Let us show that the Frobenius structure (4.25) on $M$ coincides with
the structure (1.66) (this will give us the simplest proof of that
the formulae (1.66) give an example of Frobenius manifold). It will be
convenient first to rewrite the formulae (1.66) in a slightly modified
way (cf. (2.94))
\smallskip
{\bf Lemma 4.5.} {\it \item{1.} For the example 3
of Lecture 1 the inner product $<~,~>_\lambda$ and the 3-d rank tensor
$c(\, .\, ,\, .\, ,\, .\,) = <\, .\, \cdot\, .\, ,\, .\,>_\lambda$
have the form
$$<\partial',\partial''>_\lambda = -\sum_{|\lambda |<\infty}
\res_{d\lambda = 0} {\partial'(\lambda(p)dp)\,\partial''(\lambda
(p)dp)\over d\lambda (p)}
\eqno(4.43)$$
$$c(\partial', \partial'', \partial''') =
-\sum_{|\lambda |<\infty}
\res_{d\lambda = 0} {\partial'(\lambda(p)dp)\,\partial''(\lambda
(p)dp)\, \partial'''(\lambda(p)dp)
\over dp\, d\lambda (p)}.
\eqno(4.44)$$
\item{2.} Let $q^1$,\dots, $q^n$ be the critical points of the polynomial
$\lambda(p)$,
$$\lambda'(q^i) = 0,~~i=1,\dots, n
$$
and
$$u^i = \lambda(q^i), ~~i=1,\dots, n
\eqno(4.45)$$
be the correspondent critical values. The variables $u^1$, \dots, $u^n$
are local coordinates on $M$ near the points $\lambda$ where the polynomial
$\lambda(p)$ has no multiple roots.
These are canonical coordinates for the multiplication (1.66a). The metric
(1.66b)
in these coordinates has the diagonal form
$$<~,~>|_\lambda =
\sum_{i=1}^n \eta_{ii}(u)(du^i)^2,~~\eta_{ii}(u) = {1\over \lambda''(q^i)}.
\eqno(4.46)$$
\item{3.} The metric on $M$ induced by the invariant Euclidean metric
in a point $\lambda$ where the polynomial $\lambda(p)$ has simple roots
has the form
$$(\partial',\partial'')_\lambda = -\sum_{|\lambda |<\infty}
\res_{d\lambda = 0} {\partial'(\log\lambda(p)dp)\,\partial''(\log\lambda
(p)dp)\over d\log\lambda (p)}.
\eqno(4.47)$$

}

Here $\partial',~\partial'',~\partial'''$ are arbitrary tangent vectors
on $M$ in the point $\lambda$, the derivatives
$\partial'(\lambda(p)dp)$ etc. are taken keeping $p=const$; $\lambda'(p)$
and $\lambda''(p)$ are the first and the second derivatives of the polynomial
$\lambda(p)$ w.r.t. $p$. In other words, the formulae (4.43) - (4.44)
mean that (1.65) is the LG superpotential for the Frobenius manifold (1.66)
[39].

Proof. The first formula follows immediately from (1.66b) since the sum of
residues
of a meromorphic differential $\omega$ on the Riemann $p$-sphere vanishes:
$$\res_{p=\infty} \omega + \sum \res_{|\lambda |<\infty} \omega = 0.
\eqno(4.48)$$
Here we apply the residue theorem to the meromorphic differential
$$\omega = {\partial'(\lambda(p)dp)\,\partial''(\lambda
(p)dp)\over d\lambda (p)}.
$$
{}From (4.48) it also follows that the formula (1.66a) can be rewritten as
$$c(\partial', \partial'', \partial''') =
\res_{p=\infty} {\partial'(\lambda(p)dp)\,\partial''(\lambda
(p)dp)\, \partial'''(\lambda(p)dp)
\over dp\, d\lambda (p)}.
\eqno(4.49)$$
Let
$$f(p) = \partial'(\lambda(p)),~~g(p) = \partial''(\lambda(p)),~
h(p) = \partial'''(\lambda(p)),$$
$$f(p)g(p) = q(p) + r(p) \lambda'(p)
$$
for polynomials $q(p)$, $r(p)$, $\deg q(p) < n$. In the algebra
$\Cc [p]/(\lambda'(p)$ we have then
$$f\cdot g = q.$$
On the other side, for the residue (4.49) we obtain
$$\res_{p=\infty} {\partial'(\lambda(p)dp)\,\partial''(\lambda
(p)dp)\, \partial'''(\lambda(p)dp)
\over dp\, d\lambda (p)} = \res_{p=\infty} {q(p) h(p) dp\over
d\lambda(p)} + \res_{p=\infty} r(p) h(p) dp.
$$
The second residue in the r.h.s. of the formula equals zero
while the first one coincides with the inner product
$<q,h>_\lambda = <f\cdot g,h>_\lambda$.

Let us prove the second statement of Lemma. Let $\lambda(p)$ be
a polynomial without multiple roots. Independence of the critical
values $u^1$, ..., $u^n$ as functions
of the polynomial is a standard fact
(it also follows from the explicit formula (4.55) for the Jacobi matrix).
Let us choose $\xi_1$, ..., $\xi_n$ as the coordinates on the hyperplane
(4.41). These are not orthonormal: the matrix of the (contravariant)
$W$-invariant metric in these coordinates has the form
$$g^{ab} = \delta^{ab} - {1\over n+1}.
\eqno(4.50)$$
We have
$$\lambda(p) = (p+\xi_1 +\dots +\xi_n)\prod_{a=1}^n(p-\xi_a),
{}~~\lambda'(p) = \prod_{i=1}^n(p-q^i),
\eqno(4.51)$$
$$\deli \lambda(p) ={1\over p-q^i} {\lambda'(p)\over \lambda''(q^i)}.
\eqno(4.52)$$
The last one is the Lagrange interpolation formula since
$$\deli \lambda(p)|_{p=q^j} = \delta_{ij}.
\eqno(4.53)$$
Substituting $p=\xi_a$ to the identity
$$(\deli \xi_1 +\dots +\deli \xi_n)\prod_{b=1}^n (p-\xi_b)
- \sum_{a=1}^n {\lambda(p)\over p-\xi_a}\deli \xi_a = \deli\lambda(p)
\eqno(4.54)$$
we obtain the formula for the Jacobi matrix
$$\deli\xi_a = - {1\over (\xi_a - q^i)\lambda''(q^i)},~~i,a
= 1,\dots, n.
\eqno(4.55)$$
Note that for a polynomial $\lambda(p)$ without multiple roots
we have $\xi_a\neq q^i$, $\lambda''(q^i)\neq 0$.

For the metric (4.43) from (4.53) we obtain
$$<\deli ,\delj > = -\delta_{ij}{1\over \lambda''(q^i)}.
\eqno(4.56)$$
For the tensor (4.44) for the same reasons only $c(\deli ,\deli ,\deli )$
could
be nonzero and
$$c(\deli ,\deli ,\deli )\equiv <\deli\cdot\deli ,\deli >
 = - {1\over \lambda''(q^i)}.
\eqno(4.57)$$
Hence
$$\deli\cdot\delj = \delta_{ij} \deli
\eqno(4.58)$$
in the algebra (1.66).

To prove the last statement of the lemma we observe that the metric
(4.47) also is diagonal in the coordinates $u^1$, ..., $u^n$ with
$$g_{ii}(u) := (\deli ,\deli ) = - {1\over u^i \lambda''(q^i)}.
\eqno(4.59)$$
The inner product of the gradients $(d\xi_a,d\xi_b)$ w.r.t. the metric
(4.59) is
$$\sum_{i=1}^n{1\over g_{ii}(u)}{\partial \xi_a\over \partial u^i}
{\partial \xi_b\over \partial u^i} = -
\sum_{i=1}^n {u^i\over (\xi_a -q^i)(\xi_b -q^i)\lambda''(q^i)}
$$
$$= - \sum_{i=1}^n\res_{d\lambda = 0} {\lambda(p)\over
(p-\xi_a)(p-\xi_b)\lambda'(p)} = \left[ \res_{p=\infty} + \res_{p=\xi_a}
+\res_{p=\xi_b}\right] {\lambda(p)\over
(p-\xi_a)(p-\xi_b)\lambda'(p)}
$$
$$= \delta^{ab} - {1\over n+1}.
$$
So the metric (4.47) coincides with the $W$-invariant Euclidean metric
(4.7).
Lemma is proved.
\medskip
{\bf Exercise 4.1.} Prove that the function
$$V(u) := -{1\over 2(n+1)} \left[ \xi_0^2 +\dots
+\xi_n^2\right] |_{\xi_0+\dots +\xi_n = 0}
\eqno(4.60)$$
is the potential for the metric (4.59):
$$\deli V(u) = \eta_{ii}(u).
$$
\medskip
Let us check that the curvature of
the metric
(1.66b) vanishes. I will construct explicitly the flat coordinates
for the metric (cf. [39, 124]). Let us consider the function
$p = p(\lambda)$ inverse to the polynomial $\lambda = \lambda(p)$.
It can be expanded in a Puiseaux series as $\lambda \to\infty$
$$p = p(k) = k + {1\over n+1}\left( {t^n\over k} +{t^{n-1}\over k^2}
+ \dots + {t^1\over k^n} \right) + O \left({1\over k^{n+1}
}\right)
\eqno(4.61)$$
where $k := \lambda^{1\over n+1}$, the coefficients
$$t^1 = t^1 (a_1,\dots, a_n), \dots, t^n = t^n (a_1,\dots, a_n)
\eqno(4.62)$$
are determined by this expansion. The inverse functions can be found
from the identity
$$(p(k))^{n+1} + a_n (p(k))^{n-1} + \dots + a_1 = k^{n+1}.
\eqno(4.63)$$
This gives a triangular change of coordinates of the form
$$a_i = - t^i + f_i(t^{i+1}, \dots, t^n), ~~i=1,\dots, n.
\eqno(4.64)$$
So the coefficients $t^1$, ..., $t^n$ can serve as global
coordinates on the orbit space $M$ (they give a distinguished
basis of symmetric polynomials of $(n+1)$ variables).
\smallskip
{\bf Exercise 4.2.} Show the following formula [39, 124] for the coordinates
$t^\alpha$
$$t^\alpha = -{n+1\over n-\alpha +1}\res_{p=\infty}
\left(\lambda^{n-\alpha +1
\over n+1}(p)\, dp\right) .
\eqno(4.65)$$
\medskip
Let us prove that the variables $t^\alpha$ are the flat coordinates
for the metric (1.66b),
$$<\dalpha, \dbeta > = \delta_{\alpha + \beta, n+1}.
\eqno(4.66)$$
To do this (and also in other proofs) we will use the following
\lq\lq thermodynamical identity".
\smallskip
{\bf Lemma 4.6.} {\it Let $\lambda = \lambda (p, t^1,\dots, t^n)$
and $p = p(\lambda , t^1,\dots, t^n)$ be two mutually inverse
functions depending on the parameters $t^1, \dots, t^n$. Then
$$\dalpha (\lambda \, dp)_{p = const} = - \dalpha (p\, d\lambda)_
{\lambda = const},
\eqno(4.67)$$
$\dalpha = \partial/\partial t^\alpha$.
}

Proof. Differentiating the identity
$$\lambda(p(\lambda,t),t) \equiv \lambda
$$
w.r.t. $t^\alpha$ we obtain
$${d\lambda\over dp} \dalpha p(\lambda, t)_{\lambda = const}
+ \dalpha \lambda(p(\lambda,t), t)_{p=const} = 0.
$$
Lemma is proved.
\medskip
Observe that $k = \lambda^{1\over n+1}$ can be expanded as a
Laurent series in $1/p$
$$k = p + O\left({1\over p}\right).
$$
By $[~~]_+$ I will denote the polynomial part of a Laurent series
in $1/p$. For example, $[k]_+ = p$. Similarly, for a differential
$f\, dk$ where $f$ is a Laurent series in $1/p$ we put
$[f\, dk]_+ := [f\, dk/dp]_+ dp$.
\smallskip
{\bf Lemma 4.7.} {\it The following formula holds true
$$\dalpha (\lambda\, dp)_{p=const} = - [k^{\alpha - 1} dk]_+,
{}~~\alpha = 1,\dots, n.
\eqno(4.68)$$
}

Proof. We have
$$-\dalpha (\lambda\, dp)_{p=const} = \dalpha (p\, d\lambda)_{\lambda = const}
= \left( {1\over n+1} {1\over k^{n-\alpha +1}} + O\left( {1\over k^{n+1}}
\right) \right) dk^{n+1} $$
$$= k^{\alpha -1} dk + O\left({1\over k}
\right) dk
$$
since $k=const$ while $\lambda = const$. The very l.h.s. of this chain of
equalities is a polynomial differential in $p$. And $[O(1/k)dk]_+ = 0$.
Lemma is proved.
\medskip
{\bf Corollary 4.6.} {\it The variables $t^1$, ..., $t^n$ are the flat
coordinates for the metric (1.66b). The coefficients of the
metric in these coordinates are
$$\eta_{\alpha\beta} = \delta_{\alpha + \beta, n+1}.
\eqno(4.69)$$
}

Proof. From the previous lemma we have
$$\dalpha (\lambda\, dp)_{p = const} = -k^{\alpha - 1} dk
+ O(1/k) dk.
$$
Substituting to the formula (4.43) we obtain
$$<\dalpha , \dbeta >_\lambda = \res_{p=\infty}
{k^{\alpha - 1}dk\, k^{\beta - 1} dk \over dk^{n+1}} =
{1\over n+1} \delta_{\alpha + \beta , n+1}
$$
(the terms of the form $O(1/k)dk$ do not affect the residue).
Corollary is proved.
\medskip
Now we can easyly prove that the formulae of Example 1.7
describe a Frobenius structure on the space $M$ of polynomials
$\lambda(p)$. Indeed, the critical values of $\lambda(p)$ are
the canonical coordinates $u^i$ for the multiplication in the algebra
of truncated polynomials $\Cc /(\lambda'(p)) = T_\lambda M$. The metric
(1.66b) is flat on $M$ and it is diagonal in the canonical coordinates.
{}From the flatness and from (4.60) it follows that this is a Darboux
- Egoroff metric on $M$. From Lemma 4.5 we conclude that $M$ with
the structure (1.66) is a Frobenius manifold. It also follows
that the correspondent intersection form coincides with the $A_n$-invariant
metric on $\Cc^n$. From the uniqueness part of Theorem 4.1 we conclude
that the Frobenius structure (1.66) coincides (up to an equivalence)
with
the Frobenius structure of Theorem 4.1.
\smallskip
{\bf Remark 4.4.} For the derivatives of the correspondent polynomial
$F(t)$ in [39] the following formula was obtained
$$\dalpha F = {1\over (\alpha + 1)(n+\alpha + 2)} \res_{p=\infty}
\lambda^{n+\alpha + 2\over n+1}dp.
\eqno(4.70)$$
\medskip
The constructions of this Lecture can be generalized for the case
when the monodromy group is an extension of affine Weyl groups.
The simplest solution of this type is given by the quantum multiplication
on $CP^1$ (see Example G.2 above). We will not describe here the
general construction (to be published elsewhere) but we will give two
examples of it. In these examples one obtains three-dimensional
Frobenius manifolds.
\smallskip
{\bf Exercise 4.3.} Prove that
$$F = \half t_1^2 t_3 + \half t_1t_2^2 -{1\over 24} t_2^4 + t_2 e^{t_3}
\eqno(4.71)$$
is a solution of WDVV with the Euler vector field
$$E= t_1\partial_1 +\half t_2\partial_2 + {3\over 2}\partial_3.
\eqno(4.72)$$
Prove that the monodromy group of the Frobenius manifold coinsides
with an extension of the affine Weyla group $\tilde W(A_2)$.
Hint: Prove that the flat coordinates $x$, $y$, $z$ of the intersection
form are given by
$$\eqalign{t_1 &= 2^{-{1\over 3}} e^{{2\over 3}z}\left[
e^{x+y} + e^{-x} + e^{-y}\right]\cr
t_2 &= 2^{-{2\over 3}} e^{{1\over 3}z}\left[ e^{-x-y} + e^x+e^y\right]\cr
t_3 &= z.\cr}
\eqno(4.73)$$
The intersection form is proportional to
$$ds^2 = dx^2 + dy^2 _ dz^2.
\eqno(4.74)$$
\medskip
{\bf Exercise 4.4.} Prove that
$$F = \half t_1^2t_3 + \half t_1t_2^2 -{1\over 48} t_2^4 + {1\over 4} t_2^2
e^{t_3} + {1\over 32} e^{2t_3}
\eqno(4.75)$$
is a solution of WDVV with the Euler vector field
$$E = t_1\partial_1 + \half t_2\partial_2 +\partial_3.
\eqno(4.76)$$
Prove that the monodromy group of the Frobenius manifold is
an extension of the affine Weyl group $\tilde W(B_2)$.
Hint: Show that the flat coordinates of the intersection
form are given by
$$\eqalign{t_1 &=e^{2\pi i z}\left[ \cos 2\pi x \cos 2\pi y + \half\right]\cr
t_2 &= e^{\pi i z}\left[ \cos 2\pi x + \cos 2\pi y\right]\cr
t_3 &= 2\pi z.\cr}
\eqno(4.77)$$
The intersection form in these coordinates is proportional to
$$ds^2 = dx^2 + dy^2 -\half dz^2.
$$

{\bf Remark 4.5.} The correspondent extension of the dual affine Weyl
group $\tilde W(C_2)$ gives an equivalent Frobenius 3-manifold.
\medskip
In the appendix to this Lecture we outline a generalization of our
constructions to the case of extended complex crystallographic groups.
\medskip
We obtain now an integral representation of the solution of the Riemann -
Hilbert b.v.p. of Lecture 3 for the polynomial Frobenius manifolds
on the space of orbits of a finite Coxeter group $W$.

Let us fix a system of $n$ reflections $T_1$, \dots, $T_n$ generating
the group $W$ (the order of the reflections also will be fixed). Via
$e_1$, \dots, $e_n$ I denote the normal vectors to the mirrors
of the reflections normalized as
$$T_i(e_i) = -e_i,
\eqno(4.78a)$$
$$(e_i,e_i) = 2, ~~i=1, \dots, n.
\eqno(4.78b)$$
Let $x_1$, \dots, $x_n$ be the coordinates in ${\bf R}^n$ w.r.t. the basis
(4.78).

Let us consider the system of equations for the unknowns $x_1$, \dots, $x_n$
$$\eqalign{y_1(x_1, \dots, x_n) &= y_1-\lambda\cr
y_2(x_1, \dots, x_n &= y_2\cr
\dots & \cr
y_n(x_1, \dots, x_n) &= y_n\cr}
\eqno(4.79)$$
where $y_i(x)$ are basic homogeneous $W$-invariant polynomials for the group
$W$, $\deg \, y_1 = h = {\rm max}$ (see above). Let
$$\eqalignno{x_1 & = x_1(y,\lambda) &(4.80_1)\cr
\dots & & \cr
x_n &= x_n(y,\lambda) &(4.80_n)\cr}
$$
be the solution of this system (these are algebraic functions),
$y = (y_1, \dots, y_n)\in M = \Cc /W$. (Note that these are the basic
solutions of the Gauss - Manin equations on the Frobenius manifold.)
By $\lambda = \lambda_1(x_1, y)$, \dots, $\lambda = \lambda_n(x_n,y)$
we denote the inverse functions to (4.80$_1$), \dots, (4.80$_n$) resp.
\smallskip
{\bf Proposition 4.2.} {\it The functions
$$h_a(y,z) := -z^{1\over h} \int e^{z\lambda(x,y)}dx, ~~a = 1, \dots, n
\eqno(4.81)$$
are flat coordinates of the deformed connection (3.3) on the Frobenius
manifold $\Cc /W$. Taking
$$\psi_{ia}(y,z) := {\deli h_a(u,z)\over \sqrt{\eta_{ii}(y)}},~i=1, \dots, n
\eqno(4.82)$$
where $\deli = \partial/\partial u^i$, $u_i$ are the roots of (4.29), we
obtain the solution of the Riemann - Hilbert problem of Lecture 3
for the Frobenius manifold.}

Proof. The formula (4.81) follows from (H.51). The second statement is the
inversion of Theorem H.2. Proposition is proved.
\medskip
{\bf Corollary 4.7.} {\it The nonzero off-diagonal entries of the Stokes
matrix of the Frobenius manifold $\Cc /W$ for a finite Coxeter group $W$
coincide with the entries of the Coxeter matrix of $W$.}
\medskip
So the Stokes matrix of the Frobenius manifolds is \lq\lq a half"
of the correspondent Coxeter matrix. For the simply-laced groups
(i.e., the $A$ - $D$ - $E$ series) this was obtained from physical
considerations in [27].
\medskip
To obtain the LG superpotential for the Frobenius manifold $\Cc /W$
we are to find the inverse function $\lambda = \lambda(p,y)$ to
$$p = x_1(y,\lambda) + \dots + x_n(y,\lambda)
\eqno(4.83)$$
according to (I.10).
\smallskip
{\bf Example 4.3.} We consider again the group $A_n$ acting on the hyperplane
(4.41) of the Euclidean space ${\bf R}^{n+1}$ with a standard basis
$f_0$, $f_1$, \dots, $f_n$. We chose the permutations
$$T_1:\, \xi_0\leftrightarrow\xi_1, \dots,
T_n:\, \xi_0\leftrightarrow\xi_n
\eqno(4.84)$$
as the generators of the reflection group. The correspondent root basis
(in the hyperplane (4.41)) is
$$e_1 = f_1 - f_0, \dots, e_n = f_n - f_0.
\eqno(4.85)$$
The coordinates of a vector $\xi_0 f_0 + \xi_1 f_1 + \dots
+ \xi_n f_n$ are
$$x_1 = \xi_1, \dots, x_n = \xi_n.
$$
Note that the sum
$$p = x_1 + \dots + x_n = -\xi_0
\eqno(4.86)$$
is one of the roots (up to a sign) of the equation
$$p^{n+1} + a_1 p^{n-1} + \dots + a_n = 0.
$$
The invariant polynomial of the highest degree is $a_n$. So to construct
the LG superpotential we are to solve the equation
$$p^{n+1} + a_1 p^{n-1} + \dots + a_n-\lambda = 0.
\eqno(4.87)$$
and then to invert it. It's clear that we obtain
$$\lambda = \lambda(p, a_1, \dots, a_n) =
p^{n+1} + a_1 p^{n-1} + \dots + a_n.
\eqno(4.88)$$
We obtain a new proof of Lemma 4.5.

For other Coxeter groups $W$ the above algorithm gives a universal
construction of an analogue of the versal deformation of the
correspondent simple singularity. But the calculations are more
involved.
\vfill\eject

\medskip
\centerline{\bf Appendix J.}
\smallskip
\centerline{\bf Extended complex crystallographic Coxeter groups}
\smallskip
\centerline{\bf and twisted Frobenius manifolds.}
\medskip
Complex crystallographic groups were introduced by Bernstein
and Schwarzman in [15] (implicitly they had been already used
by Looijenga in [95]). These are the groups of affine transformations of
a complex affine $n$-dimensional space $V$ with the linear part
generated by reflections.
The very important subclass is {\it complex crystallographic Coxeter
groups} (CCC grous briefly). In this case by definition $V$ is
the complexification of a real space $V_{\bf R}$; it is required
that the linear parts of the transformations of a CCC group form
a Coxeter group acting in the real linear space of translations of
$V_{\bf R}$.

CCC groups are labelled by Weyl groups of simple Lie algebras. For
any fixed Weyl group $W$ the correspondent CCC group $\tilde W$ depends
on a complex number $\tau$ in the upper half-plane as on the parameter.
Certain factorization w.r.t. a discrete group of M\"obius transformations
of the upper half-plane that we denote by $\Gamma_W$
must be done to identify equivalent CCC groups
with the given Weyl group $W$. Bernstein and Schvarzman found also
an analogue of the Chevalley theorem for CCC groups. They proved
that for a fixed $\tau$ the space of orbits of a CCC group
$\tilde W$ is a weighted projective space. The weights coincide with
the markings on the extended Dynkin graph of $W$. Note that the
discriminant locus (i.e. the set of nonregular orbits) depends on $\tau$.

We have a natural fiber bundle over the quotient $\{ Im \tau > 0 \}
/ \Gamma_W$ with the fiber $V/\tilde W$. It turns out that the space
of this bundle (after adding of one more coordinate, see below
the more precise construction) carries a natural structure of
a twisted Frobenius manifold in the sense of Apendix B (above).

For the simply-laced case $W = A_l, ~D_l, ~E_l$ the construction
[15] of CCC groups is of special simplicity. The Weyl
group $W$ acts by integer linear transformations in the space
$\Cc^l$ of the complexified root lattice ${\bf Z}^l$. This
action preserves the lattice ${\bf Z}^l \oplus \tau {\bf Z}^l$.
The CCC group $\tilde W = \tilde W(\tau)$ is the semidirect product
of $W$ and of the lattice ${\bf Z}^l \oplus \tau {\bf Z}^l$.
The groups $\tilde W(\tau)$ and $\tilde W(\tau')$ are equivalent
{\it iff}
$$\tau' = {a\tau + b\over c\tau + d}, ~~\left(\matrix{
a & b \cr c & d \cr} \right) \in SL(2, {\bf Z}).
$$

The space of orbits $\Cc^l / \tilde W(\tau)$ can be obtained in two
steps. First we can factorize over the translations
${\bf Z}^l\oplus \tau {\bf Z}^l$. We obtain the  direct product of $l$
copies of identical elliptic curves $E_\tau = \Cc /\{{\bf Z} \oplus
\tau {\bf Z}\}$. The Weyl group $W$ acts on $E_\tau^l$. After factorization
$E_\tau^l/ W$ we obtain the space of orbits.

To construct the Frobenius structure we need a $\tilde W$-invariant metric
on the space of the fiber bundle over the modular curve
$M= \left\{ Im \tau >0 \right\} /SL(2,{\bf Z})$ with the fiber
$\Cc^l/\tilde W(\tau)$. Unfortunately, such a metric does not exist.

To resolve the problem
we will consider a certain central extension
of the group $\tilde W(\tau)$ acting in an extended space
$\Cc^l \oplus \Cc$. The invariant metric we need lives on the
extended space. The modular group acts by conformal transformations of
the metric. Factorizing over the action of all these groups we
obtain the twisted Frobenius manifold that corresponds to the given
CCC group.

I will explain here the basic ideas of the construction
for the simplest example of the CCC group $\tilde A_1$ leaving more
general considerations for a separate publication. The Weyl group
$A_1$ acts on the complex line $\Cc$ by reflections
$$v\mapsto -v.
\eqno(J.1a)$$
The group $\tilde A_1(\tau)$ is the semidirect product of the
reflections and of the translations
$$v \mapsto v+m + n\tau, ~~m, n \in {\bf Z}.
\eqno(J.1b)$$
The quotient
$$\Cc /\tilde A_1(\tau) = E_\tau /\{ \pm 1\}
\eqno(J.2)$$
is the projective line. Indeed, the invariants of the group
(J.1) are even elliptic functions on $E_\tau$. It is well-known
that any even elliptic function is a rational function of the
Weierstrass $\wp$. This proves the Bernstein - Schwarzman's
analogue of the Chevalley theorem for this very simple case.

Let us try to invent a metric on the space $(v,\tau)$ being
invariant w.r.t. the transformations (J.1). We immediately
see that the candidate $dv^2$ invariant w.r.t. the
reflections does not help since under the
transformations (J.1b)
$$dv^2 \mapsto (dv + n d\tau)^2\neq dv^2.
$$

The problem of constructing of an invariant metric can be
solved by adding of one more auxiliary coordinate to the
$(v,\tau)$-space adjusting the transformation law of the new
coordinate in order to preserve invariance of the metric. The
following statement gives the solution of the problem.
\smallskip
{\bf Lemma J.1.} {\it \item{1.} The metric
$$ds^2 := dv^2 + 2 d\phi \, d\tau
\eqno(J.3)$$
remains invariant under the transformations $v\mapsto -v$ and
$$\eqalign{\phi &\mapsto \phi - nv - \half n^2 \tau + k\cr
v & \mapsto v + m + n \tau \cr
\tau & \mapsto \tau .\cr}
\eqno(J.4a)$$
\item{2.} The formulae
$$\eqalign{\phi\mapsto \tilde \phi &= \phi + \half {c v^2 \over
c\tau + d} \cr
v\mapsto \tilde v & = {v\over c\tau + d} \cr
\tau \mapsto \tilde \tau &= {a\tau + b\over c\tau +d}\cr}
\eqno(J.4b)$$
with $ad - bc = 1$ determine a conformal transformation of the metric
$ds^2$
$$d\tilde v^2 + 2 d\tilde \phi \, d\tilde \tau =
{dv^2 + 2 d\phi\, d\tau\over (c\tau+d)^2}.
\eqno(J.5)$$
}

Proof is in a simple calculation.
\medskip
We denote by $\hat A_1$ the group generated by the reflection
$v\mapsto -v$ and by the transformations (J.4) with integer
$m$, $n$, $k$, $a$, $b$, $c$, $d$.

Observe that the subgroup of the translations (J.1b) in $\tilde A_1$
becomes non-commutative as the subgroup in $\hat A_1$.

The group generated by the transformations (J.4) with integer
parameters is called {\it Jacobi group}. This name was proposed
by Eichler and Zagier [58]. The automorphic forms of subgroups
of a finite index of Jacobi
group are called {\it Jacobi forms} [{\it ibid.}]. They were
systematically studied in [58]. An analogue $\hat W$ of the
group $\hat A_1$ (the transformations (J.4)
together with the Weyl group $v\mapsto -v$)
can be constructed for any CCC group $\tilde W$ taking the Killing
form of $W$ instead of the squares $v^2$, $n^2$ and $nv$.
Invariants of these Jacobi groups were studied by Saito [126] and
Wirthm\"uller [146] (see also the relevant papers [96,
78, 7]). Particularly, Saito constructed flat
coordinates for the so-called \lq\lq codimension 1"  case. (This means
that there exists a unique maximum among the markings of the
extended Dynkin graph. On the list of our examples only the $E$-groups
are of codimension one.)
Explicit formulae of the Jacobi forms
for $\hat E_6$ that are the flat coordinates
in the sense of [126] have been obtained recently in [128].
\medskip
Let $\Cc^3 = \{ (\phi, v, \tau ), Im \tau >0\}$. I will
show that the space of orbits
$${\cal M}_{\hat A_1} := \Cc^3/\hat A_1
\eqno(J.6)$$
carries a natural structure of a twisted Frobenius manifold with
the intersection form proportional to (J.3). It turns out that this
coincides with the twisted Frobenius manifold of Appendix C.
Furthermore, it will be shown that this twisted Frobenius manifold
can be described by the LG superpotential
$$\lambda(p;\omega, \omega', c) := \wp(2\omega p;\omega, \omega')
+ c.
\eqno(J.7)$$

The factorization over $\hat A_1$ will be done in two steps. First we
construct a map
$$\Cc_0^3 = \{ (\phi, v, \tau), ~Im \tau >0, ~v \neq m + n\tau\}
\to \Cc_0^3 = \{ (z,\omega, \omega'), ~Im {\omega'\over \omega}>0,
{}~ z\neq 2m\omega + 2n\omega' \}
\eqno(J.8)$$
such that the action of the group $\hat A_1$ transforms to the action of
the group of translations
$$z\mapsto z+ 2m\omega + 2n\omega'
\eqno(J.9a)$$
reflections
$$z \mapsto -z
\eqno(J.9b)$$
and changes of the basis of the lattice
$$\eqalign{\omega' &\mapsto a\omega' +b\omega \cr
\omega &\mapsto c\omega' + d\omega\cr}, ~~\left(\matrix{
a & b \cr c & d \cr}\right) \in SL(2,{\bf Z}).
\eqno(J.9c)$$
Note that the subgroup generated by (J.9) look very similar to
the CCC group $\tilde A_1$ but the non-normalized lattice
$\{ 2m\omega + 2n\omega'\}$ is involved in the construction of
this subgroup.

We will use the Weierstrass $\sigma$-function
$$\sigma(z;\omega,\omega') = z
\prod_{m^2 + n^2 \neq 0} \left\{ \left( 1 - {z\over w}\right)
\exp \left[ {z\over w} + \half \left( {z\over w}\right)^2
\right] \right\} ,
\eqno(J.10a)$$
$$w := 2m\omega + 2n\omega'
$$
$${d\over dz}\log\sigma(z;\omega,\omega') = \zeta(z;\omega,\omega').
\eqno(J.10b)$$
It is not changed when changing the basis $2\omega$, $2\omega'$ of the
lattice while for the translations (J.9a)
$$\sigma(z+2m\omega + 2n\omega' ;\omega, \omega')
$$
$$=(-1)^{m+n+mn}\sigma(z;\omega,\omega')
\exp \left[ (z+m\omega + n\omega')(2m\eta + 2n\eta')\right]
\eqno(J.11)$$
where $\eta = \eta(\omega, \omega')$, $\eta' = \eta'(\omega, \omega')$
are defined in (C.60)
\smallskip
{\bf Lemma J.2.} {\it The equations
$$\eqalign{\phi &= {1\over 2\pi i} \left[ \log \sigma(z;\omega,\omega')
-{\eta\over 2\omega}z^2\right]\cr
v &= {z\over 2\omega}\cr
\tau &= {\omega'\over\omega}\cr}
\eqno(J.12)$$
determine a map (J.8). This map locally is a bi-holomorphic
equivalence.
}

Proof. Expressing $\sigma$-function via the Jacobi theta-functions
we obtain
$$\phi = {1\over 2\pi i}\left[ \log 2\omega +
\log {\theta_1(v;\tau)\over \theta_1'(0;\tau)}\right].
$$
This can be uniquely solved for $\omega$
$$\omega = \half {\theta_1' (0;\tau)\over \theta_1(v;\tau)}e^{2\pi i\phi}.
\eqno(J.13a)$$
Substituting to (J.12b) we obtain
$$\eqalignno{z &= v {\theta_1'(0;\tau)\over \theta_1(v;\tau)}e^{2\pi i \phi}
&(J.13b)\cr
\omega' &= {\tau\over 2} {\theta_1'(0;\tau)\over \theta_1(v;\tau)}
e^{2\pi i \phi}.
&(J.13c)\cr}
$$
Using the transformation law of $\sigma$-function and (J.4) we complete
the proof of the lemma.
\medskip
{\bf Corollary J.1.} {\it The space of orbits $\Cc^3/\hat A_1$ coincides
with the universal torus (C.51) factorized over the involution $z\mapsto
-z$.
}
\medskip
Let us calculate the metric induced by (J.3) on the universal
torus. To simplify this calculation I will use the following
basis of vector fields on the universal torus (cf. Appendix B
above)
$$\eqalign{ D_1 &= \omega {\partial\over\partial\omega}
+ \omega' {\partial\over\partial\omega'} + z{\partial\over \partial z}
\cr
D_2 &= {\partial\over\partial z} \cr
D_3 &= \eta {\partial\over\partial\omega}
+ \eta'{\partial\over\partial\omega'}
+ \zeta(z;\omega,\omega') {\partial\over\partial z}.\cr}
\eqno(J.14)$$
As it follows from Lemma the inner products of these vector fields
are functions on the universal torus (invariant elliptic functions in the
terminology of Appendix C).
\smallskip
{\bf Proposition J.1.} {\it In the basis (J.14) the metric $(D_a,D_b)
=: g_{ab}$ has the form
$$(g_{ab}) = {1\over 4\omega^2}\left(
\matrix{0 & 0 & -1 \cr
0 & 1 & 0 \cr
-1 & 0 & -\wp(z;\omega,\omega')\cr}\right).
\eqno(J.15)$$
}

Proof. By the definition we have
$$(D_a,D_b) = D_av D_bv + D_a\phi D_b\tau + D_a\tau D_b\phi.
\eqno(J.16)$$
We have
$$D_1\phi = {1\over 2\pi i},~~D_1v = D_1\tau = 0,
$$
$$D_2\phi ={1\over 2\pi i} \left[ \zeta (z;\omega,\omega') -
{\eta\over\omega}z\right] ,~~D_2v ={1\over 2\omega}, ~~D_2\tau = 0,
$$
$$D_3\phi = {1\over 4\pi i}\left[\left(\zeta(z;\omega,\omega') -
{\eta\over\omega}z\right)^2 + \wp(z;\omega,\omega')\right],
$$
$$D_3v = {1\over 2\omega} \left[ \zeta(z;\omega,\omega') -
{\eta\over\omega}z\right], D_3\tau = -{\pi i\over 2\omega^2}.
\eqno(J.17)$$
In the derivation of these formulae we used the formulae of
Appendix C and also one more formula of Frobenius and
Stickelberger
$$\eta{\partial\log \sigma\over \partial\omega}
+ \eta'{\partial\log \sigma\over \partial\omega'}=
-\half \zeta(z)^2 + \half \wp(z) - {1\over 24} g_2 z
\eqno(J.18)$$
[63, formula 30.]. From (J.17) we easily obtain (J.15). Proposition is proved.
\medskip
{\bf Remark J.1.} The metric (J.15) on the universal torus has still its values
in the line bundle $\ell$. I recall (see above Appendix C) that $\ell$
is the pull-back of the tangent bundle of the modular curve w.r.t.
the projection $(z,\omega,\omega')\mapsto \omega'/\omega$.
\smallskip
{\bf Remark J.2.} The Euler vector field $D_1$ in the coordinates
$(\phi, v, \tau)$ has the form
$$D_1 = {1\over 2\pi i}{\partial\over\partial\phi}.
\eqno(J.19)$$
\medskip
Let
$$\hat {\cal M} := \left\{ (z,\omega,\omega'), ~Im{\omega'\over\omega}>0,
{}~z\neq 0\right\} / \{z\mapsto \pm z+2m\omega + 2n\omega'\}.
\eqno(J.20)$$
The metric (J.15) is well-defined om $\hat{\cal M}$. The next step is to
construct a Frobenius structure on $\hat{\cal M}$.

The functions on $\hat{\cal M}$ are rational combinations of $\omega$,
$\omega'$ and $\wp(z;\omega,\omega')$. We define a grading in the ring
of these functions with only a pole at $z=0$ allowed putting
$$\deg \omega = \deg \omega' = -\half , ~~\deg \wp = 1.
\eqno(J.21a)$$
The correspondent Euler vector field for this grading is
$$E = -\half D_1.
\eqno(J.21b)$$
Note the relation of the grading to that defined by the vector field
$\partial /\partial\phi$.
\smallskip
{\bf Proposition J.2.} {\it There exists a unique Frobenius structure on the
manifold $\hat{\cal M}$ with the intersection form $-4\pi i \, ds^2$
(where $ds^2$ is defined in (J.3)), the Euler vector field (J.21) and
the unity vector field $e = \partial /\partial\wp$. This Frobenius
structure coincides with (C.87).
}

Proof. I introduce coordinates $t^1$, $t^2$, $t^3$ on $\hat{\cal M}$
putting
$$\eqalign{t^1 &= -{1\over \pi i} \left[ \wp(z;\omega,\omega') +
{\eta\over\omega}\right]= -{1\over \pi i}\left[
{\theta_1''(v;\tau)\theta_1(v;\tau) - {\theta_1'}^2(v;\tau)\over
{\theta_1'}^2(0;\tau)}\right] e^{-4\pi i \phi}
\cr
t^2 &= {1\over\omega}= 2 {\theta_1(v;\tau)\over \theta_1'(0;\tau)}
e^{-2\pi i \phi}
\cr
t^3 &= {\omega'\over\omega}=\tau .\cr
}\eqno(J.22)$$
Let us calculate the contravariant metric in these coordinates. To do this
I use again the vector fields (J.14):
$$(dt^\alpha, dt^\beta ) = g^{ab}D_at^\alpha D_bt^\beta
\eqno(J.23)$$
where $g^{ab} = g^{ab}(z;\omega,\omega')$ is the inverse matrix to
$-4\pi i g_{ab}$ (J.15)
$$\left( g^{ab}\right) = -{\omega^2\over \pi i}\left(\matrix{
\wp & 0 & -1 \cr
0 & 1 & 0 \cr
-1 & 0 & 0 \cr} \right).
\eqno(J.24)$$
Using the formulae of Apppendix C we obtain
$$D_1 t^1 = {2\over \pi i} \left[ \wp + {\eta\over\omega}\right],
{}~~D_1t^2 = -{1\over\omega}, ~~D_1 t^3 = 0
$$
$$D_2 t^1 = -{1\over \pi i}\wp' , ~~D_2 t^2 = D_2 t^3 = 0
$$
$$D_3 t^1 = {1\over \pi i}\left[ 2\wp^2 -{1\over 4}g_2 +{\eta^2\over
\omega^2}\right], ~~D_3 t^2 = -{\eta\over\omega^2}, ~~D_3 t^3 = -{\pi i
\over 2\omega^2}.
\eqno(J.25)$$
{}From this it follows that the matrix $g^{\alpha\beta}: =
(dt^\alpha,dt^\beta)$
has the form
$$g^{11} ={i\over\pi^3\omega^4}\left[\omega^6g_3 - \eta\omega\,\omega^4
g_2 + 4(\eta\omega)^3\right],
$$
$$g^{12} = {3\over\pi^2\omega^3}\left[(\eta\omega)^2 - {1\over 12} \omega^4
g_2\right], ~~g^{13} = {i\over \pi}\left[ \wp +{\eta\over\omega}\right],
$$
$$g^{22} ={i\over\pi}\left[\wp - 2{\eta\over\omega}\right],~~
g^{23} = {1\over 2\omega}, ~~g^{33}=0.
\eqno(J.26)$$
Differentiating $g^{\alpha\beta}$ w.r.t. $e$ we obtain a constant matrix
$$\left(\eta^{\alpha\beta}\right) = \left( -{1\over\pi i}{\partial
g^{\alpha\beta}\over\partial\wp}\right)\equiv \left({\partial g^{\alpha\beta}
\over \partial t^1}\right) =
\left(\matrix{0 & 0 & 1\cr 0 & 1 & 0 \cr 1 & 0 & 0 \cr}\right).
$$
So $t^1$, $t^2$, $t^3$ are the flat coordinates.

The next step is to calculate the matrix
$$F^{\alpha\beta} := {g^{\alpha\beta}\over\deg g^{\alpha\beta}}
\eqno(J.27)$$
and then to find the function $F$ from the condition
$${\partial^2 F\over\partial t^\alpha \partial t^\beta}
= \eta_{\alpha\alpha'}\eta_{\beta\beta'}F^{\alpha'\beta'}.
$$
(We cannot do this for the $F^{33}$ entry. But we know that $F^{33}$ must be
equal to $\partial^2F/\partial{t^1}^2$.) It is straightforward to verify
that the function $F$ of the form (C.87) satisfies (J.27). Proposition is
proved.
\medskip
{\bf Remark J.3.} Another way to prove the proposition is close to the proof
of the main theorem of Lecture 4. We verify that the entries of the
contravariant
metric on the space of orbits $\hat{\cal M}$
(J.6) and the contravariant Christoffel
simbols (3.25) of it are elliptic functions with at most
second order pole in the point $z=0$. From this it immediately follows
that the metrics $g^{\alpha\beta}$ and $\partial g^{\alpha\beta}/
\partial \wp$ form a flat pencil. This gives a Frobenius structure on
$\hat{\cal M}$. The last step is to prove analyticity of the components
of $g^{\alpha\beta}$ in the point $\omega'/\omega \, = i\infty$. This completes
the alternative proof of the proposition.
\medskip
Factorizing the Frobenius manifold we have obtained over the action
of the modular group we obtain the twisted Frobenius structure on the space
of orbits of the extended CCC group $\hat A_1$ coinciding with
this of Appendix C. Observe that the flat coordinates $t^\alpha$ are not
globally single-valued functions on the orbit space due to the transformation
law (B.19) determining the structure of the twisted Frobenius manifold.

We will consider briefly the twisted Frobenius manifolds for
the CCC groups $\hat A_n$ in Lecture 5. The twisted Frobenius
structures for the groups $\hat E_6$ and $\hat E_7$ can be obtained
using the results of [141] and [99].
\vfill\eject
\bigskip
\centerline{\bf Lecture 5.}
\medskip
\centerline{\bf Differential geometry of Hurwitz spaces.}
\medskip
Hurwitz spaces are the moduli spaces of Riemann surfaces of a given
genus $g$ with a given number of sheets $n+1$. In other words, these are
the moduli spaces of pairs $(C, \lambda)$ where $C$ is a smooth
algebraic curve of the genus $g$ and $\lambda$ is a meromorphic
function on $C$ of the degree $n+1$. Just this function realizes
$C$ as a $n+1$-sheet covering over $CP^1$ (i.e. as a $n+1$-sheet
Riemann surface). I will consider Hurwitz spaces
with an additional assumption that the type of ramification of
$C$ over the infinite point is fixed.

The idea is to take the function $\lambda= \lambda (p)$ on the Riemann
surface $p\in C$ depending on the moduli of the Riemann surface
as the superpotential in the sense of Appendix I. But we are
to be more precise to specify what is the argument of the superpotential
to be kept unchanged whith the differentiation along the moduli
(see (I.11) - (I.13)).

I will construct first a Frobenius structure on a covering of these
Hurwitz spaces corresponding to fixation of a symplectic
basis of cycles in the
homologies $H_1(C,{\bf Z})$. Factorization over the group
$Sp(2g, {\bf Z})$ of changes of the basis gives us a twisted
Frobenius structure on the Hurwitz space. It will carry a
$g$-dimensional family of metrics being sections
of certain bundle over the Hurwitz space.

We start with explicit description of the Hurwitz
spaces.

Let $M=M_{g;n_0,\dots ,n_m}$  be a moduli space of dimension
$$n=2g+n_0+\dots +n_m+2m\eqno(5.1)$$
of sets
$$(C;\infty_0,\dots ,\infty_m; \lambda )\in M_{g;n_0,\dots ,n_m}\eqno(5.2)$$
where $C$ is a Riemann surface with marked points $\infty_0$, ...,
$\infty_m$, and a marked meromorphic function
$$\lambda :C\to CP^1,~~\lambda^{-1}(\infty ) = \infty_0\cup\dots ,
\cup\infty_m \eqno(5.3)$$
having a degree $n_i+1$ near the point $\infty_i$.
(This is a connected manifold as it follows from [110].) We need the critical
values of $\lambda$
$$u^j=\lambda(P_j),~d\lambda |_{P_j}=0,~j=1,\dots ,n\eqno(5.4)$$
(i.e. the ramification points of the Riemann surface (5.3)) to be local
coordinates in  open domains in $\hat M$ where
$$u^i\neq u^j~{\rm for}~i\neq j\eqno(5.5)$$
(due to the Riemann existence theorem). Note that $P_j$ in (5.4) are the branch
points of the Riemann surface.
Another assumption is that the
one-dimensional affine group acts on $\hat M$ as
$$(C;\infty_0,\dots ,\infty_m;\lambda ;\dots )\mapsto
(C;\infty_0,\dots ,\infty_m;a\lambda +b;\dots )\eqno(5.6a)$$
$$u^i\mapsto au^i+b,~ i=1,\dots ,n.\eqno(5.6b)$$
\smallskip

{\bf Example 5.1.} For the case $g=0$, $m=0$, $n_0 =n$ the Hurwitz space
consists of all the polynomials of the form
$$\lambda(p) = p^{n+1} + a_{n} p^{n-1} + \dots + a_1, ~~a_1, \dots, a_{n}
\in \Cc .
\eqno(5.7)$$
We are to supress the term with $p^{n}$ in the polynomial $\lambda(p)$
to provide a possibility to coordinatize (5.7) by the $n$ critical values
of $\lambda$. The affine transformations $\lambda \mapsto a\lambda + b$
act on (5.7) as
$$p \mapsto a^{1\over n+1}p,
{}~~a_i \mapsto a_i a^{n-i+1\over n+2} ~{\rm for }~ i>1,
{}~ a_1 \mapsto a\, a_1 + b.
\eqno(5.8)$$
\smallskip
{\bf Example 5.2.} For $g=0$, $m = n$, $n_0 = \dots =n_m = 0$
the Hurwitz space consists of all
rational functions of the form
$$\lambda(p) = p + \sum_{i=1}^n {q_i\over p - p_i}.
\eqno(5.9)$$
The affine group $\lambda \mapsto a\lambda + b$ acts on (5.9) as
$$p\mapsto ap+b, ~~ p_i \mapsto p_i + {b\over a}, ~~
q_i \mapsto aq_i.
\eqno(5.10)$$
\smallskip
{\bf Example 5.3.} For a positive genus $g$, $m=0$, $n_0 = 1$
the Hurwitz space consists of all hyperelliptic curves
$$\mu^2 = \prod_{j=1}^{2g+1} (\lambda - u^j).
$$
The critical values $u^1$, \dots, $u^{2g+1}$
of the projection $(\lambda,\mu)\mapsto\lambda$
are the local coordinates on the moduli space. Globally they are
well-defined up to a permutation.
\smallskip
{\bf Example 5.4.} $g>0$, $m=0$, $n: = n_0\geq q$. In this case
the quotient of
the Hurwitz space over the affine group
(5.6) is obtained from the moduli space of {\it all}
smooth algebraic curves of the genus $g$ by fixation of a
non-Weierstrass point $\infty_0\in C$.
\medskip
I describe first the basic idea of introducing of a Frobenius
structure in the covering of a Hurwitz space. We define the multiplication
of the tangent vector fields declaring that the ramification points
$u^1$, \dots, $u^n$ are the canonical coordinates for the multiplication,
i.e.
$$\deli \cdot \delj = \delta_{ij}\deli
\eqno(5.11)$$
for
$$\deli := {\partial\over\partial u^i}.
$$
This definition works for Riemann surfaces with pairwise distinct
ramification points of the minimal order two. Below we will extend the
multiplication onto all the Hurwitz space.
The unity vector field $e$ and the Euler vector field $E$ are the
generators of the action (5.6) of the affine group
$$ e = \sum_{i=1}^n \deli , ~~E = \sum_{i=1}^n u^i\deli .
\eqno(5.12)$$
To complete the description of the Frobenius structure we are
to describe admissible one-forms on the Hurwitz space. I recall that
a one-form $\Omega$ on the manifold with a Frobenius algebra structure in the
tangent planes is called admissible if the invariant inner product
$$<a,b>_\Omega := \Omega(a\cdot b)
\eqno(5.13)$$
(for any two vector-fields $a$ and $b$) determines on the manifold
a Frobenius structure (see Lecture 3).

Any quadratic differential $Q$ on $C$ holomorphic for $|\lambda |<\infty$
determines a one-form $\Omega_Q$
on the Hurwitz space by the formula
$$\Omega_Q := \sum_{i=1}^n du^i \res_{P_i}{Q\over d\lambda}.
\eqno(5.14)$$
A quadratic differential $Q$ is called $d\lambda$-{\it divisible} if
it has the form
$$Q = q d\lambda
\eqno(5.15)$$
where the differential $q$ has no poles in the branch points of $C$.
For a $d\lambda$-divisible quadratic differential $Q$ the correspondent
one-form $\Omega_Q=0$. Using this observation we extend the construction
of the one-form $\Omega_Q$ to {\it multivalued} quadratic differentials.
By the definition this is a quadratic
differential $Q$ on the universal covering of
the curve $C$ such that the monodromy transformation along any cycle
$\gamma$
acts on $Q$ by
$$Q \mapsto Q + q_\gamma d\lambda
\eqno(5.16)$$
for a differential $q_\lambda$.
Multivalued quadratic differentials determine one-forms on the
Hurwitz space by the same formula (5.14).
\medskip
We go now to an appropriate covering of a Hurwitz space in order
to describe multivalued quadratic differentials for which
the one-forms (5.14) will define metrics on the Hurwitz space
according to (5.13).

The covering $\hat M=\hat M_{g;n_0,\dots ,n_m}$
will consist of the sets
$$(C;\infty_0,\dots ,\infty_m; \lambda ; k_0,\dots ,k_m;
a_1,\dots ,a_g,b_1,\dots ,b_g)\in M_{g;n_0,\dots ,n_m}\eqno(5.17)$$
with the same $C$, $\infty_0$, ...,
$\infty_m$, $\lambda$ as above and with a marked symplectic basis
$a_1,\dots ,a_g$, $b_1,\dots ,b_g\in H_1(C,{\bf Z})$, and marked
branches of roots
of $\lambda$
near $\infty_0$, ..., $\infty_m$ of the orders $n_0+1$, ..., $n_m+1$
resp.,
$$k_i^{n_i+1}(P) = \lambda(P), ~P~{\rm near}~\infty_i.\eqno(5.18)$$
(This is still a connected manifold.)

The admissible quadratic differentials on the Hurwitz space will be
constructed as squares $Q = \phi^2$ of certain differentials $\phi$
on $C$ (or on a
covering of $C$). I will call them
{\it primary} differentials. I give now the list of primary differentials.
\smallskip
Type 1. One of the
normalized Abelian differentials of the second kind
on $C$ with poles only at $\infty_0$, \dots, $\infty_m$ of
the orders less then the correspondent orders of the differential
$d\lambda$.
More explicitly,
$$\phi = \phi_{t^{i;\alpha}}, ~~i = 0,\dots, m, ~~\alpha = 1, \dots, n_i
\eqno(5.19a)$$
is the normalized Abelian differential of the second kind with a pole in
$\infty_i$,
$$\phi_{t^{i;\alpha}} = -{1\over\alpha}dk_i^\alpha
+~{\rm regular~terms~~~near~}\infty_i,
\eqno(5.19b)$$
$$\oint_{a_j}\phi_{t^{i;\alpha}}=0;\eqno(5.19c)$$
\smallskip
Type 2.
$$\phi = \sum_{i=1}^m \delta_i\phi_{v^i} ~{\rm for }~ i=1, \dots, m
\eqno(5.20a)$$
with the coefficients $\delta_1$, \dots, $\delta_i$ independent on the point
of $\hat M$. Here $\phi_{v^i}$
is one of the normalized Abelian differentials of the second kind on $C$ with
a pole only at $\infty_i$ with the principal part of the form
$$\phi_{v^i} = -d\lambda +~{\rm regular~terms~~~~near~}\infty_i,\eqno(5.20b)$$
$$\oint_{a_j}\phi_{v^i} = 0;\eqno(5.20c)$$
\smallskip
Type 3.
$$\phi = \sum_{i=1}^m \alpha_i \phi_{w^i}
\eqno(5.21a)$$
$$\oint_{a_j}\phi = 0
\eqno(5.21b)$$
with the coefficients $\alpha_1$, \dots, $\alpha_m$ independent on the
point of $\hat M$. Here
$\phi_{w^i}$ is the  normalized Abelian differential of the third kind with
simple poles at $\infty_0$ and $\infty_i$ with residues -1 and +1 resp.;
\smallskip
Type 4.
$$\phi = \sum_{i=1}^g \beta_i \phi_{r^i}
\eqno(5.22a)$$
with the coefficients $\beta_1$, \dots, $\beta_g$ independent on the point
of $\hat M$. Here
$\phi_{r^i}$ is the  normalized multivalued
differential on $C$ with increments
along the cycles $b_j$ of the form
$$\phi_{r^i} (P+b_j)-\phi_{r^i}(P)
= -\delta_{ij}d\lambda,
\eqno(5.22b)$$
$$\oint_{a_j}\phi_{r^i} = 0\eqno(5.22c)$$
without singularities but those prescribed by (5.22b);
\smallskip
Type 5.
$$\phi = \sum_{i=1}^g \gamma_i \phi_{s^i}
\eqno(5.23a)$$
with the coefficients $\gamma_1$, \dots, $\gamma_g$ independent
on the point of $\hat M$. Here
$\phi_{s^i}$ is the holomorphic differentials
on $C$ normalized by the condition
$$\oint_{a_j}\phi_{s^i} = \delta_{ij}.
\eqno(5.23b)$$
\medskip
Let $\phi$ be one of the primary differentials of the above list.
We put
$$Q = \phi^2
\eqno(5.24)$$
and we will show that the correspondent one-form on the Hurwitz space
is admissible. This will give a Frobenius structure on the covering
$\hat M$ of the Hurwitz space for any of the primary differentials.
We recall that the metric correspondent to the one-form $\Omega_{\phi^2}$
has by definition the form
$$<\partial', \partial''>_\phi := \Omega_{\phi^2}(\partial'\cdot\partial'')
\eqno(5.25)$$
for any two tangent fields $\partial'$, $\partial''$ on $\hat M$.

To construct the superpotential of this Frobenius structure we introduce
a multivalued function $p$ on $C$ taking the integral of $\phi$
$$p(P) := {\rm v.p.}\int_{\infty_0}^P \phi.
\eqno(5.26)$$
The principal value is defined by the subtraction of the divergent part
of the integral as the correspondent function on $k_0$. So
$$\phi = dp.
\eqno(5.27)$$
Now we can consider the function $\lambda$ on $C$ locally as the function
$\lambda = \lambda(p)$ of the complex variable $p$. This function also
depends on the point of the space $\hat M$ as on parameter.

Let $\hat M_\phi$ be the open domain in $\hat M$ specifying by
the condition
$$\phi(P_i)\neq 0, ~~i = 1, \dots, N.
\eqno(5.28)$$
\smallskip
{\bf Theorem 5.1.} {\it \item{1.} For any primary differential $\phi$
of the list (5.19) - (5.23) the multiplication
(5.11), the unity and the Euler
vector field (5.12), and the one-form $\Omega_{\phi^2}$ determine on
$\hat M_\phi$ a structure of Frobenius manifold. The correspondent
flat coordinates $t^A$, $A = 1, \dots, N$ consist of the five parts
$$t^A = \left( t^{i;\alpha},\, i = 0, \dots, m, \, \alpha = 1, \dots, n_i
; ~p^i, q^i, \, i = 1, \dots, m; ~r^i, s^i, \, i = 1, \dots, g
\right)
\eqno(5.29)$$
where
$$t^{i;\alpha}=\res_{\infty_i}k_i^{-\alpha} p\, d\lambda ,~
i=0,\dots ,m,~\alpha =1,\dots ,n_i;\eqno(5.30a)$$
$$p^i= {\rm v.p.}\int_{\infty_0}^{\infty_i}dp ,~i=1,\dots
,m;\eqno(5.30b)$$
$$q^i = -\res_{\infty_i} \lambda dp,~i=1,\dots ,m;\eqno(5.30c)$$
$$r^i = \oint_{b_i}dp,
\eqno(5.30d)$$
$$s^i=-{1\over 2\pi i}\oint_{a_i}\lambda dp,~
i=1,\dots ,g.\eqno(5.30e)$$
The metric (5.22) in the coordinates has the following form
$$\eta_{t^{i;\alpha}t^{i;\beta}} = {1\over n_i+1}\delta_{ij}
\delta_{\alpha +\beta ,n_i+1}\eqno(5.31a)$$
$$\eta_{v^iw^j} = {1\over n_i+1}\delta_{ij}\eqno(5.31b)$$
$$\eta_{r^is^j} = {1\over 2\pi i}\delta_{ij},\eqno(5.31c)$$
other components of the $\eta$ vanish.

The function $\lambda = \lambda(p)$ is the superpotential of this
Frobenius manifold in the sense of Appendix I.
\item{2.} For any other primary differential $\varphi$ the one-form
$\Omega_\varphi$ is an admissible one-form on the Frobenius manifold.
}

Proof. From (5.11) it follows that the metric (5.25) which we will denote
also by $ds^2_\phi$ is diagonal in the coordinates $u^1$, \dots, $u^N$
$$ds^2_\phi = \sum_{i=1}^N \eta_{ii}(u){du^i}^2
\eqno(5.32a)$$
with
$$\eta_{ii}(u) = \res_{P_i} {\phi^2\over d\lambda}.
\eqno(5.32b)$$
We first prove that this is a Darboux - Egoroff metric
for any primary differential.
\smallskip
{\bf Main lemma.} {\it For any primary differential $\phi$ on the list
(5.19) - (5.23) the metric (5.32)
is a Darboux - Egoroff metric satisfying also the
invariance conditions
$${\cal L}_e ds^2_\phi = 0,
\eqno(5.33a)$$
$${\cal L}_E ds^2_\phi ~{\rm is ~proportional ~to}~ ds_\phi^2.
\eqno(5.33b)$$
The rotation coefficients of the metric do not depend on the choice of
the primary differential $\phi$.
}

Proof. We prove first some identity relating a bilinear combination
of periods and principal parts at $\lambda = \infty$ of differentials on $C$
as functions on the moduli to the residues in the branch points.

We introduce a local parameter $z_a$ near the point $\infty_a$
putting
$$z_a = k_a^{-1}.
\eqno(5.34)$$
Let $\omega^{(1)}, ~\omega^{(2)}$ be two differentials on the universal
covering of $C\setminus (\infty_0 \cup \dots \cup\infty_m)$ holomorphic
outside of the infinity with the following properties at the infinite
points
$$\omega^{(i)} = \sum_k c_{ka}^{(i)} z^k_adz_a +
d\sum_{k>0} r_{ka}^{(i)} \lambda^k \log \lambda, ~~P \to \infty_a
\eqno(5.35a)$$
$$\oint_{a_\alpha} \omega^{(i)} = A_\alpha^{(i)}
\eqno(5.35b)$$
$$\eqalignno{\omega^{(i)} \left( P + a_\alpha\right) - \omega^{(i)}\left(
P\right)
&= dp_\alpha^{(i)}(\lambda),
{}~~p_\alpha^{(i)} (\lambda)= \sum_{s>0} p_{s\alpha}^{(i)}\lambda^s
&(5.35c)\cr
\omega^{(i)} \left( P + b_\alpha\right) - \omega^{(i)}\left( P\right)
&= dq_\alpha^{(i)}(\lambda),
{}~~q_\alpha^{(i)} (\lambda)= \sum_{s>0} q_{s\alpha}^{(i)}\lambda^s,
&(5.35d)\cr}
$$
$i = 1, 2$, where $c_{ka}^{(i)}, ~r_{ka}^{(i)}, ~A_\alpha^{(i)}, ~
p_{s\alpha}^{(i)}, ~q_{s\alpha}^{(i)}$ are some constants (i.e. independent
on the curve). We introduce also a bilinear pairing of such differentials
putting
$$<\omega^{(1)}\, \omega^{(2)}>:=
$$
$$= - \sum_{a=0}^m \left[ \sum_{k\geq 0} {c_{-k-2, a}^{(1)}\over k+1}
c_{k,a}^{(2)} + c_{-1, a} {\rm v.p.} \int_{P_0}^{\infty_a} \omega^{(2)}
+ 2\pi i {\rm v.p. }\int_{P_0}^{\infty_a} r_{k,a}^{(1)} \lambda^k \omega^{(2)}
\right]
$$
$$+ {1\over 2\pi i}\sum_{\alpha = 1}^g \left[ - \oint_{a_\alpha}
q_\alpha^{(1)}(\lambda)\omega^{(2)} + \oint_{b_\alpha}
p_\alpha^{(1)}(\lambda)\omega^{(2)} +
A_\alpha^{(1)} \oint_{b_\alpha} \omega^{(2)}
\right] .
\eqno(5.36)$$
The principal values, as above, are obtained by subtraction of the
divergent parts of the integrals as the corresponding functions on
$k_0$, \dots, $k_m$; $P_0$ is a marked point of the curve $C$ with
$$\lambda(P_0) = 0.
\eqno(5.37)$$

We will use also a natural connection in the tautological bundle
$$\matrix{\downarrow & C \cr \hat M & \cr}.
\eqno(5.38)$$
The connection is uniquely determined by the requirement that
for the horizontal lifts of the vector fields $\deli$
$$\deli \lambda = 0.
\eqno(5.39)$$
\smallskip
{\bf Lemma 5.1.} {\it The following identity holds
$$\res_{P_j}{\omega^{(1)}\omega^{(2)}\over d\lambda}
= \delj <\omega^{(1)}\, \omega^{(2)}>.
\eqno(5.40)$$
}

Proof. We realize the symplectic basis $a_1, \dots, a_g, \,
b_1, \dots, b_g$ by oriented cycles passing through the point
$P_0$. After cutting $C$ along these cycles we obtain a $4g$-gon.
We connect one of the vertices of the $4g$-gon (denoting this
again by $P_0$) with the infinite points $\infty_0, \dots,
\infty_m$ by pairwise nonintersecting paths running inside
of the $4g$-gon. Adding cuttings along these paths we obtain a domain
$\tilde C$. We assume that the $\lambda$-images of all the cuttings
do not depend on the moduli $u\in \hat M$. This can be done locally.
Then we have an identity
$${1\over 2\pi i} \oint_{\partial \tilde C} \left(
\omega^{(1)}(P) \int_{P_0}^P \delj \omega^{(2)}\right)
= -\res_{P_j}{\omega^{(i)}\omega^{(2)}\over d\lambda}.
\eqno(5.41)$$
After calculation of all the residues and of all contour integrals
we obtain (5.40). Lemma is proved.
\medskip
{\bf Corollary 5.1.} {\it The pairing (5.36) of the differentials of the form
(5.35) is symmetric up to an additive constant not depending on the moduli.}
\medskip
Proof of Main Lemma. From Lemma 5.1 we obtain
$$\eta_{jj}(u) = \delj <\phi\, \phi >, ~~j = 1, \dots, N.
\eqno(5.42)$$
{}From this we obtain the symmetry of the rotation coefficients of the
metric (5.32). To prove the identity (3.70a) for the rotation coeffients
let us consider the differential
$$\deli\delj\int \delk \phi
$$
for distinct $i, j, k$. It has poles only in the branch points
$P_i$, $P_j$, $P_k$. The contour integral of the differential along
$\partial \tilde C$ (see above the proof of Lemma 5.1) equals zero. Hence
the sum of the residues vansish. This reads
$$\delj \sqrt{\eta_{ii}}\delk \sqrt{\eta_{ii}}
+ \deli \sqrt{\eta_{jj}} \delk \sqrt{\eta_{jj}} =
\sqrt{\eta_{kk}} \deli \delj \sqrt{\eta_{kk}}.
$$
This can be written in the form (3.70a) due to the symmetry
$\gamma_{ji} = \gamma_{ij}$.

Let us prove now that the rotation coefficients do not depend
on $\phi$. Let $\varphi$ be another primary differential.
We consider the differential
$$\deli \phi \int \delj \varphi
$$
for $i\neq j$. From vanishing of the sum of the residues we obtain
$$\sqrt{\eta_{jj}^\varphi}\deli \sqrt{\eta_{jj}^\phi} =
\sqrt{\eta_{ii}^\phi} \delj \sqrt{\eta_{ii}^\varphi}.
$$
Using the symmetry $\gamma_{ji} = \gamma_{ij}$
we immediately obtain that the rotation
coefficients of the two metrics coincide.

Now we are to prove the identity (5.33a). Let us define an operator
$D_e$ on functions $f=f(P,u)$ by the formula
$$D_ef = {\partial f\over \partial \lambda} + \partial_e f.
\eqno(5.43)$$
The operator $D_e$ can be extended to differentials as the Lie derivative
(i.e. requiring $dD_e = D_ed$). We have
$$D_e\phi = 0
\eqno(5.44)$$
for any of the primary differentials $\phi$. Indeed, from the definition
of these differentials it folows that these do not change with the
transformations
$$\lambda \mapsto \lambda + b, ~~u^j \mapsto u^j + b, ~~j = 1, \dots, N.
$$
Such invariance is equivalent to (3.70b). From (5.42) we immediately
obtain
$$\partial_e \eta_{jj} = 0
$$
for the metric (5.32). Note that this implies also (3.70b).

Doing in a similar way we prove also (5.33b). Introduce the operator
$$D_E : = \lambda {\partial\over\partial\lambda} + \partial_E
\eqno(5.45)$$
we have
$$D_E \phi = [\phi ]\phi.
\eqno(5.46)$$
Here the numbers $[\phi ]$ for the differentials (5.19) - (5.23) have the form
$$\eqalign{[\phi_{t^{i;\alpha}}] &= {\alpha\over n_i+1}\cr
[\phi_{v^i}] &= 1\cr
[\phi_{w^i}] &= 0\cr
[\phi_{r^i}] &= 1\cr
[\phi_{s^i}] &= 0.\cr}
\eqno(5.47)$$
{}From this we obtain
$$\partial_E \eta_{ii}^\phi = (2[\phi ]-1)\eta_{ii}^\phi.
\eqno(5.48)$$
The equation (3.70c) also follows from (5.48). Lemma is proved.
\medskip
So we have obtained for any primary differential $\phi$ a
flat Darboux - Egoroff metric in an open domain of $\hat M$
being invariant w.r.t. the multiplication (5.11). It is easy
to see that the unity vector field is covariantly
constant w.r.t. the Levi-Civit\`a connection for the metric.
According to the results of Lecture 3, this determine a Frobenius
structure on the domain in $\hat M$. Any other metric for another
primary differential will be admissible for this Frobenius
structure due to Lemma 5.1 and Proposition 3.6.
So all the Frobenius structures
are obtained one from another by the Legendre-type transformations (B.2).

The open domains $\hat M_\phi$ cover all the universal covering of the Hurwitz
space under consideration.

We are to prove the formulae for the flat coordinates of the metric
and to establish that $\lambda = \lambda(p)$ is the superpotential
for this Frobenius manifold.

We start from the superpotential.
\smallskip
{\bf Lemma 5.2.} {\it Derivatives of $\lambda(p)dp$ along the
variables (5.19) - (5.23) have the form
$$\eqalign{\partial_{t^{i;\alpha}} \lambda(p) dp &= -\phi_{t^{i;\alpha}}\cr
\partial_{v^i} \lambda(p) dp &=- \phi_{v^i}\cr
\partial_{w^i} \lambda(p)dp &= -\phi_{w^i}\cr
\partial_{r^i} \lambda(p)dp &= -\phi_{r^i}\cr
\partial_{s^i}\lambda(p) dp &= -\phi_{s^i}\cr}
\eqno(5.49)$$
}

I recall that the differentiation in (5.49) is to be done with $p=const$.

Proof. Using the \lq\lq thermodynamic identity" (4.67) we can rewrite
any of the derivatives (5.49) as
$$\partial_{t^A} \left(\lambda(p)dp\right)_{p=const} =
-\partial_{t^A} \left(p\, d\lambda\right)_{\lambda =const} .
\eqno(5.50)$$
The derivatives like
$$\partial_{t^A} p(\lambda)_{\lambda = const}
$$
are holomorphic on the finite part of the curve $C$ outside
the branch points
$P_j$. In the branch point these derivatives have simple poles.
But the differential $d\lambda$ vanishes precisely at the branch points.
So (5.50) is holomorphic everywhere.

Let us consider now behaviour of the derivatives (5.50) at the infinity.
We have
$$p = {\rm ~singular ~part}~ -
$$
$$-(1-\delta_{i0})v^i -{1\over n_i+1} \sum_{\alpha =1}^{n_i}
t^{i;\alpha}k_i^{-(n_i-\alpha+1)} - {1\over n_i+1} w^i k_i^{-(n_i+1)}+
O\left( k_i^{-(n_i+2)}\right)
\eqno(5.51)$$
near the point $\infty_i$. The singular part by the construction
does not depend on the moduli. Also we have
$$p(P + b_\alpha) - p(P) = r^\alpha , ~~\alpha = 1, \dots, g.
\eqno(5.52)$$
For the differential $\omega := p\, d\lambda$ we will obtain
from (5.51), (5.52) and from
$$d\lambda = (n_i+1)k_i^{n_i}dk_i ~~{\rm near}~\infty_i
\eqno(5.53)$$
the following analytic properties
$$\omega = {\rm singular ~part}~ -(1-\delta_{i0})v^id\lambda
-\sum_{\alpha =1}^{n_i} t^{i;\alpha} k_i^{\alpha -1} dk_i
- w^i {dk_i\over k_i} + O\left( k_i^{-2}\right) dk_i
\eqno(5.54a)$$
$$\omega(P+b_\alpha ) - \omega(P) = r^\alpha d\lambda
\eqno(5.54b)$$
$$\oint_{a_\alpha}\omega = - s^\alpha.
\eqno(5.54c)$$
Differentiating these formulae w.r.t. one of the variables
(5.30) we obtain precisely one of the differentials
with the analytic properties (5.19) - (5.23). This proves the lemma.
\medskip
To complete the proof of Theorem we need to prove that (5.30)
are the flat coordinates\footnote{$^{*)}$}{The statement of my papers
[44, 45] that these are global coordinates on the Hurwitz space is wrong.
They are coordinates on $\hat M_\phi$ only. Changing $\phi$ we obtain
a coordinate system of the type (5.30) in a neighbourhood of any point
of the Hurwitz space.}
 for the metric $<~,~>_\phi$ and that
the matrix of this metric in the coordinates (5.30) has the form
(5.31).

Let $t^A$ be one of the coordinates (5.30). We denote
$$\phi_A := -\partial_{t^A} \lambda\, dp.
\eqno(5.55)$$
By the definition and using Lemma 5.1 we obtain
$$<\partial_{t^A}, \partial_{t^B}>_\phi
= \sum_{|\lambda |<\infty}\res_{d\lambda = 0}
{\phi_A\phi_B\over d\lambda} = \partial_e <\phi_A\, \phi_B>.
\eqno(5.56)$$
Note that in the formula (5.56) for $<\phi_A\, \phi_B>$
only the contribution of the second differential $\phi_B$
depends on the moduli. Let us define the coefficients $c_{ka}^{(A,\, B)}$,
$A_\alpha^{(A,\, B)}$, $q_{1\alpha}^{(A,\, B)}$ for the differentials
$\phi_{A,\, B}$ as in the formula (5.35). Note that by the construction
of the primary differentials all the coefficients $r_{ka}^{(A,\, B)}$,
$p_{s\alpha}^{(A,\, B)}$ vanish; the same is true for $q_{s\alpha}^{(A,\, B)}$
for $s>1$. From the equation $D_e \phi_B = 0$ (see (5.44) above) we obtain
$$\partial_e c_{ka}^{( B)} = {k+1\over n_a+1}c_{k-n_a-1, a}^{(B)},
$$
$$\partial_e \int_{P_0}^{\infty_a} \phi_B = {c_{-n_a-2,a}^{(B)}\over n_a
+1}
+ \left({\phi_B\over
d\lambda}\right)_{P_0}
$$
$$\partial_e\oint_{a_\alpha} \lambda \phi_B = \oint_{a_\alpha} \phi_B,
$$
$$\partial_e \oint_{b_\alpha}\phi_B = -{\phi_B\over d\lambda}(P+b_\alpha)
+{\phi_B\over d\lambda}(P)
\eqno(5.57)$$
(this does not depend on the point $P$).

The proof of all of these formulae is essentially the same. I will prove
for example the last one.

Let us represent locally the differential $\phi_B$ as
$$\phi_B = d\Phi(\lambda; u^1, \dots, u^N).
$$
We have for arbitrary $a$, $b$
$$\partial_e \int_a^b \phi_B = \partial_e\int_a^b d\Phi(\lambda;
u^1,\dots, u^N) = {d\over d\epsilon} \int_a^b d\Phi(\lambda;
u^1+\epsilon,\dots, u^N+\epsilon)|_{\epsilon =0}.
\eqno(5.58)$$
Doing the change of the variable $\lambda \mapsto \lambda -\epsilon$
we rewrite (5.58) as
$${d\over d\epsilon} \int_{a-\epsilon}^{b-\epsilon} d\Phi(\lambda +\epsilon ;
u^1+\epsilon,\dots, u^N+\epsilon)|_{\epsilon =0} =
-\left({d\Phi(\lambda
; u^1, \dots,u^N)\over d\lambda}\right)_{\lambda = a}^
{\lambda = b}
\eqno(5.59)$$
since
$${d\over d\epsilon} d\Phi(\lambda+\epsilon ; u^1+\epsilon,\dots,
u^N +\epsilon ) \equiv D_e \phi_B = 0
$$
(see (5.44) above). Using (5.59) for the contour integral we obtain (5.57).

{}From this and from (5.36) we obtain
$$\partial_e <\phi_A\, \phi_B>=
-\sum_{a=0}^m\left( {1\over n_a+1}\sum_{k=0}^{n_a-1}c_{-k-2,a}^{(A)}
c_{k-n_a-1,a}^{(B)}\right) -
$$
$$- \sum_{a=1}^{m}{1\over n_a+1}\left(
c_{-1,a}^{(A)}c_{-n_a-2,a}^{(B)} +c_{-1,a}^{(B)}c_{-n_a-2,a}^{(A)}
\right)
-{1\over 2\pi i} \sum_{\alpha=1}^g\left(q_{1\alpha}^{(A)} A_\alpha^{(B)}
+ A_\alpha^{(A)}q_{1\alpha}^{(B)}\right).
\eqno(5.60)$$
The r.h.s. of this expression is a constant that can be easyly
calculated using the explicit form of the differentials
$\phi_A,\, \phi_B$. This gives (5.31).

Observe now that
the structure functions $c_{ABC}(t)$ of the Frobenius manifold
can be calculated as
$$c_{ABC}(t) =\sum_{i=1}^n\res_{P_i}
{\phi_{t^A}\phi_{t^B}\phi_{t^C}\over d\lambda dp}.\eqno(5.61)$$
Extension of the Frobenius structure on {\it all} the moduli space $\hat M$ is
given by the condition that the differential
$${\phi_{t^A}\phi_{t^B}-c_{AB}^C\phi_{t^C}dp\over d\lambda}\eqno(5.62)$$
is holomorphic for $|w|<\infty$. (The Frobenius algebra on $T_tM$ will be
nilpotent for Riemann surfaces $\lambda:C\to CP^1$ with more than double branch
points.) Theorem is proved.
\medskip
{\bf Remark 5.1.} Interesting
algebraic-geometrical examples of solutions of equations
of associativity (1.14)
(not satisfying the scaling invariance) generalizing ours
were
constructed in [88]. In these examples $M$ is a moduli space of
Riemann surfaces of genus $g$ with marked points, marked germs of local
parameters near these points, and
with a marked normalized Abelian differential
of the second kind $d\lambda$ with poles at marked points and with fixed
$b$-periods
$$\oint_{b_i}d\lambda
=B_i, ~i=1,\dots ,g.
\eqno(5.63)$$
For $B_i=0$ $d\lambda$ is a perfect differential of a function $\lambda$.
So
one obtains the above Frobenius structures on
$M_{g;n_0,\dots ,n_m}$.
\medskip
{\bf Exercise 5.1.} Prove that the function $F$ for the Frobenius
structure constructed in Theorem has the form
$$F = -\half <p\, d\lambda~ p\, d\lambda>
\eqno(5.64)$$
where the pairing $<~\,~>$ is defined in (5.36).
\smallskip
{\bf Exercise 5.2.} Prove the formula
$$\partial_{t^A}\partial_{t^B} F=-<\phi_A\, \phi_B>.
\eqno(5.65)$$
\medskip
Note, particularly, that for the $t^A$-variables of the fifth
type (5.30e) the formula (5.65) reads
$$\partial_{s^\alpha}\partial_{s^\beta}F = -\tau_{\alpha\beta}
\eqno(5.66a)$$
where
$$\tau_{\alpha\beta} := \oint_{b_\beta}\phi_{s^\alpha}
\eqno(5.66b)$$
is the period matrix of holomorphic differentials on the
curve $C$.
\smallskip
{\bf Remark 5.2.} The formula (5.66) means that the Jacobians $J(C)$ of
the curve $C$
$$J(C) := \Cc^g/ \{m+\tau n\} , ~~m,\, n \in {\bf Z}^g
$$
are Lagrangian manifolds for the symplectic structure
$$\sum_{\alpha =1}^g ds^\alpha \wedge dz_\alpha
\eqno(5.67)$$
where $z_1, \dots, z_g$ are the natural coordinates
on $J(C)$ (i.e., coming from the linear coordinates in
$\Cc^g$). Indeed, the shifts
$$z \mapsto z+m+\tau n
\eqno(5.68)$$
preserve the symplectic form (5.67) due to (5.66). Conversely, if
the shifts (5.68) preserve (5.67) then the matrix $\tau$ can be
presented in the form (5.66a). So the representation (5.66)
is a manifestation of
the phenomenon that the Jacobians are {\it complex Liouville
tori} and the coordinates $z_\alpha$, $s^\alpha$ are the
{\it complex action-angle variables} on the tori.
The origin of the symplectic
structure (5.67) in the geometry of Hurwitz spaces is in realization of these
spaces as the moduli spaces [86] of the algebraic-geometrical solutions
of integrable hierarchies of the KdV type (see also the next Lecture).

In the recent paper [42] an interesting symplectic structure has been
constructed
on the fiber bundle of the intermediate Jacobians of a Calabi - Yau
three-folds $X$ over the moduli space of pairs $(X,\Omega)$, $\Omega
\in H^{3,0}(X)$. The intermediate Jacobians are also complex Liouville
tori (i.e., Lagrangian manifolds) although they are not Abelian
varieties. Their period matrix thus also can be represented in the form
(5.66a) for appropriate coordinates on the moduli space.
\medskip
{\bf Remark 5.3.} A part of the flat coordinates of the intersection form
of the Frobenius manifold can be obtained by the formulae
similar to (5.30b) - (5.30d) with the substitution $\lambda \mapsto
\log\lambda$. Another part is given, instead of (5.30a),  by the formula
$$\tilde t^a := p(Q_a), ~~a= 1, \dots, n
\eqno(5.69a)$$
where
$$n+1 := n_0+1 + n_1 + 1 + \dots + n_m+1
$$
is the number of sheets of the Riemann surface
$\lambda : C\to CP^1$, $Q_0$, $Q_1$, \dots, $Q_n$ are the zeroes of
$\lambda$ on $C$,
$$\lambda(Q_a) = 0.
\eqno(5.69b)$$
\medskip
The last step in our construction is to factorize over
the group $Sp(g,{\bf Z})$  of changes of the symplectic
basis $a_1, \dots, a_g, \, b_1, \dots, b_g$
$$\eqalign{a_i &\mapsto \sum_{j=1}^g\left( C_{ij}b_j + D_{ij} a_j
\right) \cr
b_i &\mapsto \sum_{j=1}^g\left( A_{ij} b_j + B_{ij} a_j\right)
\cr}
\eqno(5.70)$$
where the matrices $A=(A_{ij}), ~B=(B_{ij}), ~C=(C_{ij}), ~D=(D_{ij})$
are integer-valued matrices satisfying
$$\left(\matrix{A & B \cr C & D \cr}\right)
\left(\matrix{0 & -1 \cr 1 & 0 \cr}\right)
\left(\matrix{A & B \cr C & D \cr}\right)^T =
\left(\matrix{0 & -1 \cr 1 & 0 \cr}\right)
\eqno(5.71)$$
and changes of the branches $k_j$ of the roots of $\lambda$
near $\infty_j$
$$k_j \mapsto e^{2\pi i l\over n_j+1}k_j, ~~~l = 1, \dots, n_j+1.
\eqno(5.72)$$
We can calculate the transformation law of the primary differentials
(5.19) - (5.23)
 with the transformations (5.70) - (5.72).
This determines the transformation law
of the metrics (5.32). They transform like squares of the
primary differentials. All this gives a messy picture on the
quotient of the twisted
Frobenius manifold coinciding with the Hurwitz space.
An important simplification we have is that the multiplication
law of tangent vectors to the Hurwitz space stays invariant
w.r.t. the transformations.

The picture is simplified drastically if the genus of the curves
equals zero. In this case the modular group (5.70) disappears and
the action of the group of roots of unity (5.72) is very simple.
\smallskip
{\bf Exercise 5.3.} Verify that the Frobenius structure of Theorem 5.1
on the Hurwitz space (5.7) for the primary differential $\phi = dp$
coincides with the structure of Example 1.7. Check also
that the formulae (5.30a) for the flat coordinates in this case
coincide with the formulae (4.61).
\medskip
For the positive genus case the above transformation law of the
invariant metrics splits into five blocks corresponding to the
five types of the primary differentials (5.19) - (5.23). Let us consider
only the invariant metrics and the correspondent Frobenius
structures that correspond to the holomorphic primary differentials
(the type five in the previous notations). Any of these structures
for a given symplectic basis of cycles is parametrized
by $g$ constants $\gamma_1$, \dots, $\gamma_g$ in such a way that
the correspondent primary differential is
$$\phi = \sum_{i=1}^g\gamma_i \phi_{s^i}.
\eqno(5.73)$$
I recall that $\phi_{s^i}$ are the basic normalized
holomorphic differentials on the curve $C$ w.r.t. the given
symplectic basis of cycles.
The correspondent metric (5.32) we denote by $<~,~>_\gamma$.
A change (5.70) of the symplectic basis of cycles determines the
transformation law of the metric:
$$<~,~>_\gamma \mapsto <~,~>_{(C\tau + D)^{-1}\gamma}.
\eqno(5.74)$$
Here
$$\gamma = (\gamma_1, \dots, \gamma_g)^T.
$$
This is an example of twisted Frobenius
manifold: $g$-parameter family of Frobenius structures
on the Hurwitz space with the action (5.74) of the Siegel modular group
(5.70), (5.71).
\medskip
We stop now the general considerations postponing for further
publications, and we consider examples.
\smallskip
{\bf Example 5.5.} $M_{0;\, 1,0}$. This is the space of rational
functions of the form
$$\lambda = \half p^2 + a + {b\over p-c}
\eqno(5.75)$$
where $a$, $b$, $c$ are arbitrary complex parameters.
We take first $dp$ as the basic primary differential.
So $\lambda(p)$ is the LG superpotential (I.11).
Using the
definition (I.11) we immediately obtain for the metric
$$<\partial_a,\partial_a> = <\partial_b,\partial_c>=1
\eqno(5.76)$$
other components vanish,
and for the trilinear tensor (I.12)
$$\eqalign{<\partial_a,\partial_a,\partial_a>&=1\cr
<\partial_a,\partial_b,\partial_c>&=1\cr
<\partial_b,\partial_b,\partial_b>&=b^{-1}\cr
<\partial_b,\partial_c,\partial_c>&=c\cr
<\partial_c,\partial_c\partial_c>&=b\cr}
\eqno(5.77)$$
otherwise zero. The free energy and the Euler operator read
$$F = {1\over 6} a^3 + abc +\half b^2 \log b + {1\over 6} bc^3
-{3\over 4} b^2
\eqno(5.78)$$
(we add the quadratic term for a convenience later on)
$$E = a\partial_a+{3\over 2} b\partial_b + \half c\partial_c.
\eqno(5.79)$$
This is the solution of WDVV of the second type (i.e. $<e,e>\neq 0$).

To obtain a more interesting solution of WDVV
let us do the Legendre transform $S_b$ (in the notations of Appendix
B). The new flat coordinates read
$$\eqalign{\hat t^a&=c\cr
\hat t^b&= a + \half c^2\cr
\hat t^c&= \log b.\cr}
\eqno(5.80)$$
The new free energy $\hat F$
is to be determined from the equations (B.2b).
After simple calculations we obtain
$$\hat F = \half t_1^2 t_3 + \half t_1 t_2^2
-{1\over 24} t_2^4 + t_2 e^{t_3}
\eqno(5.81a)$$
where
$$\eqalign{t_1 &:=\hat t^b\cr
t_2&:= \hat t^a\cr
t_3&:= \hat t^c.\cr}
\eqno(5.81b)$$
The Euler vector field of the new solution is
$$E= t_1 \partial_1 + \half t_2\partial_2 +{3\over 2}\partial_3.
\eqno(5.82)$$
So for the solution (5.81) of WDVV $d=1$, $r=3/2$. This is just the
Frobenius manifold of Exercise 4.3.
The correspondent
primary differential is
$$d\hat p := \partial_b (\lambda dp) = {dp\over p-c}.
$$
{}From (5.26) we obtain
$$\hat p = \log{p-c\over\sqrt{2}}.
$$
Substituting to (5.75) we derive
the superpotential for the Frobenius manifold (5.81)
$$\lambda = \lambda (\hat p) =
e^{2\hat p} + t_2\sqrt{2}e^{\hat p} + t_1 + {1\over\sqrt{2}}e^{t_3-\hat p}.
\eqno(5.84)$$
The intersection form of the Frobenius manifold (5.81) is given by
the matrix
$$\left(g^{\alpha\beta}(t)\right) =
\left(\matrix{2t_2 e^{t_3} & {3\over 2}e^{t_3} & t_1\cr
{3\over 2}e^{t_3} & t_1 -\half t_2^2 & \half t_2\cr
t_1 & \half t_2 &{3\over 2}\cr}\right).
\eqno(5.85)$$
The flat coordinates $x,\, y,\, z$ of the intersection form are obtained
using Remark 5.3 in the form (4.73)
where
$$\hat p = -{1\over 6}\log 2 + {z\over 3} +x +(2m+1)\pi i~{\rm and}~
\hat p = -{1\over 6}\log 2 + {z\over 3} +y +(2n+1)\pi i
\eqno(5.86)$$
are two of the roots of the equation $\lambda(\hat p) = 0$
($m$, $n$ are arbitrary integers).
As in Exercise 4.3 the monodromy around the discriminant of the
Frobenius manifold is an affine Weyl group (this time of the type
$A_2^{(1)}$). It is generated, say, by the transformations
$$\eqalign{x&\mapsto y\cr y &\mapsto x\cr z &\mapsto z\cr}
\eqno(5.87a)$$
and
$$\eqalign{x &\mapsto x\cr y &\mapsto -x -y \cr z &\mapsto z\cr}
\eqno(5.87b)$$
and by translations
$$\eqalign{x&\mapsto x+2m\pi i\cr
y&\mapsto y+2n\pi i\cr
z&\mapsto z.\cr}
\eqno(5.87c)$$
The monodromy around the obvious closed loop along $t_3$ gives an
extension of the affine Weyl group by means of the transformation
$$\eqalign{x &\mapsto y - {2\pi i\over 3}\cr
y &\mapsto -x - y - {2\pi i\over 3}\cr
z &\mapsto z+ 2\pi i.\cr}
\eqno(5.87d)$$
This is the analogue of the gliding reflection (G.24b) since the cube of it
is just a translation
$$\eqalign{x&\mapsto x\cr
y&\mapsto y\cr
z&\mapsto z+ 6\pi i.\cr}
\eqno(5.87e)$$

We conclude that
the monodromy group of the Frobenius manifold (5.81) coincides
with the extension of the affine Weyl group of the type
$A_2^{(1)}$ by means of the group of cubic roots of the
translation (5.87e) (cf. Exercise 4.3 above).
\smallskip
{\bf Exercise 5.4.} Using an appropriate generalization of the
above example to the Hurwitz space $M_{0;n-1,\, 0}$ show, that
the monodromy group of the Frobenius manifold with the superpotential
$$\lambda (p) = e^{np} + a_1 e^{(n-1)p} + \dots + a_n + a_{n+1}e^{-p}
\eqno(5.88a)$$
($a_1, \dots, a_{n+1}$ are arbitrary complex numbers) is an extension
of the affine Weyl group of the type $A_n^{(1)}$ by means of the
group of roots of the order $n+1$ of the translation
$$\log a_{n+1} \mapsto \log a_{n+1} + 2(n+1)\pi i.
\eqno(5.88b)$$
\medskip
{\bf Example 5.6.} $M_{1;1}$. By the definition this is the space
of elliptic curves of the form
$$\mu^2 = 4\lambda^3 + a_1 \lambda^2 + a_2 \lambda + a_3
\eqno(5.89)$$
with arbitrary coefficients $a_1$, $a_2$, $a_3$ providing that
the polynomial in the r.h.s. of (5.89) has no multiple roots.
I will show that in this case the twisted Frobenius structure
of Theorem 5.1 coincides with the twisted Frobenius structure
on the space of orbits of the extended CCC group $\hat A_1$
(see above Appendix C and Appendix J).

It is convenient to use elliptic uniformization of the curves.
I will use the Weierstrass uniformization. For this I rewrite
the curve in the form
$$\mu^2 = 4(\lambda -c)^2 - g_2(\lambda -c) -g_3 =
4(\lambda -c-e_1)(\lambda -c-e_2)(\lambda -c-e_3)
\eqno(5.90)$$
where $c$ and the parameters $g_2$, $g_3$ of the Weierstrass
normal form are uniquely specified by $a_1$, $a_2$, $a_3$.
The
Weierstrass uniformization of (5.90) reads
$$\eqalign{\lambda &=\wp (z)+c\cr
\mu &=\wp '(z)\cr}
\eqno(5.91)$$
where $\wp = \wp(z;g_2, g_3)$ is the Weierstrass function.
The infinite point is $z=0$.
Fixation of a basis of cycles on the curve (5.90) corresponds
to fixation of an ordering of the roots $e_1$, $e_2$, $e_3$.
The correspondent basis of the lattice of periods is $2\omega$,
$2\omega'$ where
$$\wp(\omega) = e_1, ~~\wp(\omega') = e_3.
\eqno(5.92)$$
Let us use the holomorphic primary differential
$$dp = {dz\over 2\omega}$$
to construct a Frobenius structure on $M$.
The corresponding superpotential is
$$\lambda (p) := \wp(2\omega p; \omega, \omega') + c.
\eqno(5.93)$$
$$p \simeq p + m + n\tau.
\eqno(5.94)$$
The flat coordinates (5.30) read
$$t^1 := {1\over \pi i} \oint_a \lambda dp
= {1\over 2 \pi i\omega}\int_0^{2\omega} \left[ \wp(z;\omega,\omega')
+ c\right] dz = {1\over \pi i}\left[
-c +{\eta\over\omega}\right]
\eqno(5.95a)$$
$$t^2 := -\res_{z=0} \lambda^{-1/2} p\, d\lambda =
1/\omega\eqno(5.95b)$$
$$t^3 := \oint_b dp =
\tau ~~{\rm where~} \tau = \omega '/\omega .
\eqno(5.95c)$$
(I have changed slightly the normalization of the flat coordinates (5.30).)
The metric $<~,~>_{dp}$ in these coordinates has the form
$$ds^2 = {dt^2}^2 + 2 dt^1 dt^3.
\eqno(5.96)$$
Changes of the basis of cycles on the elliptic curve determined
by the action of $SL(2,{\bf Z})$
$$\eqalign{\omega' &\mapsto a\omega' + b\omega\cr
\omega &\mapsto c\omega' + d\omega\cr}
{}~~~\left(\matrix{a & b \cr c & d \cr}\right) \in SL(2,{\bf Z})
\eqno(5.97)$$
gives rise to the following transformation of the metric
$$ds^2 \mapsto {ds^2\over (c\tau + d)^2}.
\eqno(5.98)$$
These formulae determine a twisted Frobenius structure on
the moduli space $M_{1;1}$ of elliptic curves (5.89).

The coincidence of the formulae  for the flat coordinates
and for the action (5.97) of the modular group
with
the above formulae in the constructions in the theory
of the extended CCC group $\hat A_1$ is not accidental. We will
prove now the following statement.
\smallskip
{\bf Theorem 5.2.} {\it The space of orbits of $\hat A_1$ and the manifold
$M_{1;1}$ are isomorphic as twisted Frobenius manifolds.}
\medskip

Proof. Let us consider the intersection form of the
Frobenius manifold $M_{1;1}$. Substituting
$$c = -\wp(z;\omega, \omega')
\eqno(5.99)$$
we obtain a flat metric on the $(z; \omega, \omega')$-space.
Calculating this metric in the basis of the vector fields
$D_1$, $D_2$, $D_3$ (J.14) we obtain the metric (J.24). It is clear
that the unity and the Euler vector fields of $M_{1;1}$
coincide with $e=\partial/\partial\wp$
and (J.21) resp. So Theorem 5.1 implies
the isomorphism of the Frobenius manifolds. The action of the
modular group on these manifolds is given by the same
formulae. Theorem is proved.
\medskip
{\bf Example 5.7.} We consider very briefly the Hurwitz space
$M_{1;n}$ coinciding, as we will see, with the space of
orbits of $\hat A_n$.

This is the moduli space of all elliptic curves
$$E_L := \Cc /L, ~~L = \{ 2m\omega + 2n\omega'\}
\eqno(5.100)$$
with a marked meromorphic function of degree $n+1$ with only
one pole. We can assume this pole to coincide with $z=0$.

To realize the space of orbits of
the extended CCC group $\hat A_n$ we consider
the family of the Abelian manifolds
$E_L^{n+1}$ fibered over the space ${\cal L}$ of all lattices
 $L$ in $\Cc$. The symmetric group acts on the space of the
fiber bundle
by permutations ${(z_0, z_1, \dots, z_n)} \mapsto
(z_{i_0}, z_{i_1}, \dots, z_{i_n})$ (we denote by $z_k$ the coordinate
on the $k$-th copy of the elliptic curve). We must restrict
this onto the hyperplane
$$z_0 + z_1 + \dots + z_n = 0
\eqno(5.101)$$
After factorization over the permuations we obtain the space of
orbits of $\hat A_n$.

Note that the conformal invariant metric for $\hat A_n$ on
the space $(z_0, z_1, \dots, z_n; \omega, \omega')$ can be obtained
as the restriction of the direct sum of the invariant metrics
(J.15) onto the hyperplane (5.101).

We assign now to any point of the space of orbits of $\hat A_n$ a point
$(E_L, \lambda)$
in the Hurwitz space $M_{1;n}$ where $E_L$ is the same elliptic curve
and the function $\lambda$ is defined by the formula
$$\lambda := (-1)^{n-1} {\prod_{k=0}^n\sigma (z-z_k)\over
\sigma^{n+1}(z) \prod_{k=0}^n \sigma(z_k)}
= {1\over n!}{\det \left(\matrix{1 & \wp(z) & \wp'(z) &
\ldots & \wp^{(n-1)}(z)\cr
1 & \wp(z_1) & \wp'(z_1) &
\ldots & \wp^{(n-1)}(z_1)\cr
\vdots & \vdots & \vdots & & \vdots\cr
1 & \wp(z_n) & \wp'(z_n) &
\ldots & \wp^{(n-1)}(z_n)\cr
}\right)\over
\det \left(\matrix{1 & \wp(z_1) & \wp'(z_1) &
\ldots & \wp^{(n-2)}(z_1)\cr
1 & \wp(z_2) & \wp'(z_2) &
\ldots & \wp^{(n-2)}(z_2)\cr
\vdots & \vdots & \vdots & & \vdots\cr
1 & \wp(z_n) & \wp'(z_n) &
\ldots & \wp^{(n-2)}(z_n)\cr
}\right)}
$$
$$= {(-1)^{n-1}\over n!} \wp^{(n-1)}(z) + c_n(z_0, z_1, \dots z_n)
\wp^{(n-2)}(z) + \dots + c_1(z_0, z_1, \dots , z_n)
\eqno(5.102)$$
where the coefficients $c_1, \dots, c_n$ are defined by this equation
(I have used the classical addition formula [145, p.458] for sigma-functions).
Taking, as above, the normalized holomorphic differential
$$dp := {dz\over 2\omega}
$$
we obtain the twisted Frobenius structure on the space of functions
(5.102). By the construction, the superpotential of this Frobenius
manifold has the form (5.102) where one must substitute
$$z \mapsto 2\omega p.
$$
It is easy to check that the intersection form of this
Frobenius structure coincides with the conformal invariant metric
of $\hat A_n$.
Indeed, the flat coordinates (5.69) of the intersection form are
the zeroes of $\lambda(p)$
$$\tilde t^0 := {z_0\over 2\omega}, \dots,
 ~\tilde t^n := {z_n\over 2\omega}
\eqno(5.103a)$$
related by the linear constraint
$$\tilde t^0 + \dots + \tilde t^n = 0.
\eqno(5.103b)$$
These coordinates are not well-defined. The ambiguity in their
definition comes from two origins. The first one is the monodromy
group of the Riemann surface $\lambda : C\to CP^1$. It acts by
permutations on the set of
the zeroes of the function $\lambda$. Another origin of the
ambiguity is in the multivaluedness of the function $p$ (Abelian
integral) on the Riemann surface. So the translations of the
flat coordinates
$$\tilde t^a \mapsto m+ n\tau , ~~ m,n \in {\bf Z}
\eqno(5.104)$$
give the same point of the Hurwitz space. The permutations and
the translations just give the action of the CCC group
$\tilde A_n$. To see the extended CCC group $\hat A_n$ one should
consider all the flat coordinates of the intersection form.
They are obtained by adding two more coordinates (5.30d) and (5.30e)
(with the substitution $\lambda\mapsto\log\lambda$)
 to (5.103)
$$\tau = {\omega'\over \omega}=\oint_b dp, ~~\phi : = -{1\over 2\pi i} \oint_a
\log\lambda\, dp
\eqno(5.105)$$
(cf. (J.12)).
\medskip
For the general case $g>1$ on the twisted Fobenius manifold
$M_{g;n_0, \dots, n_m}$ with the $g$-dimensional family
$<~,~>_\gamma$ (see (5.73) above)
of the metrics corresponding to the
holomorphic primary differentials the ambiguity in the
definition of the flat coordinates (5.69) can be described
by an action of
the semidirect product of the monodromy group
of the Riemann surfaces by the lattice of periods of holomorphic
differentials. This can be considered as a higher genus generalization
of CCC groups.

Our construction even for $g=0$ does not cover the Frobenius structures
of Lecture 4 on the spaces of orbits of finite Coxeter groups besides $A_n$.
To include this class of examples into the general scheme of geometry
on Hurwitz spaces and also to cover the orbit spaces of the extended
CCC groups but $\hat A_n$ we are to consider equivariant Hurwitz
spaces. These by definition consist of the pairs $(C,\lambda)$ as
above where a finite group acts on $C$ preserving invariant the
function $\lambda$. We are going to do it in further publications.
\vfill\eject
\bigskip
\centerline{\bf Lecture 6.}
\medskip
\centerline{\bf Frobenius manifolds and integrable hierarchies.}
\smallskip
\centerline{\bf Coupling to topological gravity.}
\medskip
We start with expllanation of the following observation: all the
examples of Frobenius manifolds
constructed in Lecture 5 are finite dimensional invariant
manifolds of integrable hierarchies of the KdV type. To explain
this relation we are to explain briefly the notion of semi-classical
limit of an integrable hierarchy. In physical language the
semi-classical (more particularly, the dispersionless) limit will correspond
to the description of the quantum field theory after coupling
to topological gravity considered in the genus zero (i.e., the tree-level)
approximation.

Let
$$\partial_{t_k}y^a = f_k^a(y,\partial_xy,\partial_x^2y,\dots ),~
a=1,\dots ,l,~k=0, 1,\dots
\eqno(6.1)$$
be a commutative hierarchy of Hamiltonian integrable systems of the KdV type.
\lq\lq Hierarchy" means that the systems are ordered, say, by action of a
recursion operator. Number of recursions determine a level of a system in the
hierarchy. Systems of the level zero form a primary part of the hierarchy
(these correspond to the primary operators in TFT); others can be obtained
from the primaries by recursions.
\smallskip
{\bf Example 6.1.} The $nKdV$ (or Gelfand - Dickey) hierarchy is an infinite
system of commuting evolutionary PDEs for functions $a_1(x), \dots, a_n(x)$.
To construct the equations of the hierarchy we consider the operator
$$L = \partial^{n+1} + a_1(x) \partial^{n-1} + \dots + a_n(x),
\eqno(6.2)$$
$$\partial = d/dx.
$$
For any pair
$$(\alpha, p), ~~\alpha = 1, \dots, n, ~~p = 0, 1, \dots
\eqno(6.3)$$
we consider the evolutionary system of the  Lax form
$$\partial_{t^{\alpha,p}}L = \left[ L, \left[ L^{{\alpha \over n+1} +
p}\right]_+
\right].
\eqno(6.4)$$
The brackets $[~,~]$ stand for commutator of the operators, $\left[
L^{{\alpha\over n+1}+p}\right]_+$ denotes the differential part
of the pseudodifferential operator $L^{{\alpha\over n+1}+p}$.
The commutator in the r.h.s. is an ordinary differential operator (in $x$)
of the order at most $n-1$.
So (6.4) is a system of PDE for the functions $a_1(x, t)$, \dots, $a_n(x,t)$,
the time variable is $t = t^{\alpha, p}$. For example, for $\alpha =1, ~p=0$
we have $\left[ L^{{\alpha\over n+1}+p}\right]_+ = \partial$, so the
correspondent
PDE is the $x$-translations
$$\partial_{t^{1,0}}a_i(x) + \partial_x a_i(x) = 0.
\eqno(6.5)$$

The equations (6.4) have a bihamiltonian structure: they can be represented
in the form
$$\partial_{t^{\alpha,p}}a_i(x) = \left\{ a_i(x), H_{\alpha,p}\right\}
= \left\{ a_i(x), H_{\alpha,p-1}\right\}_1
\eqno(6.6)$$
for some
family of local
Hamiltonians $H_{\alpha,p} = H_{\alpha,p}[a(x)]$ and w.r.t.
a pair of Poisson brackets $\{ ~,~\}$ and $\{~,~\}_1$ on an appropriate space
of functionals of $a_i(x)$. There is no symmetry between the Poisson
brackets: the Casimirs (i.e. the annihilator) of the {\it first}
Poisson bracket $\{~,~\}$ only are local functionals $H_{\alpha, -1}
[a(x)]$. For the example of the KdV hierarchy ($n=1$) the two Poisson brackets
are
$$\eqalignno{\{ u(x), u(y)\} &= \delta'(x-y)
&(6.7a)\cr
\{ u(x),u(y)\}_1 &= -\half \delta'''(x-y) +2u(x)\delta'(x-y)
+ u'(x)\delta(x-y).
&(6.7b)\cr}
$$
Here $u(x) = - a_1(x)$ in the previous notations.
The annihilator of the first Poisson bracket is generated
by the local Casimir
$$H_{-1} = \int u(x)\, dx.
\eqno(6.8)$$

The Poisson brackets satisfy a very important
property of {\it compatibility}: any linear combination
$$\{ ~,~\}_1 - \lambda \{~,~\}
\eqno(6.9)$$
for an arbitrary parameter $\lambda$ is again a Poisson bracket.
This gives a possibility to construct an infinite sequence of the
commuting Hamiltonians $H_{\alpha,p}$ starting from the Casimirs
$H_{\alpha,-1}$, $\alpha = 1, \dots, n$ of the first Poisson bracket
$\{ ~,~\}$ using (6.6) as the recursion relations (see [101] for
details). For any {\it primary} Hamiltonian $H_{\alpha, -1}$ (a Casimir
of the first Poisson structure)
we obtain
an infinite chain of its {\it descendants} $H_{\alpha,p}$, $p\geq 0$
determined from the recursion relations
$$\left\{ a_i(x), H_{\alpha,p+1}\right\} =
\left\{ a_i(x), H_{\alpha,p}\right\}_1, ~~p = -1, 0, 1, 2, \dots.
\eqno(6.10)$$
They all are local functionals (i.e. integrals of polynomials of $a_i(x)$
and of their derivatives).
\medskip
The hierarchy
posesses a rich family of finite-dimensional invariant manifolds. Some of
them can be found in a straightforward way; one needs to apply
algebraic geometry methods [52] to construct more wide class of invariant
manifolds. Any of these manifolds after an extension to complex domain
turns out to be fibered over some base $M$
(a complex manifold of some dimension $n$) with $m$-dimensional tori as the
fibers (common invariant tori of the hierarchy). For $m=0$ $M$ is nothing
but the family of common stationary points of the hierarchy. For any
$m\geq 0$ $M$ is
a moduli space of Riemann  surfaces of some genus $g$ with certain additional
structures: marked points and  a marked meromorphic function with poles of
a prescribed order in these points [86]. This is just a Hurwitz space
considered in Lecture 5.
Therefore they are the
families of parameters of the
finite-gap (\lq\lq $g$-gap") solutions of the hierarchy. Our main
observation is that any such $M$ carries a natural structure of
a Frobenius manifold.
\smallskip
{\bf Example 6.2.} For $nKdV$ the family of stationary solutions
of the hierarchy consists of all operators $L$ with constant coefficients
(any two such operators commute pairwise). This coincides with
Frobenius manifold of Example 1.7.
\smallskip
{\bf Example 6.3.} For the same $nKdV$ the family of $``g$-gap" solutions
is the Hurwitz space $M_{g,n}$ in the notations of Lecture 5.
\medskip
To give an idea how an integrable Hamiltonian hierarchy of the above form
induces tensors $c_{\alpha\beta}^\gamma$, $\eta_{\alpha\beta}$ on a finite
dimensional invariant manifold $M$ I need to introduce the notion of
semiclassical limit of a hierarchy near a family $M$ of invariant tori
(sometimes it is called also a {\it dispersionless limit } or {\it Whitham
averaging} of the hierarchy; see details in [53, 54].
In the simplest case of
the family of stationary solutions the semiclassical limit is defined as
follows: one should substitute in the equations of the hierarchy
$$x\mapsto\epsilon x = X,~ t_k\mapsto\epsilon t_k = T_k
\eqno(6.11)$$
and tend $\epsilon$ to zero. For more general $M$ (family of invariant tori)
one should add averaging over the tori. As a result one obtains a new
integrable Hamiltonian hierarchy where the dependent variables are coordinates
$v^1$, ..., $v^n$ on $M$ and the independent variables are the slow
variables $X$ and $T_0$, $T_1$, ... . This new hierarchy always has a form of a
quasilinear system of PDE of the first order
$$\partial_{T_k}v^p = c_k{_q^p}(v)\partial_Xv^q,~k=0, 1, \dots\eqno(6.12)$$
for some matrices of coefficients $c_k{_q^p}(v)$. One can keep in mind the
simplest example of a semiclassical limit (just the dispersionless limit) of
the KdV hierarchy. Here $M$ is the one-dimensional family of constant
solutions of the KdV hierarchy. For example, rescaling the KdV one obtains
$$u_T=uu_X+\epsilon^2u_{XXX}\eqno(6.13)$$
(KdV with small dispersion). After $\epsilon\to 0$ one obtains
$$u_T=uu_X.\eqno(6.14)$$
The semiclassical limit of all the KdV hierarchy has the form
$$\partial_{T_k}u = {u^k\over k!}\partial_Xu,~k=0, 1,\dots .\eqno(6.15)$$

A semiclassical limit of spatialy discretized hierarchies (like Toda
system) is obtained by a similar way. It still is a system of quasilinear
PDE of the first order.

It is important to note that the commutation representation (6.4)
in the semiclassical
limit takes the form
$$\partial_{T^{\alpha,p}}\lambda(X,p) = \left\{ \lambda(X,p),\rho
_{\alpha,p}(X,p)\right\} .
\eqno(6.16)$$
In the r.h.s. $\{ ~,~\}$ stands for the standard Poisson bracket
on the $(X, p)$-plane
$$\left\{\lambda(x,p),\rho(x,p)\right\} =
{\partial\lambda\over\partial p}{\partial\rho\over\partial x}
-{\partial\rho\over\partial p}{\partial\lambda\over\partial x}.
\eqno(6.17)$$
We will call (6.16), (6.17) {\it semiclassical Lax representation}.
For the dispersionless limit the function $\lambda(x,p)$ is just the symbol
of the $L$-operator obtained by the substitution $d/dx \to p$.
The function $\rho(x,p)$ can be computed using fractional powers
as in (6.4). For the case of the semiclassical limits on a family
of oscillating finite-gap solutions the construction of the functions
$\lambda$ and $\rho$ is more complicated (roughly speaking, $\lambda(X,p)$
is the Bloch dispersion law, i.e. the dependence of the eigenvalue $\lambda$
of the operator $L$ with periodic
or quasiperiodic coefficients on the quasimomentum $p$ and on the slow
spatial variable $X=\epsilon x$).
\medskip
Let us come back to determination of tensors $\eta_{\alpha\beta}$,
$c_{\alpha\beta}^\gamma$ on $M$. Let $v_1, \dots, v^n$ be arbitrary
coordinates on $M$.
A
semiclassical limit (or \lq\lq averaging") of both the Hamiltonian structures
in the sense of general construction of S.P.Novikov and the author induces
a compatible pair of
Hamiltonian structures of the semiclassical hierarchy: a pair of Poisson
brackets of
the form
$$\{ v^p(X),v^q(Y)\}_{{\rm semiclassical}} =
\eta^{ps}(v(X))[\delta_s^q\partial_X\delta (X-Y) -
\gamma_{sr}^q(v)v_X^r\delta (X-Y)]
\eqno(6.18a)$$
$$\{ v^p(X),v^q(Y)\}_{{\rm semiclassical}}^1 =
g^{ps}(v(X))[\delta_s^q\partial_X\delta (X-Y) -
\Gamma_{sr}^q(v)v_X^r\delta (X-Y)]
\eqno(6.18b)$$
where $\eta^{pq}(v)$ and
$g^{pq}(v)$ are contravariant components of two metrics on $M$ and
$\gamma_{pr}^q(v)$ and
$\Gamma_{pr}^q(v)$ are the Christoffel symbols of the corresponding
Levi-Civit\'a connections
for the metrics $\eta^{pq}(v)$ and
$g^{pq}(v)$ resp. (the so-called {\it Poisson brackets of hydrodynamic type}).
Observe that the metric $\eta_{\alpha\beta}$ is obtained from
the semiclassical limit of
the first Hamiltonian structure of the original hierarchy.
{}From the general theory of Poisson brackets of
hydrodynamic type [53, 54] one concludes
that both the metrics on $M$
have zero curvature. [In fact, from the compatibility of
the Poisson brackets it follows that the metrics $\eta^{pq}(v)$
and $g^{pq}(v)$ form a flat pencil in the sense of Lecture 3. ]
So local flat coordinates $t^1$, ..., $t^n$ on $M$ exist
such that the metric $\eta^{pq}(v)$ in this coordinates is constant
$${\partial t^\alpha\over\partial v^p}{\partial t^\beta\over\partial v^q}
\eta^{pq}(v) = \eta_{\alpha\beta} = {\rm const}.$$
The Poisson bracket $\{ ~,~\}_{{\rm semiclassical}}$ in these coordinates has
the form
$$\{ t^\alpha (X),t^\beta (Y)\}_{{\rm semiclassical}} = \eta^{\alpha\beta}
\delta '(X-Y).\eqno(6.19)$$
The tensor $(\eta_{\alpha\beta})=(\eta^{\alpha\beta})^{-1}$ together with the
flat coordinates $t^\alpha$ is the first part of a structure we want to
construct. (The flat coordinates $t^1$, ..., $t^n$ can be expressed
via Casimirs of the original Poisson bracket and action variables and wave
numbers along the invariant tori - see details in [53, 54].)

To define a tensor $c_{\alpha\beta}^\gamma (t)$ on $M$ (or, equivalently,
the \lq\lq primary free energy" $F(t)$) we need to use a semiclassical limit
of the $\tau$-function of the original hierarchy [87, 88, 44, 45, 133].
For the dispersionless limit the definition of the semiclassical
$\tau$-function reads
$$\log\tau_{{\rm semiclassical}}(T_0,T_1,\dots ) =
\lim_{\epsilon\to 0} \epsilon^{-2}\log\tau (\epsilon t_0,\epsilon t_1,
\dots ).\eqno(6.20)$$
Then
$$F = \log \tau_{{\rm semiclassical}}\eqno(6.21)$$
for a particular $\tau$-function of the hierarchy.
Here $\tau_{{\rm semiclassical}}$ should be considered as a function only
of the $n$ primary slow variables.
The semiclassical $\tau$-function as the function of all slow variables
coincides with the tree-level partition function of the matter sector
$\eta_{\alpha\beta}$, $c_{\alpha\beta}^\gamma$ coupled to topological gravity.

Summarizing, we can say that a structure of Frobenius manifold
(i.e., a solution of WDVV) on an invariant manifold $M$ of an integrable
Hamiltonian hierarchy is induced by a semiclassical limit of the first Poisson
bracket of the hierarchy and of
a particular $\tau$-function of the hierarchy.

Now we will try to solve the inverse problem: starting from a Frobenius
manifold $M$ to construct a bi-hamiltonian hierarchy and to realize
$M$ as an invariant manifold of the hierarchy in the sense of the
previous construction. By now we are able only to construct the
semiclassical limit of the unknown hierarchy corresponding to any Frobenius
manifold $M$. The problem of recovering of the complete hierarchy
looks to be more complicated. Probably, this can be done not
for arbitrary Frobenius manifold - see an interesting discussion
of this problem in the recent preprint [57].

We describe now briefly the corresponding construction. After this we explain
why this is equivalent to coupling of the matter sector of TFT
described by the Frobenius manifold to topological gravity in the tree-level
approximation.
\medskip
Let us fix a Frobenius manifold $M$.
Considering this as the
matter sector of a 2D TFT model, let us try to calculate the
tree-level (i.e., the zero-genus) approximation of the complete model
obtained by coupling of the matter sector to topological gravity. The
idea to use hierarchies of  Hamiltonian systems of hydrodynamic type for such a
calculation was proposed by E.Witten [149] for the case of topological
sigma-models. An advantage of my approach is in effective construction
of these hierarchies for {\it any} solution of WDVV. The tree-level free
energy of the model will be identified with $\tau$-function of a
particular solution of the hierarchy. The hierarchy carries a
bihamiltonian structure under a non-resonance assumption for scaling
dimensions of the model.

So let $c_{\alpha\beta}^\gamma (t)$, $\eta_{\alpha\beta}$ be a solution of
WDVV, $t=(t^1,\dots , t^n)$. I will construct a hierarchy of the first
order PDE systems linear in derivatives ({\it systems of hydrodynamic
type}) for functions $t^\alpha (T)$, $T$ is an infinite vector of times
$$T=(T^{\alpha ,p}),~ ~\alpha =1,~\dots ,~n,~~p=0,~1,~\dots ;
{}~T^{1,0}=X,\eqno(6.22)$$
$$\partial_{T^{\alpha ,p}}t^\beta = {c_{(\alpha ,p)}}_\gamma^\beta (t)
\partial_X t^\gamma \eqno(6.23)$$
for some matrices of coefficients
${c_{(\alpha ,p)}}_\gamma^\beta (t)$.
The marked variable $X=T^{1,0}$ usualy is called {\it cosmological
constant}.

I will consider the equations (6.23) as dynamical systems (for any
$(\alpha , p)$) on the loop space ${\cal L}(M)$
of functions $t=t(X)$ with values in the
Frobenius manifold $M$.

A. Construction of the systems. I define a Poisson bracket on the
space of functions $t=t(X)$ (i.e. on the loop space ${\cal L}(M)$) by the
formula
$$\{ t^\alpha (X),t^\beta (Y)\} = \eta^{\alpha\beta}\delta '(X-Y).
\eqno(6.24)$$
All the systems (6.23) have hamiltonian form
$$\partial_{T^{\alpha ,p}}t^\beta = \{ t^\beta (X), H_{\alpha ,
p}\} \eqno(6.25)$$
with the Hamiltonians of the form
$$H_{\alpha , p} = \int h_{\alpha , p+1}(t(X))dX. \eqno (6.26)$$
The generating functions of densities of the Hamiltonians
$$h_\alpha (t,z) = \sum_{p=0}^\infty h_{\alpha , p}(t)
z^p,~\alpha =1,\dots ,n \eqno (6.27)$$
coincide with the flat coordinates $\tilde t(t,z)$ of the perturbed connection
$\tilde\nabla (z)$ (see (3.5) - (3.7)). That means that they are
determined by the system (cf. (3.5))
$$\partial_\beta\partial_\gamma h_\alpha (t,z) = z
c_{\beta\gamma}^\epsilon (t) \partial_\epsilon h_\alpha (t,z).
\eqno(6.28)$$
This gives simple recurrence relations for the densities $h_{\alpha
,p}$. Solutions of (6.28) can be normalized in such a way that
$$h_\alpha (t,0) = t_\alpha =\eta_{\alpha\beta}t^\beta ,\eqno(6.29a)$$
$$<\nabla h_\alpha (t,z),\nabla h_\beta (t,-z)> =
\eta_{\alpha\beta} \eqno(6.29b)$$
$$\partial_1 h_\alpha(t,z) = z h_\alpha(t,z) +\eta_{1\alpha}.
\eqno(6.29c)$$
Here $\nabla$ is the gradient (in $t$) w.r.t. the metric $<~,~>$.
\smallskip
{\bf Example 6.4.} For a massive Frobenius manifold with $d<1$
the functions $h_\alpha(t,z)$ can be determined uniquely
from the equation
$$\deli h_\alpha = z^{-\mu_\alpha} \sqrt{\eta_{ii}(u)}
\psi^0_{i\alpha}(u,z)
\eqno(6.30)$$
where the solution $\psi^0_{i\alpha}(u,z)$ of the system (3.122)
has the form (3.124). Thus these functions can be continued analytically
in any sector of the complex $z$-plane.
\medskip
{\bf Exercise 6.1.} Prove the following identities for the gradients
of the generating functions $h_\alpha(t,z)$:
$$\nabla <\nabla h_\alpha(t,z),\nabla h_\beta (t,w)> =
(z+w)\nabla h_\alpha(t,z)\cdot \nabla h_\beta (t,w)
\eqno(6.31)$$
$$\left[ \nabla h_\alpha (t,z), \nabla h_\beta(t,w)\right] =
(w-z) \nabla h_\alpha (t,z) \cdot \nabla h_\beta (t,w).
\eqno(6.32)$$
There is the product of the vector fields on the Frobenius manifold
in the r.h.s. of the
formulae.
\medskip
Let us show that the Hamiltonians (6.26) are in involution. So all
the systems of the hierarchy (6.23) commute pairwise.
\smallskip
{\bf Lemma 6.1.} {\it The Poisson brackets (6.24) of the functionals $h_\alpha
(t(X),z)$ for any fixed $z$
have the form
$$\left\{ h_\alpha (t(X),z_1),h_\beta (t(Y),z_2)\right\} =
\left[ q_{\alpha\beta}(t(Y), z_1, z_2) + q_{\beta\alpha}(t(X),z_2,z_1)
\right]
\eqno(6.33a)$$
where
$$q_{\alpha\beta}(t,z_1,z_2) := {z_2\over z_1+z_2} <\nabla h_\alpha
(t,z_1),\nabla h_\beta (t,z_2)>.
\eqno(6.33b)$$}

Proof. For the derivatives of $q_{\alpha\beta}(t,z_1,z_2)$ one has from
(6.31)
$$\nabla q_{\alpha\beta}(t,z_1,z_2) = z_2 \nabla h_\alpha (t,z_1)\cdot
\nabla h_\beta (t,z_2).
$$
The l.h.s. of (6.33a) has the form
$$\eqalign{\left\{ h_\alpha (t(X),z_1), h_\beta (t(Y),z_2)\right\} &=
<\nabla h_\alpha (t(X), z_1), \nabla h_\beta (t(Y), z_2)>
\delta'(X-Y)\cr
&= <\nabla h_\alpha (t(X), z_1), \nabla h_\beta (t(X), z_2)>
\delta'(X-Y)\cr
&+ z_2<\nabla h_\alpha (t, z_1)\cdot \nabla h_\beta (t, z_2),
\partial_Xt>
\delta(X-Y).\cr}
$$
This completes the proof.
\medskip
{\bf Exercise 6.2.} For any solution $h(t,z)$ of the equation (6.28)
prove that
$$\partial_{T^{\alpha,k}}h(t,z) = \res_{w=0}{w^{-k-1}\over z+w}
<\nabla h(t,z), \nabla h_\alpha (t,w)>.
\eqno(6.34)$$
(The denominator $z+w$ was lost in the formula (3.53) of [46].)
\medskip
{\bf Corollary 6.1.} {\it The Hamiltonians (6.26) commute pairwise.}
\medskip
Observe that the functionals
$$H_{\alpha,-1}= \int t_\alpha (X)\, dX
$$
span the annihilator of the Poisson bracket (6.24).
\smallskip
{\bf Exercise 6.3.} Show that the equations (6.25) for $p=0$
(the primary part of the hierarchy)
have the form
$$\partial_{T^{\alpha,0}} t^\gamma = c_{\alpha\beta}^\gamma(t)\partial_X
t^\beta.
\eqno(6.35)$$
For the equations with $p>0$ obtain the recursion relation
$$\partial_{T^{\alpha,p}} = \nabla^\epsilon h_{\alpha,p}
\partial_{T^{\epsilon,0}}.
\eqno(6.36)$$
\medskip
{\bf Exercise 6.4.} Prove that for a massive Frobenius manifold all
the systems of the hierarchy (6.25) are diagonal in the canonical
coordinates $u^1$, \dots, $u^n$.

In this case from the results of Tsarev [136] it follows completeness
of the family of the conservation laws (6.26) for any of the systems
in the hierarchy (6.25).
\medskip
\bigskip
B. Specification of a solution $t=t(T)$. The hierarchy (6.25) admits an
obvious scaling group
$$T^{\alpha ,p} \mapsto cT^{\alpha ,p},~~t\mapsto t.\eqno(6.37)$$
Let us take the nonconstant invariant solution for the symmetry
$$(\partial_{T^{1,1}} - \sum T^{\alpha ,p}\partial_{T^{\alpha ,p}})
t(T) = 0 \eqno(6.38)$$
(I identify $T^{1,0}$ and $X$. So the variable $X$ is supressed in the
formulae.) This solution can be found without quadratures from a fixed
point equation for the gradient map
$$t=\nabla \Phi_T(t),\eqno(6.39)$$
$$\Phi_T(t) = \sum_{\alpha ,p}T^{\alpha ,p}h_{\alpha ,p}(t).\eqno(6.40)$$
\smallskip
{\bf Proposition 6.1.} {\it The hierarchy (6.25) in the domain
$$T^{1,1} = \epsilon, ~~X, ~T^{\alpha,p} = o(\epsilon) ~{\rm for}~
(\alpha,p)\neq (1,1), ~\epsilon \to 0
\eqno(6.41)$$
has a unique nonconstant solution $t^\beta = t^\beta(X,T)$ invariant
w.r.t. the transformations (6.38). This can be found from the variational
equations
$${\rm grad}_t \left[\Phi_T(t) +X t_1\right] = 0
\eqno(6.42)$$
or, equivalently, as the fixed point of the gradient map
$$t = \nabla \Phi_T(t)
\eqno(6.43)$$
where
$$\Phi_T(t) = \sum_{\alpha,\, p}T^{\alpha, p}h_{\alpha,p}(t).
\eqno(6.44)$$}

Proof. For the invariant solutions of (6.23) one has
$$\left( X\delta_\gamma^\beta +\sum c_{(\alpha,p)\, y}^\beta(t)\right)
\partial_X t^\gamma = 0, ~~\beta = 1, \dots, n.
\eqno(6.45)$$
Using the recursion (6.36) this system can be represented in the form (6.72).
In the domain (6.41) one has
$$\partial_\mu\partial_\nu\left[\Phi_T(t)+Xt_1\right] = \epsilon \eta_{\mu\nu}
+ o(\epsilon).
$$
Hence the solution is locally unique. Therefore it satisfies (6.25).
Proposition
is proved.
\medskip
\bigskip
C. $\tau$-function. Let us define functions
$<\phi_{\alpha ,p}\phi_{\beta,q}>(t)$ from the expansion
$$(z+w)^{-1}(<\nabla h_\alpha (t,z), \nabla h_\beta
(t,w)> - \eta_{\alpha\beta})$$
$$ = \sum_{p,q=0}^\infty <\phi_{\alpha ,p}
\phi_{\beta ,q}>(t)z^p w^q =: <\phi_{\alpha}(z)\phi_\beta(w)>(t).
\eqno(6.46)$$
The infinite matrix of the coefficients $<\phi_{\alpha ,p}\phi_{\beta ,q}>
(t)$ has a
simple meaning: it is the energy-momentum tensor of the commutative
Hamiltonian hierarchy (6.25). That means that
the matrix entry $<\phi_{\alpha ,p}\phi_{\beta ,q}>(t)$
is the density of flux of the Hamiltonian $H_{\alpha ,p}$ along the flow
$T^{\beta ,q}$:
$$\partial_{T^{\beta ,q}}h_{\alpha ,p+1}(t) = \partial_X
<\phi_{\alpha ,p}\phi_{\beta ,q}>(t).\eqno(6.47)$$
Then we define
$$\log\tau (T) = {1\over 2}\sum <\phi_{\alpha ,p}\phi_{\beta ,q}>(t(T))
T^{\alpha, p}T^{\beta ,q}$$
$$ + \sum <\phi_{\alpha ,p}\phi_{1,1}>(t(T))
T^{\alpha,p} + {1\over 2}<\phi_{1,1}\phi_{1,1}> (t(T)).\eqno(6.48)$$
\smallskip
{\bf Exercise 6.5.} Prove that the $\tau$-function satisfies the
identity
$$\partial_{T^{\alpha,p}}\partial_{T^{\beta,q}} \log\tau
= <\phi_{\alpha,p}\phi_{\beta,q}>.
\eqno(6.49)$$

The equations (6.47), (6.49) are the main motivation for the name
\lq\lq$\tau$-function" (cf. [127]).
\smallskip
{\bf Exercise 6.6.} Derive the following formulae
$$\eqalign{<\phi_{\alpha,0}\phi_{\beta,0}> &= \dalpha\dbeta F\cr
<\phi_{\alpha,p}\phi_{1,0}> &= h_{\alpha,p}\cr
<\phi_{\alpha,p}\phi_{\beta,0}> &= \dbeta h_{\alpha,p+1}\cr
<\phi_{\alpha,p}\phi_{1,1}> &= \left( t^\gamma \dgamma - 1\right)
h_{\alpha,p+1}\cr
<\phi_{\alpha,0}\phi_{1,1}> &= t^\lambda \dalpha\partial_\lambda F
-\dalpha F\cr
<\phi_{1,1}\phi_{1,1}> &= t^\alpha t^\beta \dalpha\dbeta F -
2 t^\alpha \dalpha F + 2 F.\cr}
\eqno(6.50)$$
\smallskip
{\bf Exercise 6.7.} Show that
$$<\phi_\alpha(z)\phi_1(w)> = h_\alpha(t,z) + O(w).
\eqno(6.51)$$
\medskip
{\bf Remark 6.1.} More general solutions of (6.25) has the form
$$\nabla [\Phi_T(t)-\Phi_{T_0}(t)] = 0 \eqno(6.52)$$
for arbitrary constant vector $T_0 = T_0^{\alpha ,p}$. For massive Frobenius
manifolds these form a dense subset in the space of all solutions of (6.25)
(see [136, 46]). Formally
they can be obtained from the solution (6.42) by a shift of
the arguments $T^{\alpha ,p}$.
$\tau$-function of the solution (6.52) can be formaly obtained from
(6.48) by the same shift. For the example of topological gravity [148,
149]
such a shift is just the operation that relates the tree-level free energies
of the topological phase of 2D gravity and of the matrix model. It should be
taken into account that the operation of such a time shift in systems of
hydrodynamic type is a subtle one: it can pass through a point of gradient
catastrophe where derivatives become infinite. The correspondent solution of
the KdV hierarchy has no gradient catastrophes but oscillating zones arise
(see [69] for details).
\medskip
{\bf Theorem 6.1}. {\it Let
$${\cal F}(T)=\log \tau (T),\eqno(6.53a)$$
$$<\phi_{\alpha ,p} \phi_{\beta ,q}\dots >_0 = \partial_{T^{\alpha
,p}} \partial_{T^{\beta ,q}}\dots {\cal F}(T).\eqno(6.53b)$$
Then the following relations hold
$${\cal F}(T)|_{_{T^{\alpha ,p}=0 ~{\rm for}~p>0,~T^{\alpha
,0}=t^\alpha}} = F(t) \eqno(6.54a)$$
$$\partial_X{\cal F}(T) = \sum T^{\alpha ,p}\partial_{T^{\alpha ,p-1}}
{\cal F}(T)+ {1\over 2}
\eta_{\alpha\beta}T^{\alpha ,0}T^{\beta ,0} \eqno(6.54b)$$
$$<\phi_{\alpha ,p}\phi_{\beta ,q}\phi_{\gamma ,r}>_0 = <\phi_{\alpha
,p-1}\phi_{\lambda ,0}>_0 \eta^{\lambda\mu} <\phi_{\mu ,0} \phi_{\beta
,q} \phi_{\gamma ,r}>_0.\eqno(6.54c)$$}

Proof can be obtained using the results of the Exercises (6.5) - (6.7) (see
[46] for details).
\medskip
Let me establish now a 1-1 correspondence between the statements of
the theorem and the axioms of Dijkgraaf and Witten of coupling
to topological gravity. In a complete model
of 2D TFT (i.e. a matter sector coupled to topological gravity) there
are infinite number of operators. They usualy are denoted by
$\phi_{\alpha ,p}$ or $\sigma_p(\phi_\alpha )$. The operators
$\phi_{\alpha ,0}$ can be identified with the primary operators
$\phi_\alpha$; the operators $\phi_{\alpha ,p}$ for $p>0$ are called
{\it gravitational descendants} of $\phi_\alpha$. Respectively one has
infinite number of coupling constants $T^{\alpha ,p}$. For example, for the
$A_n$-topological minimal models (see Lecture 2 above) the gravitational
descendants come from the Mumford - Morita - Miller classes of the
moduli spaces $\moduli$. A similar description of gravitational
descendants can be done also for topological sigma-models [149, 38].

The formula (6.53a)
expresses the tree-level (i.e. genus zero)
partition function of the model of 2D TFT
via logarythm of
the $\tau$-function (6.48). Equation (6.53b) is the standard relation between
the correlators (of genus zero) in the model and the free energy.
Equation (6.54a) manifests that before coupling to gravity the partition
function (6.53a) coincides with the primary partition function of the
given matter sector. Equation (6.54b) is the string equation for the free
energy [148, 149, 38].
And equations (6.54c) coincide with the genus zero recursion
relations for correlators of a TFT [40, 149].

We conclude  with another formulation of Theorem 6.1.
\smallskip
{\bf Theorem 6.2.} {\it For any Frobenius manifold the formulae
give a solution of the Dijkgraaf - Witten relations defining
coupling to topological gravity at the tree level.}
\medskip
I recall that the solution of these relations for a given matter sector
of a TFT (i.e., for a given Frobenius manifold) is unique, according to
[40].
\medskip
Particularly, from (6.53) one obtains
$$<\phi_{\alpha,p}\phi_{\beta,q}>_0 = <\phi_{\alpha,p}\phi_{\beta,q}>
\eqno(6.55a)$$
where the r.h.s. is defined in (6.46),
$$<\phi_{\alpha ,p}\phi_{1,0}>_0 =h_{\alpha ,p}(t(T)),\eqno(6.55b)$$
$$<\phi_{\alpha ,p}\phi_{\beta ,q}\phi_{\gamma ,r}>_0 =
<\nabla h_{\alpha ,p}\cdot\nabla h_{\beta ,q}\cdot\nabla h_{\gamma
,r},[e-\sum T^{\alpha ,p}\nabla h_{\alpha ,p-1}]^{-1}>.\eqno(6.55c)$$
The second factor of the inner product in the r.h.s. of (6.55c) is
an invertible element (in the Frobenius algebra of vector fields on $M$) for
sufficiently small $T^{\alpha ,p}$, $p>0$. From the last formula one obtains

{\bf Proposition 6.2.} {\it The coefficients
$$c_p,_{\alpha\beta}^\gamma (T) = \eta^{\gamma\mu}
\partial_{T^{\alpha ,p}}
\partial_{T^{\beta ,p}}
\partial_{T^{\mu ,p}}\log \tau (T) \eqno(6.56)$$
for any $p$ and any $T$ are structure constants of a commutative
associative algebra with the invariant inner product $\eta_{\alpha\beta}$.}

As a rule such an algebra has no unity.

In fact the Proposition holds also for a $\tau$-function of an arbitrary
solution of the form (6.52).

We see that the hierarchy (6.25) determines a family of B\"acklund
transforms of the associativity equations (1.14)
$$F(t)\mapsto \tilde F(\tilde t),$$
$$\tilde F = \log\tau ,~\tilde t^\alpha = T^{\alpha ,p}
\eqno(6.57)$$
for a fixed $p$ and for arbitrary $\tau$-function of (6.25). So it is
natural to consider equations of the hierarchy as Lie -- B\"acklund
symmetries of WDVV.

Up to now I even did not use the scaling invariance (1.9). It turns out
that this gives rise to a bihamiltonian structure of the hierarchy
(6.25) under certain nonresonancy conditions.

We say that a pair $\alpha ,p$ is {\it resonant} if
$${d+1 \over 2} -q_\alpha +p = 0.\eqno(6.58)$$
Here $p$ is a nonnegative integer. The Frobenius manifold with the scaling
dimensions $q_\alpha, ~d$
is {\it nonresonant}
if all pairs $\alpha ,p$ are nonresonant. For example, manifolds
satisfying the inequalities
$$0=q_1\leq q_2\leq\dots\leq q_n=d<1 \eqno(6.59)$$
all are nonresonant.
\smallskip
{\bf Theorem 6.3}. {\it 1) For a Frobenius manifold
 with the scaling dimensions $q_\alpha$ and
$d$ the formula
$$\{t^\alpha (X), t^\beta (Y)\}_1 =g^{\alpha\beta}(t(X))\delta'(X-Y)
+ \left({d+1\over 2}-q_\beta \right) c_\gamma^{\alpha\beta}(t(X))
\partial_Xt^\gamma(X)
\delta (X-Y) \eqno(6.60)$$
where $g^{\alpha\beta}(t)$ is the intersection form of the Frobenius
manifold
determines a Poisson bracket compatible with the Poisson bracket (6.24).
2) For a nonresonant TCFT model all the equations of the hierarchy (6.25)
are Hamiltonian equations also with respect to the Poisson bracket
(6.60).}

Proof follows from (H.12).
\medskip
The nonresonancy condition is essential: equations (6.25) with resonant
numbers $(\alpha ,p)$ do not admit another Poisson structure.
\medskip
{\bf Example 6.5.} Trivial Frobenius manifold corresponding
to a graded $n$-dimensional
Frobenius algebra $A$.
Let
$${\bf t}=t^\alpha e_\alpha \in A.\eqno(6.61)$$
The linear system (6.28) can be solved easily:
$$h_\alpha (t,z)=z^{-1}<e_\alpha , e^{z{\bf
t}}
-1>.
\eqno(6.62)$$
This gives the following form of the hierarchy (6.25)
$$\partial_{T^{\alpha ,p}}{\bf t} = {1\over p!}e_\alpha {\bf
t}^p\partial_X{\bf t}.\eqno(6.63)$$
The solution (6.43) is specified as the fixed point
$$G({\bf t})={\bf t},\eqno(6.64a)$$
$$G({\bf t}) = \sum_{p=0}^\infty {{\bf T}_p\over p!}{\bf t}^p.
\eqno(6.64b)$$
Here I introduce $A$-valued coupling constants
$${\bf T}_p = T^{\alpha ,p}e_\alpha\in A,~p=0,~1,\dots .
\eqno(6.65)$$
The solution of (6.64a) has the well-known form
$${\bf t} =G(G(G(\dots ))) \eqno(6.66)$$
(infinite number of iterations). The $\tau$-function of the
solution (6.66) has the form
$$\log\tau = {1\over 6}<1,{\bf t}^3> - \sum_p{<{\bf T}_p, {\bf
t}^{p+2}>\over (p+2)p!} + {1\over 2}\sum_{p,q} {<{\bf T}_p
{\bf T}_q,{\bf t}^{p+q+1}>\over (p+q+1)p!q!}. \eqno(6.67)$$
For the tree-level correlation functions of a TFT-model with constant
primary correlators one immediately obtains
$$<\phi_{\alpha ,p}\phi_{\beta ,q}>_0 = {<e_{\alpha}e_{\beta },
{\bf t}^{p+q+1}>\over (p+q+1)p!q!}, \eqno(6.68a)$$
$$<\phi_{\alpha ,p}\phi_{\beta ,q}\phi_{\gamma ,r}>_0 =
{1\over p!q!r!}<e_\alpha e_\beta e_\gamma ,{{\bf
t}^{p+q+r}\over 1 - \sum_{s\geq 1}{{\bf T}_s{\bf t}^{s-1}
\over (s-1)!}}>. \eqno(6.68b)$$

Let us consider now the second hamiltonian structure (6.60). I
start with the most elementary case $n=1$ (the pure gravity).
Let me redenote the coupling constant
$$u=t^1.$$
The Poisson bracket (6.60) for this case reads
$$\{ u(X),u(Y)\}_1 ={1\over 2}(u(X)+u(Y))\delta '(X-Y).
\eqno(6.69)$$
This is nothing but the Lie -- Poisson bracket on the dual
space to the Lie algebra of one-dimensional vector fields.

For arbitrary graded Frobenius algebra $A$ the Poisson
bracket (6.60) also is linear in the coordinates $t^\alpha$
$$\{ t^\alpha (X),t^\beta (Y)\}_1 = [({d+1\over 2} - q_\alpha
)c^{\alpha\beta}_\gamma t^\gamma (X) + ({d+1\over 2} - q_\beta
)c^{\alpha\beta}_\gamma t^\gamma (Y)]\delta '(X-Y). \eqno(6.70)$$
It determines therefore a structure of an infinite dimensional
Lie algebra on the loop space ${\cal L}(A^*)$ where $A^*$ is
the dual space to the graded Frobenius algebra $A$. Theory
of linear Poisson brackets of hydrodynamic type and of
corresponding infinite dimensional Lie algebras was
constructed in [12] (see also [54]). But the class of examples
(6.70) is a new one.
\medskip
Let us come back to a general (i.e. nontrivial) Frobenius manifold.
I will assume that the scaling dimensions
are ordered in such a way that
$$0=q_1< q_2\leq\dots\leq q_{n-1}<q_n=d.\eqno(6.71)$$
Then from (6.60) one obtains
$$\{ t^n(X),t^n(Y)\}_1 = {1-d\over 2}(t^n(X)+t^n(Y))\delta
'(X-Y). \eqno(6.72)$$
Since
$$\{ t^\alpha (X), t^n(Y)\}_1 = [({d+1\over 2}-q_\alpha )t^\alpha
(X) +{1-d\over 2}t^\alpha (Y)]\delta '(X-Y), \eqno(6.73)$$
the functional
$$P = {2\over 1-d}\int t^n(X)dX \eqno(6.74)$$
generates spatial translations.
We see that for $d\neq 1$ the Poisson bracket (6.60) can be
considered as a nonlinear extension of the Lie algebra of
one-dimensional vector fields. An interesting question is to
find an analogue of the Gelfand -- Fuchs cocycle for this
bracket. I found such a cocycle for a more particular class of
Frobenius manifolds. We say that a Frobenius manifold is {\it graded} if for
any $t$ the Frobenius algebra $c_{\alpha\beta}^\gamma (t)$,
$\eta_{\alpha\beta}$ is graded.
\smallskip
{\bf Theorem 6.4}. {\it For a graded Frobenius manifold the formula
$$\{ t^\alpha (X),t^\beta (Y) \}^{\hat{}} _1 =
\{ t^\alpha (X),t^\beta (Y) \} _1 + \epsilon^2\eta^{1\alpha}\eta^{1\beta}
\delta '''(X-Y) \eqno(6.75)$$
determines a Poisson bracket compatible with (6.24) and (6.60) for
arbitrary $\epsilon^2$ (the central charge). For a generic graded
Frobenius manifold
this is the only one deformation of the Poisson bracket
(6.60) proportional to $\delta '''(X-Y)$.}
\medskip
For $n=1$ (6.75) determines nothing but the Lie -- Poisson
bracket on the dual  space to the Virasoro algebra
$$\{ u(X),u(Y)\} _1^{\hat{}}  = {1\over 2}[u(X)+u(Y)] \delta '(X-Y) +\epsilon^2
\delta '''(X-Y)\eqno(6.76)$$
(the second Poisson structure of the KdV hierarchy).
For $n>1$ and constant primary correlators (i.e. for a
constant graded Frobenius algebra $A$) the Poisson bracket
(6.75) can be considered as a vector-valued extension (for $d\neq
1$) of the Virasoro.

The compatible pair of the Poisson brackets (6.24) and (6.75)
generates an integrable hierarchy of PDE for a non-resonant
graded Frobenius manifold
using the standard machinery of the bihamiltonian
formalism (see above)
$$\partial_{T^{\alpha ,p}}t^\beta =\{ t^\beta (X),\hat H_{\alpha ,p}\} =
\{ t^\beta (X), \hat H_{\alpha ,p-1}\}_1^{\hat{}}.\eqno(6.77)$$
Here the Hamiltonians have the form
$$\hat H_{\alpha ,p} = \int \hat h_{\alpha ,p+1}dX,\eqno(6.78a)$$
$$\hat h_{\alpha ,p+1 }=[{d+1\over 2} -q_\alpha + p]^{-1}
 h_{\alpha ,p+1}(t)
+\epsilon^2 \Delta\hat h_{\alpha ,p+1}(t, \partial_Xt,
\dots , \partial_X^pt;\epsilon^2)\eqno(6.78b)$$
where $\Delta\hat h_{\alpha ,p+1}$ are some polynomials determined
by (6.77). They are graded-homogeneous of degree 2 where deg$\partial^k_Xt
=k$, deg$\epsilon = -1$. The small dispersion parameter $\epsilon$ also
plays the role of the string coupling constant.
It is clear that the hierarchy (6.25) is the
zero-dispersion limit of this hierarchy. For $n=1$ using the
pair (6.24) and (6.76) one immediately recover the KdV hierarchy.
Note that this describes the topological gravity.
For a trivial manifold (i.e. for a graded
Frobenius algebra $A$) the first nontrivial equations of the
hierarchy are
$$\partial_{T^{\alpha ,1}}{\bf t} = e_\alpha {\bf t}{\bf t}_X +
{2\epsilon^2\over 3-d}e_\alpha e_n{\bf t}_{XXX}. \eqno(6.79)$$

For non-graded Frobenius manifolds it could be of interest to find nonlinear
analogues of the cocycle (6.75). These should be differential geometric
Poisson brackets of the third order [54] of the form
$$\{ t^\alpha (X),t^\beta (Y)\}_1^{\hat{}}
= \{ t^\alpha (X),t^\beta (Y)\}_1 +$$
$$\epsilon^2\{ g^{\alpha\beta}(t(X))\delta '''(X-Y) +
b^{\alpha\beta}_\gamma (t(X))t^\gamma_X\delta ''(X-Y)+$$
$$[f^{\alpha\beta}_\gamma (t(X))t_{XX}^\gamma +h_{\gamma\delta}^{\alpha\beta}
(t(X))t_X^\gamma t_X^\delta ]\delta '(X-Y)+$$
$$[p^{\alpha\beta}_\gamma (t)t_{XXX}^\gamma +
q_{\gamma\delta}^{\alpha\beta}(t)t_{XX}^\gamma t_X^\delta
+r_{\gamma\delta\lambda}^{\alpha\beta}(t)t_X^\gamma t_X^\delta
t_X^\lambda ]\delta (X-Y)\}.\eqno(6.80)$$
I recall (see [54]) that the form (6.80) of the Poisson bracket should be
invariant with respect to nonlinear changes of coordinates in the
manifold $M$. This implies that the leading term $g^{\alpha\beta}(t)$
transforms like a metric (may be, degenerate) on the cotangent bundle $T_*M$,
$b^{\alpha\beta}_\gamma (t)$ are contravariant components of a connection on
$M$ etc. The Poisson bracket (6.80) is assumed to be compatible with (6.24).
Then the compatible pair (6.24), (6.80) of the Poisson brackets generates an
integrable hierarchy of the same structure (6.77), (6.78). The hierarchy (6.25)
will be the dispersionless limit of (6.77).
\medskip
{\bf Example 6.6.} Let me describe the hierarchies (6.25) for
two-dimensional Frobenius manifolds.
 Let us redenote the
coupling constants
$$t^1 = u,~t^2 = \rho .\eqno(6.81)$$
For $d\neq -1, 1,  3$ the
primary free energy $F$ has the form
$$F = {1\over 2}\rho u^2 + {g\over a(a+2)}\rho^{a+2},\eqno(6.82)$$
$$a={1+d\over 1-d} \eqno(6.83)$$
where we introduce an arbitrary constant
$g$.
Let me
give an example of equations of the hierarchy (6.25) (the $T=T^{1,1}$-flow)
$$u_T+uu_X+g\rho^a\rho_X  =0 \eqno (6.84a)$$
$$\rho_T+(\rho u)_X = 0.\eqno(6.84b)$$
These are the equations of isentropic motion of one-dimensional fluid with
the dependence of the pressure on the density of the form
$p={g\over a+2}\rho^{a+2}$. The Poisson structure (6.24) for these equations
was proposed in [116].
For $a=0$ (equivalently $d=-1$) the system coincides with the
equations of  waves on shallow water (the dispersionless limit [156] of the
nonlinear Schr\"odinger equation (NLS)).

For $d=1$ the primary free energy has the form
$$F = {1\over 2}\rho u^2 + ge^\rho .\eqno(6.85)$$
This coincides with the free energy of the topological sigma-model
with  $CP^1$ as the target space.
The corresponding $T=T^{2,0}$-system of the hierarchy (6.25) reads
$$u_T=g(e^\rho )_X$$
$$\rho_T=u_X.$$
Eliminating $u$ one obtains the long wave limit
$$\rho_{TT} =g(e^\rho )_{XX} \eqno(6.86)$$
of the Toda system
$${\rho_n}_{tt} = e^{\rho_{n+1}} -2e^{\rho_n} + e^{\rho_{n-1}}. \eqno(6.87)$$
\medskip
{\bf Example 6.7.} For the Frobenius manifold of Example 1.7 the hierarchy
(6.25) is just the dispersionless limit of the $nKdV$-hierarchy. This
was essentially obtained in [39] and elucidated by Krichever in [87].
The two metrics on the
Frobenius manifold (I recall that this coincides with the
space of orbits of the group $A_n$) are just obtained
from the two hamiltonian structures of the $nKdV$ hierarchy:
the Saito metric is obtained by the semiclassical limit of
from the first Gelfand-Dickey Poisson bracket of $nKdV$ and the
Euclidean metric is obtained by the same semiclassical limit from the second
Gelfand-Dickey Poisson bracket. The Saito and the Euclidean coordinates
on the orbit space are the Casimirs for the corresponding Poisson brackets.
The factorization map $V\to M=V/W$ is the semiclassical limit of the Miura
transformation.
\medskip
{\bf Example 6.8.} The Hurwitz spaces $M_{g; n_0, \dots, n_m}$ parametrize
\lq\lq $g$-gap" solutions of certain integrable hierarchies. The Lax
operator $L$ for these hierarchies must have a form of a $(m+1)\times
(m+1)$-matrix. The equation $L\, \psi = 0$ for a $(m+1)$-component
vector-function $\psi$ must read as a system of ODE (in $x$) of the
orders $n_0+1$, \dots, $n_m+1$ resp. Particularly, for $m=0$ one
obtains the scalar operator $L$ of the order $n = n_0$. So the
Hurwitz spaces $M_{g;n}$ parametrize algebraic-geometrical solutions
(of the genus $g$) of the $nKdV$ hierarchy.

To describe the hierarchy (6.25) we are to solve the recursion system (6.28).
\smallskip
{\bf Proposition 6.3.} {\it
The generating functions $h_{t^A}(t;z)$ (6.27) (where $t^A$ is one of the flat
coordinates (5.30) on the Hurwitz space)
have the form
$$\eqalign{h_{t^{i;\alpha}} (t;z) &=-{n_i+1\over \alpha}\res_{p=\infty_i}
k_i^\alpha {}_1F_1(1;1+{\alpha\over n_i+1};z\lambda (p))dp\cr
h_{p^i} &={\rm v.p.}\int_{\infty_0}^{\infty_i}e^{\lambda(p) z}dp\cr
h_{q^i} &= \res_{\infty_i}{e^{\lambda z}-1\over z}dp\cr
h_{r^i} &= \oint_{b_i}e^{\lambda z}dp\cr
h_{s^i} &={1\over 2\pi i}\oint_{a_i}pe^{\lambda z}d\lambda.\cr}
\eqno(6.88)$$}

Here $_1F_1(a,b,z)$ is the Kummer confluent hypergeometric
function (see [100]).

We leave the proof as an exercise for the reader (see [44]).
\smallskip
{\bf Remark 6.2.} Integrals of the form (6.88) seem to be interesting functions
on the moduli space of the form $M=M_{g;n_0,\dots ,n_m}$. A simplest example of
such an integral for a family of elliptic curves reads
$$\int_0^\omega e^{\lambda\wp (z)}dz\eqno(6.89)$$
where $\wp (z)$ is the Weierstrass function with periods $2\omega$, $2\omega
'$.
For real negative $\lambda$ a degeneration of the elliptic curve ($\omega
\to\infty$) reduces (6.89) to the standard probability integral
$\int_0^\infty e^{\lambda x^2}dx$. So the integral (6.89) is an analogue
of the probability integral as a function on $\lambda$ and on moduli of
the elliptic curve. I recall that dependence on these parameters is specified
by the equations (6.28).
\medskip
Gradients of this functions on the Hurwitz space $M$ have the form
$$\eqalign{\partial_{t^A}h_{t^{i;\alpha}} &= \res_{\infty_i}
k_i^{\alpha - n_i-1}
{}_1F_1(1;{\alpha\over n_i+1};z\lambda (p))\phi_{t^A}\cr
\partial_{t^A}h_{p^i} &=\eta_{t^Ap^i}-z {\rm v.p.}
\int_{\infty_0}^{\infty_i}e^{\lambda z}\phi_{t^A}\cr
\partial_{t^A}h_{q^i} &=
\res_{\infty_i} e^{\lambda z}\phi_{t^A}\cr
\partial_{t^A}h_{r^i} &=\eta_{t^Ar^i}-z\oint_{b_i}e^{\lambda z}
\phi_{t^A}\cr
\partial_{t^A}h_{s^i} &={1\over 2\pi i}\oint_{a_i}
e^{\lambda z}\phi_{t^A}.\cr}
\eqno(6.90)$$
The pairing (6.46) $<\phi_{\alpha}(z)\phi_\beta(w)>$ involved in the definition
of the $\tau$-function (6.48) coincides with (5.36).
\smallskip
{\bf Remark 6.3.}
For any Hamiltonian $H_{A ,k}$ of the form (6.26), (6.88)  one can
construct a differential $\Omega_{A,k}$ on $C$ or on a covering $\tilde C$
with singulariries only at the marked infinite points such that
$${\partial\over\partial u^i}h_{t^A,k}=\res_{P_i}
{\Omega_{A,k}dp\over d\lambda},~ i=1,\dots ,n.\eqno(6.91)$$
We give here, following [44],
an explicit form of these differentials (for $m=0$ also see
[45]). All these will be normalized (i.e. with vanishing $a$-periods)
differentials on $C$ or on the universal covering of $C$ with no
other singularities or multivaluedness but those indicated in
(6.92) - (6.97)
$$\Omega_{t^{i;\alpha},k} = -{1\over n_i+1}
\left[ \left({\alpha\over n_i+1}\right)_{k+1}\right]^{-1}
d\lambda^{{\alpha\over n_i+1} + k} + {\rm regular ~ terms}
\eqno(6.92)$$
$$\Omega_{v^i, k} = -d\left( {\lambda^{k+1}\over (k+1)!}\right)
+ {\rm regular ~terms}, ~i = 1, \dots, m
\eqno(6.93)$$
$$\Omega_{w^i, k} = \cases{-{1\over n_i+1} d\psi_k(\lambda)
+ {\rm reg. ~terms} & near $\infty_i$\cr
{1\over n_0+1} d\psi_k(\lambda) + {\rm reg. ~terms} & near $\infty_0$\cr}
\eqno(6.94)$$
where
$$\psi_k(\lambda) := {\lambda^k\over k!} \left[\log \lambda
-
\left( 1 + \half + \dots + {1\over k}\right)\right], ~k>0
\eqno(6.95)$$
$$\Omega_{r^i, k}(P+ b_j) - \Omega_{r^i,k}(P) = -\delta_{ij}
{\lambda^k\over k!} d\lambda
\eqno(6.96)$$
$$\Omega_{s^i, k}(P+a_j) - \Omega_{s^i, k}(P) =
\delta_{ij} {\lambda^{k-1}\over (k-1)!} d\lambda.
\eqno(6.97)$$

Using these differentials the hierarchy (6.25) can be written in the
Flaschka -- Forest --McLaughlin form [59]
$$\partial_{T^{A,p}}dp = \partial_X\Omega_{A,p}\eqno(6.98)$$
(derivatives of the differentials are to be calculated with $\lambda=$const.).

Integrating (6.98) along the Riemann surface $C$ one obtains for the
the Abelian integrals
$$q_{A,k} := \int \Omega_{A,k}
$$
a similar representation
$$\partial_{T^{A,p}}p(\lambda) = \partial_Xq_{A,p}(\lambda).
$$
Rewriting this for the inverse function $\lambda = \lambda(p)$
we obtain the semiclassical Lax representation of the hierarchy
(6.25) for the Hurwitz space (see details below in a more general setting)
$$\partial_{T^{A,p}}\lambda(p) = \left\{ \lambda, \rho_{A,k}\right\}
$$
where
$$\rho_{A,k} := q_{A,k}(\lambda(p)).
$$

The matrix $<\phi_{A,p}\phi_{B,q}>$ determines a pairing of these differentials
with values in functions on the moduli space
$$<\Omega_{A,p}\,\Omega_{B,q}> =<\phi_{A,p}\phi_{B,q}>(t)\eqno(6.99)$$
This pairing coincides with the two-point correlators (6.55a).
Particularly, the primary free energy $F$ as a function on $M$ can be written
in the form
$$F=-{1\over 2}<pd\lambda\,pd\lambda>.\eqno(6.100)$$
Note that the differential $pd\lambda$ can be written in the form
$$pd\lambda =\sum{n_i+1\over n_i+2}\Omega_{\infty_i}^{(n_i+2)}+
\sum t^A\phi_{t^A}\eqno(6.101)$$
where
$\Omega_{\infty_i}^{(n_i+2)}$ is the Abelian differential of the second kind
with a pole at $\infty_i$ of the form
$$\Omega_{\infty_i}^{(n_i+2)} = dk_i^{n_i+2} +
{}~{\rm regular~terms~~~~near~}\infty_i.\eqno(6.102)$$
For the pairing (6.99) one can obtain from [88] the following formula
$$<f_1d\lambda\, f_2d\lambda>={1\over 2}
\int\int_C(\bar\partial f_1\partial f_2
+\partial f_1\bar\partial f_2)\eqno(6.103)$$
where the differentials $\partial$ and $\bar\partial$ along the Riemann surface
should be understood in the distribution sense. The meromorphic differentials
$f_1dw$ and $f_2dw$ on the covering $\tilde C$ should be considered as
piecewise
meromorphic differentials on $C$ with jumps on some cuts.
\smallskip
{\bf Exercise 6.8.} Prove that the three-point correlators
$\partial_{T^{A,p}}\partial_{T^{B,q}}\partial_{T^{C,r}} {\cal F}$
as functions on $T^{\alpha, 0} = t^\alpha $ with $T^{\alpha,p} = 0$
for $p>0$ can be written in the form
$$<\phi_{A,p}\phi_{B,q}\phi_{C,r}>
= \sum \res_{d\lambda = 0}{\Omega_{A,p}\Omega_{B,q}\Omega_{C,r}
\over d\lambda dp}.
\eqno(6.104)$$
Here $A$, $B$, $C$ denote the labels of one of the flat coordinates
(5.29), the numbers $p$, $q$, $r$ take values $0, 1, 2, \dots$.
[Hint: use (6.55c) and (6.91).]

For the space of polynomials $M_{0;n}$ (the Frobenius manifold
of the $A_n$-topological minimal model) the formula (6.104) was obtained
in [97, 56].
\medskip
The corresponding hierarchy (6.25) is obtained by averaging along invariant
tori
of a family of $g$-gap solutions of a KdV-type hierarchy related to a matrix
operator $L$ of the matrix order $m+1$ and of orders $n_0$, ..., $n_m$
in $\partial /\partial x$.
The example $m=0$ (the averaged Gelfand -- Dickey hierarchy)
was considered in more details in [45].

Also for $g+m>0$ one needs to extend the
KdV-type hierarchy to obtain (6.25) (see [45]). To explain the nature of such
an extension let us consider the simplest example of $m=0$, $n_0=1$. The moduli
space $M$ consists of hyperelliptic curves of genus $g$ with marked homology
basis
$$y^2 = \prod_{i=1}^{2g+1}(\lambda-u_i).\eqno(6.105)$$
This parametrizes the family of $g$-gap solutions of the KdV. The $L$ operator
has the well-known form
$$L= -\partial^2_x+u.\eqno(6.106)$$
In real smooth periodic case $u(x+T)=u(x)$ the quasimomentum $p(\lambda)$ is
defined by the formula
$$\psi (x+T,\lambda) = e^{ip(\lambda)T}\psi (x,\lambda)\eqno(6.107)$$
for a solution $\psi (x,\lambda)$ of the equation
$$L\psi = \lambda\psi \eqno(6.108)$$
(the Bloch -- Floquet eigenfunction). The differential $dp$ can be extended
onto the family of all (i.e. quasiperiodic complex meromorphic)
$g$-gap operators (6.105) as a
normalized Abelian differential of the second kind with a double pole at the
infinity $\lambda
=\infty$. (So the superpotential $\lambda = \lambda(p)$ has the sense of
the Bloch dispersion law, i.e. the dependence of the energy $\lambda$ on
the quasimomentum $p$.) The Hamiltonians of the KdV hierarchy can be
obtained as coefficients of expansion of $dp$ near the infinity. To
obtain a complete family of conservation laws of the averaged hierarchy
(6.25) one needs to extend the family of the KdV integrals by adding
nonlocal functionals of $u$ of the form
$$\oint_{a_i}\lambda^kdp,~~\oint_{b_i}\lambda^{k-1}dp,~k=1,2,\dots .
\eqno(6.109)$$
\medskip
We will obtain now the semiclassical Lax representation
for the equations of the hierarchy (6.25) for {\it arbitrary}
Frobenius manifold.

Let $h(t,z)$ be any solution of the equation
$$\dalpha\dbeta h(t,z) = zc_{\alpha\beta}^\gamma (t) \dgamma
h(t,z)
\eqno(6.110)$$
 normalized by the homogeneity condition
$$z\partial_z h = \LE h.
\eqno(6.111)$$
By $p(t,\lambda)$ I will denote the correspondent flat coordinate
of the pencil (H.1) given by the integral (H.11)
$$p(t,\lambda) = \oint z^{d-3\over 2} e^{-\lambda z} h(t,z)\, dz.
\eqno(6.112)$$
We introduce the functions
$$q_{\alpha, k}(t,\lambda) :=\res_{w=0}\oint
{z^{d-3\over 2}w^{-k-1}\over z+w}
<\nabla h(t,z), \nabla h_\alpha (t,w)>\, dz.
\eqno(6.113)$$
\smallskip
{\bf Lemma 6.2.} {\it The following identity holds
$$\partial_T^{\alpha, k}p(t,\lambda)= \partial_Xq_{\alpha,k}(t,\lambda).
\eqno(6.114)$$}

Proof. Integrating the formula (6.34) with the weight $z^{d-3\over 2}
e^{-\lambda z}$ we obtain (6.114). Lemma is proved.
\medskip
{\bf Exercise 6.9.} Let $p(\lambda)$, $q(\lambda)$ be two functions of
$\lambda$ depending also on parameters $x$ and $t$ in such a way that
$$\partial_t p(\lambda)_{\lambda = const} = \partial_x q(\lambda)_{\lambda
= const}.
\eqno(6.115)$$
Let $\lambda = \lambda(p)$ be the function inverse to $p = p(\lambda)$
and
$$\rho(p) := q(\lambda(p)).
\eqno(6.116)$$
Prove that
$$\partial_t \lambda(p)_{p=const} =
\left\{ \lambda, \rho\right\}
:= {\partial\lambda\over \partial x} {\partial \rho\over
\partial p} -
{\partial\rho\over \partial x} {\partial \lambda\over
\partial p}.
\eqno(6.117)$$
\medskip
{\bf Theorem 6.5.} {\it The hierarchy (6.25) admits the semiclassical Lax
representation
$$\partial_{T^{\alpha, k}} = \left\{ \lambda, \rho_{\alpha, k}
\right\}
\eqno(6.118)$$
where
$$\rho_{\alpha,k} := q_{\alpha, k}(t, \lambda(p,t))
\eqno(6.119)$$
and the functions $\lambda = \lambda (p,t)$ is the inverse to (6.112).}

Proof follows from (6.114) and (6.117).
\medskip
In fact we obtain many semiclassical Lax representations of the
hierarchy (6.25): one can take any solution of (6.114) and the correspondent
flat coordinate of the intersection form and apply the above procedure.
The example of $A_n$ Frobenius manifold suggests that for $d<1$ one should
take in (6.118) the flat coordinate $p = p(\lambda, t)$
(I.10) inverse to the LG superpotential constructed in Appendix I
for any massive Frobenius manifold with $d<1$.

The next chapter in our story about Frobenius manifolds could be a
quantization of the dispersionless Lax pairs (6.118). We are to
substitute back $p \to d/dx$ and to obtain a hierarchy
of the KdV type. We hope to address the problem of quantization in
subsequent publications.

\vfill\eject
{\bf References.}
\medskip
\item{1.} Ablowitz M., Chakravarty S., and Takhtajan L., Integrable systems,
self-dual Yang-Mills equations and connections with modular forms, Univ.
Colorado Preprint PAM \# 113 (December 1991).
\medskip
\item{2.}
Arnol'd V.I., Normal forms of functions close to degenerate critical
points. The Weyl groups $A_k$, $D_k$, $E_k$, and Lagrangian singularities,
{\sl Functional Anal.} {\bf 6} (1972) 3 - 25.
\medskip
\item{3.} Arnol'd V.I., Wave front evolution and equivariant Morse lemma,
{\sl Comm. Pure Appl. Math.} {\bf 29} (1976) 557 - 582.
\medskip
\item{4.} Arnol'd V.I., Indices of singular points of 1-forms on a manifold
with boundary, convolution of invariants of reflection groups, and singular
projections of smooth surfaces, {\sl Russ. Math. Surv.} {\bf 34} (1979)
1 - 42.
\medskip
\item{5.} Arnol'd V.I., Gusein-Zade S.M., and Varchenko A.N., Singularities
of Differentiable Maps, volumes I, II, Birkh\"auser, Boston-Basel-Berlin,
1988.
\medskip
\item{6.} Arnol'd V.I., Singularities of Caustics and Wave Fronts,
Kluwer Acad. Publ., Dordrecht - Boston - London, 1990.
\medskip
\item{7.} van Asch A., Modular forms and root systems, {\sl Math. Anal.}
{\bf 222} (1976), 145-170.
\medskip
\item{8.} Aspinwall P.S., Morrison D.R., Topological field theory and
rational curves, {\sl Comm. Math. Phys.} {\bf 151} (1993) 245 - 262.
\medskip
\item{9.} Astashkevich A., and Sadov V., Quantum cohomology of partial flag
manifolds, hep-th/9401103.
\medskip
\item{10.} Atiyah M.F., Topological quantum field theories, {\sl Publ. Math.
I.H.E.S.} {\bf 68} (1988) 175.
\medskip
\item{11.} Atiyah M.F., and Hitchin N., The Geometry and Dynamics of Magnetic
Monopoles, Princeton Univ. Press, Princeton, 1988.
\medskip
\item{12.} A. Balinskii and S. Novikov, {\sl Sov. Math. Dokl.} {\bf 32} (1985)
228.
\medskip
\item{13.} Balser W., Jurkat W.B., and Lutz D.A., Birkhoff invariants
and Stokes multipliers for meromorphic linear differential equations,
{\sl J. Math. Anal. Appl.} {\bf 71} (1979), 48-94.
\medskip
\item{14.} Batyrev V.V., Quantum cohomology rings of toric manifolds,
{\sl Ast\'erisque} {\bf 218} (1993), 9-34.
\medskip
\item{15.} Bernstein J.N., and Shvartsman O.V., Chevalley theorem for complex
crystallographic group, {\sl Funct. Anal. Appl.} {\bf 12} (1978), 308-310;
\item{}Chevalley theorem for complex crystallographic groups, Complex
crystallographic Coxeter groups and affine root systems, In: {\sl Seminar
on Supermanifolds, 2}, ed. D.Lei\-tes, Univ. Stockholm (1986).
\medskip
\item{16.} Bershadsky M., Cecotti S., Ooguri H., and Vafa C., Kodaira - Spencer
theory of gravity and exact results for quantum string amplitudes,
Preprint HUTP-93/A025, RIMS-946, SISSA-142/93/EP.
\medskip
\item{17.} Beukers F., and Heckman G., Monodromy for hypergeometric
function $_nF_{n-1}$, {\sl Invent. Math.} {\bf 95} (1989), 325-354.
\medskip
\item{18.} Birkhoff G.D., Singular points of ordinary linear differential
equations, {\sl Proc. Amer. Acad.} {\bf 49} (1913), 436-470.
\medskip
\item{19.} Birman J., Braids, Links, and Mapping Class Groups,
Princeton Univ. Press, Princeton, 1974.
\medskip
\item{20.} Blok B. and Varchenko A., Topological conformal field theories
and the flat coordinates, {\sl Int. J. Mod. Phys.} {\bf A7} (1992) 1467.
\medskip
\item{21.} Bourbaki N., Groupes et Alg\`ebres de Lie, Chapitres 4, 5 et 6,
Masson, Paris-New York-Barcelone-Milan-Mexico-Rio de Janeiro, 1981.
\medskip
\item{22.}Brezin E., Itzykson C., Parisi G., and Zuber J.-B., {\sl Comm.
Math. Phys.} {\bf 59} (1978) 35.
\item{} Bessis D., Itzykson C., and Zuber J.-B., {\sl Adv. Appl. Math.}
{\bf 1} (1980) 109.
\item{} Mehta M.L., {\sl Comm. Math. Phys.} {\bf 79} (1981) 327;
\item{} Chadha S., Mahoux G., and Mehta M.L., {\sl J.Phys.} {\bf A14}
(1981) 579.
\medskip
\item{23.} Brieskorn E. Singular elements of semisimple algebraic groups,
In: Actes Congres Int. Math., {\bf 2}, Nice (1970), 279 - 284.
\medskip
\item{24.} Candelas P., de la Ossa X.C., Green P.S., and Parkes L.,
A pair of Calabi - Yau manifolds as an exactly soluble superconformal
theory, {\sl Nucl. Phys.} {\bf B359} (1991), 21-74.
\medskip
\item{25.} Cassels J.W., An Introduction to the Geometry of Numbers,
Springer, N.Y. etc., 1971.
\medskip
\item{26.} Cecotti S. and Vafa C., Topological-antitopological fusion,
{\sl Nucl. Phys.} {\bf B367} (1991) 359-461.
\medskip
\item{27.} Cecotti S. and  Vafa C., On classification of $N=2$ supersymmetric
theories, Preprint HUTP-92/A064 and SISSA-203/92/EP, December 1992.
\medskip
\item{28.} Ceresole A., D'Auria R., and Regge T., Duality group for
Calabi - Yau 2-modulii space, Preprint POLFIS-TH. 05/93, DFTT 34/93.
\medskip
\item{29.} Chazy J., Sur les \'equations diff\'erentiellles dont
l'int\'egrale g\'en\'erale poss\`ede un coupure essentielle mobile,
{\sl C.R. Acad. Sc. Paris} {\bf 150} (1910), 456-458.
\medskip
\item{30.} Chakravarty S., Private communication, June 1993.
\medskip
\item{31.} Coxeter H.S.M., Discrete groups generated by reflections,
{\sl Ann. Math.} {\bf 35} (1934) 588 - 621.
\medskip
\item{32.} Coxeter H.S.M., The product of the generators of a finite group
generated by reflections,
\medskip
\item{33.} Darboux G., {\it Le\c{c}ons sur les syst\`emes ortogonaux et les
cordonn\'ees curvilignes}, Paris, 1897. \item{}
Egoroff D.Th., {\it Collected papers on differential geometry}, Nauka, Moscow
(1970) (in Russian).
\medskip
\item{34.} Date E., Kashiwara M, Jimbo M., and Miwa T., Transformation groups
for soliton equations, In: {\sl Nonlinear Integrable Systems -- Classical
Theory and Quantum Theory}, World Scientific, Singapore, 1983, pp. 39-119.
\medskip
\item{35.} Deligne P., and Mumford D., The irreducibility of the space of
curves
of given genus, {\sl Inst. Hautes \'Etudes Sci. Publ. Math.}
{\bf 45} (1969), 75.
\medskip
\item{36.} Di Francesco P., Lesage F., and Zuber J.B., Graph rings
and integrable perturbations of N=2 superconformal theories,
{\sl Nucl. Phys.} {\bf B408} (1993), 600-634.
\medskip
\item{37.} Dijkgraaf R., {\it A Geometrical Approach to Two-Dimensional
Conformal
Field Theory}, Ph.D. Thesis (Utrecht, 1989).
\medskip
\item{38.} Dijkgraaf R., Intersection theory, integrable hierarchies and
topological field theory, Preprint IASSNS-HEP-91/91, December 1991.
\medskip
\item{39.} Dijkgraaf R., E.Verlinde, and H.Verlinde, {\sl Nucl. Phys.}
{\bf B 352} (1991) 59;
\item{} Notes on topological string theory and 2D quantum gravity,
Preprint PUPT-1217, IASSNS-HEP-90/80, November 1990.
\medskip
\item{40.} Dijkgraaf  R., and Witten E., {\sl Nucl. Phys.} {\bf B 342} (1990)
486.
\medskip
\item{41.} Drinfel'd V.G. and Sokolov V.V., {\sl J. Sov. Math.} {\bf 30} (1985)
1975.
\medskip
\item{42.} Donagi R., and Markman E., Cubics, Integrable systems,
and Calabi - Yau threefolds, Preprint (1994).
\medskip
\item{43.} Dubrovin B., On differential geometry of strongly integrable
systems of hydrodynamic type, {\sl Funct. Anal. Appl.} {\bf 24}
(1990).
\medskip
\item{44.} Dubrovin B., Differential geometry of moduli spaces and its
application to soliton equations and to topological field theory,
Preprint No.117,  Scuola Normale Superiore, Pisa (1991).
\medskip
\item{45.} Dubrovin B., Hamiltonian formalism of Whitham-type hierarchies
and topological Landau - Ginsburg models, {\sl Comm. Math. Phys.}
{\bf 145} (1992) 195 - 207.
\medskip
\item{46.} Dubrovin B., Integrable systems in topological field theory,
{\sl Nucl. Phys.} {\bf B 379} (1992) 627 - 689.
\medskip
\item{47.} Dubrovin B., Geometry and integrability of
topological-antitopological fusion, {\sl Comm. Math. Phys.}{\bf 152}
(1993), 539-564.
\medskip
\item{48.} Dubrovin B., Integrable systems and classification of
2-dimensional topological field theories,
In \lq\lq Integrable Systems", Proceedings
of Luminy 1991 conference dedicated to the memory of J.-L. Verdier.
Eds. O.Babelon, O.Cartier, Y.Kosmann-Schwarbach, Birkh\"auser, 1993.
\medskip
\item{49.} Dubrovin B., Topological conformal field theory from the point
of view of integrable systems, In: Proceedings of 1992 Como
workshop \lq\lq Quantum Integrable Field Theories". Eds. L.Bonora, G.Mussardo,
A.Schwimmer, L.Girardello, and M.Martellini, Plenum Press, 1993.
\medskip
\item{50.} Dubrovin B., Differential geommetry of the space of orbits
of a Coxeter group, Preprint SISSA-29/93/FM (February 1993).
\medskip
\item{51.} Dubrovin B., Fokas A.S., and Santini P.M., Integrable functional
equations and algebraic geometry, Preprint INS \# 208 (May 1993), to
appear in {\sl Duke Math. J.}.
\medskip
\item{52.}
Dubrovin B., Krichever I.,  and Novikov S., {\it Integrable Systems}. I.
Encyclopaedia of Mathematical Sciences, vol.4 (1985) 173,
Springer-Verlag.
\medskip
\item{53.} Dubrovin B. and Novikov S.P., The Hamiltonian formalism
of one-dimensional systems of the hydrodynamic type and
the Bogoliubov - Whitham averaging method,{\sl Sov. Math. Doklady}
{\bf 27} (1983) 665 - 669.
\medskip
\item{54.} Dubrovin B. and Novikov S.P., Hydrodynamics of weakly deformed
soliton lattices. Differential geometry and Hamiltonian theory,
{\sl Russ. Math. Surv.} {\bf 44:6} (1989) 35 - 124.
\medskip
\item{55.} Dubrovin B., Novikov S.P., and Fomenko A.T., Modern Geometry,
Parts 1 - 3, Springer Verlag.
\medskip
\item{56.} Eguchi T., Kanno H., Yamada Y., and Yang S.-K., Topological
strings, flat coordinates and gravitational descendants, {\sl Phys. Lett.}{\bf
B305} (1993), 235-241;
\item{} Eguchi T., Yamada Y., and Yang S.-K., Topological field theories
and the period integrals, {\sl Mod. Phys. Lett.} {\bf A8} (1993), 1627-1638.
\medskip
\item{57.} Eguchi T., Yamada Y., and Yang S.-K., On the genus expansion
in the topological string theory, Preprint UTHEP-275 (May 1994).
\medskip
\item{58.} Eichler M., and Zagier D., The Theory of Jacobi Forms,
Birkh\"auser, 1983.
\medskip
\item{59.} Flaschka H., Forest M.G., and McLaughlin D.W., {\sl Comm. Pure Appl.
Math.} {\bf 33} (1980) 739.
\medskip
\item{60.} Flaschka H., and Newell A.C., {\sl Comm. Math. Phys.} {\bf 76}
(1980)
65.
\medskip
\item{61.} Fokas A.S., Ablowitz M.J., On a unified approach to transformations
and elementary solutions of Painlev\'e equations, {\sl J. Math. Phys.}
{\bf 23} (1982), 2033-2042.
\medskip
\item{62.} Fokas A.S., Leo R.A., Martina L., and Soliani G., {\sl Phys. Lett.}
{\bf A115} (1986) 329.
\medskip
\item{63.} Frobenius F.G., and Stickelberger L., \"Uber die Differentiation
der ellipyischen Functionen nach den Perioden und Invarianten,
{\sl J. Reine Angew. Math.} {\bf 92} (1882).
\medskip
\item{64.} Gelfand I.M., and Dickey L.A., {\it
A Family of Hamilton Structures Related
to Integrable Systems}, preprint IPM/136 (1978) (in Russian).\item{}
Adler M., {\sl Invent. Math.} {\bf 50} (1979) 219.\item{}
Gelfand I.M., and Dorfman I., {\sl Funct. Anal. Appl.} {\bf 14} (1980) 223.
\medskip
\item{65.} Givental A.B., Sturm theorem for hyperelliptic integrals,
{\sl Algebra and Analysis} {\bf 1} (1989), 95-102 (in Russian).
\medskip
\item{66.} Givental A.B., Convolution of invariants of groups generated
by reflections, and connections with simple singularities of functions,
{\sl Funct. Anal.} {\bf 14}  (1980) 81 - 89.
\medskip
\item{67.} Givental A., Kim B., Quantum cohomology of flag manifolds and
Toda lattices, Preprint UC Berkeley (December 1993).
\medskip
\item{68.} Gromov M., Pseudo-holomorphic curves in symplectic
manifolds, {\sl Invent. Math.} {\bf 82} (1985), 307.
\medskip
\item{69.} Gurevich A.V., and Pitaevskii L.P., {\sl Sov. Phys. JETP} {\bf 38}
(1974) 291; {\sl ibid.}, {\bf 93} (1987) 871; {\sl JETP Letters} {\bf 17}
(1973) 193.\item{}
Avilov V., and Novikov S., {\sl Sov. Phys. Dokl.} {\bf 32} (1987) 366.\item{}
Avilov V., Krichever I., and Novikov S., {\sl Sov. Phys. Dokl.} {\bf 32}
(1987) 564.
\medskip
\item{70.} Halphen G.-H., Sur un syst\`emes d'\'equations diff\'erentielles,
{\sl C.R. Acad. Sc. Paris} {\bf 92} (1881), 1001-1003, 1004-1007.
\medskip
\item{71.} Hanay A., Oz Y., and Plesser M.R., Topological Landau - Ginsburg
formulation and integrable strucures of 2d string theory, IASSNS-HEP 94/1.
\medskip
\item{72.} Hitchin N., Talk at ICTP (Trieste), April, 1993.
\medskip
\item{73.} Hosono S., Klemm A., Theisen S., and Yau S.-T.,
Mirror symmetry, mirror map and applications to complete intersection
Calabi - Yau spaces, Preprint HUTMP-94/02, CERN-TH.7303/94,
LMU-TPW-94-03 (June 1994).
\medskip
\item{74.} Hurwitz A., \"Uber die Differentsialglechungen dritter
Ordnung, welchen die Formen mit linearen Transformationen in
sich gen\"ugen, {\sl Math. Ann.} {\bf 33} (1889), 345-352.
\medskip
\item{75.} Ince E.L., Ordinary Differential Equations, London
- New York etc., Longmans, Green and Co., 1927.
\medskip
\item{76.}
Its A.R.,  and Novokshenov V.Yu., {\it The Isomonodromic Deformation
Method in the Theory of Painlev\'e Equations}, {\sl Lecture Notes in
Mathematics}
1191, Springer-Verlag, Berlin 1986.
\medskip
\item{77.} Jacobi C.G.J., {\sl Crelle J.} {\bf 36} (1848), 97-112.
\medskip
\item{78.} Kac V.G., and Peterson D., Infinite-dimensional Lie algebras,
theta functions and modular forms, {\sl Adv. in Math.} {\bf 53} (1984),
125-264.
\medskip
\item{79.} Kim B., Quantum cohomology of partial flag manifolds
and a residue formula for their intersection pairing, Preprint
UC Berkeley (May 19914).
\medskip
\item{80.} Koblitz N., Introduction to Elliptic Curves and Modular
Forms, Springer, New York etc., 1984.
\medskip
\item{81.} Kodama Y. A method for solving the dispersionless KP equation
and its exact solutions, {\sl Phys. Lett.} {\bf 129A} (1988), 223-226;
\item{} Kodama Y. and Gibbons J., A method for solving the dispersionless
KP hierarchy and its exact solutions, II, {\sl Phys. Lett.} {\bf 135A} (1989)
167-170;
\item{}Kodama Y., Solutions of the dispersionless Toda equation,
{\sl Phys. Lett.} {\bf 147A} (1990), 477-482.
\medskip
\item{82.} Kontsevich M.,  {\sl Funct. Anal.} {\bf 25} (1991) 50.
\medskip
\item{83.} Kontsevich M., Intersection theory on the moduli space of curves
and the matrix Airy function,
{\sl Comm. Math. Phys.} {\bf 147} (1992) 1-23.
\medskip
\item{84.} Kontsevich M.,
Enumeration of rational curves via torus action, Preprint MPI
(May 1994).
\medskip
\item{85.} Kontsevich M., Manin Yu.I., Gromov - Witten classes, quantum
cohomology and enumerative geometry, MPI preprint (1994).
\medskip
\item{86.} Krichever I.M., Integration of nonlinear equations by the methods
of algebraic geometry, {\sl Funct. Anal. Appl.} {\bf 11} (1977), 12-26.
\medskip
\item{87.} Krichever I.M., The dispersionless Lax equations and topological
minimal
models, {\sl Commun. Math. Phys.} {\bf 143} (1991), 415-426.
\medskip
\item{88.} Krichever I.M.,
 The $\tau$-function of the universal Whitham hierarchy, matrix models
and topological field theories, Preprint LPTENS-92-18.
\medskip
\item{89.} Lerche W., Generalized Drinfeld - Sokolov hierarchies, quantum
rings, and W-gravity, Preprint CERN-TH.6988/93.
\medskip
\item{90.} Lerche W., Chiral rings and integrable systems for models
of topological gravity, Preprint CERN-TH.7128/93.
\medskip
\item{91.} Lerche W., On the Landau - Ginsburg realization of topological
gravities, Preprint CERN-TH.7210/94 (March 1994).
\medskip
\item{92.} Lerche W., Vafa C., and Warner N.P., Chiral rings in N=2
superconformal
theories, {\sl Nucl. Phys.} {\bf B324} (1989), 427-474.
\medskip
\item{93.} Li K., {\it Topological gravity and minimal matter},
CALT-68-1662, August
1990; {\it Recursion relations in topological gravity with minimal matter},
CALT-68-1670, September 1990.
\medskip
\item{94.} Looijenga E., A period mapping for certain semiuniversal
deformations, {\sl Compos. Math.} {\bf 30} (1975) 299 - 316.
\medskip
\item{95.} Looijenga E., Root systems and elliptic curves, {\sl Invent. Math.}
{\bf 38} (1976), 17-32.
\medskip
\item{96.} Looijenga E., Invariant theory for generalized root systems,
{\sl Invent. Math.} {\bf 61} (1980), 1-32.
\medskip
\item{97.} Losev A., Descendants constructed from matter field in topological
Landau - Ginsburg theories coupled to topological gravity, Preprint ITEP
(September 1992).
\medskip
\item{98.} Losev A.S., and Polyubin I., On connection between topological
Landau - Ginsburg gravity and integrable systems, hep-th/9305079
(May, 1993).
\medskip
\item{99.} Maassarani Z., {\sl Phys. Lett.} {\bf 273B} (1992) 457.
\medskip
\item{100.} Magnus W., Oberhettinger F., and Soni R.P., {\it
Formulas and Theorems for
the Special Functions of Mathematical Physics}, Springer-Verlag, Berlin -
Heidelberg - New York, 1966.
\medskip
\item{101.} Magri F., {\sl J. Math. Phys.} {\bf 19} (1978) 1156.
\medskip
\item{102.} Malgrange B., \'Equations Diff\'erentielles \`a
Coefficients Polynomiaux, Birkh\"auser, 1991.
\medskip
\item{103.} McCoy B.M., Tracy C.A., Wu T.T.,
{\sl J. Math. Phys.} {\bf 18} (1977)
1058.
\medskip
\item{104.} McDuff D., and Salamon D., J-Holomorphic curves and quantum
cohomology,
Pre\-print SUNY and Math. Inst. Warwick (April 1994).
\medskip
\item{105.} Miller E., The homology of the mapping class group,
{\sl J. Diff. Geom.} {\bf 24} (1986), 1.
\item{} Morita S., Characteristic classes of surface bundles, {\sl
Invent. Math.} {\bf 90} (1987), 551.
\item{} Mumford D., Towards enumerative geometry of the moduli space of
curves, In: {\sl Arithmetic and Geometry}, eds. M.Artin and J.Tate,
Birkh\"auser, Basel, 1983.
\medskip
\item{106.} Miwa T., Painlev\'e property of monodromy presereving
equations and the analyticity of $\tau$-functions, {\sl Publ. RIMS}
{\bf 17} (1981), 703-721.
\medskip
\item{107.} Morrison D.R., Mirror symmetry and rational curves on quintic
threefolds: a guide for mathematicians, {\sl J. Amer. Math. Soc.} {\bf 6}
(1993), 223-247.
\medskip
\item{108.} Nagura M., and Sugiyama K., Mirror symmetry of K3 and torus,
Preprint UT-663.
\medskip
\item{109.} Nakatsu T., Kato A., Noumi M., and Takebe T., Topological strings,
matrix integrals, and singularity theory, {\sl Phys. Lett.} {\bf B322}
(1994), 192-197.
\medskip
\item{110.} Natanzon S., {\sl Sov. Math. Dokl.} {\bf 30} (1984) 724.
\medskip
\item{111.} Noumi M., Expansions of the solution of a Gauss - Manin system
at a point of infinity, {\sl Tokyo J. Math.} {\bf 7:1} (1984), 1-60.
\medskip
\item{112.} Novikov S.P. (Ed.),
{\it The Theory of Solitons: the Inverse Problem
Method}, Nauka, Moscow, 1980. Translation: Plenum Press, N.Y., 1984.
\medskip
\item{113.} Novikov S.P., and Veselov A.P., {\sl Sov. Math. Doklady} (1981);
\item{} {\sl Proceedings of Steklov Institute} (1984).
\medskip
\item{114.} Okamoto K., Studies on the Painlev\'e equations, {\sl
Annali Mat. Pura Appl.} {\bf 146} (1987), 337-381.
\medskip
\item{115.} Okubo K., Takano K., and Yoshida S., A connection problem for the
generalized hypergeometric equation, {\sl Fukcial. Ekvac.} {\bf 31}
(1988), 483-495.
\medskip
\item{116.} Olver P.J., {\sl Math. Proc. Cambridge Philos. Soc.} {\bf 88}
(1980) 71.
\medskip
\item{117.} Piunikhin S., Quantum and Floer cohomology have the same ring
structure,
Preprint MIT (March 1994).
\medskip
\item{118.} Procesi C. and Schwarz G., Inequalities defining orbit spaces,
{\sl Invent. Math.} {\bf 81} (1985) 539 - 554.
\medskip
\item{119.} Rankin R.A., The construction of automorphic forms from the
derivatives of a given form, {\sl J. Indian Math. Soc.} {\bf 20} (1956),
103-116.
\medskip
\item{120.} Ruan Y., Tian G., A mathematical theory of quantum cohomology,
{\sl Math. Res. Lett.} {\bf 1} (1994), 269-278.
\medskip
\item{121.} Rudakov A.N., Integer valued bilinear forms and vector bundles,
{\sl Math. USSR Sbornik} {\bf 180} (1989), 187-194.
\medskip
\item{122.} Sadov V., On equivalence of Floer's and quantum cohomology,
Preprint HUTP-93/A027.
\medskip
\item{123.} Saito K., On a linear structure of a quotient variety by a finite
reflection group, Preprint RIMS-288 (1979).
\medskip
\item{124.} Saito K., Yano T., and Sekeguchi J., On a certain generator system
of the ring of invariants of a finite reflection group,
{\sl Comm. in Algebra} {\bf 8(4)} (1980) 373 - 408.
\medskip
\item{125.} Saito K., Period mapping associated to a primitive form,
{\sl Publ. RIMS} {\bf 19} (1983) 1231 - 1264.
\medskip
\item{126.} Saito K., Extended affine root systems II (flat invariants),
{\sl Publ. RIMS} {\bf 26} (1990) 15 - 78.
\medskip
\item{127.} Sato M., Miwa T., and Jimbo A., {\sl Publ. RIMS} {\bf 14} (1978)
223;
{\bf 15} (1979) 201, 577, 871; {\bf 16} (1980) 531.
\item{} Jimbo A., Miwa T., Mori Y., Sato M., Physica {\bf 1D} (1980) 80.
\medskip
\item{128.} Satake I., Flat structure of the simple elliptic
singularity of type $\tilde E_6$ and Jacobi form, hep-th/9307009.
\medskip
\item{129.} Shcherbak O.P., Wavefronts and reflection groups, {\sl Russ. Math.
Surv.} {\bf 43:3} (1988) 149 - 194.
\medskip
\item{130.} Schlesinger L., \"Uber eine Klasse von Differentsialsystemen
beliebliger Ordnung mit festen kritischer Punkten, {\sl J. f\"ur Math.}
{\bf 141} (1912), 96-145.
\medskip
\item{131.} Slodowy P., Einfache Singularitaten und Einfache Algebraische
Gruppen,
Preprint, Regensburger Mathematische Schriften {\bf 2}, Univ. Regensburg
(1978).
\medskip
\item{132.} Spiegelglass M., Setting fusion rings in topological Landau
- Ginsburg, {\sl Phys. Lett.} {\bf B274} (1992), 21-26.
\medskip
\item{133.} Takasaki K. and Takebe T., SDiff(2) Toda equation - hierarchy, tau
function and symmetries, {\sl Lett. Math. Phys.} {\bf 23} (1991), 205-214;
\item{} Integrable hierarchies
and dispersionless limi, Preprint UTMS 94-35.
\medskip
\item{134.} Takhtajan L., Modular forms as tau-functions for
certain integrable reductions of the Yang-Mills equations, Preprint
Univ. Colorado (March 1992).
\medskip
\item{135.} Todorov A.N., Global properties of the moduli of Calabi - Yau
manifolds-II (Teichm\"uller theory), Preprint UC Santa Cruz (December 1993);
\item{}Some ideas from mirror geometry applied to the moduli space of K3,
Preprint UC Santa Cruz (April 1994).
\medskip
\item{136.} Tsarev S., {\sl Math. USSR Izvestija} {\bf 36} (1991); {\sl Sov.
Math. Dokl.} {\bf 34} (1985) 534.
\medskip
\item{137.} Vafa C., {\sl Mod. Phys. Let.} {\bf A4} (1989) 1169.
\medskip
\item{138.} Vafa C., Topological mirrors and quantum rings, in [154].
\medskip
\item{139.} Varchenko A.N. and Chmutov S.V., Finite irreducible groups,
generated
by reflections, are monodromy groups of suitable singularities,
{\sl Func. Anal.} {\bf 18} (1984)
171 - 183.
\medskip
\item{140.} Varchenko A.N. and Givental A.B., Mapping of periods and
intersection form, {\sl Funct. Anal.} {\bf 16} (1982) 83 - 93.
\medskip
\item{141.} Verlinde E. and Warner N., {\sl Phys. Lett.} {\bf 269B} (1991) 96.
\medskip
\item{142.} Voronov A.A., Topological field theories, string backgrounds,
and homotopy algebras, Preprint University of Pennsylvania (December 1993).
\medskip
\item{143.} Wall C.T.C., A note on symmetry of singularities, {\sl Bull. London
Math. Soc.} {\bf 12} (1980) 169 - 175;
\item{} A second note on symmetry of singularities, {\sl ibid.}, 347 - 354.
\medskip
\item{144.} Wasow W., Asymptotic expansions for ordinary differential
equations, Wiley, New York, 1965.
\medskip
\item{145.} Whittaker E.T., and Watson G.N., A Course of Modern Analysis,
N.Y. AMS Press, 1979.
\medskip
\item{146.} Wirthm\"uller K., Root systems and Jacobi forms, {\sl
Compositio Math.} {\bf 82} (1992), 293-354.
\medskip
\item{147.} Witten E., {\sl Comm. Math. Phys.} {\bf 117} (1988) 353;
\item{} {\sl ibid.}, {\bf 118} (1988) 411.
\medskip
\item{148.} Witten E., On the structure of the topological phase of
two-dimeensional
gravity,
{\sl Nucl. Phys.} {\bf B 340} (1990) 281-332.
\medskip
\item{149.} Witten E., Two-dimensional gravity and intersection theory on
moduli
space,
{\sl Surv. Diff. Geom.} {\bf 1}  (1991) 243-210.
\medskip
\item{150.} Witten E., Algebraic geometry associated with matrix models of
two-dimensional gravity, Preprint IASSNS-HEP-91/74.
\medskip
\item{151.} Witten E., Lectures on mirror symmetry, In [154].
\medskip
\item{152.} Witten E., On the Kontsevich model and other models of two
dimensional gravity, Preprint IASSNS-HEP-91/24.
\medskip
\item{153.} Witten E., Phases of N=2 theories in two dimensions, {\sl Nucl.
Phys.}
{\bf B403} (1993), 159-222.
\medskip
\item{154.} Yau S.-T., ed., Essays on Mirror Manifolds, International
Press Co., Hong Kong, 1992.
\medskip
\item{155.} Yano T., Free deformation for isolated singularity, {\sl Sci.
Rep. Saitama Univ.} {\bf A9} (1980), 61-70.
\medskip
\item{156.} Zakharov V.E., On the Benney's equations, {\sl Physica} {\bf 3D}
(1981), 193-202.
\medskip
\item{157.} Zuber J.-B., On Dubrovin's topological field theories, Preprint
SPhT 93/147.
\medskip

\vfill\eject\end